\documentclass[11pt,a4paper,twoside,titlepage]{book}
\usepackage[english,slovene]{babel}
\usepackage{graphicx}
\usepackage{feynarts}
\usepackage{fancyhdr}
\usepackage{makeidx}
\usepackage{braket}
\usepackage{psfrag}
\usepackage{amsmath}
\usepackage{color}
\usepackage{longtable}
\usepackage{tabularx}
\usepackage{array}
\usepackage{graphics}
\usepackage{epsfig}
\usepackage{amssymb}

\def\bi#1{\hbox{\boldmath{$#1$}}}
\def\slashed#1{\displaystyle{\not}#1}

\newcommand{\nn}{\nonumber}
\newcommand{\bea}{\begin{eqnarray}}
\newcommand{\eea}{\end{eqnarray}}
\newcommand{\beq}{\begin{equation}}
\newcommand{\eeq}{\end{equation}}

\newcommand{\gev}{{\rm GeV}}

\newcommand{\pdir}{p\kern -5.2pt\raise 0.2ex\hbox {/}}
\newcommand{\vdir}{v\kern -5.75pt\raise 0.15ex\hbox {/}}
\newcommand{\kdir}{k\kern -5.75pt\raise 0.15ex\hbox {/}}
\newcommand{\epsdir}{\epsilon\kern -5.0pt\raise 0.15ex\hbox {/}}
\newcommand{\bvdir}{\bar{v}\kern -5.75pt\raise 0.15ex\hbox {/}}
\newcommand{\Ddir}{D\kern -7.75pt\raise 0.20ex\hbox {/}}
\newcommand{\ldir}{l\kern -5.0pt\raise 0.2ex\hbox{/}}
\newcommand{\varepsdir}{\varepsilon\kern -5.5pt\raise 0.15ex\hbox{/}}

\newcommand{\bbarq}{B^0_q-\overline B^0_q}
\newcommand{\bbars}{B^0_s-\overline B^0_s}
\newcommand{\bbard}{B^0_d-\overline B^0_d}

\newcommand {\vek}[1]{\mathbf{#1}}
\newcommand {\E}[1]{\times 10^{#1}} 
\newcommand {\e}[1]{\mathrm{~#1}}       

\newcommand{\mc}[1]{\mathcal{#1}}

\definecolor{Gray}{rgb}{0.5,0.5,0.5}
\definecolor{DarkGray}{rgb}{0.25,0.25,0.25}
\definecolor{LightGray}{rgb}{0.94,0.94,0.94}
\definecolor{Blue}{rgb}{0.,0.,1.}
\definecolor{VeryLightBlue}{rgb}{0.9,0.9,1}
\definecolor{LightBlue}{rgb}{0.8,0.8,1}
\definecolor{DarkBlue}{rgb}{0,0,0.6}
\definecolor{LightGreen}{rgb}{0.88,1,0.88}
\definecolor{Green}{rgb}{0.,0.9,0.}
\definecolor{DarkGreen}{rgb}{0,0.6,0}
\definecolor{VeryLightYellow}{rgb}{1,1,0.9}
\definecolor{LightYellow}{rgb}{1,1,0.6}
\definecolor{MidYellow}{rgb}{1,1,0.5}
\definecolor{VeryLightRed}{rgb}{1,0.9,0.9}
\definecolor{LightRed}{rgb}{1,0.8,0.8}
\definecolor{DarkRed}{rgb}{0.75,0.,0.}
\definecolor{Red}{rgb}{1.,0.,0.}
\definecolor{Black}{rgb}{0.,0.,0.}
\definecolor{White}{rgb}{1.,1.,1.}

\newcommand{\Blue}[1]{{\color{Blue}{#1}}}

\newcommand{\Red}[1]{{\color{Red}{#1}}}

\fancypagestyle{plain}{
   \fancyhead{} 
   \fancyfoot[CE,CO]{\bfseries\thepage}
}

\makeindex
\begin{document}

\selectlanguage{english}

\pagestyle{empty}

\frontmatter

\begin{center}

\vfill

\begin{sc}
{\LARGE University of Ljubljana
\\
\vspace{0.1cm}
Faculty of Mathematics and Physics}
\end{sc}

\vfill

\begin{LARGE}
Jernej Fesel Kamenik
\\
\vspace{0.5cm}
{\huge\bf
Role of Resonances in Heavy Meson Processes within Standard Model and Beyond
}
\\
\vspace{0.5cm}
Ph.D. Thesis

\vfill

Advisor: Prof. Svjetlana Fajfer

\vfill

Ljubljana, 2007

\end{LARGE}
\end{center}

\newpage
\begin{quotation}
\end{quotation}
\newpage

\begin{center}

\vfill

\begin{sc}
{\LARGE Univerza v Ljubljani
\\
\vspace{0.1cm}
Fakulteta za matematiko in fiziko}
\end{sc}

\vfill

\begin{LARGE}
Jernej Fesel Kamenik
\\
\vspace{0.5cm}
{\huge\bf
Vpliv resonanc na procese te\5kih mezonov znotraj standardnega modela in njegovih raz\3iritev
}
\\
\vspace{0.5cm}
Disertacija

\vfill

Mentorica: prof. dr. Svjetlana Fajfer

\vfill

Ljubljana, 2007

\end{LARGE}
\end{center}

\newpage
\begin{quotation}
\end{quotation}
\newpage

\begin{quotation}
\end{quotation}
\vfill
{\hfill \LARGE \bfseries MAJI}
\vfill
\begin{quotation}
\end{quotation}

\newpage

\begin{quotation}
\end{quotation}

\newpage

\begin{quotation}
For the conception and completion of this text I am much indebted to my advisor Svjetlana Fajfer, who has always managed to fine-tune and guide confidently my passage through the crevasses of studies and research. I am also especially grateful to Damir Be\'cirevi\'c, whose insight, energy and enthusiasm for the problems in the field continues to excite and inspire me. Not the least I thank him for his many insightful and valuable comments and suggestions he has given to the manuscript. I would also like to thank all the other collaborators, with whom parts of this work have been done, namely Nejc Ko\v snik, Jan O. Eeg and the sadly desist Paul Singer. With this I must not omit to mention the generosity of Doris Kim and Jim Wiss from the FOCUS collaboration, who have willingly shared parts of their experimental data with me. Thanks are also due to Miha Nemev\v sek and my father, Borut Kamenik, for carefully proof reading parts of the manuscript on a very short notice and calling my attention to numerous typos and other linguistic errors.

Most of this work was done at the Department of Theoretical Physics at the Jo\v zef Stefan Institute and I am indebted to the colleagues there for many enlightening discussions, especially to Miha Nemv\v sek, Nejc Ko\v snik, Jure Zupan, Borut Bajc and Jernej Mravlje. I am also grateful to the Laboratoire de Physique Th\'eorique at Universit\'e Paris Sud, Centre d'Orsay for the hospitality during spring 2005, where part of this work was done.  However, many thanks also go to numerous excellent and welcoming hosts, devoted lecturers and stimulating student colleagues at numerous summer and winter schools which contributed enormously by deepening my understanding and broadening my horizons during the past four years.

It seems to me impossible to thank enough to those that I hold dear, Maja, mine and her parents, and our closest friends. For they have expressed patience and understanding sometimes beyond to me reasonable amount for all my absences and hours spent at work, while completing this text.

I would like to acknowledge that this work was supported in part by the Slovenian Research Agency. Support through the PAI project
{\sl ``Proteus''} and the European Commission RTN network,
Contract No. MRTN-CT-2006-035482 (FLAVIAnet) is also kindly acknowledged.
\end{quotation}
\newpage

\begin{quotation}
\end{quotation}

\newpage
\begin{quotation}
Za zasnovo in neprecenljivo pomo\v c pri izvedbi pri\v cujo\v cega dela sem hvale\v zen mentorici Svjetlani Fajfer, ki je vedno znala izbrati in natan\v cno umeriti mojo pot mimo mnogih previs tekom \v studija in raziskav. Velika zahvala gre tudi Damirju Be\v cirevi\v cu, \v cigar vpogled, energija in zanesenja\v stvo nad problemi na najinem skupnem podro\v cju me vedno znova navdu\v sujejo in navdihujejo. Nenazadnje sem mu hvale\v zen za njegove premnoge temeljite in ob\v sirne komentarje ter predloge na predhodno verzijo tega teksta.  Rad bi se zahvalil tudi preostalim sodelavcem na uspe\v snih skupnih projektih, Nejcu Ko\v sniku, Janu O. Eegu ter \v zal prezgodaj preminulemu Paulu Singerju. Ob tem pa ne smem izpustiti omembe velikodu\v snih Doris Kim in Jimu Wiss iz kolaboracije FOCUS, ki sta mi voljno ponudila del njihovih eksperimentalnih rezultatov v analizo. Hvale\v zen sem tudi Mihi Nemev\v sku in mojemu o\v cetu, Borutu Kameniku, da sta v zelo kratkem roku skrbno prebrala dele besedila in me opozorila na mnoge tipkarske in druge jezikovne spodrsljaje.

Pri\v cujo\v ce delo je v veliki meri nastalo na Odseku za teoreti\v cno fiziko In\v stituta Jo\v zef Stefan in zahvala gre vsem sodelavcem za premnoge diskusije, \v se posebej pa bi ob tem rad izpostavil Miho Nemev\v ska, Nejca Ko\v snika, Jureta Zupana, Boruta Bajca in Jerneja Mravljeta. Hvale\v zen sem tudi Laboratoire de Physique Th\'eorique na Universit\'e Paris Sud, Centre d'Orsay, za gostoljubje spomladi leta 2005, kjer sem opravil del tukaj predstavljenih raziskav. Nenazadnje sem dol\v zan zahvale mnogim gostiteljem, predanim u\v citeljem in so\v studentom na mnogih poletnih in zimskih \v solah, ki so mi omogo\v cili poglobiti moje razumevanje in raz\v siriti obzorja v zadnjih \v stirih letih.

Nikakor se ne morem dovolj zahvaliti mojim najdra\v zjim, Maji, mojim ter njenim  star\v sem ter najinim najo\v zjim prijateljem, ki so v\v casih izkazali \v se preveliko mero potrpljenja in razumevanja ob mojih \v stevilnih odsotnostih in predvsem urah porabljenih med pripravo pri\v cujo\v cega teksta.

Rad bi izpostavil, da je delo delno financirala Javna agencija za raziskovalno dejavnost Republike Slovenije. Stro\v ski raziskav so bili delno kriti tudi iz projekta PAI
{\sl ``Proteus"} ter s pomo\v cjo mre\v ze RTN evropske komisije, pogodba \v st. MRTN-CT-2006-035482 (FLAVIAnet).
\end{quotation}
\newpage

\newpage
\begin{quotation}
\end{quotation}
\newpage

\newpage
\begin{quotation}
\end{quotation}
\vfill
\hfill
\parbox {9.cm}{
\it Great beauty seems invariably to portend\\some tragic fate.
\\
\\
\hfil\rm Michel Houellebecq, Les particules el\'ementaires
%
%
%
%
%
}

\chapter*{Abstract}

The effective theory based on combined chiral and heavy quark symmetry, the heavy meson chiral perturbation theory, is applied to studying the role of resonances in various processes of heavy mesons within and beyond the Standard Model.

Chiral corrections including both positive and negative parity heavy meson doublets are calculated to the effective strong couplings featuring in the effective theory leading order interaction Lagrangian. Bare values of the chirally corrected couplings are extracted from the measured decay widths of charmed resonances. Chiral behavior of the couplings is studied in the leading logarithmic approximation. The mass splitting between heavy mesons of opposite parities spoils the chiral limit of the amplitudes. We restore a well behaved chiral limit by expanding the relevant loop integral expressions in inverse powers of the mass splitting.

In semileptonic heavy to light decays we determine resonance contributions to the various form factors within an effective theory inspired model at zero recoil. We employ a form factor parameterization based on effective theory limits to extrapolate our results to the whole kinematical region in charm decays. We compare our results with experimental data and lattice calculations, and conclude that for a consistent description of the heavy to light semileptonic form factors, one needs to go beyond a single resonance pole approximation in the form factor parameterization.

In semileptonic decays of $B$ mesons to charm resonances we calculate chiral corrections to the relevant Isgur-Wise functions. We evaluate loop contributions of both positive and negative parity heavy mesons to the chiral running of the amplitudes. A well defined chiral limit is only restored after an appropriate loop integral expansion is performed.

We calculate chiral loop corrections to the complete set of supersymmetric four-quark operators mediating heavy neutral meson mixing. The impact of heavy scalar meson contributions in the chiral loops on the chiral behavior of the bag parameters is studied and a well defined chiral extrapolation procedure is defined.

Very rare nonleptonic decays of the $B_c$ meson are studied within the Standard Model where they are mediated by box loop diagrams, and within a number of Standard Model extensions. Based on existing experimental searches for related $B$ meson decays, limits are imposed on some of the models studied. The most promissing nonleptonic two- and three-body decay channels of the $B_c$ meson in the search for such new physics contributions are identified.

\vspace{0.5cm}

\noindent\textsf{Key Words:} heavy meson chiral perturbation theory, decays of charmed mesons, weak decays of heavy mesons, hadronic decays of heavy mesons, heavy neutral meson oscillations, new physics searches, lattice quantum chromodynamics

\vspace{0.5cm}

\noindent\textsf{PACS:} 12.39.Fe, 13.20.Fc, 13.25.Ft, 13.25.Hw, 12.39.Hg, 12.38.Gc



\chapter*{Povzetek}

V doktorskem delu uporabimo efektivno teorijo, ki vklju\v cuje tako kiralno simetrijo, kot simetrijo te\v zkih kvarkov, za \v studij vplivov resonanc na procese te\v zkih mezonov znotraj in izven standardnega modela.

V mo\v cnih razpadih te\v zkih mezonov izra\v cunamo kiralne popravke k efektivnim sklopitvenim konstantam, kjer upo\v stevamo prispevke te\v zkih mezonov tako pozitivne kot negativne parnosti. Gole vrednosti efektivnih sklopitev dolo\v cimo iz razpadnih \v sirin \v carobnih resonanc. Analiziramo tudi kiralno obna\v sanje efektivnih sklopitev v pribli\v zku vodilnih logaritmov. Opazimo, da masna re\v za med te\v zkimi mezoni pozitivne in negativne parnosti pokvari kiralno limito amplitud. S pomo\v cjo razvoja zan\v cnih integralov po obratni vrednosti masne re\v ze ponovno vzpostavimo dobro dolo\v ceno kiralno limito.

Znotraj efektivnega modela prou\v cujemo prispevke resonanc k semileptonskim razpadom \v carobnih v lahke mezone. Napovedi v limiti ni\v ctega odboja ekstrapoliramo na celotno kinematsko podro\v cje s pomo\v cjo splo\v sne parametrizacije oblikovnih funkcij, ki temelji na limitah efektivnih teorij kvantne kromodinamike. Na\v se rezultate primerjamo z eksperimentalnimi podatki in izra\v cuni na mre\v zi. Zaklju\v cimo, da enostavni pribli\v zek enega pola ne more ve\v c zadovoljivo opisati semileptonskih oblikovnih funkcij.

V semileptonskih razpadih mezonov $B$ v \v carobne mezonske resonance izra\v cunamo kiralne popravke k funkcijam Isgur-Wise. Pri tem upo\v stevamo prispevke te\v zkih mezonov obeh parnosti h kiralnemu obna\v sanju amplitud. Dobro definirano kiralno limito dobimo le po primernem razvoju zan\v cnih integralov.

Izra\v cunamo kiralne popravke k celotnemu naboru kvarkovskih operatorjev, ki povzro\v cajo oscilacije te\v zkih nevtralnih mezonov. Obravnavamo prispevke te\v zkih skalarnih mezonov h kiralnemu obna\v sanju parametrov ``vre\v ce'' ter predpi\v semo dobro definiran postopek njihove kiralne ektrapolacije.

Zelo redke neleptonske razpade mezonov $B_c$ obravnavamo znotraj standardnega modela, kjer potekajo le preko \v skatlastih zank, ter znotraj nekaterih njegovih raz\v siritev. Na podlagi obstoje\v cih eksperimentalnih iskanj sorodnih razpadov mezona $B$, postavimo meje na parametre nekaterih obravnavanih modelov. Nato predlagamo najobetavnej\v se dvo- in trodel\v cne razpadne kanale mezona $B_c$ za bodo\v ca iskanja signalov nove fizike.

\vspace{0.5cm}

\noindent\textsf{Klju\1ne besede:} kiralna perturbacijska teorija s te\v zkimi mezoni, razpadi \v carobnih mezonov, \v sibki razpadi te\v zkih mezonov, hadronski razpadi te\v zkih mezonov, oscilacije nevtralnih te\v zkih mezonov, signali nove fizike, izra\v cuni kvantne kromodinamike na mre\v zi

\vspace{0.5cm}

\noindent\textsf{Stvarni vrstilec - PACS:} 12.39.Fe, 13.20.Fc, 13.25.Ft, 13.25.Hw, 12.39.Hg, 12.38.Gc

\chapter*{Notation}

\begin{list}{}{\setlength{\itemsep}{0.5cm}\setlength{\listparindent}{0pt}\setlength{\leftmargin}{0pt}}

\item The characters from the middle fo the Greek alphabet $\mu$, $\nu$,\ldots in general run over space-time indices 0, 1, 2, 3, while the Latin indices $i$, $j$, $k$,\ldots tun over spatial indices 1, 2, 3.

\item The characters from the beginning of the Latin alphabet $a$, $b$,\ldots in general run over light quark flavor indices 1, 2,\ldots,N in case of $SU(N)$ chiral flavor theory.

\item Spatial vector quantities are denoted with bold Latin letters e.g.~$\bf p$, while indices of Lorentz covariant quantities are writen explicitely e.g~$p^{\mu}$.

\item The metric used in the thesis is $\eta^{\mu\nu}=\mathrm{diag}(1,-1,-1,-1)$, where the indices run over 0, 1, 2, 3, with 0 the temporal index.

\item The Levi-Civita tensor $\epsilon^{\mu\nu\rho\sigma}$ is defined as a totally antisymmetric tensor with $\epsilon^{0123}=1$.

\item The Einstein summation over repeated indices is assumed unless stated otherwise. The dot-product $p\cdot k$ denotes $p^{\mu} k_{\mu}$.

\item The Dirac matrices are defined so that $\gamma_{\mu} \gamma_{\nu} + \gamma_{\nu} \gamma_{\mu} = 2 \eta_{\mu\nu}$. Also, $\gamma_{5} = i \gamma_{0}\gamma_{1}\gamma_{2}\gamma_{3}$.
The matrix $\sigma^{\mu\nu} = \frac{i}{2}[\gamma^{\mu},\gamma^{\nu}]$. The slash on a character denotes $\slashed p = p^{\mu} \gamma_{\mu}$. The trace Tr runs over Dirac matrix indices.

\item The Hermitian adjoint of a vector, matrix or operator $O$ is denoted $O^{\dagger}$. A bar on a Dirac bispinor $u$ denotes $\bar u = u^{\dagger} \gamma_0$.

\item The imaginary and real part of a complex number $z$ are deonted $\Im(z)$ and $\Re(z)$ respectively.

\item Natural units with $\hbar$ and the speed of light taken to be unity are used. The fine structure constant is thus $\alpha_{\mathrm{e.m.}}=e^2/4\pi\simeq1/137$.

\end{list}

\tableofcontents
\addcontentsline{toc}{chapter}{Contents}
\listoftables
\addcontentsline{toc}{chapter}{List of Tables}
\listoffigures
\addcontentsline{toc}{chapter}{List of Figures}

\pagestyle{fancy}

\renewcommand{\thesection}{\arabic{section}}

\selectlanguage{slovene}

\chapter*{Povzetek doktorskega dela}
\addcontentsline{toc}{chapter}{Povzetek doktorskega dela}
\section{Uvod}

Standardni model (SM) fizike osnovnih delcev je kvantna teorija umeritvenih polj, ki opisuje temeljne elektromagnetne, \v sibke in mo\v cne interakcije. Izoblikoval se je v \v sestdesetih letih prej\v snjega stoletja in je vse odtlej popolnoma obvladoval podro\v cje~\cite{Donoghue:1992dd}. Osnovni gradniki SM so fermioni -- leptoni in kvarki -- ki so uvr\v s\v ceni v tri dru\v zine. Umeritvena grupa SM je $SU(3)_c \times SU(2)_L \times U(1)_Y$, kjer $SU(3)_c$ zaznamuje umeritveno grupo kvantne kromodinamike (ang. quantum chromodynamics -- QCD), $SU(2)_L$ je umeritvena grupa \v sibkega izospina, medtem ko je $U(1)_Y$ umeritvena grupa \v sibkega hipernaboja. Samo levoro\v cni kiralni ferminoi se transformirajo kot izospinski dubleti pod $SU(2)_L$, medtem ko kvarki hkrati tvorijo fundamentalno tripletno reprezentacijo $SU(3)_c$. Mase leptonov in kvarkov v SM generiramo s pomo\v cjo Higgsovega mehanizma -- s spontanim zlomom simetrije, ko (kiralne) simetrije teorije njen vakuum ne spo\v stuje. V ta namen se v teorijo doda skalarni \v sibko-izospinski dublet. Njegova vakuumska pri\v cakovana vrednost zlomi umeritveno invarianco na podgrupo $SU(3)_c\times U(1)_{EM}$ in inducira mase \v sibkim $W^{\pm}$ in $Z$ umeritvenim bozonom.

\par

Kvarkovska polja v $SU(2)_L$ bazi v splo\v snem niso lastna stanja mase. Zato jih obi\v cajno s pomo\v cjo unitarne matrike zavrtimo v masno bazo. Po konvenciji rotacijo izvedemo na poljih spodnjih kvarkov in rotacijsko matriko imenujemo Cabibbo-Kobayashi-Maskawa (CKM). V celoti jo lahko opi\v semo s pomo\v cjo treh realnih kotov in ene kompleksne faze, ki kr\v si simetrijo $CP$.

\par

SM se lahko pohvali z mnogimi uspe\v sno prestanimi testi opisa osnovnih interakcij.
Njegove napovedi so bile izdatno preverjene v pospe\v sevalni\v skih laboratorijih in se dobro ujemajo z meritvami do najvi\v sjih energij dosegljivih do sedaj: precizni elektro\v sibki testi so v splo\v snem v izjemnem ujemanju z napovedmi SM~\cite{:2005em}, medtem ko meritve kr\v sitev simetrije $CP$  v sistemih z mezoni $K$, $D$ in $B$ podpirajo CKM opis z eno univerzalno fazo~\cite{Charles:2004jd,Bona:2006ah}. Zadnji osnovni gradnik, ki trenutno \v se \v caka na svojo eksperimentalno odkritje je Higgsov bozon.

\par

Kljub velikim uspehom SM pa iz opazovanj \v ze vemo, da SM ne predstavlja popolne slike na najmanj\v sih prostorskih skalah. Tako na primer SM ne vsebuje gravitacije. Navkljub izrednim naporom, ki so jih v zadnjih desetljetjih teoreti\v cni fiziki namenili tej temi, je napredek po\v casen in izsledki neprepri\v cljivi. Predvsem tudi zaradi skoraj popolne odsotnosti eksperimentalnih namigov na tem podro\v cju. Po drugi strani pa SM prav tako ne pojasni nedavno izmerjenih nevtrinskih oscilacij~\cite{GonzalezGarcia:2003qf}. Te ka\v zejo na neni\v celne mase nevtrinov, v nasprotju z opisom, ki ga ponuja SM. Hkrati vse ve\v c astrofizikalnih opazovanj nakazuje, da ve\v cina materije v vesolju ni ne svetilna, ne barionske sestave~\cite{Astier:2005qq}. Hkrati je relativno po\v casna oziroma ``hladna''. SM ne ponuja kandidatov za nebarionsko hladno temno snov.  Nenazadnje na\v se trenutno razumevanje bariogeneze -- tvorbe merjene asimetrije med barioni in anti-barioni -- v zgodnjem vesolju zahteva mnogo ve\v cje kr\v sitve simetrije $CP$, kot so dovoljene znotraj SM~\cite{Strumia:2006qk}.

\par

Pravilna interpretacija eksperimentalnih podatkov in morebitna potrditev napovedi SM oziroma odkritje signalov nove fizike zahtevajo zanesljive izra\v cune relevantnih hadronskih procesov, temelje\v c na fundamentalnem kvarkovskem opisu teorije. Neperturbativna narava QCD pri nizkih energijah, ki hkrati kvarke in gluone dr\v zi ujete znotraj hadronov, nam pri tem povzro\v ca obilico preglavic. Razvoj po sklopitveni konstanti v tem re\v zimu namre\v c ni ve\v c mogo\v c. Neposredni izra\v cuni opazljivk na podlagi osnovnih principov QCD so \v se vedno mogo\v ci s pomo\v cjo simulacij QCD na mre\v zi, vendar so te ra\v cunsko izredno zahtevne~\cite{Montvay:1994cy}. Ena od mo\v znosti, ki nam preostanejo je uporaba simetrij Lagrangevega operatorja, na podlagi katerih skonstruiramo efektivne teorije~\cite{Ecker:1994gg}. Neznane parametre v efektivni teoriji dolo\v cimo iz eksperimentov ali, kadar je to mogo\v ce, s pomo\v cjo neposredne primerjave z napovedmi polne teorije QCD. Tak\v sne efektivne teorije lahko potem uporabimo neposredno za napovedi nekaterih eksperimentalnih procesov ali za oporo izra\v cunom QCD na mre\v zi pri pravilnem upo\v stevanju napak in aproksimacij.

\par

Ena pomembnih manifestacij mo\v cne dinamike QCD pri nizkih energijah je pojav resonanc v spektru delcev. Zaznane so bile pred mnogimi leti v procesih pionov in kaonov, kjer so bile tudi podrobno raziskane~\cite{Donoghue:1992dd}. Izkazale so se kot izredno vplivne v mnogih nizkoenergijskih procesih. Po eni strani omejujejo veljavnost dolo\v cenih efektivnih teorij, ki resonan\v cnih pojavov niso sposobne zadovoljivo opisati. Hkrati je znano, da njihova prisotnost skoraj popolnoma zabri\v se prispevke redkih procesov znotraj SM oziroma nove fizike k oscilacijam mezona $D$ in njegovim redkim razpadom~\cite{Prelovsek:2000rj}. Po drugi strani pa so fiziki dolgo predvidevali, da so zaradi relativno velikih mas kvarkov $c$ in $b$ prispevki resonanc te\v zkih mezonov v procesih teh dveh kvarkov manj pomembni.

\par

V zadnjih nekaj letih pa so mnogi eksperimenti poro\v cali o prvih opa\v zanjih resonanc v spektru \v carobnih mezonov~\cite{Abe:2003zm, Link:2003bd, Aubert:2003fg, Vaandering:2004ix, Besson:2003jp, Krokovny:2003zq}. \v Studije osnovnih lastnosti teh novih stanj so bile \v se posebej stimulirane zaradi dejstva, da mase resnanc v nasprotju s teoreti\v cnimi napovedi kvarkovskih modelov~\cite{Godfrey:1985xj, Godfrey:1986wj} in izra\v cunov na mre\v zi~\cite{Hein:2000qu, Dougall:2003hv} niso le\v zale dale\v c nad masami osnovnih stanj. To hkrati namiguje na potencialno velik vpliv resonanc v procesih $D$ in $D_s$ mezonov in nam zastavlja naslednja vpra\v sanja: Ali lahko ocenimo pomembne vplive najni\v zje le\v ze\v cih resonanc te\v zkih mezonov v procesih osnovnih stanj te\v zkih mezonov? Ali lahko ohranimo nadzor nad temi efekti, \v se posebej znotraj efektivnih teorij QCD? Ali nam lahko morda pomagajo razumeti nekatere vidike opa\v zenih in izmerjenih procesov osnovnih stanj te\v zkih mezonov? In kon\v cno, katere zaklju\v cke pridobljene v \v carobnem sektorju lahko prenesemo in apliciramo v procesih mezonov $B$ in $B_s$, katerih resonance so trenutno \v se izven dosega eksperimentalnih laboratorijev.

\par

V tej disertaciji bomo raziskali mnogo aspektov resonanc v procesih te\v zkih mezonov~\cite{Fajfer:2004mv,Fajfer:2005ug,Fajfer:2005mk,Fajfer:2006uy,Fajfer:2006hi,Eeg:2007ha,Becirevic:2006me,Fajfer:2004fx,Fajfer:2006av}. Njihovi poglavitni prispevki bodo analizirani v relevantnem pristopu efektivnih teorij QCD. Znotraj tega ogrodja bomo izra\v cunali hadronske parametre, ki nastopajo v mnogih nizkornergijskih procesih in preu\v cili vpliv resonanc te\v zkih mezonov na opazljivke. Te vsebujejo mo\v cne in semileptonske razpadne \v sirine te\v zkih mezonov, kot tudi parametre me\v sanja nevtralnih te\v zkih mezonov. Mo\v cni razpadni kanali, kadar so dovoljeni, ponavadi prevladujejo v izmerjenih razpadnih \v sirinah, zato jih lahko uporabimo kot kriterije veljavnosti izbranega efektivno-teoretskega pristopa ter hkrati iz njih dolo\v cimo osnovne parametre efektivnih teorij. Semileptonski razpadi, ki potekajo preko nabitih kvarkovskih in leptonskih tokov, SM opisuje \v ze v drevesnem redu. V teh procesih zato potrjeno prevladujejo prispevki SM. Poglobljene raziskave tega podro\v cja zato predvsem preverjajo konsistentnosti znotraj SM, kot so meritve razli\v cnih matri\v cnih elementov CKM ter testi unitarnosti matrike CKM. Po drugi strani pa me\v sanje te\v zkih nevtralnih mezonov znotraj SM poteka v redu ene \v skatlaste zanke. Tako se v teh procesih odpira okno za iskanje prispevkov nove fizike, ki so lahko, ne pa nujno, obte\v zeni s faktorji zank. Znotraj na\v sega pristopa bomo obravnavali vse mogo\v ce hadronske amplitude, ki nastopajo v me\v sanju te\v zkih nevtralnih mezonov znotraj SM in izven. Nazadnje bomo obravnavali tudi zelo redke hadronske razpade dvojno-te\v zkega mezona $B_c$, ki potekajo, tako kot mezonsko me\v sanje, znotraj SM \v sele v redu \v skatlaste zanke. Uporabili bomo nekaj pridobljenega znanja o vplivu resonanc na izra\v cune relevantnih hadronskih razpadnih amplitud. Tako bomo s pomo\v cjo obstoje\v cih meritev postavili nekatere nove meje na mnoge predloge nove fizike in hkrati predlagali perspektivne nove smeri iskanja nove fizike.

\section{Efektivne teorije te\v zkih in lahkih kvarkov}

Trdovraten problem fenomenolo\v skih ra\v cunov v hadronski fiziki predstavlja neperturbativna narava mo\v cne interakcije. Pristop efektivnih teorij se je v minulih desetletjih izkazal kot izredno koristno orodje v tovrstnih obravnavah. Kot je obi\v caj v sodobni fiziki, tudi tu uporabljamo simetrije za poenostavitev zahtevnih problemov.

\par

Lagrangev operator QCD ima v limiti brezmasnih $N_f$ kvarkov kiralno simetrijo $SU(N_f)_R\times SU(N_f)_L$, za katero na podlagi mnogih eksperimentalnih in teoreti\v cnih argumentov predpostavimo, da je spontano zlomljena v vektorsko podgrupo $SU(N_f)_V$. Posledica tak\v sne sponatne zlomitve je pojav brezmasnih Goldstonovih bozonov, ki parametrizirajo faktorski prostor $SU(N_f)_R\times SU(N_f)_L/SU(N_f)_V$ in so tudi edine prostostne stopnje v nizkoenergijskih procesih. Za najbolj pogost primer $N_f=3$ Goldstonova polja zapi\v semo v obliki matrike
\begin{equation}
\Pi = \begin{pmatrix}
    \frac{1}{\sqrt 6}\eta_8 + \frac{1}{\sqrt 2} \pi^0 & \pi^+ & K^+ \\
   \pi^- & \frac{1}{\sqrt 6}\eta_8 - \frac{1}{\sqrt 2} \pi^0 & K^0 \\
   K^- & \overline K^0 & -\sqrt{\frac{2}{3}}\eta_8
    \end{pmatrix},
    \label{eq:2.3}
\end{equation}
medtem ko v primeru $N_f=2$ upo\v stevamo le pionska polja. Njihove efektivne interakcije ne vsebujejo prispevkov z manj kot dvema odvodoma kar omogo\v ca razvoj po prenosih gibalnih koli\v cin $p$, kjer je npr. vsak odvod reda $p$. Lagrangev operator v vodilnem redu tak\v snega kiralnega razvoja je~\cite{Donoghue:1992dd,Gasser:1984gg}
\begin{equation}
\mathcal L^{(2)}_{\chi} = \frac{f^2}{8} \partial_{\mu} \Sigma_{ab} \partial^{\mu} \Sigma^{\dagger}_{ba} + \lambda_0 \left[(m_q)_{ab} \Sigma_{ba} + (m_q)_{ab} \Sigma^{\dagger}_{ba}\right],
\end{equation}
kjer je $\Sigma = \exp 2 i \Pi/f$. Mase psevdo-Goldstonovih bozonov, ki so posledica mas kvarkov $u$, $d$ in predvsem $s$, pogosto parametriziramo v obliki Gell-Mannovih formul~\cite{Becirevic:2004uv}
\begin{equation}
\begin{array}{rclrclrcl}
m^2_{\pi} & = & \frac{8 \lambda_0 m_s}{f^2} r, & m_K^2 & = & \frac{8 \lambda_0 m_s}{f^2} \frac{r + 1}{2}, &
m^2_{\eta_8} & = & \frac{8 \lambda_0 m_s}{f^2} \frac{r+2}{3},
\end{array}
\label{eq:2.6}
\end{equation}
kjer je $r=m_{u,d}/m_s$ in $8 \lambda_0 m_s / f^2 = 2m_K^2-m_{\pi}^2$.

\par

Nekoliko druga\v cna je simetrija te\v zkih kvarkov, ki je posledica asimptotske svobode QCD. Pri dovolj velikih energijah, ki so povezane z masami te\v zkih kvarkov, je narava QCD perturbativna in v mnogih pogledih podobna QED. Hkrati v interakcijah te\v zkih kvarkov njihov spin prispeva le v obliki relativisti\v cnih kromomagnetnih u\v cinkov. V limiti neskon\v cno te\v zkih kvarkov ti u\v cinki izginejo in dobimo efektivno $SU(2)$ spinsko simetrijo. Nenazadnje QCD lo\v ci med okusi kvarkov le po njihovih masah. V limiti, ko mase $N_Q$ te\v zkih kvarkov hkrati po\v sljemo proti neskon\v cnosti, postanejo ti efektivno nerazlo\v cljivi in dobimo novo $SU(N_Q)$ okusno simetrijo, stanja pa namesto po njihovi gibalni koli\v cini razlikujemo po hitrosti $v$. Efektivno teorijo, ki upo\v steva omenjene simetrije te\v zkih kvarkov imenujemo efektivna teorija te\v zkih kvarkov (ang. heavy quark effective theory -- HQET) Simetrije te\v zkih kvarkov so izredno uporabne tudi v kombinaciji s kiralno simetrijo lahkih kvarkov in sicer v opisu interakcij mezonov, ki vsebujejo par te\v zkega in lahkega kvarka. Spinska simetrija te\v zkih kvarkov tu zahteva, da so hadronska stanja neodvisna od spina te\v zkega kvarka, kar te\v zko-lahke mezone uredi v masno degenerirane pare glede na parnost in spin lahkih prostostnih stopenj znotraj hadrona. Osnovna tak\v sna dubleta negativne in pozitivne parnosti sta $H_v = (1+\slashed{v})/2 [ \slashed P_v^{*} - P_v \gamma_5 ]$ in $S_v = (1+\slashed{v})/2 [ \slashed P_{1v}^{*}\gamma_5 - P_{0v} ]$, ter vsebujeta osnovna psevdoskalarna ($P_v$), vektorska ($P_v^{*\mu}$), skalarna ($P_{0v}$) in aksialna ($P_{1v}^{*\mu}$) stanja te\v zkih mezonov. Njihove interakcije dolo\v cata kiralna simetrija in simetrije te\v zkih kvarkov. V prvem redu obeh razvojev zapi\v semo Lagrangev operator tak\v sne efektivne teorije te\v zkih mezonov in psevdo-Goldstonovih bozonov (ang. heavy meson chiral perturbation theory -- HM$\chi$PT)~\cite{Casalbuoni:1996pg,Manohar:2000dt}
\begin{eqnarray}
    \mathcal L^{(1)}_{\mathrm{HM}\chi\mathrm{PT}} &=& \mathcal L^{(1)}_{\frac{1}{2}^-} + \mathcal L^{(1)}_{\frac{1}{2}^+} + \mathcal L^{(1)}_{\mathrm{mix}}, \nonumber\\
    \mathcal L^{(1)}_{\frac{1}{2}^-} &=& - \mathrm{Tr}\left[ \overline H_a (i v \cdot \mathcal{D}_{ab} - \delta_{ab} \Delta_H ) H_b\right] + g \mathrm{Tr} \left[ \overline H_b H_a \slashed{\mathcal A}_{ab} \gamma_{5} \right], \nonumber\\
    \mathcal L^{(1)}_{\frac{1}{2}^+} &=& \mathrm{Tr} \left[\overline S_a ( i v \cdot \mathcal{D}_{ab} - \delta_{ab} \Delta_S) S_b \right] + \widetilde g \mathrm{Tr} \left[ \overline S_b S_a \slashed{\mathcal A}_{ab} \gamma_{5}  \right], \nonumber\\
    \mathcal L^{(1)}_{\mathrm{mix}} &=& h \mathrm{Tr} \left[ \overline H_b S_a \slashed{\mathcal A}_{ab} \gamma_{5}  \right] + \mathrm{h.c.}.
    \label{eq:2_13}
\end{eqnarray}
Z h.c. smo ozna\v cili dodatni hermitsko konjugiran operator, Tr pa ozna\v cuje sled \v cez Diracove indekse. Uvedli smo \v se operator $\xi = \sqrt{\Sigma}$ s katerim definiramo operatorja kiralnega vektorskega $\mathcal V_{\mu} = (\xi \partial_{\mu} \xi^{\dagger} + \xi^{\dagger} \partial_{\mu} \xi)/2$ in aksialnega $\mathcal A_{\mu} = i (\xi^{\dagger} \partial_{\mu} \xi- \xi \partial_{\mu} \xi^{\dagger})/2 = i \xi^{\dagger} \partial_{\mu} \Sigma \xi^{\dagger}/2$ toka. Prvi nastopa v kovariantnem odvodu kineti\v cnega \v clena Lagrangevega operatorja $\mathcal{D}^{\mu}_{ab} = \delta_{ab}\partial^{\mu} -\mathcal{V}^{\mu}_{ab}$, drugi pa definira interakcije med pari te\v zko-lahkih mezonov in lihim \v stevilom psevdo-Goldstonovih bozonov, ki jih parametrizirajo efektivne sklopitvene konstante $g$, $h$ in $\widetilde g$. Prosta masna parametra te\v zkih mezonov $\Delta_H$ in $\Delta_S$ bi lahko v primeru, da bi obravnavali le interakcije te\v zkih mezonov ene parnosti, postavili na ni\v c s primerno redefinicijo hitrosti. V na\v sem primeru pa to ni ve\v c mogo\v ce in v izra\v cunih se nam pojavi nova invariantna koli\v cina -- razlika obeh \v clenov, ki jo ozna\v cimo s $\Delta_{SH} \equiv \Delta_S - \Delta_H$. Videli bomo, da ta koli\v cina pomembno vpliva na interpretacijo in veljavnost izra\v cunov znotraj HM$\chi$PT. V prvem redu razvoja v kiralni simetriji in simetriji te\v zkih kvarkov zapi\v simo \v se operator \v sibkega toka
\begin{equation}
J^{(0)\mu}_{(V-A)\mathrm{HM}\chi\mathrm{PT}} = \frac{i \alpha}{2} \mathrm{Tr} [ \gamma^{\mu}(1-\gamma_5) H_b] \xi_{ba}^{\dagger} - \frac{i \alpha'}{2} \mathrm{Tr} [ \gamma^{\mu}(1-\gamma_5) S_b] \xi_{ba}^{\dagger} + \mathcal O \left( 1/{m_Q} \right),
\label{eq:2_17}
\end{equation}
ki ga bomo potrebovali pri izra\v cunu \v sibkih procesov te\v zkih mezonov. $\alpha$ in $\alpha'$  sta prosta parametra, ki ju lahko identificiramo z razpadnima konstantama te\v zkih mezonov lihe in sode parnosti.

\section[Hadronske amplitude -- pristopi in resonance]{Hadronske amplitude -- efektivni pristopi in resonance}

Pri fenomenolo\v ski obravnavi \v sibkih interakcij v hadronskih sistemih pogosto uporabljamo nekatere standardne metode in matemati\v cne pripomo\v cke. Tako nam npr. metode razvoja v operatorsko vrsto omogo\v cajo raz\v clenitev problema v perturbativen izra\v cun visokoenergijskih prispevkov z asimptotsko prostimi kvarki na eni strani, ter na temeljnem nivoju neperturbativen izra\v cun hadronskih matri\v cnih elementov operatorjev, ki pa vsebujejo le lahke prostostne stopnje QCD. Na kratko bomo oplazili nekatere splo\v sne lastnosti, pribli\v zke in relacije med tak\v snimi hadronskimi amplitudami.

\par

Osnovna ideja razvoja v operatorsko vrsto je raz\v clenitev poljubnega nelokalnega produkta operatorjev v vsoto lokalnih operatorjev pomno\v zenih z efektivnimi t.i. Wilsonovimi parametri
\begin{equation}
T\{A_1(x_1) A_2(x_2)\ldots A_k(x_k) \xrightarrow[x_i\to x]{} \sum_{n} C_{n}^{A_1\ldots A_k}(x-x_1,\ldots, x-x_k) \mathcal O_n (x),
\end{equation}
kjer $T$ ozna\v cuje operator \v casovne ureditve. Mo\v c tak\v sne raz\v clenitve je dvojna: prvi\v c velja na operatorski ravni, je neodvisna od zunanjih stanj, na katero jo apliciramo in nam zato slu\v zi za izgradnjo efektivnih Hamiltonovih operatorjev; drugi\v c pa nam omogo\v ca raz\v clenitev skal v problemih, kjer lahko izmenjavo virtualnih prostostnih stopenj pri visokih energijah zakodiramo v Wilsonove koeficiente, fiziko nizkih energij pa opi\v semo z efektivnimi operatorji. Pogosto lahko tako vrednosti Wilsonovih koeficientov izra\v cunamo analiti\v cno oz. perturbativno. Preostane nam izvrednotenje matri\v cnih elementov efektivnih operatorjev med zunanjimi stanji ($\bra{f}$ in $\ket{i}$), ki opisujejo verjetnostno amplitudo za proces $\mathcal M_{fi}$
\begin{equation}
\mathcal M_{fi} = \sum_i C_i \bra{f}\mathcal O_i \ket{i}.
\end{equation}
V tem izvrednotenju le\v zi sr\v z vseh te\v zav povezanih z izra\v cunom \v sibkih prehodov med hadronskimi stanji. Trenutno najbolj\v sa metoda za tak\v sne ra\v cune so simulacije QCD na mre\v zi. Vendar je naloga tako te\v zavna, \v se posebno v prehodih med te\v zkimi in lahkimi hadronskimi stanji, da morajo tudi ``eksaktne'' simulacije na mre\v zi uporabljati mnoge pribli\v zke. Eden tak\v snih fenomenolo\v sko in teoreti\v cno motiviranih pribli\v zkov je zelo enostaven a izjemno uporaben pribli\v zek vakuumskega zasi\v cenja (ang. vacuum saturation approximation -- VSA) oz. popolne faktorizacije. Formalno ga izrazimo tako, da med produkte operatorjev, ki jih lahko identificiramo s kvazi-stabilnimi hadronskimi stanji vstavimo celoten nabor kvantnih stanj, nato pa zavr\v zemo vsa razen vakuuma.
Soroden je pribli\v zek zasi\v cenja z resonancami, kjer vmesna stanja modeliramo z izmenjavo resonanc znotraj nekega efektivnega pristopa oz. modela.

\par

Na primeru semileptonskih prehodov si oglejmo \v se nekaj splo\v snih lastnosti hadronskih matri\v cnih elementov, ki so nam pogosto v pomo\v c pri analizi hadronskih procesov. Hadronski matri\v cni element, ki opisuje semileptonske prehode med (psevdo)skalarnimi mezoni ($P_i \to P_f$) lahko v splo\v snem parametriziramo s pomo\v cjo primernih Lorentzovih kovariant na podlagi gibalnih koli\v cin s katerima ozna\v cimo za\v cetno in kon\v cno stanje ($p_i$ in $p_f$), pomno\v zenih z oblikovnimi funkcijami -- skalarnimi funkcijami kvadrata izmenjane gibalne koli\v cine $s=(p_i-p_f)^2$. Matri\v cni element toka $J_{V-A}$ ima tako le dva \v clena
\begin{eqnarray}
\bra{ P_f (p_f) } J_{V-A}^{\mu} \ket{ P_i (p_i)} &=& F_+(s) \left( (p_i+p_f)^{\mu} -
    \frac{m_{P_i}^2-m_{P_f}^2}{s}(p_i-p_f)^{\mu} \right) \nonumber\\*
    &&+ F_0(s) \frac{m_{P_i}^2-m_{P_f}^2}{s}(p_i-p_f)^{\mu},
\label{eq:def-ff}
\end{eqnarray}
kjer sta $F_{+,0}$ vektorska in skalarna oblikovna funkcija. Posebnost tak\v sne izbire parametrizacije je, da v te\v znostnem sistemu za\v cetnega stanja natan\v cno razlo\v cuje prispevke stanj razli\v cnih vrtilnih koli\v cin k amplitudi. Kot nakazuje ze samo poimenovanje, $F_{+}$ vsebuje le prispevke stanj z vrtilno koli\v cino $1$ -- vektorske prispevke, medtem ko $F_0$ opisuje prispevke skalarnih stanj z vrtilno koli\v cino $0$. Zahteva po kon\v cni vrednosti amplitude pri $s=0$ nam da \v se dodatno kinematsko omejitev
\begin{equation}
F_+(0)=F_0(0)~.
\label{eq:PP_form_factor_relations}
\end{equation}

\par

Podobno raz\v clembo lahko naredimo tudi v primeru, ko imamo v kon\v cnem stanju vektorski mezon ($P\to V$). Takrat lahko zapi\v semo matri\v cna elementa aksialnega ($J_A$) in vektorskega ($J_V$) toka kot
\begin{eqnarray}
    \bra{ V (\epsilon,p_V) } J^{\mu}_{V} \ket{P (p_P)} &=& \frac{2 V(s)}{m_P + m_V} \varepsilon^{\mu\nu\alpha\beta} \epsilon_{\nu}^* p_{P\alpha} p_{V\beta}, \nonumber\\*
    \bra{ V (\epsilon,p_V) } J^{\mu}_{A} \ket{P (p_P)} \rangle &=&  - i \epsilon^* \cdot (p_P-p_V) \frac{2m_V}{s} (p_P-p_V)^{\mu} A_0(s) \nonumber\\*
    &&- i(m_P + m_V) \left[\epsilon^{*\mu} - \frac{\epsilon^{*} \cdot (p_P-p_V)}{s} (p_P-p_V)^{\mu}\right] A_1(s) \nonumber\\*
    &&+ i \frac{\epsilon^{*}\cdot (p_P-p_V)}{(m_P + m_V)}\left[(p_P+p_V)^{\mu} - \frac{m_P^2-m_V^2}{s} (p_P-p_V)^{\mu}\right] A_2(s),\nonumber\\
\label{eq:PtoV_slo}
\end{eqnarray}
kjer smo z $\epsilon$ ozna\v cili polarizacijo kon\v cnega vektorskega stanja. Oblikovna funkcija $V$ bo sedaj vsebovala vse prispevke vektorskih stanj, $A_1$ in $A_2$ opisujeta izmenjave aksialnih prostostnih stopenj, $A_0$ pa psevdoskalarne prispevke. Tudi tokrat nam dodatna omejitev zagotavlja kon\v cnost matri\v cnega elementa pri $s=0$
\begin{equation}
A_0(0)-\frac{m_P+m_V}{2m_V}A_1(0)+\frac{m_P-m_V}{2m_V}A_2(0)=0~.
\label{eq:PV_form_factor_relations}
\end{equation}
V razpadih psevdoskalarnih v vektorske mezone je pogosto uporabna parametrizacija razpadne amplitude v obliki t.i. su\v cnostnih amplitud
\begin{eqnarray}
H_{\pm}(y) &=&  (m_P+m_V) A_1(m_P^2 y) \mp \frac{2 m_P |{\bf p}_V(y)|}{m_P+m_V} V(m_P^2 y),\nonumber\\*
H_0(y) &=& \frac{m_P+m_V}{2 m_P m_V \sqrt y} [ m_P^2 (1-y) -m_V^2] A_1(m_P^2 y) - \frac{2 m_P |{\bf p}_V(y)|^2}{m_V(m_P+m_V) \sqrt y } A_2(m_P^2 y),\nonumber\\
\label{eq:helicity}
\end{eqnarray}
kjer je $y=s/m_P^2$ in vektor gibalne koli\v cine kon\v cnega stanja je podan kot
\begin{equation}
|{\bf p}_V (y)|^2 = \frac{[m_P^2 (1-y) + m_V^2]^2}{4 m_P^2} -m_V^2.
\end{equation}

\section{Mo\v cni razpadi te\v zkih mezonov}

%

Natan\v cno poznavanje efektivnih mo\v cnih sklopitev v prvem redu HM$\chi$PT je bistveno za teoreti\v cne izra\v cune \v sibkih procesov te\v zkih mezonov znotraj HM$\chi$PT, saj te sklopitve nastopajo v vseh zan\v cnih kiralnih korekcijah k kateremu koli operatorju znotraj HM$\chi$PT. Trenutna najzaneslivej\v sa metoda za oceno hadronskih matri\v cnih elementov so numeri\v cne simulacije QCD na mre\v zi. Zaradi ra\v cunskih te\v zav ob pribli\v zevanju kiralni limiti, \v studije na mre\v zi uporavljajo velike vrednosti mas lahkih kvarkov. Fizikalne rezultate potem dobijo s pomo\v cjo kiralne ektrapolacije. Ta v postopek vnese nove sistemati\v cne napake, ki jih je izredno te\v zko nadzorovati. Z ni\v zanjem mas kvarkov namre\v c pri\v cakujemo vse bolj izrazite u\v cinke spontanega zloma kiralne simetrije~\cite{Yamada:2001xp,Kronfeld:2002ab,Becirevic:2002mh}. HM$\chi$PT nam omogo\v ca vzpostaviti sistemti\v cno kontrolo nad tak\v snimi efekti saj napoveduje kiralno obna\v sanje hadronskih koli\v cin v procesih te\v zko-lahkih kvarkovskih sistemov. Njene napovedi lahko neposredno uporabimo kot vodilo pri kiralni ektrapolaciji rezultatov na mre\v zi. Znotraj HM$\chi$PT lahko izra\v cunamo kiralne logaritemske popravke (imenovane tudi ne-analiti\v cni \v cleni). Pri\v cakujemo, da bodo najbolj izraziti v limiti izredno majhnih energij oz. mas $m_q\ll \Lambda_{\rm QCD}$. \v Ce je pogoj zagotovo izpolnjen za kvarke $u$ in $d$, je situacija v primeru kvarka $s$ precej manj jasna~\cite{Descotes-Genon:2003cg,Descotes-Genon:2002yv}. Pravtako nejasna je velikost skale kiralne zlomitve $\Lambda_{\chi}$. Nekateri avtorji uporabljajo vrednosti okrog $4\pi f_\pi \simeq 1$~GeV~\cite{Manohar:1983md}, medtem ko jo drugi raje enotijo z maso prve vektorske resonance $m_\rho=0.77$~GeV~\cite{Pich:1995bw,Gasser:1984gg}. Ob\v casno se uporabljajo tudi \v se manj\v se vrednosti. V sistemih te\v zko-lahkih mezonov postane situacija \v se bolj zapletena. Prva orbitalno vzbujena stanja ($j_{\ell}^P=1/2^+$) namre\v c le\v zijo nenavadno blizu najni\v zje le\v ze\v cih stanj ($j_{\ell}^P=1/2^-$). Nedavna eksperimentalna odkritja mezonov $D_{0s}$ in $D_{1s}$ postavljajo velikost masne re\v ze le na pribli\v zno $\Delta_{S_s} \equiv m_{D_{0s}^{\ast}}-m_{D_s}=350$~MeV~\cite{Aubert:2003fg,Vaandering:2004ix,Besson:2003jp,Abe:2003jk}. Malce ve\v cja je v primeru stanj brez \v cudnosti $\Delta_{S_q}=430(30)$~MeV~\cite{Abe:2003zm, Link:2003bd}. Hkrati ra\v cuni QCD na mre\v zi v limiti stati\v cnih te\v zkih kvarkov~\cite{Green:2003zz} dajejo slutiti, da so masne re\v ze majhne tudi v sektorju $b$ kvarkov. Takoj opazimo, da sta tako $\Delta_{S_s}$ kot $\Delta_{S_q}$ manj\v si od $\Delta_{\chi}$, $m_{\eta}$ in celo $m_K$, kar zahteva ponovni premislek o napovedih HM$\chi$PT.

\par

\begin{figure}
\begin{center}
\epsfxsize9cm\epsffile{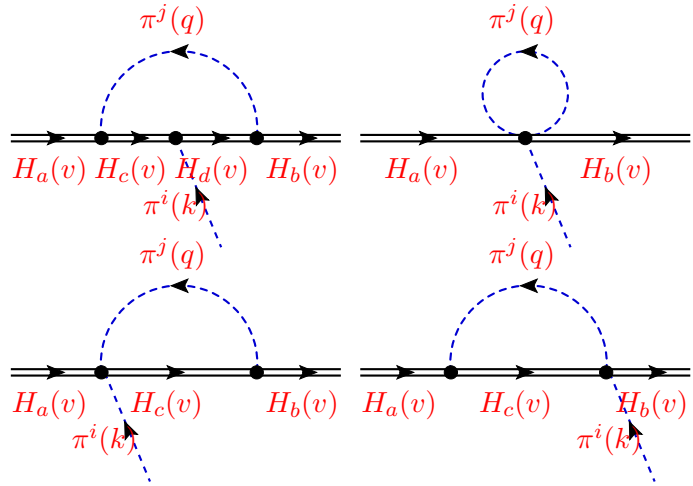}
\end{center}
\caption{\small\it \label{fig:mocne_zanke} Enozan\v cni diagrami, ki prispevajo h kiralnim popravkom efektivnega mo\v cnega vozli\v s\v ca.}
\end{figure}

V na\v sem izra\v cunu kiralnih popravkov  k mo\v cnim razpadom te\v zkih mezonov upo\v stevamo prispevke te\v zko-lahkih mezonov pozitivne in negativne parnosti. Uporabimo Lagrangev operator~(\ref{eq:2_13}) in izpeljemo Feynmanova pravila za izra\v cun Feynmanovih zan\v cnih diagramov na sliki~\ref{fig:mocne_zanke}.

\psfrag{pijl}[bc]{$\Red{ \pi^i(q)}$}
\psfrag{Ha}[cc]{${\Red{H_a(v)}}$}
\psfrag{Hb}[cc]{$\Red{H_b(v)}$}
\psfrag{pi}[bc]{$\Red {\pi^i(k)}$}
\psfrag{Hc}[cc]{$\Red{H_c(v)}$}
\psfrag{Hd}[cc]{$\Red{H_d(v)}$}
\psfrag{pij}[bc]{$\Red{\pi^j(q)}$}

Zapi\v semo in razlo\v cimo tudi vse potrebne kontra\v clene v redu $\mathcal O(m_q)$, v katere lahko pospravimo neskon\v cne prispevke zan\v cnih izra\v cunov.
Nato iz merjenih razpadnih \v sirin \v carobnih resonanc izlu\v s\v cimo gole vrednosti sklopitvenih konstant. Zaradi velikega \v stevila kontra\v clenov v redu ene zanke, njihovih vrednosti ne moremo dolo\v citi. V osnovni analizi, ki jo opravimo pri fiksni skali renormalizacije $\mu = 1~\mathrm{GeV}$, njihove prispevke zanemarimo~\cite{Stewart:1998ke}. V tabeli~\ref{table:povzetek} med seboj primerjamo rezultate izra\v cunov v drevesnem redu, ter v redu ene zanke z in brez prispevkov te\v zko-lahkih mezonov pozitivne parnosti.
\begin{table}[!t]
\begin{center}
\begin{tabular}{|l|c|c|c|}
\hline
 Ra\v cunska shema & $g$ & $|h|$ & $\widetilde g$ \\
 \hline
 \hline
 Drevesni red & $0.61$~\cite{Anastassov:2001cw} & $0.52$ & $-0.15$\\
 \hline
 Red ene zanke brez stanj pozitivne parnosti & $0.53$ & & \\
 \hline
 Red ene zanke s stanji pozitivne parnosti  & $0.66$ & $0.47$ & $-0.06$ \\
 \hline
\end{tabular}
\end{center}
\caption{\small\it \label{table:povzetek} Povzetek na\v sih rezultatov za efektivne sklopitve, kot je razlo\v zeno v besedilu. Vse vrednosti v redu ene zanke so dobljene ob zanemaritvi prispevkov kontra\v clenov na regularizacijski skali $\mu = 1~\mathrm{GeV}$.}
\end{table}
Prispevke kontr\v clenov k izra\v cunom v redu ene zanke nato ocenimo posredno preko odvisnosti na\v sih rezultatov od skale renormalizacije, kot tudi neposredno s pomo\v cjo Monte-Carlo \v zrebanja nakju\v cnih vrednosti kontra\v clenov in njihovih prispevkov k analizi razpadnih \v sirin. Dodatno preverimo tudi ob\v cutljivost na\v sih rezultatov na vhodne podatke mas in predvsem masnih razlik te\v zko-lahkih mezonov.

\par

Nato obravnavamo prispevke dodatnih resonanc znotraj kiralnih zank h kiralni ekstrapolaciji, ki jo uporabljajo simulacije QCD na mre\v zi~\cite{Abada:2003un, McNeile:2004rf}. Kaoni in mezoni $\eta$ znotraj zank k tak\v sni ekstrapolaciji prakti\v cno ne prispevajo, medtem ko poglavitna nelinearnost izhaja iz prispevkov pionov. Rezultati iz prej\v snjega odstavka dopu\v s\v cajo mo\v znost relativno velikih popravkov k renormalizaciji sklopitvenih konstant. Masna razlika med te\v zko-lahkimi mezoni obratnih parnosti $\Delta_{SH}$ je namre\v c velika v primerjavi z masami pionov in povzro\v ci, da imajo le-ti znotraj kiralnih zank veliko gibalno koli\v cino. To postavi veljavnost tak\v sne raz\v sirjene ra\v cunske sheme pod vpra\v saj, saj dajo navidezno najve\v cje popravke prav zanke v katerih nastopajo vzbujena stanja te\v zko-lahkih mezonov. Oglejmo si torej tipi\v cni zan\v cni integral v raz\v sirjeni shemi, ki bo v splo\v snem sedaj vseboval dve dimenzijski skali ($m$ in $\Delta$). Kiralna teorija dodatno zahteva, da morajo biti vse gibalne koli\v cine pseudo-Goldstonovih bozonov (pionov) majhne v primerjavi s skalo kiralne zlomitve $\Lambda_{\chi}$. Prvo skalo, ki nastopa znotraj zan\v cnih integralov indentificiramo z masami pseudo-Goldstonovih bozonov znotraj zank. \v Studije QCD na mre\v zi lahko to koli\v cino spreminjajo in uporabljajo vrednosti vse do $m\sim 1~\mathrm{GeV}$, \v ceprav jo kiralna teorija varuje pred velikimi popravki in ima napovedno vrednost le za $m\ll\Lambda_{\chi}$. Po drugi strani pa lahko druga koli\v cina $\Delta$ v raz\v sirjeni ra\v cunski shemi vsebuje tudi masne razlike med te\v zko-lahkimi mezoni obratnih parnosti. V tem primeru vrednost $\Delta$ ni ve\v c varovana ne s strani kiralne simetrije in ne simetrij te\v zkih kvarkov in lahko zavzame vrednosti reda tipi\v cne hadronske skale $\mathcal O(\Lambda_{QCD})$. Ko torej integriramo gibalne koli\v cine pionov znotraj zank tudi preko te skale, se napovedna mo\v c in perturbativnost sheme poru\v sita.

\par

\v Ce smo lahko fenomenolo\v ske sklopitve iz eksperimentalno merjenih razpadnih \v sirin izlu\v s\v cili ne ozirajo\v c se na tak\v sne probleme (rezultati niso bili kriti\v cno odvisni od izbrane vrednosti $\Delta_{SH}$), pa je situacija v kiralnih ektrapolacijah popolnoma druga\v cna. Tu namre\v c pri\v cakujemo, da bodo ne-analiti\v cni logaritemski prispevki prevladovali medtem ko lahko vse analiti\v cne prispevke enostavno pri\v stejemo k relevantnim kontra\v clenom. V teoriji z le eno skalo ($m$) so tako prevladujo\v ci popravki vedno oblike $m^2 \log m^2$ in imajo dobro dolo\v ceno kiralno limito, ko gre $m\to 0$. V na\v si raz\v sirjeni shemi pa dobimo med drugimi tudi nove prispevke oblike $\Delta_{SH}^2 \log m^2$, ki v kiralni limiti divergirajo. Takoj razumemo, da bo situacija najslab\v sa prav v primeru pionov, katerih mase moramo iz vrednosti, simuliranih na mre\v zi, ekstrapolirati najdlje proti kiralni limiti.

\par

Te\v zavo poskusimo re\v siti pri izvoru, zato se osredoto\v cimo na kiralno limito teorije in poskusimo narediti razvoj zan\v cnih integralov po majhnem parametru. V izbrani limiti so to ravno potence obratne vrednosti nove skale $1/\Delta$. Postopek bo legitimen ob predpostavki, da le\v zi relevantno obmo\v cje integracije stran od te skale, torej za majhne mase in gibalne koli\v cine pionov znotraj zank in dovolj velike vrednosti $\Delta\sim \Delta_{\chi}$. Tako pridelamo vsoto integralov oblike
\begin{equation}
\left.\frac{\mu^{4-D}}{(2\pi)^D} \int \mathrm{d}^D q \frac{q^{\mu} q^{\nu}}{(q^2-m^2)(v\cdot q - \Delta)}\right|_{\Delta = \mathrm{large}} = \frac{\mu^{4-D}}{(2\pi)^D} \int \mathrm{d}^D q \frac{q^{\mu} q^{\nu}}{(q^2-m^2)} \frac{-1}{\Delta} (1+\frac{q\cdot v}{\Delta} + \ldots),
\label{eq_int1a}
\end{equation}
kjer smo s tremi pikami ozn\v cili vi\v sje \v clene v razvoju po $1/\Delta$. Postopek lahko razumemo tudi kot razvoj okrog limite, v kateri se vzbujena stanja razklopijo, njihovi prispevki k teoriji pa se preobrazijo v vrsto lokalnih operatorjev, du\v senih s potencami $1/\Delta$. Interpretiramo jih kot nove prispevke h kontra\v clenom teorije brez dinami\v cnih vzbujenih stanj. Kakr\v sna koli ve\v cja odstopanja tak\v snega pristopa od napovedi teorije brez vzbujenih stanj s primerno zamaknjenimi parametri, bi signalizirala zlom razvoja. To bi pomenilo, da prispevkov dinami\v cnih te\v zko-lahkih mezonov pozitivne parnosti v procesih osnovnih stanj ne moremo zanemariti. Pri\v cakujemo, da bo razvoj dobro deloval na primeru kiralne teorije s simetrijsko grupo $SU(2)$, ki vsebuje le dinami\v cne pione kot pseudo-Goldstonove bozone, katerih mase so mnogo manj\v se od fenomenolo\v ske vrednosti $\Delta_{SH}$. Za ilustracijo lahko skiciramo relevantne energijske skale v efektivni teoriji
\begin{equation}
m_{u,d} \sim \frac{m_{\pi}^2}{\Lambda_{\chi}} < \Delta_{SH} \lesssim m_s \sim \frac{m_{K,\eta}^2}{\Lambda_{\chi}} < \Lambda_{\chi} \ll m_Q.
\end{equation}
Znotraj celotne $SU(3)$ invariantne kiralne teorije s dinami\v cnimi te\v zko-lahkimi mezoni obeh parnosti razvijamo po potencah
$\{m_{\pi,K,\eta},\Delta_{SH}\}/\Lambda_{\chi}$ in $\{m_{\pi,K,\eta},\Delta_{SH},\Lambda_{\chi}\}/m_{Q}$, medtem ko v $SU(2)$ kiralni teoriji z $1/\Delta_{SH}$ razvojem zan\v cnih integralov razvijamo po $m_{\pi}/\{\Lambda_{\chi},\Delta_{SH}\}$ in $\{m_{\pi},\Delta_{SH},\Lambda_{\chi}\}/m_{Q}$.

\par

\begin{figure}
\begin{center}
\hspace*{-2cm}{\includegraphics{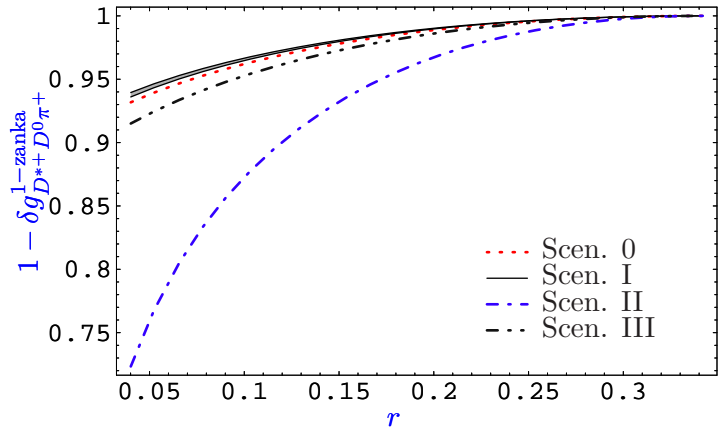}}
\end{center}
\caption{\small\it \label{fig:coupling_1}Renormalizacija sklopitve $g$ v procesu $D^{*+}\to D^0\pi^+$. Primerjava kiralne ekstrapolacije v (I) $SU(2)$ limiti z vodilnimi \v cleni v razvoju zan\v cnih integralov (\v crna neprekinjena \v crta), (II) celotni logaritemski prispevki v $SU(3)$ teoriji s te\v zko-lahkimi multipleti obeh parnosti (modra \v crtasto-pik\v casta \v crta), (III) njihova degenerirana limita (siva \v crtasto-dvojno pik\v casta \v crta), in (0) $SU(3)$ logaritemski prispevki stanj negativne parnosti (rde\v ca \v crtasta \v crta), kot je razlo\v zeno v besedilu.}
\end{figure}

Zgoraj opisan pristop uporabimo za ektrapolacijo efektivnih mo\v cnih sklopitev $g$, $h$ in $\widetilde g$ v redu ene zanke. Najprej zapi\v semo vodilne logaritemske prispevke v $SU(2)$ kiralni teoriji skupaj z vodilnimi $1/\Delta_{SH}$ popravki
\begin{subequations}
\begin{eqnarray}
\label{eq_g_extra1_slo}
g^{\mathrm{eff.}}_{P^{*}_a P_b \pi^{\pm}} &=& g \left\{1 + \frac{1}{(4\pi f)^2} m_{\pi}^2 \log\frac{m_{\pi}^2}{\mu^2} \left[ - 4 g^2 - \frac{m_{\pi}^2}{8\Delta_{SH}^2} h^2 \left( 3 + \frac{\widetilde g}{g}\right) \right]\right\}, \\
g^{\mathrm{eff.}}_{P^{*}_a P_a \pi^0} &=& g \left\{1 + \frac{1}{(4\pi f)^2} m_{\pi}^2 \log\frac{m_{\pi}^2}{\mu^2} \left[ - 5 g^2 - \frac{m_{\pi}^2}{8\Delta_{SH}^2} h^2 \left( 3 - \frac{\widetilde g}{g}\right) \right]\right\},\\
h^{\mathrm{eff.}}_{P^{'}_{a0} P_b \pi^{\pm}} &=& h \left\{1 + \frac{1}{(4\pi f)^2} m_{\pi}^2 \log\frac{m_{\pi}^2}{\mu^2} \left[ \frac{3}{4} (2g\widetilde g - 3 g^2 - 3\widetilde g^2) - \frac{m^2_{\pi}}{2\Delta_{SH}^2} h^2 \right]\right\},\\
h^{\mathrm{eff.}}_{P^{'}_{a0} P_a \pi^0} &=& h \left\{1 + \frac{1}{(4\pi f)^2} m_{\pi}^2 \log\frac{m_{\pi}^2}{\mu^2} \left[ \frac{3}{4} (- 2 g \widetilde g - 3 g^2 - 3 \widetilde g^2) - \frac{m^2_{\pi}}{4\Delta_{SH}^2} h^2 \right]\right\},\\
\widetilde g^{\mathrm{eff.}}_{P^{*}_{a1} P^{'}_{b0} \pi^{\pm}} &=& \widetilde g \left\{1 + \frac{1}{(4\pi f)^2} m_{\pi}^2 \log\frac{m_{\pi}^2}{\mu^2} \left[ - 4 \widetilde g^2 + \frac{m_{\pi}^2}{8\Delta_{SH}^2} h^2 \left( 3 + \frac{ g}{\widetilde g}\right) \right]\right\}, \\
\widetilde g^{\mathrm{eff.}}_{P^{*}_{a1} P^{'}_{a0} \pi^0} &=& \widetilde g \left\{1 + \frac{1}{(4\pi f)^2} m_{\pi}^2 \log\frac{m_{\pi}^2}{\mu^2} \left[ - 5 \widetilde g^2 + \frac{m_{\pi}^2}{8\Delta_{SH}^2} h^2 \left( 3 - \frac{ g}{\widetilde g}\right) \right]\right\},
\label{eq_g_extra_slo}
\end{eqnarray}
\end{subequations}
kjer lo\v cimo med razpadi z nevtralnim ali nabitim pionom v kon\v cnem stanju. Te formule bomo primerjali s celotnimi $SU(3)$ kiralnimi prispevki tudi v teoriji z dinami\v cnimi stanji obeh parnosti. V slede\v ci analizi bomo uporabili fenomenolo\v ske vrednosti sklopitev iz tabele~\ref{table:povzetek} in primerjali:
\begin{description}
\item[(I)] Razvoj zan\v cnih integralov v limiti $SU(2)$. Vodilne prispevke da teorija brez vzbujenih stanj, mi pa bomo upo\v setavali tudi vodilne poravke clenov reda $1/\Delta_{SH}^2$.
\item[(II)] Celotna $SU(3)$ logaritemska ekstrapolacija s prispevki te\v zko-lahkih multipletov obeh parnosti.
\item[(III)] Enako kot (II) vendar v degenerirani limiti $\Delta_{SH}=0$,
\item[(0)] Kiralna $SU(3)$ ektrapolacija brez $1/\Delta_{SH}^2$ prispevkov v ena\v cbah~(\ref{eq_g_extra1}-\ref{eq_g_extra}).
\end{description}
Predpostavimo eksaktno $SU(2)$ izospinsko limito in parametriziramo mase pseudo-Goldstonovih bozonov s pomo\v cjo Gell-Mannovih formul~(\ref{eq:2.6}). Posledi\v cno v kiralni ektrapolaciji variiramo le razmerje $r$ -- med masama lahkih kvarkov in maso \v cudnega kvarka, ki jo dr\v zimo na njeni fizikalni vrednosti. Ker nas zanima le ne-analiti\v cna $r$-odvisnost na\v sih amplitud, lahko od\v sejemo skupno odvisnost od skale renormalizacije skupaj z vsemi prispevki, analiti\v cnimi  v $r$. Na\v se rezultate normaliziramo na skupno vrednost v vseh primerih pri skali $8 r_{\mathrm{ab}} \lambda_0 m_s / f^2= \Delta_{SH}^2 $ in jih od tu ekstrapoliramo proti kiralni limiti. Za primer si oglejmo primer efektivne sklopitve v procesu brez \v cudnosti $D^{*+}\to D^0\pi^+$ na sliki~\ref{fig:coupling_1}.
\psfrag{xk1}[bc]{\Blue{$r$}}
\psfrag{xm1}[tc][tc][1][90]{\Blue{$1-\delta g^{1\mathrm{-zanka}}_{D^{*+}D^0\pi^+}$}}
\psfrag{s0}{Scen. 0}
\psfrag{s1}{Scen. I}
\psfrag{s2}{Scen. II}
\psfrag{s3}{Scen. III}
Takoj opazimo, da vklju\v citev celotnih $SU(3)$ logaritemskih prispevkov vzbujenih stanj v zankah vnese velika ($\gtrsim 30\%$) odstopanja od ekstrapolacije brez teh stanj. \v Ce pa namesto tega uporabimo razvoj zan\v cnih integralov, se odstopanja drasti\v cno zmanj\v sajo. Ekstrapolaciji znotraj $SU(2)$ in $SU(3)$ teorij brez vzbujenih stanj sta skoraj identi\v cni saj v obeh glavnino prispevajo pionske zanke. Vodilne prispevke izintegriranih vzbujenih stanj ocenimo s pomo\v cjo sivega podro\v cja med obema krivuljama $SU(2)$ teorije v scenariu (I). Razlika nanese komaj $0.5\%$, kar ka\v ze na dobro konvergenco razvoja.

\par

Za popolnost izri\v semo \v se diagram kiralne ektrapolacije efektivne sklopitve $h$ v procesu $D_0^+ \to D^0 \pi^+$ (ekstrapolacija sklopitve $\widetilde g$ poteka v istem slogu kot $g$ ob zamenjavi obeh sklopitev v ektrapolacijskih formulah). \psfrag{xk1}[bc]{\Blue{$r$}}
\psfrag{xm1}[tc][tc][1][90]{\Blue{$1-\delta h^{1\mathrm{-zanka}}_{D^{*+}_0 D^0\pi^+}$}}
\begin{figure}
\begin{center}
\hspace*{-2cm}{\includegraphics{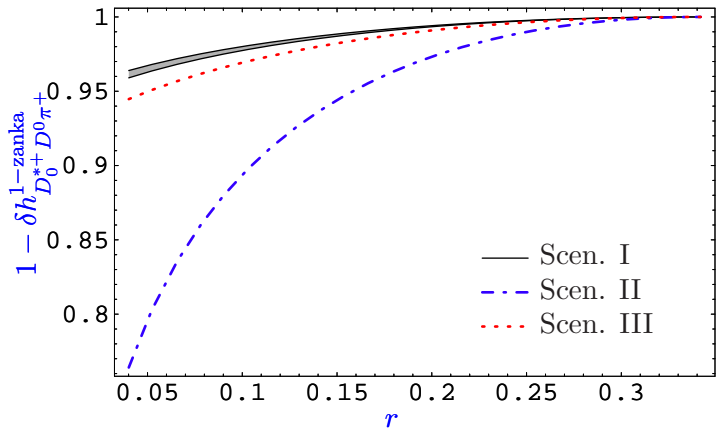}}
\end{center}
\caption{\small\it \label{fig:coupling_2} Kiralna ekstrapolacija sklopitve $h$ v razpadu $D^{*+}_0\to D^0\pi^+$. Primerjava kiralne ekstrapolacije z (I) razvojem zan\v cnih integralov v $SU(2)$ limiti (\v crna neprekinjena \v crta), (II) celotni $SU(3)$ logaritemski popravki (modra \v crtasto-pik\v casta \v crta), in (III) njihova degenerirana limita (rde\v ca \v crtasta \v crta), kot je razlo\v zeno v besedilu.}
\end{figure}
Tukaj scenarij (0) nima pomena saj v zunanjih stanjih nastopajo te\v zko-lahki mezoni obeh parnosti. Po drugi strani pa je razvoj zan\v cnih integralov v scenariju (I) \v se vedno dobro definiran, \v ceprav je njegove fizikalna interpretacija sedaj manj jasna. Namre\v c iz teorije ne integriramo ve\v c stanj posamezne parnosti temve\v c dejansko re\v zemo visokoenergijske prispevke posami\v cnih zan\v cnih integralov, ki nastanejo kot posledica kinematike vmesnih in zunanjih te\v zko-lahkih stanj. Tudi tukaj so vplivi vodilnih popravkov tak\v snega razvoja reda velikosti $0.5\%$.

\par

Na\v sa analiza kiralne ektrapolacije sklopitve $g$ je pokazala, da lahko prispevki vzbujenih te\v zko-lahkih stanj pomembno vplivajo na naklon in krivino ektrapolacij v limiti $m_{\pi}\to 0$. Zagovarjamo tezo, da je to posledica velikih gibalnih koli\v cin pionov v zan\v cnih integralih, kjer nastopajo masne razlike med te\v zko-lahkimi multipleti razli\v cnih parnosti $\Delta_{SH}$, ki v kiralni limiti ne gredo proti ni\v c. \v Ce pa uporabimo fizikalno motiviranih pribli\v zkov -- tak\v sne integrale razvijemo po potencah $1/\Delta_{SH}$ -- se njihovi efekti drasti\v cno zmanj\v sajo in prispevajo k ektrapolaciji le \v se reda $0.5\%$. Posledi\v cno lahko sklepamo o dobri konvergenci izbranega razvoja. Torej lahko zaklju\v cimo, da je mogo\v ce dr\v zati kiralne popravke zank v mo\v cnih razpadih te\v zkih mezonov pod nadzorom v kolikor ekstrapolacije izvajamo pod skalo $\Delta_{SH}$, vsekakor pa ostajajo pomembni za natan\v cno dolo\v citev efektivnih mo\v cnih sklopitev $g$, $h$ in $\widetilde g$.

\section{Semileptonski razpadi te\v zkih mezonov}

Eden izmed trenutno najpomembnej\v sih programov v hadronski fiziki se ukvarja z dolo\v citvijo parametrov CKM iz ekskluzivnih razpadov. Bistvena sestavina tega pristopa je natan\v cno poznavanje oblikovnih funkcij v te\v zko-te\v zkih in te\v zko-lahkih \v sibkih prehodih hadronov. Tradicionalno so najve\v c pozornosti po\v zeli razpadi bezonov $B$ ter z njimi povezano izlu\v s\v cenje faze CKM in absolutnih vrednosti elementov CKM $V_{ub}$ in $V_{cb}$. Hkrati pa v sektorju \v carobnih mezonov absolutne vrednoste $V_{cs}$ in $V_{cd}$ najnatan\v cneje dolo\v ca unitarnost CKM, medtem ko neposredne meritve v ekspereimentih zavira slabo teoreti\v cno poznavanje velikosti relevantnih oblikovnih funkcij. V obeh sektorjih lahko prisotnost blizu le\v ze\v cih resonanc vzbujenih te\v zko-lahkih mezonov trenutno sliko precej spremeni.

\subsection{Te\v zko -- lahki prehodi}

V tem razdelku bomo na kratko obnovili splo\v sno parametrizacijo oblikovnih funkcij v prehodih $H\to P$, ki sta jo prva zasnovala Be\'cirevi\'c in Kaidalov~\cite{Becirevic:1999kt}, ter izoblikovali podobno parametrizacijo tudi za vse oblikovne funkcije v prehodih $H\to V$. Tak\v sna parametrizacija mora upo\v stevati znana eksperimentalna dejstva o prisotnosti resonanc v relevantnih razpadnih kanalih ter tudi teoreti\v cne limite teorij HQET in teorije kolinearnih prostostnih stopenj (ang. soft colinear effective theory -- SCET), ko so te relevantne za dan problem. Potem bomo analizirali prispevke nedavno odkritih \v carobnih resonanc k oblikovnim funkcijam prehodov $H\to P$ in $H\to V$, za kar bomo uporabili efektivni model na podlagi HM$\chi$PT v katerega bomo dodali nova polja, ter splo\v snih parametrizacij oblikovnih funkcij. V na\v si diskusiji se bomo omejili na prispevke v prvem redu kiralnega razvoja ter razvoja po obratnih vrednostih mas te\v zkih kvarkov $1/m_{Q}$.

\par

V limiti majhnega odboja, ko izhajajo\v ci lahki mezon $P$ v masnem sistemu za\v cetnega te\v zkega mezona $H$ skorajda miruje HQET napoveduje dobro znana umeritvena pravila za vse oblikovne funkcije $H\to P$ in $H\to V$~\cite{Isgur:1990kf}. Hkrati v obratni limiti, ko mezon $P$ v masnem sistemu za\v cetnega te\v zkega mezona $H$ izide z maksimalno energijo za oblikovne funkcije veljajo nekoliko druga\v cna umeritvena pravila~\cite{Charles:1998dr}. Na\v sa naloga je, da poi\v semo primerno konsistentno parametrizacijo poteka oblikovnih funkcij, ki bo zvezno povezala obe limitni obmo\v cji. Pri tem nam nekoliko pomaga dejstvo, da \v ze poznamo prevladujo\v ce prispevke k nekaterim oblikovnim funkcijam v limiti majhnega odboja. V bli\v zini tega kinemtatskega obmo\v cja le\v zijo namre\v c znane \v carobne resonance negativne parnosti, ki bodo prispevale bli\v znje pole v konfiguracijski ravnini. Le-te lahko v primeru vektorskih oblikovnih funkcij $F_+$ v $H\to P$ ter $V$ v $H\to V$ (kot tudi psevdoskalarne oblikovne funkcije $A_0$) izoliramo in dobro dolo\v cimo, saj pripadajo najni\v zje le\v ze\v cim vektorskim (psevdoskalarnim) \v carobnim resonancam $H^*$ ($H'$). Upo\v stevajo\v c \v se vse umeritvene limite ter kinematske relacije med oblikovnimi funkcijami, lahko njihovo energijsko odvisnost kompaktno parametriziramo kot
\begin{eqnarray}
\label{eq:ff_general1}
F_+(s) &=& \frac{F_+(0)}{(1-x)(1-ax)}, \nonumber\\
\label{eq:ff_general2}
F_0(s) &=& \frac{F_0(0)}{(1-bx)}, \nonumber\\
\label{eq:ff_general3}
V(s) &=& \frac{V(0)}{(1-x)(1-a'x)}, \nonumber\\
A_0(s) &=& \frac{A_0(0)}{(1-y)(1-a''y)}, \nonumber\\
A_1(s) &=& \frac{A_1(0)}{1-b'x}, \nonumber\\
A_2(s) &=& \frac{A_2(0)}{(1-b'x)(1-b''x)},
\label{eq:ff_general}
\end{eqnarray}
kjer smo ozna\v cili $x=s/m_{H^*}^2$ in $y=s/m_{H'}^2$ in hkrati veljajo znane kinematske relacije med $F_+(0)=F_0(0)$ ter $V(0)$, $A_0(0)$, $A_1(0)$ in $A_2(0)$.

\par

\begin{figure}[!t]
\begin{center}
\epsfxsize8cm\epsffile{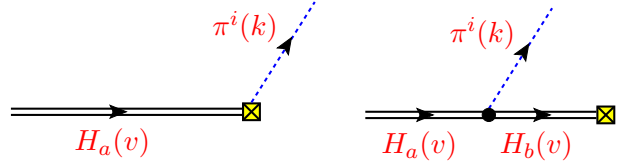}
\end{center}
\caption{\small\it \label{fig:HP}Diagrama, ki prispevata k oblikovnim funkcijam $H\to P$.}
\end{figure}

Proste parametre gornje parametrizacije, kamor \v stejemo tudi parametre polov $a$, $a'$, $b'$ in $b''$, dolo\v cimo s pomo\v cjo efektivnega modela, ki temelji na HM$\chi$PT. Feynmanova pravila HM$\chi$PT veljajo v podro\v cju majhnega odboja in dajo prispevke k oblikovnim funkcijam $H\to P$ v prvem redu razvoja na sliki~\ref{fig:HP}.
\psfrag{Ha}[cc]{${\Red{H_a(v)}}$}
\psfrag{Hb}[cc]{$\Red{H_b(v)}$}
\psfrag{pi}[bc]{$\Red {\pi^i(k)}$}
Opazimo, da desni diagram na sliki \v ze spominja na resonan\v cni prispevek. Vendar pa ob primerjavi s parametrizacijo vektorske oblikovne funkcije $F_+$ opazimo, da lahko z danimi $1/2^+$ in $1/2^-$ polji HM$\chi$PT identificiramo le prvega izmed obeh polov, ki prispevata k oblikovni funkciji. Nerodnost razre\v simo, tako da v efektivno teorijo vnesemo nov set $j_{\ell}^P=1/2^-$ polj $\widetilde H$, ki predstavljajo radialno vzbujena stanja psevdoskalarnih in vektorskih mezonov. Potrebne spremembe Lagrangevega operatorja HM$\chi$PT so enostavne
\begin{eqnarray}
    \mathcal L^{(1)}_{\mathrm{HM}\chi\mathrm{PT}} &+=& \widetilde{\mathcal L}^{(1)}_{\frac{1}{2}^-} + \widetilde{\mathcal L}^{(1)}_{\mathrm{mix}}, \nonumber\\
    \widetilde{\mathcal L}^{(1)}_{\frac{1}{2}^-} &=& - \mathrm{Tr}\left[ \overline {\widetilde H}_a (i v \cdot \mathcal{D}_{ab} - \delta_{ab} \Delta_{\widetilde H} ) \widetilde H_b\right], \nonumber\\
    \widetilde{\mathcal L}^{(1)}_{\mathrm{mix}} &=& \widetilde h \mathrm{Tr} \left[ \overline H_b \widetilde H_a \slashed{\mathcal A}_{ab} \gamma_{5}  \right] + \mathrm{h.c.},
    \label{eq:Lradial}
\end{eqnarray}
in
\begin{equation}
J^{(0)\mu}_{a(V-A)\mathrm{HM}\chi\mathrm{PT}} += \frac{i \widetilde \alpha}{2} \mathrm{Tr} [ \gamma^{\mu}(1-\gamma_5) \widetilde H_b] \xi_{ba}^{\dagger},
\end{equation}
kjer smo vnesli tudi tri nove parametre: $\Delta_{\widetilde H}$ je masni popravek novega multipleta, $\widetilde h$ je efektivna sklopitev med osnovnimi in radialno vzbujenimi stanji negativne parnosti, $\widetilde \alpha$ pa je efektivna \v sibka sklopitev novih stanj, povezana z njihovo razpadno konstanto. Sedaj \v stevilo polov v parametrizaciji oblikovnih funkcij $H\to P$ ustreza \v stevilu resonan\v cnih prispevkov v HM$\chi$PT modelu.

\par

Podobno igro lahko poskusimo tudi v primeru prehodov $H\to V$. Ker pa HM$\chi$PT ne vsebuje lahkih vektorskih mezonov, moramo teorijo spet raz\v siriti z fenomenolo\v skim modelom. Poslu\v zimo se pogosto uporabljenega principa skrite simetrije~\cite{Casalbuoni:1996pg}, po katerem lahke vektorske mezone vpeljemo kot umeritvena polja neke raz\v sirjene $SU(3)_V$ simetrije in jih opi\v semo z operatorjem $\hat\rho_{\mu} = i \frac{g_V}{\sqrt{2}} \rho_{\mu}$, kjer je $\rho_{\mu}$ matrika lahkih vektorskih mezonov
\begin{equation}
\rho_{\mu} =
   \begin{pmatrix}
    \frac{1}{\sqrt 2} (\omega_{\mu} + \rho^0_{\mu}) & \rho^+_{\mu} & K^{*+}_{\mu} \\
    \rho^-_{\mu} & \frac{1}{\sqrt 2} (\omega_{\mu} - \rho^0_{\mu}) & K^{*0}_{\mu} \\
    K^{*-}_{\mu} & \overline K^{*0}_{\mu} & \phi_{\mu}
   \end{pmatrix}.
\end{equation}
Kineti\v cni in masni Lagrangev operator tak\v snih polj sta potem
\begin{equation}
\mathcal L_V = \frac{1}{2g_V^2}[ F_{\mu\nu}(\hat\rho)_{ab} F^{\mu\nu}(\hat\rho)_{ba}] - \frac{a f^2}{2} (\mathcal V^{\mu}_{ab} - \hat \rho^{\mu}_{ab})(\mathcal V_{\mu,ba} - \hat \rho_{\mu,ba}),
\end{equation}
kjer smo definirali $F_{\mu\nu}(\hat\rho) = \partial_{\mu} \hat\rho_{\nu} - \partial_{\nu} \hat\rho_{\mu} + [\hat \rho_{\mu} , \hat \rho_{\nu}]$. V prvem redu razvoja $1/m_H$ lahko potem zapi\v semo tudi interakcijski Lagrangev operator med lahkimi vektorskimi mezoni in te\v zko-lahkimi mezoni~\cite{Casalbuoni:1996pg,Bajc:1995km}
\begin{eqnarray}
\mathcal L^{\mathrm{int}}_{1/2^-} &=& - i \beta \mathrm{Tr} [H_b v_{\mu} \hat\rho^{\mu}_{ba} \overline H_a ] + i \lambda \mathrm{Tr} [ H_b \sigma^{\mu\nu} F_{\mu\nu}(\hat\rho)_{ba} \overline H_a ],\\
\mathcal L^{\mathrm{int}}_{\mathrm{mix}} &=&  - i \zeta \mathrm{Tr} [ H_b v_{\mu} \hat\rho^{\mu}_{ba} \overline S_a ] + \mathrm{h.c.} \nonumber\\*
&& + i \mu \mathrm{Tr} [ H_b \sigma^{\mu\nu} F_{\mu\nu}(\hat\rho)_{ba} \overline S_a ] + \mathrm{h.c.},\\
\widetilde{\mathcal L}^{\mathrm{int}}_{\mathrm{mix}} &=&  - i \widetilde \zeta \mathrm{Tr} [ H_b v_{\mu} \hat\rho^{\mu}_{ba} \overline {\widetilde{H}}_a ] + \mathrm{h.c.} \nonumber\\*
&& + i \widetilde \mu \mathrm{Tr} [ H_b \sigma^{\mu\nu} F_{\mu\nu}(\hat\rho)_{ba} \overline {\widetilde H}_a ] + \mathrm{h.c.},
\label{eq:L_even_odd}
\end{eqnarray}
ter
\begin{eqnarray}
J^{(0)\mu}_{a(V-A)\mathrm{HM}\chi\mathrm{PT}} &+=& \alpha_1 \mathrm{Tr} [\gamma^5 H_b \hat\rho^{\mu}_{ba} ] + \alpha_2 \mathrm{Tr} [ \gamma^{\mu} \gamma^{5} H_b v_{\alpha} \hat\rho^{\alpha}_{ba} ].
\end{eqnarray}
S temi dodatnimi gradniki lahko sestavimo Feynmanove diagrame, ki prispevajo k prehodom $H\to V$, in so topolo\v sko ekvivalentni tistim na sliki~\ref{fig:HP}, z zamenjavo pionskih z lahko-vektorskimi linijami. Ti diagrami lepo reproducirajo strukturo polov v splo\v sni parametrizaciji oblikovnih funkcij s prispevki resonanc s primernimi kvantnimi \v stevili. Izjema je oblikovna funkcija $A_2$, katere parametrizacija vsebuje dva efektivna pola, medtem ko na\v s model napoveduje le prispevek ene (edine) aksialne resonance iz $j_{\ell}^P=1/2^+$ multipleta.

\par

Na\v s pristop \v zelimo primerjati z izmerjenimi razpadnimi \v sirinami in kotnimi porazdelitvami semileptonskih razpadov \v carobnih mezonov v lahke psevdoskalarne in vektorske mezone. Zato moramo napovedi HM$\chi$PT modelov ektrapolirati \v cez celotno kinematsko podro\v cje. Poslu\v zimo se splo\v snih parametrizacij oblikovnih funkcij~(\ref{eq:ff_general1}) ter hkrati uporabiti \v cimve\v c obstoje\v cih eksperimentalnih informacij. Efektivne parametre polov oblikovnih funkcij ($a$, $a'$, $b'$ in $b''$) tako zasi\v cimo z merjenimi oz. teoreti\v cno napovedanimi masami \v carobnih resonanc, katerih prispevke smo identificirali v HM$\chi$PT modelu. Preostale parametre dolo\v cimo s prilagajanjem napovedi za razpadne \v sirine znotraj na\v sega pristopa z eksperimentalno izmerjenimi vrednostmi. Rezultati tak\v snega postopka so vrednosti parametrov v tabelah~\ref{table:params} in~\ref{table:params_V}.
\begin{table*}[!t]
\begin{center}
\begin{tabular}{|l|cc|cc|}
\hline
Razpad & $F_+(0)$ & $F_0(0)$ & $a$ & $b$ \\
\hline\hline
    ${D_0\to \pi^-}^{\dagger}$ & $0.60$ & $0.60$ & $0.55$ & $0.76$\\
    ${D^0\to K^{-}}$ & $0.72$ & $0.72$ & $0.57$ & $0.83$\\
    $D^+\to\pi_0$ & $0.60$ & $0.62$ & $0.55$ & $0.76$\\
    $D^+\to \overline K_0$ & $0.72$ & $0.71$ & $0.57$ & $0.83$\\
    $D_s\to\eta$ & $0.73$ & $0.81$ & $0.57$ & $0.83$\\
    $D_s\to\eta'$ & $0.87$  & $0.66$ & $0.57$ & $0.83$\\
    $D^+\to\eta$ & $0.60$ & $0.62$ & $0.55$ & $0.76$\\
    $D^+\to\eta'$ & $0.60$ & $0.62$ & $0.55$ & $0.76$\\
    $D_s\to\overline K_0$ & $0.60$ & $0.62$ & $0.55$ & $0.76$\\
\hline
\end{tabular}
\end{center}
\caption{\small\it \label{table:params} Napovedi na\v sega modela za vrednosti parametrov, ki nastopajo v formulah splo\v sne parametrizacije oblikovnih funkcij~(\ref{eq:ff_general1}) za obravnavane razpadne kanale $D\to P\ell\nu_{\ell}$. Razpadni kanal $D^0\to \pi^-$ ozna\v cen s kri\v zcem (${\dagger}$) smo uporabili za prilagajanje novih parametrov.}
\end{table*}
\begin{table*}[!t]
\begin{center}
\begin{tabular}{|l|cccc|cc|}
\hline
Razpad & $V(0)$ & $A_0(0)$ & $A_1(0)$ & $A_2(0)$ & $a''=a'$ &  $b'$ \\
\hline
\hline
    ${D^0\to \rho^-}^{\dagger}$ & $1.05$ & $1.32$ & $0.61$ & $0.31$ & $0.55$ & $0.76$ \\
    ${D^0\to K^{-*}}^{\dagger}$ & $0.99$ & $1.12$ & $0.62$ & $0.31$ & $0.57$ & $0.83$  \\
    ${D^+\to \rho^0}^{\dagger}$ & $1.05$ & $1.32$ & $0.61$ & $0.31$ & $0.55$ & $0.76$ \\
    ${D^+\to K^{0*}}^{\dagger}$ & $0.99$ & $1.12$ & $0.62$ & $0.31$ & $0.57$ & $0.83$ \\
    $D^+\to \omega$ & $1.05$ & $1.32$ & $0.61$ & $0.31$ & $0.55$ & $0.76$ \\
    ${D_s\to \phi}^{\dagger}$ & $1.10$ & $1.02$ & $0.61$ & $0.32$ & $0.57$ & $0.83$ \\
    $D_s\to K^{0*}$ & $1.16$ & $1.19$ & $0.60$ & $0.33$ & $0.55$ & $0.76$ \\
\hline
\end{tabular}
\end{center}
\caption{\small\it \label{table:params_V} Napovedi na\v sega modela za vrednosti parametrov, ki nastopajo v formulah splo\v sne parametrizacije oblikovnih funkcij~(\ref{eq:ff_general1}) za obravnavane razpadne kanale $D\to P\ell\nu_{\ell}$ ($b''=0$ za vse razpadne kanale). Razpadne kanale ozna\v cene s kri\v zcem (${\dagger}$) smo uporabili za prilagajanje novih parametrov.}
\end{table*}

\par

\begin{figure}[!t]
\begin{center}
\scalebox{1}{\includegraphics{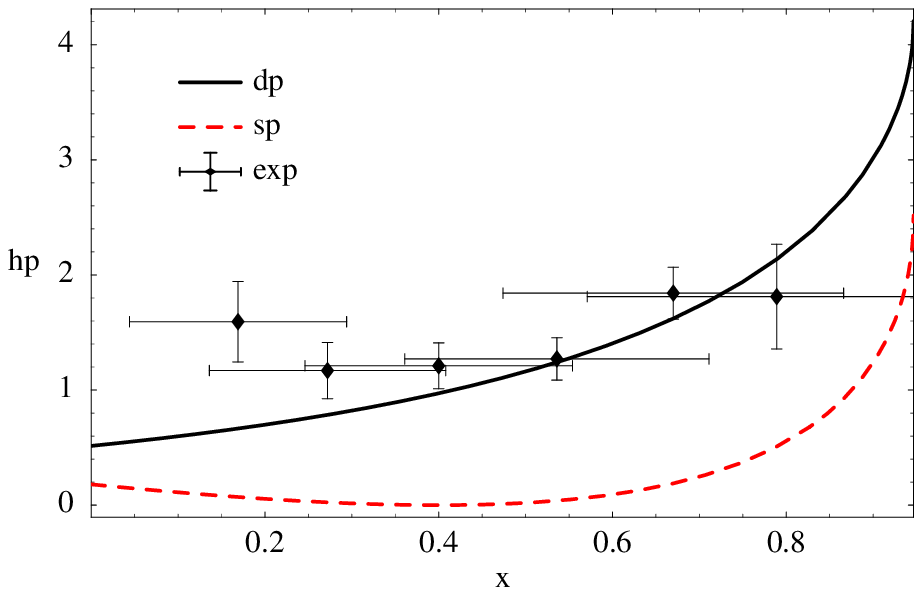}}
\end{center}
\caption{\small\it \label{fig:1} Napovedi na\v sega modela (dva pola v \v crni neprekinjeni \v crti in en pol v rde\v ci prekinjeni \v crti) za porazdelitev su\v cnostne amplitude $H_+^2(s)$ v primerjavi s podatki kolaboracije FOCUS za semileptonski razpad $D^+ \to \overline K^{*0}$.}
\end{figure}

\begin{figure}[!t]
\begin{center}
\scalebox{1}{\includegraphics{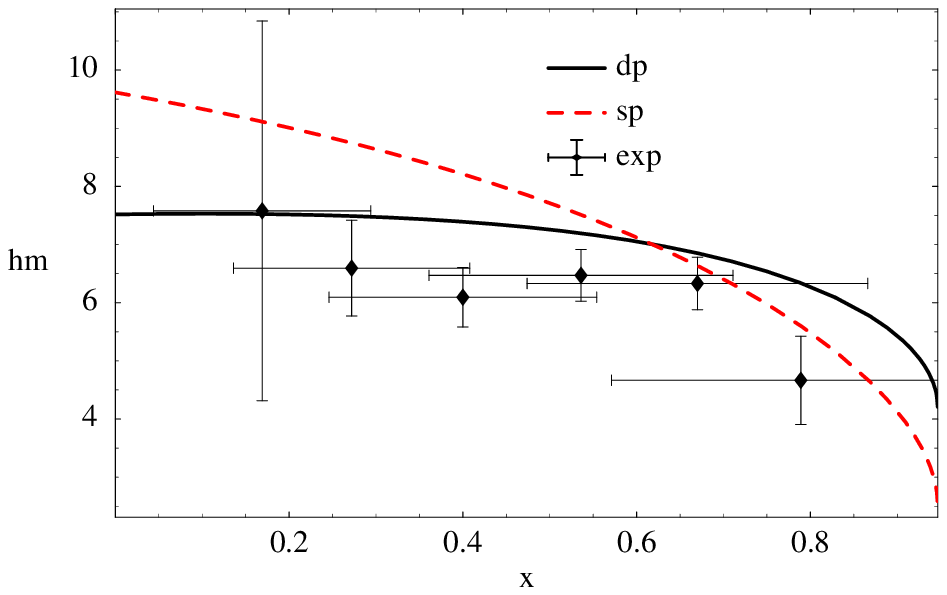}}
\end{center}
\caption{\small\it \label{fig:2} Napovedi na\v sega modela (dva pola v \v crni neprekinjeni \v crti in en pol v rde\v ci prekinjeni \v crti) za porazdelitev su\v cnostne amplitude $H_-^2(s)$ v primerjavi s podatki kolaboracije FOCUS za semileptonski razpad $D^+ \to \overline K^{*0}$.}
\end{figure}

\begin{figure}[!t]
\begin{center}
\scalebox{1}{\includegraphics{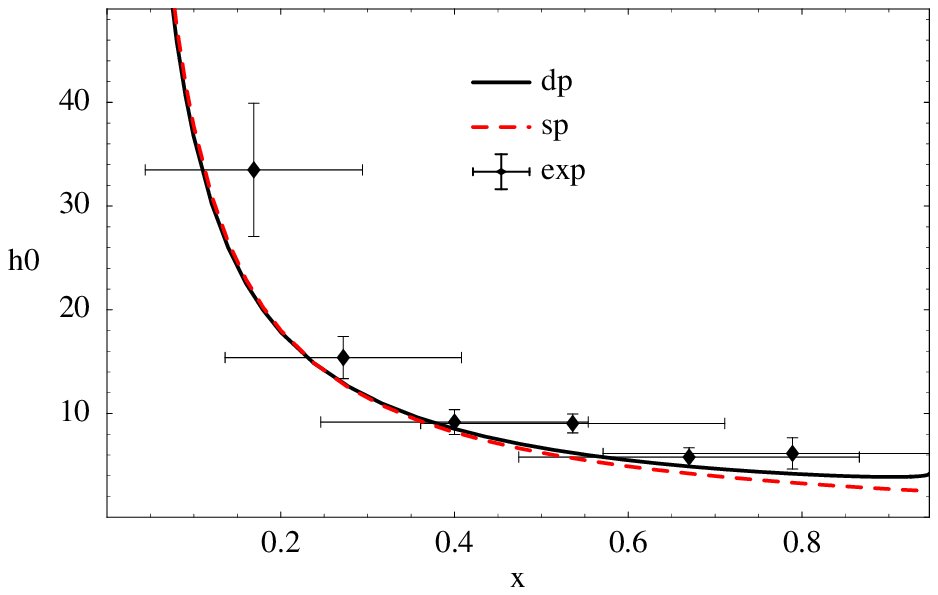}}
\end{center}
\caption{\small\it \label{fig:3} Napovedi na\v sega modela (dva pola v \v crni neprekinjeni \v crti in en pol v rde\v ci prekinjeni \v crti) za porazdelitev su\v cnostne amplitude $H_0^2(s)$ v primerjavi s podatki kolaboracije FOCUS za semileptonski razpad $D^+ \to \overline K^{*0}$.}
\end{figure}

Na podlagi tako dobljenih parametrov najprej primerjamo napovedi na\v sega pristopa z nedavno eksperimentalno analizo su\v cnostnih amplitud, ki jo je naredila kolaboracija FOCUS~\cite{Link:2005dp}. Primerjava porazdelitev posami\v cnih su\v cnostnih amplitud je na slikah~\ref{fig:1},~\ref{fig:2} in~\ref{fig:3}.
\psfrag{hp}[bc][bc][1][90]{\Blue{$\left|H_+^{(D^+ \to \overline K^{*0})}\right|^2~[\mathrm{GeV}^2]$}}
\psfrag{dp}[cl]{Napoved}
\psfrag{sp}[cl]{Napoved (en pol)}
\psfrag{exp}[cl]{FOCUS~\cite{Link:2005dp}}
\psfrag{hm}[bc][bc][1][90]{\Blue{$\left|H_-^{(D^+ \to \overline K^{*0})}\right|^2~[\mathrm{GeV}^2]$}}
\psfrag{sp}[cl]{Napoved (en p.)}
\psfrag{h0}[bc][bc][1][90]{\Blue{$\left|H_0^{(D^+ \to \overline K^{*0})}\right|^2~[\mathrm{GeV}^2]$}}
\psfrag{sp}[cl]{Napoved (en pol)}
Opazimo, da je ujemanje napovedi na\v sega modela z eksperimentalnimi podakti dobro, \v ceprav pa so eksperimentalne napake \v se velike.
Poleg splo\v sne parametrizacije primerjamo tudi napovedi na\v sega modela z enostavnim nastavkom enega pola za vse oblikovne funkcije. Iz slik~\ref{fig:1} in ~\ref{fig:2} je razvidno, da se tak\v sen pristop ne ujema dobro z eksperimentalnimi podatki.

\par

Na koncu podamo \v se napovedi za razvejtivena razmerja vseh relevantnih semileptonskih prehodov $D\to P$ in $D\to V$ in primerjamo na\v se napovedi z merjenimi vrednostmi iz PDG~\cite{Eidelman:2004wy}. Rezultati so povzeti v tabelah~\ref{table:PP_results} in \ref{table:results}.
\begin{table}[!t]
\begin{center}
\begin{tabular}{|c|c|c|c|}
   \hline
    Razpad & $\mathcal{B}\mathrm{(dva~pola)}~[\%]$ & $\mathcal{B}\mathrm{(en~pol)}~[\%]$ & $\mathcal{B}\mathrm{(Eksp.)}~[\%]$\\
    \hline
    \hline
   ${D^0\rightarrow \pi^-}^{\dagger}$ & $0.36$ & $0.36$ & $0.36\pm0.06$\\
   $D^0\rightarrow K^-$ & $3.8$ & $0.43$ & $3.43\pm0.14$ \\
   $D^+\rightarrow \pi^0$ & $0.46$ & $0.51$ & $0.31\pm0.15$\\
   $D^+\rightarrow \overline K^0$ & $9.7$ & $1.1$ & $6.8\pm0.8$ \\
   $D^+_s\rightarrow \eta$ & $2.6$ & $0.38$ & $2.5\pm0.7$ \\
   $D^+_s\rightarrow \eta'$ & $0.86$ & $0.03$ & $0.89\pm0.33$ \\
   $D^+\rightarrow \eta$ & $0.11$ & $0.006$ & $<0.5$ \\
   $D^+\rightarrow \eta'$ & $0.016$ & $0.0003$ & $<1.1$ \\
   $D^+_s\rightarrow K^0$ & $0.33$ & $0.06$ & \\
    \hline
\end{tabular}
\end{center}
\caption{\small\it \label{table:PP_results} Razvejitvena razmerja za semilptonske razpade $D\rightarrow P$. Primerjava napovedi modela z eksperimentom. Razpadni kanal $D^0\to \pi^-$ ozna\v cen s kri\v zcem (${\dagger}$) smo uporabili za prilagajanje novih parametrov.}
\end{table}
\begin{table}[!t]
\begin{center}
\begin{tabular}{|l|cc|cc|cc|}
\hline
Razpad & $\mathcal{B}$ [\%] & $\mathcal{B}$ (Eksp.) [\%] & $\Gamma_L/\Gamma_T$ &  $\Gamma_L/\Gamma_T$ (Eksp.) &  $\Gamma_+/\Gamma_-$ &  $\Gamma_+/\Gamma_-$ (Eksp.) \\\hline
    \hline
     ${D^0\to \rho^-}^{\dagger}$ & $0.20$ & $0.194(41)$~\cite{Blusk:2005fq} & $1.10$ & & $0.13$ & \\
    ${D^0\to K^{-*}}^{\dagger}$ & $2.2$ & $2.15(35)$~\cite{Eidelman:2004wy} & $1.14$ & & $0.22$ & \\
    ${D^+\to \rho^0}^{\dagger}$ & $0.25$ & $0.25(8)$~\cite{Eidelman:2004wy} & $1.10$ & & $0.13$ & \\
    ${D^+\to K^{0*}}^{\dagger}$ & $5.6$ & $5.73(35)$~\cite{Eidelman:2004wy} & $1.13$ & $1.13(8)$~\cite{Eidelman:2004wy} & $0.22$ & $0.22(6)$~\cite{Eidelman:2004wy} \\
    ${D_s\to \phi}^{\dagger}$ & $2.4$ & $2.0(5)$~\cite{Eidelman:2004wy} & $1.08$ & & $0.21$ & \\
    $D^+\to \omega$ & $0.25$ & $0.17(6)$~\cite{Blusk:2005fq} & $1.10$ & & $0.13$ & \\
    $D_s\to K^{0*}$ & $0.22$ &  & $1.03$ & & $0.13$ &\\
    \hline
\end{tabular}
\end{center}
\caption{\small\it \label{table:results} Razvejitvena razmerja ter razmerja delnih razpadnih \v sirin za semilptonske razpade $D\rightarrow V$. Primerjava napovedi modela z eksperimentom. Razpadne kanale ozna\v cene s kri\v zcem (${\dagger}$) smo uporabili za prilagajanje novih parametrov.}
\end{table}
Za primerjavo smo v tabelo~\ref{table:PP_results} vklju\v cili tudi rezultate, dobljene z na\v sim pristopom a le enim polom v parametrizaciji oblikovne funkcije $F_+$. Na\v s model ektrapoliran s splo\v sno parametrizacijo da v splo\v snem rezultate zdru\v zljive s trenutnimi eksperimentalnimi rezultati, medtem ko ekstrapolacija z enim samim polom popolnoma odpove. V principu bi lahko na\v s pristop posplo\v sili tudi na razpade mezonov $B$. Vendar pa so ti, zaradi mnogo ve\v cjega kinematskega podro\v cja mnogo bolj ob\v cutljivi na vrednosti oblikovnih funkcij pri $s\approx 0$ in zato zahtevajo pristop, ki presega le upo\v stevanje najni\v zjih resonanc.

\subsection{Te\v zko -- te\v zki prehodi}

Na na\v si misiji k natan\v cni dolo\v citvi matri\v cnega elementa CKM $V_{cb}$ igrajo pomembno vlogo \v studije razpadov mezona $B$ v \v carobne resonance. Eksperimenti z namenom dolo\v citi vrednost $V_{cb}$ dejansko izlu\v s\v cijo produkt $|V_{cb}\mathcal F(1)|$, kjer je $\mathcal F(1)$ hadronska oblikovna funkcija prehodov $B \to D$ ali $B \to D^*$ pri ni\v ctem odboju. Pomankanje natan\v cnih informacij o obliki in velikosti teh oblikovnih funkcij je tako \v se vedno glavni vir napak. V obravnavi hadronskih lastnosti s pomo\v cjo QCD na mre\v zi najve\v cje te\v zave nastanejo zaradi majhnih mas lahkih kvarkov. \v Studije na mre\v zi morajo uporabljati ve\v cje mase in rezultate naknadno ekstrapolirati k njihovim fizikalnim vrednostim. V teh \v studijah je kiralno obna\v asnje amplitud \v se posebej pomembno. HM$\chi$PT je v tem pogledu zelo uporabna, saj nam omogo\v ca nekaj kontrole nad napakami, ki se pojavijo ob pribli\v zevanju kiralni limiti. Ogledali si bomo torej popravke kiralnih zank znotraj HM$\chi$PT v semileptonskih razpadih $B$ mezonov v \v carobne mezone obeh parnosti in dolo\v cili njihov vpliv na kiralno ekstrapolacijo, kot jo uporabljajo \v studije QCD na mre\v zi pri obravnavi relevantnih oblikovnih funkcij.

\par

Ponovno uporabimo formalizem Lagrangevih operatorjev v efektivni kiralni teoriji te\v zko-lahkih mezonov in pseudo-Goldstonovih bozonov. \v Sibki del Lagrangevega operatorja, ki opisuje prehode med te\v zkimi kvarki lahko zapi\v semo s \v sibkimi tokovi te\v zkih mezonov v HM$\chi$PT~\cite{Falk:1993iu,Wise:1992hn}
\begin{eqnarray}
\overline c_{v'} \Gamma b_{v} &\to& C_{cb} \Huge\{ \,-\, \xi(w) \mathrm{Tr} \left[ \overline{H}_a(v') \Gamma H_a(v) \right] \nonumber\\
&&\phantom{C_{cb} \{}-\widetilde \xi(w) \mathrm{Tr} \left[ \overline{S}_a(v') \Gamma S_a (v) \right] \nonumber\\
&&\phantom{C_{cb} \{}-\tau_{1/2}(w) \mathrm{Tr} \left[ \overline{H}_a(v') \Gamma S_a (v) \right] + \mathrm{h.c.} \Huge\}
\end{eqnarray}
v prvem redu kiralnega razvoja in razvoja po obratnih vrednostih mas te\v zkih kvarkov $1/m_Q$. Ob tem smo ozna\v cili $\Gamma=\gamma_{\mu}(1-\gamma_5)$ in $w= v \cdot v'$. Simetrija te\v zkih kvarkov zapoveduje enakost $\xi(1)=\widetilde \xi(1) =1$, ki je imuna na vsakr\v sne kiralne korekcije. Po drugi strani pa vrednosti $\tau_{1/2}(w)$ niso tako omejene.

\par

\begin{figure}[!t]
\begin{center}
\epsfxsize5cm\epsffile{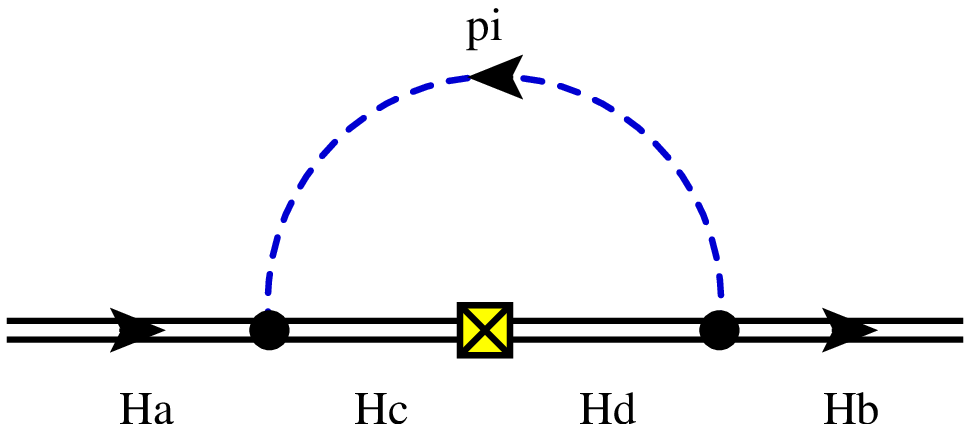}
\end{center}
\caption{\small\it\label{fig:diagram_xi}Diagram zan\v cnega popravka \v sibkega vozli\v s\v ca.}
\end{figure}

\begin{figure}
\begin{center}
\hspace*{-0.4cm}\scalebox{1}{\includegraphics{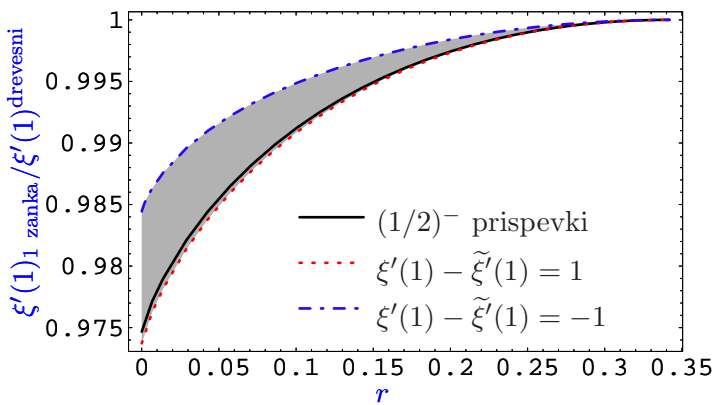}}
\end{center}
\caption{\small\it\label{fig:plot_1} Kiralna ekstrapolacija naklona funkcije IW
pri $w=1$ ($\xi'(1)$). Prispevki stanj negativne parnosti (\v crna crta) in domet mo\v znih prispevkov stanj pozitivne parnosti, kadar razliko naklonov $\xi(1)$ in $\tilde \xi(1)$ variiramo med $1$ (rde\v ca prekinjena \v crta)
in $-1$ (modra pik\v casto-prekinjena \v crta).}
\end{figure}

\begin{figure}
\begin{center}
\hspace*{-0.4cm}\scalebox{1}{\includegraphics{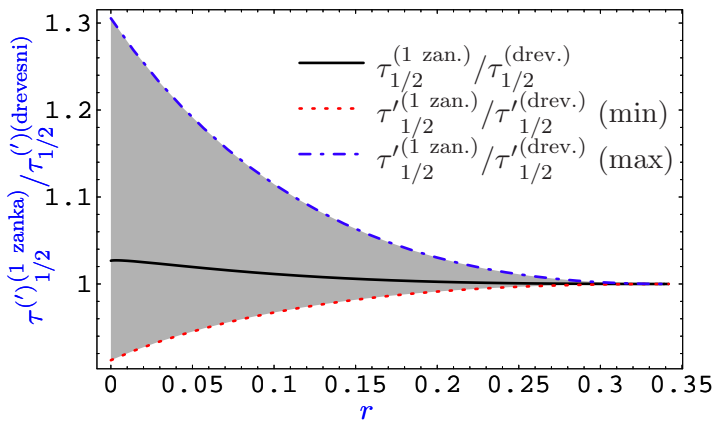}}
\end{center}
\caption{\small\it\label{fig:plot_2} Kiralna ekstrapolacija naklona funkcije $\tau_{1/2}$
in njenega naklona pri $w=1$. Ekstrapolacija $\tau_{1/2}(1)$ skupaj s  $1/\Delta_{SH}$ prispevki (\v crna neprekinjena \v crta), in domet mo\v znih prispevkov k njenemu naklonu -- $\tau_{1/2}'(1)$ -- (sivo obmo\v cje) kadar variiramo razliko naklonov $\xi'(1)$, $\tilde \xi'(1)$ in $\tau_{1/2}'(1)$ med $1$(rde\v ca prekinjena \v crta) in $-1$ (modra pik\v casto-prekinjena \v crta).}
\end{figure}

Kiralne zan\v cne popravke k funkcijam Isgur-Wise $\xi(\omega)$, $\widetilde \xi (\omega)$ in $\tau_{1/2}(\omega)$ izra\v cunamo na podlagi Feynmanovih diagramov oblike na sliki~\ref{fig:diagram_xi}.
\psfrag{pi}[bc]{$\Red{\pi^i(q)}$}
\psfrag{Ha}[cc]{$\Red{H_a(v)}$}
\psfrag{Hb}[cc]{$\Red{H_b(v')}$}
\psfrag{Hc}[cc]{$\Red{H_c(v)}$}
\psfrag{Hd}[cc]{$\Red{H_c(v')}$}
Te\v zki mezoni v za\v cetnem in kon\v cnem stanju si lahko namre\v c izmenjajo en psevdo-Goldstonov bozon, medtem ko znotraj zanke te\v ce \v se par te\v zko-lahkih mezonov pozitivne ali negativne parnosti. Prispevke vseh tak\v snih konfiguracij apliciramo na kiralne ekstrapolacije, ki jih uporabljajo \v studije QCD na mre\v zi, da prestavijo mase lahkih mezonov iz velikih vrednosti v simulacijah v bli\v zino kiralne limite~\cite{Abada:2003un, McNeile:2004rf}. \v Ze v prej\v snjih poglavjih smo opozorili na probleme s kiralno limito amplitud, ki vsebujejo masno re\v zo med te\v zko-lahkimi mezoni obeh parnosti $\Delta_{SH}$. Spet uporabimo razvoj po $1/\Delta_{SH}$, s katerim umirimo logaritemske popravke majhnih mas pionov. Kot smo \v ze razlo\v zili, tak\v sen razvoj dobro deluje na teoriji $SU(2)$, v kateri kaoni in ete, katerih mase bi konkurirale $\Delta_{SH}$, ne nastopajo v zankah. Zato zapi\v simo le izraze zan\v cno popravljenih funkcij Isgur-Wise za zunanja stanja te\v zko-lahkih mezonov brez \v cudnosti v teoriji $SU(2)$:
\begin{eqnarray}
\xi_{aa}(w) &=&  \xi(w) \Bigg\{ 1 + \frac{3}{32\pi^2 f^2} m^2_{\pi} \log \frac{m^2_{\pi}}{\mu^2} \Bigg[ g^2 2 (r(w)-1) \nonumber\\
&&\hskip1cm - h^2  \frac{m^2_{\pi}}{4\Delta^2} \left(1-w\frac{\widetilde \xi(w)}{\xi(w)}\right) - h g \frac{m^2_{\pi}}{\Delta^2} w(w-1)\frac{\tau_{1/2}(w)}{\xi(w)}\Bigg] \Bigg\},
\end{eqnarray}
in
\begin{eqnarray}
\tau_{1/2 aa}(w) &=&  \tau_{1/2}(w) \Bigg\{ 1 + \frac{3}{32\pi^2 f^2} m^2_{\pi} \log \frac{m^2_{\pi}}{\mu^2} \Bigg[ - g\widetilde g(2r(w)-1) - \frac{3}{2} (g^2+\widetilde g^2) \nonumber\\
&&\hskip1cm + h^2  \frac{m^2_{\pi}}{4\Delta^2} \left(w-1\right) - h g \frac{m^2_{\pi}}{2\Delta^2} \frac{\xi(w)}{\tau_{1/2}(w)} w(1+w) + h\widetilde g \frac{m^2_{\pi}}{2\Delta^2} \frac{\widetilde \xi(w)}{\tau_{1/2}(w)} w(1+w) \Bigg] \Bigg\}.\nonumber\\
\end{eqnarray}
Sedaj nari\v semo kiralno ekstrapolacijo funkcij Isgur-Wise pod skalo $\Delta_{SH}$, na kateri funkcije tudi normiramo (sliki~\ref{fig:plot_1} in~\ref{fig:plot_2}).
\psfrag{xk1}[bc]{\Blue{$r$}}
\psfrag{xm1}[tc][tc][1][90]{\Blue{$\xi'(1)_{\mathrm{1~zanka}}/\xi'(1)^{\mathrm{drevesni}}$}}
\psfrag{s1}[cl]{$(1/2)^-$ prispevki}
\psfrag{s3}[cl]{$\xi'(1)-\widetilde \xi'(1)=1$}
\psfrag{s4}[cl]{$\xi'(1)-\widetilde \xi'(1)=-1$}
\psfrag{xk1}[bc]{\Blue{$r$}}
\psfrag{xm1}[tc][tc][1][90]{\Blue{${\tau^{(')}}^{\mathrm{(1~zanka)}}_{1/2}/\tau^{(')\mathrm{(drevesni)}}_{1/2}$}}
\psfrag{s1}[cl]{$\tau^{\mathrm{(1~zan.)}}_{1/2}/\tau^{\mathrm{(drev.)}}_{1/2}$}
\psfrag{s3}[cl]{${\tau'}^{\mathrm{(1~zan.)}}_{1/2}/{\tau'}^{\mathrm{(drev.)}}_{1/2}$ (min)}
\psfrag{s4}[cl]{${\tau'}^{\mathrm{(1~zan.)}}_{1/2}/{\tau'}^{\mathrm{(drev.)}}_{1/2}$ (max)}
Trenutno ne poznamo zanesljivih ocen velikosti $\widetilde \xi'(1)$ in $\tau_{1/2}'(1)$, ki nastopata v formulah za kiralno ekstrapolacijo, ko upo\v stevamo prispevke stanj pozitivne parnosti. Zato njune mo\v zne prispevke ocenimo tako, da njuni vrednosti variiramo glede na vrednost $\xi'(1)$ med $1$ in $-1$. Opazimo, da so prispevki stanj pozitivne parnosti h kiralni ekstrapolaciji funkcije Isgur-Wise $\xi'(1)$ pod skalo $\Delta_{SH}$ majhni (okrog enega odstotka po na\v si oceni). Podobno splo\v sno obna\v sanje lahko pripi\v semo funkciji $\widetilde \xi'(1)$ ob zamenjavi $g \leftrightarrow \widetilde g$, $\Delta_{SH}
\leftrightarrow -\Delta_{SH}$ in $\xi'(1) \leftrightarrow \widetilde \xi'(1)$. Tudi kiralna ekstrapolacija vrednosti $\tau_{1/2}(1)$ (skupaj z $1/\Delta_{SH}$ popravki) deluje zelo polo\v zno in nakazuje, da je linearna ekstrapolacija v tem primeru lahko dober pribli\v zek. Po drugi strani pa so potencialni prispevki h kiralni ekstrapolaciji odvoda $\tau'_{1/2}(1)$ po trenutnih grobih ocenah lahko precej\v snji, do $30\%$.

\par

Na\v si rezultati so \v se posebej pomembni za izlu\v s\v cenje oblikovnih funkcij s pomo\v cjo QCD na mre\v zi. Trenutne napake na parameter $V_{cb}$ v ekskluzivnih kanalih so \v ze reda le nekaj odstotkov. To zahteva zelo natan\v cen nadzor nad teoreti\v cnimi napakami, ki lahko vplivajo na njegovo dolo\v citev. Razumevanje kiralnih popravkov je bistveno za zagotavljanje verodostojnosti izlu\v s\v cenja oblikovnih funkcij in ocene napak na mre\v zi. Na\v se ocene vodilnih $1/\Delta_{SH}$ popravkov postavljajo mejo na natan\v cnost tak\v snih ekstrapolacij.

\section{Me\v sanje te\v zkih nevtralnih mezonov}

Oscilacije v sistemih mezonov $B^0_{d,s}-\overline B^0_{d,s}$ posredujejo nevtralni tokovi, ki spreminjajo okus. Znotraj SM so prepovedane na drevesnem nivoju, zato nam njihove meritve omogo\v cajo dostop do delcev znotraj relevantnih zan\v cnih diagramov. Dandanes se natan\v cno merjene vrednosti $\Delta m_{B_d}=0.509(5)(3)\ {\rm ps}^{-1}$~\cite{Barberio:2006bi}, in $\Delta m_{B_s}=17.31(^{33}_{17})(7)\ {\rm ps}^{-1}$~\cite{Abulencia:2006mq} uporabljajo za omejitev oblike unitarnostnega trikotnika CKM in tako dolo\v cajo koli\v cino kr\v sitev CP znotraj SM~\cite{Charles:2004jd,Bona:2006ah}. Cilj nam ote\v zujejo teoreti\v cne napake v izra\v cunih vrednosti razpadnih konstant $f_{B_{s,d}}$ in parametrov ``vre\v ce'' (ang. bag parameters) $B_{B_{s,d}}$. Te koli\v cine lahko v principu izra\v cunamo na mre\v zi. Veliko oviro pri tem pa predstavlja majhna masa kvarka $d$, ki je v simulacijah ne moremo dose\v ci neposredno, temve\v c le preko kiralne ekstrapolacije. Oglejmo si torej vplive te\v zko-lahkih mezonov pozitivne parnosti na izra\v cun kiralnih popravkov razpadnih konstant in parametrov vre\v ce, ki nastopajo v \v studijah prispevkov SM in supersimetri\v cnega SM k me\v salnim amplitudam $B^0_{d,s}-\overline B^0_{d,s}$.

\par

Prispevki supersimetri\v cnega SM k me\v salni amplitudi  $B^0_{d,s}-\overline B^0_{d,s}$ se ponavadi obravnava v tako imenovani supersimetri\v cni bazi $\Delta B=2$ operatorjev~\cite{Gabbiani:1996hi}:
\begin{eqnarray}
 \label{eq:baseS}
{\phantom{{l}}}\raisebox{-.16cm}{\phantom{{j}}}
O_1 &=& \ \bar b^i \gamma_\mu (1- \gamma_{5} )  q^i \,
 \bar b^j  \gamma^\mu (1- \gamma_{5} ) q^j \,  ,
  \nonumber \\
{\phantom{{l}}}\raisebox{-.16cm}{\phantom{{j}}}
O_2&=& \ \bar b^i  (1- \gamma_{5} ) q^i \,
\bar b^j  (1 - \gamma_{5} )  q^j \, ,  \nonumber  \\
{\phantom{{l}}}\raisebox{-.16cm}{\phantom{{j}}}
O_3&=& \ \bar b^i  (1- \gamma_{5} ) q^j \,
 \bar b^j (1 -  \gamma_{5} ) q^i \, ,  \\
{\phantom{{l}}}\raisebox{-.16cm}{\phantom{{j}}}
O_4 &=& \ \bar b^i  (1- \gamma_{5} ) q^i \,
 \bar b^j   (1+ \gamma_{5} ) q^j \,  ,  \nonumber \\
{\phantom{{l}}}\raisebox{-.16cm}{\phantom{{j}}}
O_5 &=& \ \bar b^i  (1- \gamma_{5} ) q^j \,
 \bar b^j   (1+ \gamma_{5} ) q^i \,  ,  \nonumber
 \end{eqnarray}
kjer sta $i$ in $j$ barvna indeksa. Znotraj SM k me\v salni amplitudi pomembno prispeva le operator $O_1$. Matri\v cni elementi  gornjih operatorjev so ponavadi parametrizirani s pomo\v cjo parametrov vre\v ce $B_{{1}{\mathrm -}{5}}$, ki predstavljajo merilo odstopanja od pribli\v zka VSA
\begin{eqnarray}
{\langle \overline B^0_a\vert O_{{1}{\mathrm -}{5}}(\nu)\vert B^0_a\rangle
\over
\langle \overline B^0_a\vert O_{{1}{\mathrm -}{5}}(\nu)\vert B^0_a\rangle_{\rm VSA} } = B_{{1}{\mathrm -}{5}}(\nu)\,,
\end{eqnarray}
kjer smo z $\nu$ ozna\v cili skalo renormalizacije, pri kateri lo\v cimo nizko od visokoenergijskih prispevkov k amplitudam.

\par

Za opis nizkoenergijskega obna\v sanja matri\v cnih elementov operatorjev~(\ref{eq:baseS}) uporabimo HM$\chi$PT. Preden se podamo v podrobnosti naj omenimo, da obstaja v eksaktni limiti stati\v cnih te\v zkih kvarkov (v tej limiti bomo operatorje ozna\v cevali s tildo) zveza med operatorji~(\ref{eq:baseS}), in sicer $\langle  \overline B^0_a\vert  \widetilde O_3 + \widetilde O_2 +\frac{1}{2} \widetilde O_1 \vert   B^0_a \rangle = 0$, ki nam omogo\v ca, da iz obravnave izlo\v cimo enega izmed njih. V isti limiti definiramo tudi razpadne konstante te\v zko-lahkih mezonov $\hat f$ namre\v c kot
\begin{eqnarray}
\lim_{m_b\to \infty}{  \langle 0 \vert A_\mu \vert B^0_a(p) \rangle_{\rm QCD} \over \sqrt{2m_B}} =\lim_{m_b\to \infty}{ \langle 0 \vert P \vert B^0_a(p) \rangle_{\rm QCD} \over \sqrt{2m_B}}  =    \langle 0 \vert \widetilde A_0 \vert B^0_a(v) \rangle_{\rm HQET} = i \hat f_a v_\mu\,,
\end{eqnarray}
kjer smo z $A_\mu = \bar b \gamma_\mu\gamma_5 q$ in $P = \bar b \gamma_5 q$ ozna\v cili aksialni in psevdoskalarni tok. Nazadnje upo\v stevamo \v se izsledke \v studije kiralnih popravkov v kaonski fiziki~\cite{Becirevic:2004qd}, po katerih se operatorja $O_4$ in $O_5$ razlikujeta le po barvnih indeksih -- lokalni izmenjavi gluona, ki pa ne more vplivati na nizkoenergijske lastnosti amplitud. Ta dva operatorja morata torej utrpeti identi\v cne kiralne popravke. Tako nam za analizo kiralnih popravkov ostanejo le trije matri\v cni elementi operatorjev: $\widetilde O_1$, $\widetilde O_2$ in $\widetilde O_4$, ki jih s polji HM$\chi$PT zapi\v semo kot
\begin{eqnarray}\label{eq:9}
\widetilde O_{1}&= & \sum_X \beta_{1 X} {\rm Tr}\left[ \left(\xi^{\dagger} H\right)_a \gamma_{\mu }(1-\gamma_5) X\right]
 {\rm Tr}\left[ \left(\xi^{\dagger} H\right)_a \gamma^{\mu }(1-\gamma_5) X \right] \nonumber \\
&&\hspace*{5mm}  +
 \beta_{1 X}^\prime
 {\rm Tr}\left[ \left(\xi^{\dagger} H\right)_a \gamma_{\mu }(1-\gamma_5) X  \right]
 {\rm Tr}\left[ \left(\xi^{\dagger} S\right)_a  \gamma^{\mu }(1-\gamma_5) X  \right]    \nonumber \\
&&\hspace*{5mm}+  \beta_{1 X}^{\prime\prime}  {\rm Tr}\left[ \left(\xi^{\dagger} S\right)_a \gamma_{\mu }(1-\gamma_5) X \right]
 {\rm Tr}\left[ \left(\xi^{\dagger} S\right)_a  \gamma^{\mu }(1-\gamma_5) X\right] \,,\nonumber \\
\widetilde O_{2}&= & \sum_X \beta_{2 X} {\rm Tr}\left[ \left(\xi^{\dagger}  H\right)_a  (1-\gamma_5) X\right]
 {\rm Tr}\left[ \left(\xi^{\dagger} H\right)_a  (1-\gamma_5) X \right] \nonumber \\
&&\hspace*{5mm}  +
 \beta_{2 X}^\prime
 {\rm Tr}\left[ \left(\xi^{\dagger} H\right)_a  (1-\gamma_5) X  \right]
 {\rm Tr}\left[ \left(\xi^{\dagger} S\right)_a   (1-\gamma_5) X  \right]    \nonumber \\
&&\hspace*{5mm}+  \beta_{2 X}^{\prime\prime}  {\rm Tr}\left[ \left(\xi^{\dagger}  S\right)_a  (1-\gamma_5) X \right]
 {\rm Tr}\left[ \left(\xi^{\dagger} S\right)_a  (1-\gamma_5) X\right] \,, \nonumber \\
\widetilde O_{4}
&= & \sum_X \beta_{4 X} {\rm Tr}\left[ \left(\xi^{\dagger} H\right)_a  (1-\gamma_5) X\right]
 {\rm Tr}\left[ \left(\xi H\right)_a  (1+\gamma_5) X \right] \nonumber \\
&&\hspace*{5mm}  +
 \beta_{4 X}^\prime
 {\rm Tr}\left[ \left(\xi^{\dagger} H\right)_a  (1-\gamma_5) X  \right]
 {\rm Tr}\left[ \left(\xi S\right)_a   (1+\gamma_5) X  \right]   \nonumber \\
&&\hspace*{5mm}  +
\overline \beta_{4 X}^\prime
 {\rm Tr}\left[ \left(\xi^{\dagger}  S\right)_a  (1-\gamma_5) X  \right]
 {\rm Tr}\left[ \left(\xi H\right)_a   (1+\gamma_5) X  \right]   \nonumber \\
&&\hspace*{5mm}+  \beta_{4 X}^{\prime \prime} {\rm Tr}\left[ \left(\xi^{\dagger}  S\right)_a  (1-\gamma_5) X\right]
 {\rm Tr}\left[ \left(\xi S\right)_a  (1+\gamma_5) X \right] \,,
\end{eqnarray}
kjer smo ozna\v cili $X\in\{ 1, \gamma_5, \gamma_{\nu},
\gamma_{\nu}\gamma_5, \sigma_{\nu\rho}\}$.

\par

\begin{figure}
\vspace*{-0.3cm}
\begin{center}
\begin{tabular}{@{\hspace{-0.25cm}}cc}
\epsfxsize4cm\epsffile{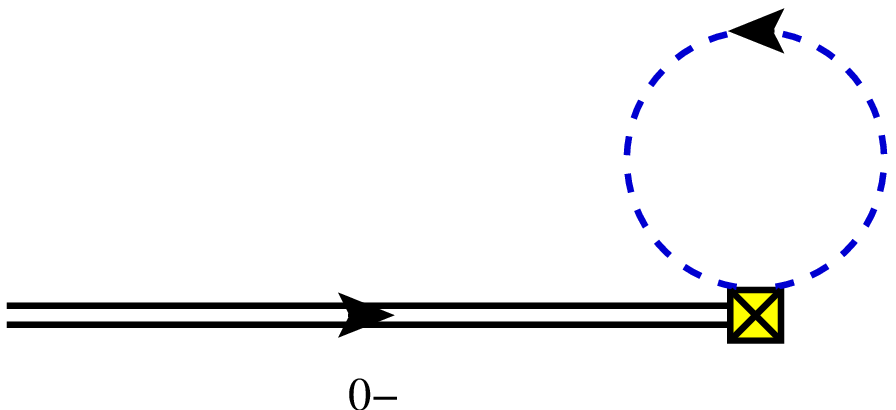}    &
\epsfxsize4cm\epsffile{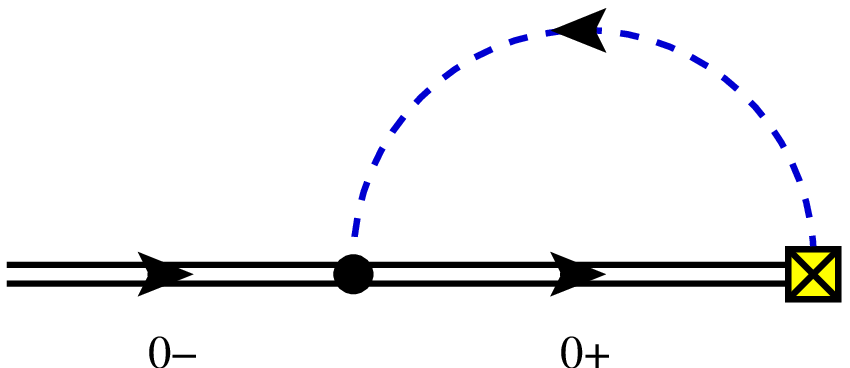}    \\
\end{tabular}
\caption{\small\it\label{fig:0}{Diagrama, ki prispevata neni\v celne kiralne popravke k psevdoskalarni razpadni konstanti te\v zko-lahkih mezonov.}}
\end{center}
\end{figure}

Najprej se osredoto\v cimo na kiralne popravke k (psevdoskalarnim) razpadnim konstantam te\v zkih mezonov. Prispevke dobimo iz zan\v cnih diagramov na sliki~\ref{fig:0}.
\psfrag{0-}[cc]{$\color{red}  H_a(v)$}                                              \psfrag{1-}[cc]{$\color{red}  H_b(v)$}                                              \psfrag{0+}[cc]{$\color{red}  H_b(v)$}
Diagram na desni prispeva le v teoriji s te\v zko-lahkimi mezoni obeh parnosti, saj lahko znotraj zanke te\v cejo le (psevdo)skalarni mezoni. Izra\v cuni popravkov na podlagi teh diagramov ponovno poka\v zejo, da prisotnost stanj pozitivne parnosti ne prizadane pionskih popravkov k razpadni konstanti, se pa popravki teh stanj kosajo s kaonskimi in eta popravki. Relavantni kiralni logaritemski popravki tako ponovno prihajajo iz teorije s simetrijo $SU(2)_L\otimes SU(2)_R \to SU(2)_V$ in jih lahko zapi\v semo kot
\begin{eqnarray}\label{eq:fB-correct}
\hat f_{q} = \alpha \left[ 1-  {1+3g^2\over 2 (4\pi f)^2} {3\over 2} m_\pi^2 \log{m_\pi^2\over \mu^2} + c_f(\mu) m_\pi^2 \right]\,,
\end{eqnarray}
kjer smo z $c_f$ ozna\v cili relevantne kontra\v clene. Hkrati lahko enostavno preverimo, da so kiralni popravki k skalarni razpadni konstanti te\v zko-lahkih mezonov identi\v cni, z zamenjavo sklopitev $g$ in $\widetilde g$
\begin{eqnarray}\label{eq:su2f}
\widetilde f_{q} = \alpha' \left[ 1-  {1+3\widetilde g^2\over 2 (4\pi f)^2} {3\over 2} m_\pi^2
\log{ m_\pi^2\over \mu^2} + \widetilde  c_f(\mu) m_\pi^2 \right]\,.
\end{eqnarray}
Kot smo \v ze pokazali v prej\v snjih razdelkih, velja $\widetilde g^2/g^2 \ll 1$, zato pri\v cakujemo da bodo odstopanja od linearne ekstrapolacije za $\widetilde f_{q}$ manj\v sa kot za $\hat f_{q}$.

\par

\begin{figure}
\vspace*{-0.3cm}
\begin{center}
\begin{tabular}{@{\hspace{-0.25cm}}c}
\epsfxsize12.cm\epsffile{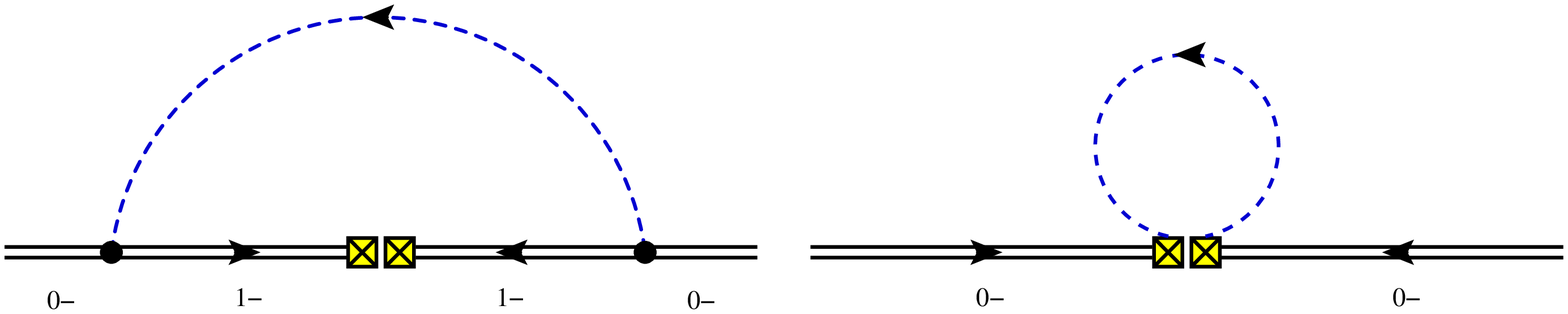}    \\
\epsfxsize12cm\epsffile{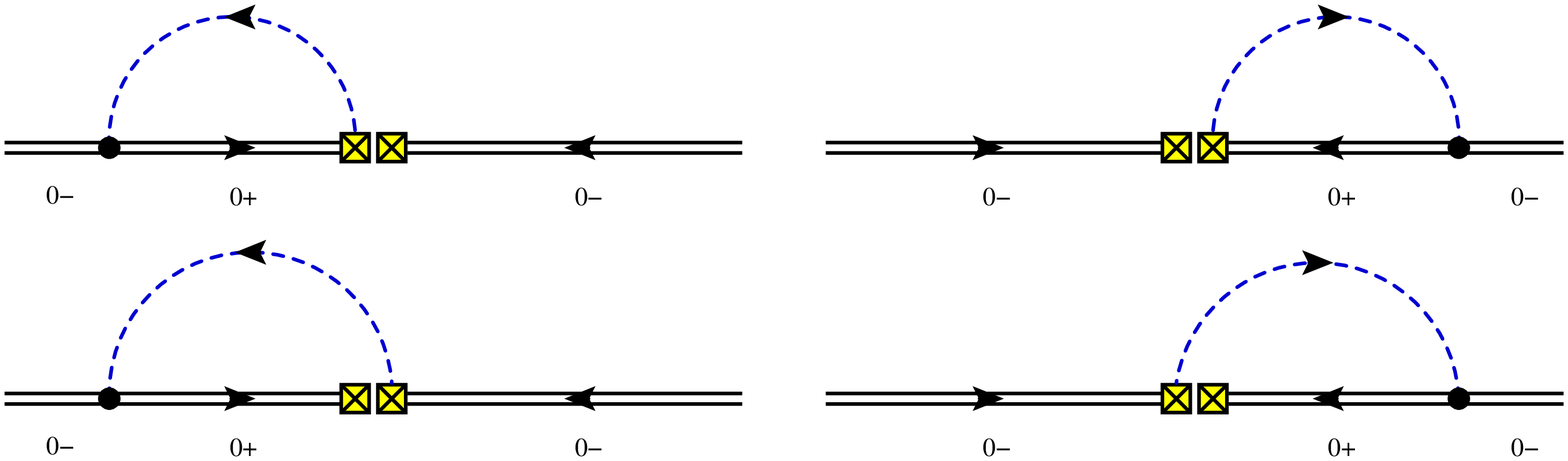}
\end{tabular}
\caption{\small\it\label{fig:4}{Diagrami, ki nastopajo v izra\v cunu kiralnih popravkov k operatorjem
$\langle \widetilde {\cal O}_{1,2,4}\rangle$.}}
\end{center}
\end{figure}

Kon\v cno se posvetimo me\v salnim amplitudam, ki dobijo kiralne popravke iz diagramov na sliki~\ref{fig:4}.
\psfrag{0-}[cc]{$\color{red}  H_a(v)$}                                              \psfrag{1-}[cc]{$\color{red}  H_b(v)$}                                              \psfrag{0+}[cc]{$\color{red}  H_b(v)$}
\psfrag{1-0+}[cc]{{$ \color{red}{\quad H_b(v)}$}}
\psfrag{0-}[cc]{$\color{red}  H_a(v)$}                                              \psfrag{1-}[cc]{$\color{red}  H_b(v)$}                                              \psfrag{0+}[cc]{$\color{red}  H_b(v)$}
Ponovno diagrami v spodnjih dveh vrsticah prispevajo le ob upo\v stevanju te\v zko-lahkih stanj obeh parnosti. Osredoto\v cimo se na obmo\v cje $m_\pi \ll \Delta_{SH}$ in postopamo podobno kot doslej. Zan\v cne integrale namre\v c razvijemo po obratnih potencah masne re\v ze med te\v zko-lahkimi mezoni obeh parnosti $1/\Delta_{SH}$ in ponovno poka\v zemo, da ostanejo vodilni pionski logaritemski popravki neprizadeti in kiralna ektrapolacija znotraj $SU(2)$ teorije dobro definirana. Zapi\v semo
\begin{subequations}
\begin{eqnarray}\label{eq:B1-correct}
B_{1q}  &=&B_1^{\rm drevesni}
 \left[
1 - {1-3 g^2  \over 2 (4\pi f)^2} m_\pi^2\log{m_\pi^2\over \mu^2} + c_{{B}_1}(\mu)m_\pi^2\right]\,,\\
\label{eq:B24-correct}
B_{2,4q}  & = &  B_{2,4}^{\rm drevesni}  \left[
 1 + {3 g^2 Y \mp 1 \over 2 (4\pi f)^2} m_\pi^2\log{m_\pi^2\over \mu^2}
 + c_{{B}_{2,4}}(\mu)m_\pi^2\right]\,,
\end{eqnarray}
\end{subequations}
kjer smo ozna\v cili $Y=(\widehat \beta^*_{2,4}/\widehat \beta_{2,4})$,
$\widehat \beta^\ast_2 = \beta_{2\gamma_{\nu}} +
\beta_{2\gamma_{\nu}\gamma_5} - 4 \beta_{2\sigma_{\nu\rho}}$ in $ \widehat
\beta^\ast_4 = - \beta_{4\gamma_{\nu}} + \beta_{4\gamma_{\nu}\gamma_5}$.

\par

Morda je na tem mestu dobro povdariti, da je diskusija tega razdelka pomembna predvsem tudi za fenomenolo\v ske pristope, ki ozna\v cujejo logaritemske popravke kaonov in et kot napovedi in hkrati dolo\v cijo relevantne kontra\v clene iz limite velikega \v stevila barv ali kak\v snega drugega modela. Pokazali smo namre\v c, da so prispevki blizu le\v ze\v cih skalarnih resonanc po velikosti konkuren\v cni prispevkom kaonov in et, ter jih zato v tak\v snih diskusijah ne moremo zanemariti ali iz njih izlo\v citi. Vendar pa je dejstvo, da bli\v znja skalarna stanja ne pokvarijo poglavitnih pionskih logaritemskih prispevkov, zelo dobrodo\v slo za \v studije QCD na mre\v zi, saj lahko te \v se vedno uporabljajo formule HM$\chi$PT za ektrapolacijo njihovih rezultatov. Tak\v sni postopki pa morajo biti omejeni na teorijo $SU(2)$ in pod skalo $\Delta_{SH}$.

\par

Izra\v cunali smo torej vodilne kiralne popravke k celotni bazi supersimetri\v cnih operatorjev, ki prispevajo k me\v sanju te\v zkih nevtralnih mezonov. Pokazali smo, da bli\v znje skalarne resonance ne vplivajo na vodilne pionske logaritemske prispevke, ki zato ostajajo zanesljivo vodilo pri kiralnih ektrapolacijah rezultatov simulacij na mre\v zi. Hkrati smo preverili, da se vodilni kiralni popravki k razpadnim konstantam te\v zko-lahkih mezonov sode in lihe parnosti ujemajo ob zamenjavi efektivnih sklopitvenih konstant $g \to \widetilde g$.

\section[Redki hadronski razpadi te\v zkih mezonov]{Redki hadronski razpadi $\Delta S=2$ in $\Delta S = -1$ mezonov $B_c$}

Redki razpadi mezonov $B$ veljajo za eno najobetavnej\v sih podro\v cij za odkritje nove fizike izven SM~\cite{Grossman:2003qi,Isidori:2004rd, Buras:2004sc}. Pri\v cakovati je namre\v c, da bodo novi virtualni delci vplivali na te procese. To \v se posebej velja za razpade, ki potekajo preko nevtralnih tokov, ki spreminjajo okus, saj ti znotraj SM potekajo le preko zank. Ekstremni primer ka\v snega pristopa je iskanje razpadov, ki so znotraj SM izredno redki, in \v ze samo opa\v zanje katerih bi pomenilo jasen signal nove fizike. Huitu, Lu, Singer in Zhang so pred leti predlagali razpade $b \to s
s \bar d$ in $b \to d d \bar s$ (v katerih se \v cudnost spremeni za $\Delta S= -1$ oz. $\Delta S= 2$)~\cite{Huitu:1998vn, Huitu:1998pa} kot prototipne v tak\v snem iskanju. Njihov predlog temelji na dejstvu, da so tak\v sni razpadi znotraj SM izredno redki, saj potekajo preko izmenjave gornjih kvarkov in bozonov $W$ znotraj \v skatlastih zank in imajo posledi\v cno razvejitvena razmerja reda $10^{-11}$ do $10^{-13}$.

\par

Prihajajo\v ci pospe\v sevalnik LHC bo med drugim izredno produktivna tovarna mezonov $B_c$, Pri\v cakovanja za njihovo proizvodnjo se namre\v c gibljejo okoli $5 \times 10^{10}$ dogodkov na leto ob luminoznosti $10^{34}\mathrm{~cm}^{-2} s^{-1}$~\cite{Gouz:2002kk}. \v Cetudi bi bila dejanska \v stevilka nekaj redov velikosti manj\v sa, bo omogo\v cala \v studij redkih razpadov mezonov $B_c$, ki bo morda osvetlil prispevke fizike izven SM. Posvetili se bomo torej izra\v cunom redkih prehodov $b \to s
s \bar d$ in $b \to d d \bar s$ v dvo- in trodel\v cnih razpadnih kanalih mezona $B_c$ znotraj SM ter nekaj najpopularnej\v sih okvirov nove fizike. Na podlagi znanih eksperimentalnih omejitev na prispevke obravnavanih modelov bomo podali napovedi za razvejitvena razmerja ter identificirali najperspektivnej\v se kanale za iskanje signalov nove fizike.

\par

Efektivni \v sibki Hamiltonov operator, ki zaobjema tudi procese $b \to s
s \bar d$ in $b \to d d \bar s$ zapi\v semo kot
\begin{equation}
\mathcal H_{\mathrm{eff.}} = \sum_{n=1}^5 \left[ C^s_n \mathcal O^s_n + \widetilde C^s_n \widetilde {\mathcal O}^s_n + C^d_n \mathcal O^d_n + \widetilde C^d_n \widetilde {\mathcal O}^d_n\right],
\end{equation}
kjer $C^q_i$ in $\widetilde C^q_i$ ozna\v cujeta efektivne Wilsonove sklopitve, s katerimi pomno\v zimo celotno bazo operatorjev, ki prispevajo k procesom $b\to d d \bar s$ (za $q=s$) in $b\to s s \bar d$ (za $q=d$). Izberemo
\begin{equation}
\begin{array}{c}
  \mathcal O^s_1 = \bar d^i_L \gamma^{\mu} b^i_L \bar d^j_R \gamma_{\mu} s^j_R,\quad
  \mathcal O^s_2 = \bar d^i_L \gamma^{\mu} b^j_L \bar d^j_R \gamma_{\mu} s^i_R,\quad
  \mathcal O^s_3 = \bar d^i_L \gamma^{\mu} b^i_L \bar d^j_L \gamma_{\mu} s^j_L, \\
  \\
  \mathcal O^s_4 = \bar d^i_R b^i_L \bar d^j_L s^j_R,\quad
  \mathcal O^s_5 = \bar d^i_R b^j_L \bar d^j_L s^i_R,

\end{array}
\end{equation}
skupaj z dodatnimi operatorji ${\mathcal {\widetilde O}^s_{1,2,3,4,5}}$, ki jih dobimo s primernimi kiralnimi transformacijami gornjih ($L\leftrightarrow R$), ter vse skupaj \v se z obrnjenimi okusi kvarkov $s$ in $d$.

\par

Znotraj SM k procesu $b \to dd \bar s$ ($b \to ss \bar d$) prispevata le operatorja $\mathcal O^{s(d)}_3$. Glavne prispevke k Wilsonovim sklopitvam dajo top in \v carobni kvarki ter bozoni $W$ znotraj \v skatlastih zank
\begin{subequations}
\begin{eqnarray}
    C^{d,SM}_3 &=& \frac{G_F^2}{4\pi^2} m_W^2 V_{tb} V^*_{ts} \Bigg[ V_{td} V_{ts}^* f\left( \frac{m_W^2}{m_t^2} \right)  + V_{cd} V_{cs}^* \frac{m_c^2}{m_W^2} g\left( \frac{m_W^2}{m_t^2},\frac{m_c^2}{m_W^2} \right) \Bigg], \\*
   C_3^{s,SM} &=& \frac{G_F^2}{4 \pi^2} m_W^2 V_{tb} V_{td}^* \Bigg[ V_{ts} V_{td}^*
  f\left(\frac{m_W^2}{m_t^2}\right) + V_{cs} V_{cd}^* \frac{m_c^2}{m_W^2} g\left(\frac{m_W^2}{m_t^2},\frac{m_c^2}{m_W^2}\right)\Bigg],
\end{eqnarray}
\end{subequations}
kjer je
\begin{subequations}
\begin{eqnarray}
f(x)&=&\frac{1-11x+4x^2}{4x(1-x)^2}
-\frac{3}{2(1-x)^3}{\rm ln}x,\\
~g(x,y)&=& \frac{4x-1}{4(1-x)} +\frac{8x-4x^2-1}{4(1-x)^2}\ln x -\ln y .
\end{eqnarray}
\end{subequations}
Z uporabo numeri\v cnih vrednosti matri\v cnih elementov CKM iz PDG~\cite{Eidelman:2004wy} lahko postavimo mejo $\left|C_3^{s,SM}\right| \leq 3\E{-13}\e{GeV}^{-2}$ in $\left|C^{d,SM}_{3}\right|\leq 4\times10^{-12} \mathrm{~GeV}^{-2}$.

\par

Obravnavamo tudi prispevke nekaterih modelov fizike izven SM: minimalni supersimetri\v cni SM (MSSM) z in brez parnosti R (RPV) ter model z generi\v cnim dodatnim vektorskim bozonom $Z'$. Znotraj MSSM prispevata, podobno kot v SM, le operatorja  $\mathcal O^{q}_3$, medtem ko sta pripadajo\v ca Wilsonova koeficienta iz izmenjave parov gluinov ($\widetilde g$) in spodnjih skvarkov ($\widetilde d$)~\cite{Gabbiani:1996hi}
\begin{subequations}
\begin{eqnarray}
C_3^{s,MSSM} &=& -\frac{\alpha_S^2 \delta_{21}^* \delta_{13}}{216
    m_{\widetilde{d}}^2} \left[24 x f_6(x) + 66 \widetilde{f}_6 (x)\right],\\
C_3^{d,MSSM} &=& -\frac{\alpha_S^2 \delta^{*}_{12}
  \delta_{23}}{216m_{\widetilde
  d}^2}\left[ 24x f_6 (x) + 66 \widetilde f_6 (x) \right],
\label{eq:Cmssm_slo}
\end{eqnarray}
\end{subequations}
kjer sta
\begin{subequations}
\begin{eqnarray}
f_6(x)=\frac{6(1+3x)\ln x +x^3-9x^2-9x+17}{6(x-1)^5}\; ,  \\
\widetilde{f}_6(x)=\frac{6x(1+x)\ln x -x^3-9x^2+9x+1}{3(x-1)^5}\;,
\end{eqnarray}
\end{subequations}
in smo definirali $x=m_{\widetilde g}^2/m_{\widetilde d}^2$. Z uporabo obstoje\v cih mej na parametre MSSM ($\delta_{ij}$, $m_{\widetilde d}$, $m_{\widetilde g}$) iz \v studij oscilacij mezono $K$, $B$ in $B_s$ ter drugih redkih razpadov mezonov $K$ in $B$ lahko spet omejimo vrednosti $|C_3^{d,MSSM}| \lesssim 5 \times 10^{-12} \mathrm{~GeV}^{-2}$ in $\left|C_3^{s,MSSM}\right| \leq 2 \E{-12}\e{GeV}^{-2}$. V primeru, da znotraj MSSM dopustimo interakcije tipa RPV, dobimo poglavitne prispevke \v ze v drevesnem redu preko izmenjave snevtrinov ($\widetilde \nu$). K efektivnemu Hamiltonovemu operatorju potem prispevajo predvsem operatorji $\mathcal O^q_{4}$ in $\mathcal{\widetilde O}^q_{4}$ z Wilsonovimi sklopitvami
\begin{eqnarray}
  C_4^{s,RPV} = -\sum_{n=1}^3 \frac{\lambda_{n31}'\lambda_{n12}'^*}{m_{\widetilde{\nu}_n}^2}, &&
  \widetilde C_4^{s,RPV} = -\sum_{n=1}^3 \frac{\lambda_{n21}'\lambda_{n13}'^*}{m_{\widetilde{\nu}_n}^2},\nonumber\\
  C_4^{d,RPV} = -\sum_{n=1}^3 \frac{\lambda_{n32}'\lambda_{n21}'^*}{m_{\widetilde{\nu}_n}^2}, &&
  \widetilde C_4^{d,RPV} = -\sum_{n=1}^3 \frac{\lambda_{n12}'\lambda_{n23}'^*}{m_{\widetilde{\nu}_n}^2}.\nonumber\\
  \label{eq:RPVcouplings_slo}
\end{eqnarray}
Obstoje\v ce \v studije ne omejujejo vseh parametrov ($\lambda'_{ijk}$, $m_{\widetilde \nu}$), ki nastopajo v teh procesih, zato bomo omejitve podali iz napovedi za merjene ekskluzivne razpadne kanale.

\par

Mnoge raz\v siritve SM vsebujejo dodatne nevtralne vektorske bozone $Z'$~\cite{Langacker:2000ju,Erler:1999nx}. Ti lahko prispevajo k efektivnemu Hamiltonovemu operatorju v drevesnem redu preko operatorjev $\mathcal O^q_{1,3}$ ter tudi $\mathcal {\widetilde O}^q_{1,3}$. Pripadajo\v ce Wilsonove koeficiente lahko zapi\v semo kot
\begin{equation} \label{eq:zPrimeOperators_slo}
\begin{array}{cc}
C_1^{s,Z'} = -\frac{4G_F y}{\sqrt{2}} B_{12}^{d_L} B_{13}^{d_R}, &
\widetilde C_1^{s,Z'} = -\frac{4G_F y}{\sqrt{2}} B_{12}^{d_R} B_{13}^{d_L}, \\
C_3^{s,Z'} = -\frac{4G_F y}{\sqrt{2}} B_{12}^{d_L} B_{13}^{d_L}, &
\widetilde C_3^{s,Z'} = -\frac{4G_F y}{\sqrt{2}} B_{12}^{d_R} B_{13}^{d_R}, \\
C_1^{d,Z'} = -\frac{4G_F y}{\sqrt{2}} B_{21}^{d_L} B_{23}^{d_R}, &
\widetilde C_1^{d,Z'} = -\frac{4G_F y}{\sqrt{2}} B_{21}^{d_R} B_{23}^{d_L}, \\
C_3^{d,Z'} = -\frac{4G_F y}{\sqrt{2}} B_{21}^{d_L} B_{23}^{d_L}, &
\widetilde C_3^{d,Z'} = -\frac{4G_F y}{\sqrt{2}} B_{21}^{d_R} B_{23}^{d_R},
\end{array}
\end{equation}
kjer je $y = (g_2/g_1)^2 (\rho_1 \sin^2 \theta+ \rho_2 \cos^2 \theta)$
in $\rho_i = m_W^2/m_i^2 \cos^2\theta_W$. Z $g_1$,
$g_2$, $m_1$ in $m_2$ smo ozna\v cili umeritvene sklopitvene konstane in mase
bozonov $Z$ in $Z'$, medtem ko $\theta$ ozna\v cuje me\v salni kot.

\par

V izra\v cunu pogostosti razpadnih kanalov mezona $B_c$, ki potekajo preko kvarkovskih prehodov $b \to dd\bar{s}$ in $b \to ss\bar{d}$ moramo izvrednotiti matri\v cne elemente efektivnih Hamiltonovih operatorjev med hadronskimi stanji. V prvem pribli\v zku uporabimo popolno faktorizacijo oz. VSA, ki v ve\v cini primerov zadovoljivo opi\v se glavne lastnosti razpadnih kanalov, ki jih obravnavamo. Izjeme so kanali, v katerih razpadne amplitude v faktorizacijskem pribli\v zku izginejo. O tak\v snih razpadih na\v sa metoda mol\v ci; potrebni bi bili bolj natan\v cni pristopi. Obravnavamo dvodel\v cne razpadne kanale $B_c^- \to D^{*-}_s \overline K^{*0}$, $D^{*-}_s \overline K^0, D_s^- \overline K^{*0}$ in $D_s^- \overline K^0$, kot tudi trodel\v cne kanale  $B_c^- \to D_s^- K^- \pi^+$, $D^{*-}_s K^- \pi^+$, $D_s^-  D^{*-}_s D^+$, $D_s^-  D_s^- D^{*+}$, $D_s^-  D_s^- D^+$, $D^0 \overline K^0 K^-$ in $D^{*0} \overline K^0 K^-$. Pri tem uporabimo teoreti\v cne uvide iz hadronskih in semileptonskih \v studij te\v zkih mezonov iz prej\v snjih razdelkov, kot tudi iz drugih virov. Predvsem se izka\v ze pomebna vloga vmesnih resonanc pri modeliranju hadronskih oblikovnih funkcij, ki nastopajo v faktorizacijskem pribli\v zku, pomagajo pa nam tudi natan\v cno dolo\v cene vrednosti nekaterih sklopitvenih konstant znotraj HM$\chi$PT.

\par

Na podlagi izra\v cunanih amplitud za izbrane hadronske kanale, najprej omejimo proste parametre modelov RPV in $Z'$, tako da na\v se napovedi primerjamo z obstoje\v cimi eksperimentalnimi mejami na razvejtiveni razmerji razpadov $\mathcal{BR}(B^- \to K^- K^- \pi^+) <$ $2.4 \times 10^{-6}$ in ${\rm BR} (B^- \to \pi^- \pi^- K^+
)< 4.5\E{-6}$, ki so jih izmerili v eksperimentu Belle~\cite{Garmash:2003er}. V primeru modela RPV normaliziramo mase snevtrinov na skupno masno skalo $100~\mathrm{GeV}$ in dobljene meje zapi\v semo kot
\begin{subequations}
\begin{eqnarray}
\label{eq:par-R_slo}
  \left| \sum_{n=1}^3 \left(\frac{100~\mathrm{GeV}}{m_{\widetilde\nu_n}}\right)^2 \left(\lambda_{n31}' \lambda_{n12}'^* + \lambda_{n21}' \lambda_{n13}'^*\right)\right| &<& 9.5\E{-5},\\
  \left| \sum_{n=1}^3 \left(\frac{100~\mathrm{GeV}}{m_{\widetilde\nu_n}}\right)^2 \left(\lambda_{n32}' \lambda_{n21}'^* + \lambda_{n21}' \lambda_{n13}'^*\right)\right| &<& 9.5\E{-5}.
\end{eqnarray}
\end{subequations}
Lahko pa predpostavimo, da poglavitni prispevki nove fizike prihajajo v obliki dodatnih bozonov $Z'$. V tem primeru dobimo meje na njihove sklopitve oblike
\begin{subequations}
  \begin{align}
    y^2 \left| B_{12}^{s_L}\, B_{13}^{s_R}+
      B_{12}^{s_R}\,  B_{13}^{s_L}\right| &<  2.7\E{-4},\\
    y^2 \left| B_{12}^{s_L}\, B_{13}^{s_L}+ B_{12}^{s_R}\,
      B_{13}^{s_R}\right| &< 5.6\E{-4},
\end{align}
\end{subequations}
in
\begin{subequations}
  \begin{align}
    y^2 \left| B_{21}^{d_L}\, B_{23}^{d_R}+
      B_{21}^{d_R}\,  B_{32}^{d_L}\right| &<  2.4\E{-4},\\
    y^2 \left| B_{21}^{d_L}\, B_{32}^{d_L}+ B_{21}^{d_R}\,
      B_{32}^{d_R}\right| &< 5.3\E{-4}.
  \label{eq:par-Z_slo}
\end{align}
\end{subequations}
Meje~(\ref{eq:par-R_slo}-\ref{eq:par-Z_slo}) so zanimive, predvsem ker omejujejo kombinacije parametrov RPV oz. $Z'$, v ortogonalni smeri od obstoje\v cih meritev oscilacij mezonov $K$, $B$, $B_s$, ter drugih redkih procesov. Na podlagi teh omejitev lahko kon\v cno podamo tudi napovedi za razvejtivena razmerja mnogih mo\v znih dvo- in trodel\v cnih razpadnih kanalov mezona $B_c$. Na\v si rezultati so povzeti v tabeli~\ref{tab:1}.
\begin{table}[!t]
\begin{center}
\begin{tabular}{|l|cccc|}
  \hline
  Razpad & SM & MSSM & RPV & $Z'$ \\\hline
  \hline
  $B_c^- \to D^- D^- D_s^+$ & $1\E{-21}$ & $5\E{-20}$ & $7\E{-9}$ & $9\E{-10}$\\
  $B_c^- \to D_s^- D_s^- D^+$ & $4\E{-19}$ & $5\E{-19}$ & $1\E{-8}$ & $1\E{-9}$\\
  $B_c^- \to D^- K^+ \pi^-$ & $2 \times 10^{-16}$ & $5\times 10^{-15}$ & $4 \times 10^{-7}$ & $2\E{-6}$ \\
  $B_c^- \to D_s K^- \pi^{+}$ & $7 \times 10^{-14}$ & $1\times 10^{-13}$ & $8 \times 10^{-7}$ & $3\E{-6}$ \\
  $B_c^- \to \overline D^0 \pi^- K^0$ & $4 \times 10^{-20}$ & $2\times 10^{-18}$ & $2 \times 10^{-8}$ & $1\times 10^{-9}$ \\
  $B_c^- \to \overline D^0 K^- \overline K^0$ & $4 \times 10^{-18}$ & $7\times 10^{-18}$ & $9 \times 10^{-9}$ & $6\times 10^{-10}$ \\
  $B_c^- \to D^- K^0$ & $4\E{-17}$ & $2\E{-15}$ & $4\E{-8}$ & $3\E{-7}$\\
  $B_c^- \to D_s^- \overline K_0$ & $1 \times 10^{-14}$ & $2 \times 10^{-14}$ & $7 \times 10^{-8}$ & $4 \times 10^{-7}$ \\
  $B_c^- \to D^{*-} K^0$ & $4\E{-17}$ & $2\E{-15}$ & $4\E{-8}$ & $3\E{-7}$\\
  $B_c^- \to D_s^{*-} \overline K_0$ & $1 \times 10^{-14}$ & $2\times 10^{-14}$ & $6 \times 10^{-8}$ & $4 \times 10^{-7}$ \\
  $B_c^- \to D^- K^{*0}$ & $8\E{-17}$ & $3\E{-15}$ & $3\E{-9}$ & $5\E{-7}$\\
  $B_c^- \to D_s^- \overline K_0^*$ & $3 \times 10^{-14}$ & $4 \times 10^{-14}$ & $6\E{-9}$ & $9\E{-7}$ \\

  $B_c^- \to D^{*-} K^{*0}$ & $6\E{-18}$ & $3\E{-16}$ & $2\E{-10}$ & $4\E{-8}$\\
  $B_c^- \to D_s^{*-} \overline K_0^*$ & $2 \times 10^{-15}$ & $3\times 10^{-15}$ & $4\E{-10}$ & $5\E{-8}$ \\
  \hline
\end{tabular}
\end{center}
\caption{\small\it Razvejitvena razmerja razpadov $\Delta S= -1$ in $\Delta S= 2$ mezona $B_c^-$ izra\v cunana znotraj modelov SM, MSSM, RPV in $Z'$.  Za dolo\v citev neznanih kombinacij parametrov RPV (\v cetrti stolpec) in $Z'$ (peti stolpec) smo uporabili eksperimentalne gornje meje $BR(B^- \to \pi^- \pi^- K^+)$ $<
  1.8 \times 10^{-6}$ in $BR(B^- \to K^- K^- \pi^+)$ $<
  2.4 \times 10^{-6}$.
}
\label{tab:1}
\end{table}
Napovedi SM in MSSM so zanemerljivo majhne. Na podlagi omejitev na parametre RPV iz razpadnih kanalov $B^- \to \pi^- \pi^- K^+$ in $B^- \to K^- K^- \pi^+$ dobimo v tem modelu najve\v cja razvejtivena razmerja za trodel\v cne razpadne kanale $B_c^- \to D^- K^+ \pi^-$ in
$B_c^- \to D_s^- K^- \pi^+$, ter dvodel\v cne razpadne kanale $B^- \to D^- K^0$,
$B^- \to D_s^- \overline K^0$,
$B_c^- \to D^{*-} K^0$ in
$B^- \to D_s^{*-} \overline K^0$.
Poleg teh, pa znotraj modela $Z'$ dobimo velika razvejitvena razmerja \v se v kanalih $B_c^- \to D^- K^{*0}$ in $B_c^- \to D^{*-} K^{*0}$. Eksperimenti namesto mezonov $K^0$ ali $\overline K^0$, dejansko izmerijo stanja mezonov $K_S$ oz. $K_L$. Posledi\v cno bo v razpadnih kanalih, ki vsebujejo nevtralne psevdoskalarne kaone, zaradi prispevkov pingvinskih diagramov znotraj SM te\v zko zaznati vplive nove fizike~\cite{Grossman:1999av}. V tem pogledu so bolj perspektivni razpadi v nabite kaone oziroma njihova vektorska stanja.

\par

V na\v sem izra\v cunu smo se naslanjali na pribli\v zek naivne faktorizacije, ki je kot prvi pribli\v zek zadostna za opis grobih lastnosti prispevkov nove fizike. Tudi v primeru, da bi morebitni nefaktorizabilni prispevki znatno spremenili vrednosti hadronskih amplitud, je razkorak med napovedmi SM in nove fizike trenutno tolik\v sen, da v vsakem primeru ohranja relevantnost obravnavanih razpadnih kanalov v iskanju nove fizike, in to nemudoma, ko bodo na voljo ve\v cje koli\v cine mezonov $B_c$.

\section{Zaklju\v cki}

Neperturbativna narava QCD je trdovraten problem ra\v cunov v hadronski fiziki. Ena izmed njegovih manifestacij je pojav resonanc v hadronskem spektru. V procesih, kjer so izmenjane gibalne koli\v cine majhne v primerjavi s skalo zlomitve kiralne simetrije $\sim 1~\mathrm{GeV}$, lahko uporabimo pristop efektivnih teorij, ki temelji na pribli\v zni kiralni simetriji lahkih kvarkov ter pribli\v zni simetriji okusov in spina  te\v zkih kvarkov. V tak\v snem okviru lahko sistemati\v cno analiziramo vpliv najni\v zje le\v ze\v cih resonanc v procesih te\v zkih mezonov.

\par

HM$\chi$PT smo uporabili na primeru mo\v cnih, semileptonskih in redkih procesih te\v zkih mezonov. V ogrodje efektivne teorije smo sistemti\v cno vklju\v cili najni\v zje le\v ze\v ce spinske multiplete te\v zkih mezonov pozitivne in negativne parnosti.

\par

V prvem redu kiralnega razvoja smo pokazali, da lahko bli\v znje le\v ze\v ca vzbujena stanja te\v zkih mezonov pomagajo razlo\v ziti nekatere lastnosti semileptonskih oblikovnih funkcij v razpadih te\v zkih v lahke mezone. Namre\v c, z uporabo omejene parametrizacije oblikovnih funkcij, ki temelji na pribli\v znih limitah efektivnih teorij QCD, smo uspeli zasi\v citi prispevke celotnega stolpa vmesnih stanj samo z najbli\v zje le\v ze\v cimi stanji primernih kvantnih \v stevil. Parametrizacijo smo napeli na izr\v cun razpadne \v sirine v kinematskem obmo\v cju majhnih izmenjav gibalne koli\v cine znotraj HM$\chi$PT. Tak\v sen model je uspe\v sno reproduciral ve\v cino oblikovnih funkcij $H\to P$ in $H\to V$ prehodov znotraj trenutnih eksperimentalnih napak in v ujemanju z obstoje\v cimi izra\v cuni QCD na mre\v zi.

\par

V drugih procesih, ki smo jih obravnavali, prispevajo vzbujena stanja te\v zkih mezonov \v sele v drugem redu kiralnega razvoja -- k tako imenovanim kiralnim popravkom. Obravnavali smo mo\v cne razpade te\v zkih mezonov ter izra\v cunali efektivne mo\v cne sklopitvene konstante  med pari te\v zkih mezonov sode ali lihe parnosti ter lahkimi psevdoskalarnimi mezoni v drugem redu kiralnega razvoja. Iz merjenih razpadnih \v sirin $D^*\to D\pi$ in $D'_0\to D \pi$ smo izlu\v s\v cili efektivne sklopitvene konstante v prvem in drugem redu kiralnega razvoja. Vpliv velikega \v stevila novih neznanih parametrov, ki nastopajo v izra\v cunih drugega reda, smo ocenili s pomo\v cjo variacij skale kiralne zlomitve ter otipanjem prostora parametrov ob prilagajanju izra\v cunov na eksperimentalne meritve. Nato smo \v studirali ekstrapolacijo sklopitvenih konstant v limiti, ko gredo mase lahkih psevdoskalarnih mezonov proti ni\v c. Ugotovili smo, da dajo naivni izra\v cuni kiralnih popravkov z upo\v stevanjem vzbujenih te\v zkih stanj slabo definirano kiralno limito. Namesto tega lahko izvedemo razvoj v obratni vrednosti masnih razlik med osnovnimi in vzbujenimi stanji te\v zkih mezonov in tako re\v simo kiralno limito izra\v cunov. Tak\v sen razvoj je zanesljiv za majhne mase lahkih psevdoskalarnih mezonov, manj\v se od masnih razlik med osnovnimi in vzbujenimi stanji te\v zkih mezonov. Potem se prispevki vzbujenih stanj te\v zkih mezonov formalno izra\v zajo kot popravki vi\v sjih redov v kiralnem razvoju teorije brez dinami\v cnih vzbujenih stanj. Ti rezultati so \v se posebej pomembni za \v studije QCD na mre\v zi, ki uporabljajo kiralno ektrapolacijo za dosego fizikalne limite simuliranih mas lahkih kvarkov. Po na\v sih ugotovitvah je relevantna kiralna limita takih ektrapolacij $SU(2)$ izpospinska limita, zanesljivo pa se lahko izvedejo le za pionske mase manj\v se od masnih razlik med osnovnimi in vzbujenimi stanji te\v zkih mezonov. Hkrati lahko ocenimo zanesljivost ektrapolacij tak\v sne simetrije v redu vodilnih logaritmov z uporabo vodilnih prispevkov vzbujenih te\v zkih stanj vi\v sjega reda.

\par

Razklopitev vzbujenih resonanc in njihove poglavitne prispevke smo preverili tudi na primeru semileptonskih oblikovnih funkcij v te\v zko-te\v zkih prehodih med mezoni sode in lihe parnosti, kjer smo izra\v cunali kiralne popravke k funkcijam Isgur-Wise. Za izlu\v s\v cenje matri\v cnega elementa CKM $V_{bc}$ namre\v c poleg izredno natan\v cne dolo\v citve razpadnih \v sirin iz eksperimentov in oblikovnih funkcij iz izra\v cunov QCD na mre\v zi potrebujemo natan\v cno poznavanje kiralne limite. Ugotovili smo, da so efekti vzbujenih resonanc te\v zkih mezonov primerljivi s trenutnimi ocenami teoreti\v cnih napak in jih bo zato v prihodnjih \v studijah potrebno upo\v stevati.

\par

Redke procese te\v zkih mezonov analiziramo predvsem z namenom iskanja signalov nove fizike izven SM.  Vendar pa lahko upamo na uspeh le ob dobrem poznavanju in nadzoru nad hadronskimi efekti. V ta namen smo obravnavali kiralno obna\v sanje celotne supersimetri\v cne baze efektivnih operatorjev $\Delta B = 2$, ki so odgovorni za oscilacije nevtralnih te\v zkih mezonov. Izra\v cunali smo popravke kiralnih zank v drugem redu kiralnega in prvem redu razvoja po masah te\v zkih kvarkov ter vklju\v cili vplive te\v zkih mezonov sode parnosti. Potrdili smo razklopitev vzbujenih stanj in podali izraze za kiralno ekstrapolacijo celotne baze efektivnih operatorjev v redu vodilnih logaritmov. Na\v s rezultat bodo tako lahko uporabile prihodnje \v studije teh procesov s simulacijami QCD na mre\v zi. Kot pomo\v zni rezultat smo izra\v cunali vodilne kiralne logaritemske popravke k razpadnim konstantam te\v zkih mezonov sode parnosti.

\par

Nazadnje smo izvrednotili zelo redke prehode $b\to ss\bar d$ in $b \to dd\bar s$ mezona $B_c$ v pristopu efektivnih teorij. Hadronske razpadne amplitude smo ocenili s pomo\v cjo pribli\v zkov faktorizacije in zasi\v cenja z resonancami. Prehode smo analizirali znotraj ve\v cih modelov nove fizike. Na podlagi obstoje\v cih eksperimentalnih mej na pogostosti razpadov $B\to K K \pi$ in $B \to \pi \pi K$ smo lahko omejili relevantne kombinacije parametrov nove fizike. Kon\v cno smo na podlagi teh omejitev identificirali najobetavnej\v se dvo- in trodel\v cne neleptonske razpade mezona $B_c$, v katerih bi lahko s pomo\v cjo prihodnjih del\v cnih trkalnikov iskali signale teh redkih prehodov.

\par

Tekom ra\v cuna smo razre\v sili tudi nekaj tehni\v cnih podrobnosti. Morali smo izlu\v s\v citi celoten nabor kontra\v clenov v drugem redu kiralnega razvoja, ki prispevajo k mo\v cnim prehodom med te\v zkimi mezoni sode in lihe parnosti, ter lahkimi psevdoskalarnimi mezoni. Vklju\v citev vzbujenih stanj te\v zkih mezonov je nato pokvarila kiralno limito izra\v cunov v redu vodilnih logaritmov. Problem smo razre\v sili s pomo\v cjo odrezanega razvoja zan\v cnih integralov po obratni vrednosti masnih razlik med osnovnimi in vzbujenimi stanji te\v zkih mezonov, na ra\v cun zmanj\v sanja intervala zanesljivosti izra\v cunov znotraj HM$\chi$PT. V primeru semileptonskih prehodov med te\v zkimi in lahkimi mezoni smo morali pravilno reproducirati limiti HQET in SCET, da smo lahko dobili veljavno parametrizacijo oblikovnih funkcij. Hkrati smo morali med seboj pravilno napeti bazi oblikovnih funkcij znotraj QCD in HQET, ter identificirati prispevke izra\v cunov znotraj HM$\chi$PT k posami\v cnim oblikovnim funkcijam. Ugotovili smo, da le tak\v sno pravilno napenjanje oblikovnih funkcij verno reproducira prispevke resonanc pravilnih kvantnih \v stevil k oblikovnim funkcijam. Nenazadnje smo morali zaradi strukture polov v parametrizaciji oblikovnih funkcij v na\v se HM$\chi$PT izra\v cune vklju\v citi prispevke radialno vzbujenih stanj te\v zkih mezonov. V izra\v cunih kiralnih popravkov k me\v sanju te\v zkih nevtralnih mezonov smo morali predpisati pravilen postopek bozonizacije efektivnih operatorjev. Izkazalo se je, da je mogo\v ce ogromen nabor vseh mo\v znih struktur znotraj HM$\chi$PT s pomo\v cjo spinske simetrije te\v zkih kvarkov in identitet matrik $4\times 4$ znantno skr\v citi. Podobno smo morali na primeru prehodov $b\to ss\bar d$ in $b \to dd\bar s$ identificirati celotno bazo kvarkovskih operatorjev, kot tudi njihov tok in me\v sanje v prvem redu ena\v cb renormalizacijske grupe. Le tako smo lahko ohranili nadzor nad vodilnimi popravki QCD v visokoenergijskem re\v zimu. Nenazadnje smo za oceno mnogih hadronskih amplitud v dvo- in trodel\v cnih razpadih mezona $B_c$ morali izvesti vhodne HM$\chi$PT izra\v cune ter po potrebi vklju\v citi tudi prispevke lahkih vektorskih ter skalarnih resonanc. Ob tem smo smiselne fenomenolo\v ske rezultate dobili le s pravilnimo predpisanimi postopki resonan\v cnega zasi\v cenja amplitud.

\mainmatter

\renewcommand{\thesection}{\arabic{chapter}.\arabic{section}}

\selectlanguage{english}

\chapter{Introduction}
\index{introduction}

The Standard Model (SM)\index{Standard Model}
 of elementary particle physics is a quantum gauge-field\index{gauge field theory} theoretical description of fundamental electromagnetic, weak and strong interactions. It emerged in the 1960's and has completely dominated the field ever since~\cite{Donoghue:1992dd}. The building blocks of the SM are fermions -- leptons and quarks -- which come in three families. The SM gauge\index{gauge group} group is $SU(3)_c \times SU(2)_L \times U(1)_Y$, where the $SU(3)_c$ is the gauge\index{gauge group} group of Quantum Chromodynamics (QCD)\index{QCD}\index{Quantum Chromodynamics|see{QCD}}, $SU(2)_L$ is the gauge\index{gauge group} group of weak isospin, while $U(1)_Y$ is the gauge\index{gauge group} group of weak hypercharge\index{hypercharge}. Only the left-handed chiral\index{chiral fermions} fermions transform as weak isospin doublets under the $SU(2)_L$, while quarks also form the fundamental triplet representation of $SU(3)_c$. The masses of leptons and quarks\index{quarks} are generated via the Higgs\index{Higgs mechanism} mechanism -- spontaneous symmetry breaking, where the (chiral) symmetry\index{chiral symmetry breaking} of the theory is not respected by the vacuum. For this purpose an additional scalar weak isospin doublet is introduced. Its vacuum expectation value also breaks gauge\index{gauge invariance} invariance of the theory to the subgroup $SU(3)_c\times U(1)_{EM}$, inducing masses for the weak $W^{\pm}$ and $Z$\index{gauge boson!$Z$} gauge bosons.

\par

The quark fields in the $SU(2)_L$ basis are not the mass eigenstates in general. Therefore it is customary to rotate them to the mass eigenbasis by means of a unitary matrix. The rotation is conventionally conveyed to the down-quark fields and the rotation matrix is called the Cabibbo-Kobayashi-Maskawa (CKM)\index{CKM}\index{Cabibbo-Kobayashi-Maskawa|see{CKM}} matrix. It can be fully described by three real mixing angles and a complex CP\index{CP violation} violating phase.

\par

The successes of the SM description are abundant. Its predictions have been extensively tested in accelerator facilities and agree well with the data measured up to energies available at present: the electroweak precision tests are generally in impressive agreement with SM predictions~\cite{:2005em} while the CP\index{CP violation} violation experiments in $K$, $D$ and $B$ meson\index{meson!$K$}\index{meson!$B$}\index{meson!$D$} systems support the CKM\index{CKM} description with one universal phase~\cite{Charles:2004jd,Bona:2006ah}. The only elementary building block presently lacking experimental detection is the Higgs boson\index{Higgs boson}.

\par

However we also know from observations, that the SM cannot be the ultimate fundamental theory. For once, it does not include gravity. Although colossal theoretical efforts have been spent on the subject in the last few decades, the progress has been slow and the results inconclusive. Mainly also due to the lack of almost any experimental hints in the area. On the other hand, the SM also does not account for the recently measured neutrino\index{neutrino} oscillations\index{oscillations!of neutrinos}~\cite{GonzalezGarcia:2003qf}. Explanation of these requires non-zero neutrino\index{neutrino} masses, contrary to the SM prescription\footnote{In fact, the matter contents of the SM can easily be extended to include right-handed neutrinos\index{neutrino} and thus allowing for Dirac neutrino masses via the Higgs\index{Higgs mechanism} mechanism. However the observed smallness of the neutrino masses\index{neutrino} and the fact that right-handed neutrinos\index{neutrino} must be singlets under the SM gauge\index{gauge group} group seem to prefer alternative mechanisms which lie beyond the SM.}. Thirdly, a growing number of astrophysical observations suggest that most of the matter in the universe is neither luminous nor baryonic~\cite{Astier:2005qq}. In addition, most of it must be slowly moving or ``cold''. The SM does not provide a candidate for nonbaryonic cold dark matter. Finally, our current understanding of baryogenessis -- the generation of the measured baryon - antibaryon asymmetry -- in the early universe requires levels of CP\index{CP violation} and baryon number violation much higher than allowed for in the SM~\cite{Strumia:2006qk}.

\par

There are also some conceptual and ``aesthetic'' problems with the SM. The running of the gauge\index{gauge coupling} couplings suggests a unification scale at $10^{14}-10^{16}~\mathrm{GeV}$ although precise unification does not occur if one takes into account only SM fields~\cite{Masiero:2005ua}. Neither does the SM describe the dynamics of such unification. In fact, even the electroweak symmetry breaking has no dynamical explanation within the SM. It is imposed by construction and renders the masses of the elementary particles as free parameters. Compared to the large unification scale the electroweak scale $1/\sqrt{G_F} \sim 250~\mathrm{GeV}$\index{G$_F$} also appears to be very small. The large scale hierarchy manifests itself in form of large quantum corrections to the mass of the Higgs\index{Higgs boson} boson, which are quadratically divergent and thus sensitive to the UV completion of the theory\footnote{This has to be compared to the logarithmic divergences of fermionic fields due to chiral\index{chiral symmetry} symmetries and gauge\index{gauge boson} boson fields due to gauge\index{gauge invariance} invariance.}. Even more appealing is the ``fine-tuning'' required for the vacuum energy density when compared to the measured present critical density of the universe $\rho_c\sim 10^{-14}~\mathrm{eV}^4$~\cite{Astier:2005qq}. Its classical value, which is a free parameter in any quantum field theory (QFT)\index{QFT}\index{Quantum Field Theory|see{QFT}}, has to cancel corrections due to spontaneous symmetry breaking and the resulting vacuum condensates in the SM at energy scales from a few $100~\mathrm{MeV}$ to a few $100~\mathrm{GeV}$. Another similar issue is related to the strong CP\index{CP violation!in QCD} phase of the QCD\index{QCD!vacuum} vacuum. Its value, a free parameter of the SM, is severely constrained by the measurements of the electric dipole moment of the neutron~\cite{Donoghue:1992dd}.

\par

All of these reasons call for physics beyond SM\index{Beyond the Standard Model|see{new physics}}, and many proposals exist on what the physical reality ought to look like above the electroweak scale. For example, supersymmetric (SUSY)\index{SUSY}\index{supersymmetric|see{SUSY}} extensions of the SM attempt to resolve the Higgs\index{Higgs hierarchy problem} hierarchy problem and provide suitable dark matter candidates~\cite{Nilles:1983ge, Weinberg:2000cr}. The simplest and most studied of these is the Minimal SUSY SM (MSSM)\index{MSSM}\index{Standard Model!minimal supersymmetric|see{MSSM}}\index{Minimal Supersymmetric Standard Model|see{MSSM}} which adds a bosonic partner to each SM fermion and a fermionic counterpart to each boson, but also doubles the Higgs sector. Possible alternative proposals come in the form of (large) extra-dimensions in which some or all of the SM fields may propagate or extended SM symmetries. In order for any of these extensions\index{new physics} to address the dynamics at the electroweak scale, new physical degrees of freedom have to appear at the TeV scale. On the other hand, many of these ``low scale'' new physics scenarios can be embedded into high scale unification theories, such as Grand Unified Theories (GUTs)\index{GUT}\index{Grand Unified Theory|see{GUT}} attempting to describe the unification of the SM couplings at large scales in terms of incorporating the SM gauge\index{gauge group} group into larger symmetries~\cite{Weinberg:1996kr}. Even more ambitious are the various String theories unified under the name of M-theory, which also attempt to address the quantization of gravity and the cosmological evolution of the very early universe~\cite{Green:1987sp, Green:1987mn}. In order to get a handle on this plethora of high energy phenomenologies one must often rely on the so-called ``bottom-up'' approach to new physics\index{new physics}. One constructs effective low energy theories by systematically parameterizing possible new physics\index{new physics!contributions} contributions to low energy processes based on symmetry principles of the expected underlying theory. Within this framework, the SM itself is regarded as an effective low energy description of a grander theory, containing the SM particle content and gauge\index{gauge symmetry} symmetry at low energies. A similar reasoning lies behind the MSSM\index{MSSM}, which is often regarded as the low energy effective theory of a high scale (and/or dimensional) GUT\index{GUT} or String theory, containing a (slightly broken) SUSY\index{SUSY} SM particle content at energies close to the electroweak scale. Alternatively, one can focus on specific low energy aspects, common to various SM extensions. A common characteristic of various new physics\index{new physics} models (including MSSM\index{MSSM}) is the appearance of a doubling of the Higgs sector, which can be put in the general form of a Two Higgs Doublet Model (THDM)\index{THDM}\index{Two Higgs Doublet Model|see{THDM}}. On the other hand many GUT\index{GUT} and String theories also predict additional low energy $U(1)$ gauge extensions to the SM -- the appearance of additional $Z'$ bosons\index{gauge boson!$Z'$}. By focusing on such common aspects, one can extract important general signatures of various new physics proposals.

\par

The experimental challenge of finding new physics\index{new physics!searches} follows two main directions. In direct searches the idea is to produce the new particles and detect them directly (often through their decay products). This requires high enough energies at particle colliders such as the Tevatron\index{Tevatron}, the upcoming LHC\index{LHC} or the planned ILC\index{ILC}. A complementary idea is to measure the effects of new particles in processes where they enter as intermediate virtual states. In this approach it is crucial to be able to disentangle the effects of new physics\index{new physics!contributions} from those conveyed by the SM particles. One may then employ the ``top-down'' approach as prototyped by the Wilsonian Operator Product Expansion (OPE)~\cite{Wilson:1969zs}\index{OPE}\index{Operator Product Expansion|see{OPE}}. Namely one may represent low energy Green's functions\index{Green's function} or scattering amplitudes\index{scattering amplitude} in terms of products of local operators, which are in term computed (matched to\index{matching}) the full original formulation of the SM and possible additional higher energy extensions. In this way SM and new physics\index{new physics!contributions} contributions are clearly separated on an amplitude-by-amplitude basis. The task is then to compute low energy scattering cross sections and decay  rates and compare them to precision measurements. This approach both tests SM predictions as well as probes possible new physics\index{new physics!contributions} contributions. Experimentally it requires high statistics and precision measurements, such as those provided in the last years by the B\index{meson!$B$} and D\index{meson!$D$} meson factories\index{B-factories}\index{meson factories} at Belle\index{meson factories!Belle|see{Belle}}\index{Belle}, BaBar\index{meson factories!BaBar|see{BaBar}}\index{BaBar}, CLEO-c\index{meson factories!CLEO|see{CLEO}}\index{CLEO}\index{CLEO-c|see{CLEO}} and others. Among their successes are the by now established neutral meson oscillations\index{oscillations!$K^0-\overline K^0$}\index{meson!$K$!oscillations} in all neutral $K-\overline K$~\cite{Donoghue:1992dd}, $D-\overline D$\index{oscillations!$D^0-\overline D^0$}\index{meson!$D$!oscillations}~\cite{Staric:2007dt,Aubert:2007wf}, $B-\overline B$\index{oscillations!$B^0_d-\overline B^0_d$}\index{meson!$B$!oscillations}~\cite{Barberio:2006bi} and $B_s -\overline B_s$\index{oscillations!$B^0_s-\overline B^0_s$}~\cite{Abulencia:2006mq,Abazov:2006dm} \index{meson!$B_s$!oscillations}\index{meson!$B$!oscillations}meson systems, as well as ever tightening consistency constraints on the CKM\index{CKM!unitarity} unitarity and the CP\index{CP violation} violating phase. So far, no clear indications of new physics\index{new physics} in these phenomena have been observed and several stringent experimental bounds on various new physics proposals have been imposed.

\par

In order to correctly interpret experimental results and justify the consistency with the SM or claim new physics signals, one first has to reliably calculate the relevant hadronic processes based on the quark picture of the OPE\index{OPE}. Due to the nonperturbative and confining nature of low energy QCD\index{QCD}, this turns out to be a daunting task. Namely, the expansion in the coupling constant is not applicable in this regime. Ab initio\index{ab initio calculations in QCD} calculations, i.e., by starting with the QCD\index{QCD!Lagrangian} Lagrangian and finishing up with predictions for physical observables are still possible, through the use Lattice QCD\index{QCD!on the lattice|see{lattice QCD}}\index{lattice QCD} techniques, but are computationally very challenging~\cite{Montvay:1994cy}. Lattice methods\index{lattice QCD} also have their own limitations. To get meaningful results, computations have to be done in Euclidean space-time, which makes calculations of processes with more than one hadron in the final state very difficult. Also, in order to make numerical difficulties tractable, a number of approximations have to be made, e.g., by working at relatively large pion masses. Another option that has been commonly used in the past, is to use symmetries of the QCD\index{QCD!Lagrangian} Lagrangian to construct effective theories~\cite{Ecker:1994gg}. Unknown parameters in the effective theory are fixed from experiments or, if possible, from perturbative comparison (matching\index{matching}) to full QCD. These effective theories may then be employed to either predict some experimental processes directly, or to assist nonperturbative Lattice QCD\index{lattice QCD} calculations making them more tractable and keeping control of the used approximations.

\par

One important manifestation of the strong QCD\index{QCD} dynamics at low energies is the appearance of resonances\index{resonance} in the particle spectrum. They have been detected long ago and studied extensively in the processes of pions and kaons\index{meson!$K$}~\cite{Donoghue:1992dd}. Their effects proved to be critical in many low energy processes. On one hand they restrict the validity of effective theory approaches, which are not able to fully include their effects, e.g., in (resonant) $\pi \pi$ scattering. Also, their dominant (long distance\index{long distance effects}) effects are known to almost completely obscure contributions due to (short distance) SM OPE\index{OPE} or possible new physics\index{new physics!contributions} contributions in $D$\index{meson!$D$!oscillations} meson oscillations\index{oscillations!$D^0-\overline D^0$} and rare decays~\cite{Prelovsek:2000rj}\index{meson!$D$!decay}. On the other hand, due to the relatively large $c$ and $b$ quark masses, heavy meson resonance effects were long believed to be less significant in processes among hadrons involving these two quarks.

\par

In the last couple of years however, many experiments have reported first observations of resonances\index{resonance!of charmed meson} in the charm spectrum.
In 2003, Belle~\cite{Abe:2003zm}\index{Belle} and FOCUS~\cite{Link:2003bd}\index{FOCUS} experiments reported the observation of broad resonances $D_0^{*+}$\index{resonance!of charmed meson}\index{meson!$D$!resonances} and $D_0^{*0}$, ca. $400-500\mathrm{~MeV}$ higher above the usual $D$\index{meson!$D$} states and with opposite parity. In the same year BaBar~\cite{Aubert:2003fg}\index{BaBar} announced a narrow meson $D_{sJ}(2317)^+$\index{meson!$D_s$!resonances}. This was confirmed by FOCUS~\cite{Vaandering:2004ix}\index{FOCUS} and CLEO~\cite{Besson:2003jp}\index{CLEO} which also noticed another narrow state, $D_{sJ}(2463)^+$\index{meson!$D_s$!resonances}. Both states were also confirmed by Belle~\cite{Krokovny:2003zq}\index{Belle}. The basic properties of the relevant charmed mesons together with the dominating hadronic decay modes are listed in table~\ref{table_input}.
\begin{table*}
\begin{center}
\begin{tabular}{|c|c|c|c|c|}
\hline
Meson & $J^P$ & Mass~$[\mathrm{GeV}]$ & Width~$[\mathrm{GeV}]$ & $Br.~[\%]$ (final states) \\
\hline
\hline
\multicolumn{5}{|c|}{$c \bar d$}\\
\hline
$D^{+}$ & $0^-$ & $1.869\pm0.001$ & & \\
\hline
$D^{*+}$ & $1^-$ & $2.010\pm0.001$ & $(9.6 \pm 2.2)\times 10^{-5}$ & $67.7\pm0.5~(D^0 \pi^+)$, \\
& & & & $30.7\pm0.5~(D^+ \pi^0)$ \\
\hline
$D^{*+}_0$~\cite{Link:2003bd} & $0^+$ & $2.403\pm 0.014 \pm 0.035$ & $0.283 \pm 0.024 \pm 0.034$ & $(D^0 \pi^+)$\footnotemark[3] \\
\hline
\hline
\multicolumn{5}{|c|}{$c \bar u$}\\
\hline
$D^{0}$ & $0^-$ & $1.865\pm0.001$ & & \\
\hline
$D^{*0}$ & $1^-$ & $2.007\pm0.001$  & $<0.002$ & $61.9\pm2.9~(D^0 \pi^0)$ \\
\hline
$D^{*0}_0$ & $0^+$ & $2.350\pm 0.027$\footnotemark[4] & $0.262 \pm 0.051$\footnotemark[4] & $(D^+ \pi^-)$\footnotemark[3]\\
\hline
$D^{'0}_1$ & $1^+$ & $2.438\pm 0.030$\footnotemark[5] & $0.329 \pm 0.084$\footnotemark[5] & $(D^{*+} \pi^-)$\footnotemark[3]\\
\hline
\hline
\multicolumn{5}{|c|}{$c \bar s$}\\
\hline
$D_s$ & $0^-$ & $1.968\pm0.001$ & & \\
\hline
$D_s^{*}$ & $1^-$ & $2.112\pm 0.001$ & $<1.9\times 10^{-3}$ & $5.8\pm2.5~(D_s \pi^0)$\\
\hline
$D_{sJ}(2317)^+$ & $0^+$ & $2.317\pm 0.001$ & $<0.005$ & $(D_s^{0} \pi^+)$\footnotemark[3]\\
\hline
$D_{sJ}(2463)^+$ & $1^+$ & $2.459\pm 0.001$ & $<0.006$ & $(D_s^{*0} \pi^+)$\footnotemark[3]\\
\hline
\end{tabular}
\caption{\small\it Experimentally measured properties of the relevant charmed mesons and their dominant hadronic decay modes. The pseudoscalar ground states are listed for completeness. Unless indicated otherwise, the values are taken from PDG.\label{table_input}}
\end{center}
\end{table*}

\footnotetext[3]{Observed channel.}
\footnotetext[4]{Average of Belle~\cite{Abe:2003zm}\index{Belle} and FOCUS~\cite{Link:2003bd}\index{FOCUS} values from~\cite{Colangelo:2004vu}.}
\footnotetext[5]{Average of Belle~\cite{Abe:2003zm}\index{Belle} and CLEO~\cite{Anderson:1999wn}\index{CLEO} values from~\cite{Colangelo:2004vu}.}

\setcounter{footnote}{5}

\par

Studies of the basic properties of these states have been triggered particularly by the fact that the  $D_{sJ}(2317)^+$\index{meson!$D_s$!resonances} and  $D_{sJ}(2463)^+$ states' masses are below threshold for the decay into ground state charmed mesons and kaons\index{meson!$K$}, as suggested by quark model studies \cite{Godfrey:1985xj, Godfrey:1986wj} and lattice calculations\index{lattice QCD} \cite{Hein:2000qu, Dougall:2003hv}. Their relative closeness to the ground state charmed mesons suggests possibly significant effects in the processes of the lowest $D$\index{meson!$D$} and $D_s$\index{meson!$D_s$} states and poses the following questions: Can we estimate the relevant effects of the lowest heavy meson resonances\index{resonance!of heavy meson} to the processes of heavy meson ground states? Can we keep their effects under control, especially within effective theories of QCD? Can they possibly help us to understand certain aspects of observed and measured ground state heavy meson processes? And finally what conclusions, drawn for the charm sector can we apply to the processes of $B$\index{meson!$B$} and $B_s$\index{meson!$B_s$} mesons, whose resonances are currently still beyond the reach of experimental facilities\footnote{During the final stages of preparation of this thesis D0\index{D0} collaboration has reported the first observation of axial resonances in the $B$\index{meson!$B$!resonances} spectrum~\cite{Collaboration:2007vq}. Their properties and interpretation are yet to be analyzed in detail.}.

\par

In this thesis we will explore several aspects of resonances\index{resonance} in the heavy meson processes~\cite{Fajfer:2004mv,Fajfer:2005ug,Fajfer:2005mk,Fajfer:2006uy,Fajfer:2006hi,Eeg:2007ha,Becirevic:2006me,Fajfer:2004fx,Fajfer:2006av}. Their leading order contributions, either at tree level or at one loop, will be analyzed in the relevant effective theory approach to QCD. Within this framework we will calculate hadronic parameters entering various low energy processes and study the impact of heavy meson resonances\index{resonance!of heavy meson} on observables. These include strong, semileptonic decay rates of heavy mesons as well as neutral heavy meson mixing parameters\index{mixing!of heavy neutral mesons}. Since strong decay channels, if open, usually dominate the measured decay widths, one may use these as benchmarks on the validity of the chosen effective theory approach and also determine from them basic parameters of the effective theory. Semileptonic decays, mediated by quark and charged lepton weak currents proceed at tree level in the SM and are confirmed to be dominated by SM contributions. Their detailed study may therefore produce important consistency checks within the SM, such as the determination of the various CKM\index{CKM!matrix elements} matrix elements and testing its unitarity, provided the relevant hadronic effects are well understood. Heavy neutral meson mixing\index{mixing!of heavy neutral mesons}, on the other hand, is mediated by box\index{box diagrams} diagrams in the SM. This makes it an important arena for studying possible new physics\index{new physics!contributions} contributions, which may or may not be suppressed by loop factors\index{loop suppression}. Within our approach we will analyze all possible hadronic amplitudes entering heavy neutral meson mixing\index{mixing!of heavy neutral mesons} within the SM or beyond. Finally, we will also analyze very rare hadronic decays of the doubly heavy $B_c$ meson\index{meson!$B_c$!decay}, which are, like the neutral meson mixing\index{mixing!of neutral mesons}, mediated by box\index{box diagrams} diagrams in the SM. There we will make use of some of the knowledge on the impact of resonances\index{resonance} in the calculation of the relevant hadronic decay amplitudes in order to constrain various new physics proposals based on existing experimental searches and also propose prospecting new search directions.

\par

The outline of the thesis is as follows. In the first two chapters we introduce the prerequisites for the phenomenological studies in the subsequent chapters. In chapter 2 we introduce the concept of effective field theories with a focus on the effective theory approaches to QCD in the limits of small and large quark masses. In chapter 3 we review some commonly used tools in hadronic calculations, such as the OPE\index{OPE}, general hadronic matrix element parameterizations, and some of their approximations. In chapter 4 we analyze strong decays\index{strong decay} of heavy mesons within an effective theory approach, including loop\index{loop contributions} contributions of excited heavy meson resonances\index{resonance!of heavy meson}. We attempt to extract the relevant effective strong meson couplings from the measured decay  rates and study the impact of the resonances\index{resonance contribution} on the coupling extraction from Lattice QCD\index{lattice QCD} calculations. In chapter 5 we analyze the leading contributions of the heavy meson resonances\index{resonance!of heavy meson} to semileptonic decays\index{semileptonic decay}. Both heavy to light as well as heavy to heavy meson transitions are analyzed. While in the former, heavy resonances may contribute already at tree level, in the latter their contributions are loop\index{loop suppression} suppressed. Similar, loop suppressed contributions to heavy neutral meson\index{mixing!of heavy neutral mesons} mixing hadronic amplitudes are studied in chapter 6. Finally, chapter 7 contains our analysis of the very rare hadronic decays\index{decay of $B_c$ meson} of the $B_c$\index{meson!$B_c$!decay} meson within the SM and some of its extensions. The \index{conclusions}conclusions are gathered in Chapter 6, while some further technicalities of our calculations as well as brief descriptions of studied SM extensions are relegated to the appendices.

\chapter{Effective theories of heavy and light quarks}

\section{What is an effective field theory?}

The content of quantum theory is encoded in its Green's functions\index{Green's function}, which in general depend in a complicated way on the properties (e.g. particle momenta) of the initial and final states. In particular they exhibit nonanalytic behavior such as cuts and poles in the configuration variables, which arise when the kinematics allow for physical intermediate states. Conversely, when the kinematics are far from being able to produce a certain propagating intermediate state, the contribution of that state to the Green's function\index{Green's function} of interest will be relatively simple, well approximated by the first few terms in a Taylor expansion (e.g. of the incoming momenta of the scattering problem). Instead of Taylor expanding each amplitude it turns out to be much more profitable to expand the Lagrangian in local operators that only involve the {\it light} degrees of freedom, where the expansion is in the powers of the generalized momenta of the light fields (appearing as derivatives in the Lagrangian) divided by the scale of {\it heavy} physics. Such a Lagrangian is called an effective field theory. Although the heavy modes do not appear explicitly anymore, their contributions are encoded through the parameters of the effective theory\footnote{This aspect of effective theories is not unique to quantum phenomena. Integrating out\index{integrating out} certain regions or scales of the phase space in order to simplify the description of certain phenomena has also been found to be of high value in other fields such as (classical) statistical mechanics or (classical) field theories.}. There are many situations in which effective field theories are of utility~\cite{Kaplan:2005es}:
\begin{itemize}

\item They allow one to compute low energy scattering amplitudes without having a detailed understanding of the short distance physics, or to avoid wasting effort calculating tiny effects from known short distance physics (such is the OPE\index{OPE} and the effective weak Hamiltonian\index{effective weak Hamiltonian}).

\item In nonperturbative theories (such as low energy QCD\index{QCD}) one can construct a predictive effective field theory for low energy phenomena by combining power counting of operators with symmetry constraints of the underlying theory (such as the $\chi$PT\index{$\chi$PT} and HM$\chi$PT\index{HM$\chi$PT}).

\item By regarding theories of known physics as effective field theory descriptions of more fundamental underlying physics, one can work bottom up, extrapolating from observed rare processes to a more complete theory of short distance physics (this approach is taken in many studies of BSM physics, such as MFV or grand unification).

\end{itemize}
At present, the general approach of effective field theory is followed in many contexts of the SM and even in more speculative theories like grand unification, supergravity, extra dimensions or superstrings.
\par

\section{Exploring the Chiral symmetry of QCD\label{sec_2_2}}
\index{$\chi$PT}
\index{Chiral Perturbation Theory|see{$\chi$PT}}
\index{QCD!chiral symmetry}
\index{symmetries!of light quarks}
\index{chiral symmetry}

One of the earliest and also one of the most successful examples of effective theories is the chiral perturbation theory ($\chi$PT\index{$\chi$PT}) which builds upon the approximate chiral\index{chiral symmetry} symmetry of QCD\index{QCD!chiral symmetry} at low energies. We will briefly review it in this section. The QCD\index{QCD!Lagrangian} Lagrangian with $N_f$ ($N_f=2, 3$) massless quarks $q^{(n)}=(u,d,\ldots)$
\begin{eqnarray}
\mathcal L^0_{QCD} &=& \sum_{n=1}^{N_f} \overline q^{(n)} i \slashed{D}  q^{(n)} + \mathcal L_{\mathrm{gauge}} + \mathcal L_{\mathrm{heavy~quarks}} \nonumber\\
&=& \sum_{n=1}^{N_f} \left[ \overline q_L^{(n)} i \slashed{D}  q_L^{(n)} + \overline q_R^{(n)} i \slashed{D}  q_R^{(n)} \right] + \mathcal L_{\mathrm{gauge}} + \mathcal L_{\mathrm{heavy~quarks}},
\label{eq_2_1}
\end{eqnarray}
where $\slashed{D} = \gamma^{\mu} D_{\mu}$ is the QCD\index{covariant derivative!in QCD} covariant derivative and $q_{R,L} = (1\pm \gamma_5) q/2$, has a global symmetry
\begin{equation}
\underbrace{SU(N_f)_R \times SU(N_f)_L}_{\mathrm{chiral~group}~G} \times U(1)_V \times U(1)_A\,.
\end{equation}
At the effective hadronic level, the quark number symmetry $U(1)_V$ is realized as baryon number. The axial\index{axial symmetry} $U(1)_A$ is anomalous\index{anomalous $U(1)_A$} and is broken by nonperturbative effects. Theoretical and phenomenological evidence suggests that the chiral\index{chiral symmetry group} group $G$ on the other hand is spontaneously broken to the vector subgroup $H=SU(N_f)_V$. The axial generators\index{axial group generators} of $G$ are realized non-linearly and associated with them are the $N_f^2-1$ massless pseudoscalar Goldstone\index{pseudo-Goldstone boson} bosons $\Pi(x) = \lambda^i \pi_i(x)$ parameterizing the $G/H$ right coset space. Here $\lambda^i$ are the broken generators of $G$ and $\pi_i(x)$ are the Goldstone\index{pseudo-Goldstone boson} fields. For the $N_f=3$ case, the $\Pi$ can be written as
\begin{equation}
\Pi = \begin{pmatrix}
    \frac{1}{\sqrt 6}\eta_8 + \frac{1}{\sqrt 2} \pi^0 & \pi^+ & K^+ \\
   \pi^- & \frac{1}{\sqrt 6}\eta_8 - \frac{1}{\sqrt 2} \pi^0 & K^0 \\
   K^- & \overline K^0 & -\sqrt{\frac{2}{3}}\eta_8
    \end{pmatrix},
    \label{eq_2.3}
\end{equation}
while in the $N_f=2$ only the pion fields remain.
To preserve all the symmetries of the fundamental theory in the effective Lagrangian\index{effective Lagrangian}, it is essential to construct it out of the Goldstone\index{pseudo-Goldstone boson} field functions which transform linearly under $G$ (see e.g.~\cite{Zupan:2002rz} for details). A customary choice is $\Sigma = \exp 2 i \Pi(x)/f$, which transforms as $\Sigma \to R \Sigma L^{\dagger}$, where $R$ and $L$ are the corresponding generators $SU(N_f)_R$ and $SU(N_f)_L$ respectively. $f$ is an undetermined constant of energy dimension one, which can be identified with the Goldstone\index{pseudo-Goldstone boson!decay constant} boson decay\index{decay constant} constant. We continue by factoring out the broken generators of $G$ from the quark fields $q =  \zeta(\Pi) \widetilde q$, where $\zeta(\Pi)$ transforms under $G$ as $\zeta(\Pi) \to \zeta(\Pi') U(x)$. Here $\Pi'(x)$ is the transformed Goldstone\index{pseudo-Goldstone boson} matrix and we demand that $U(x)$ be an element of $H$. In general it will also be a function of $\Pi(x)$. Consequently, $\widetilde q$ transforms as $\widetilde q \to U(x) \widetilde q$ and we have to modify its covariant derivative\index{covariant derivative!in $\chi$PT} to account for the coordinate dependence $ \slashed{\widetilde D} \widetilde q = (\slashed{D} + \slashed{\mathcal V})\widetilde q$ where the vector field $\mathcal V_{\mu} = (\xi \partial_{\mu} \xi^{\dagger} + \xi^{\dagger} \partial_{\mu} \xi)/2$ and $\xi = \sqrt \Sigma$ transforming as $\xi \to L \xi U^{\dagger} = U \xi R^{\dagger}$. It can be easily checked that $\mathcal V_{\mu}$ transforms under $G$ as $\mathcal V_{\mu} \to U \mathcal V_{\mu} U^{\dagger} + U \partial_{\mu} U^{\dagger}$. There exists another operator which can be built up of $\Pi(x)$, has the properties of an axial vector field $\mathcal A_{\mu} = i (\xi^{\dagger} \partial_{\mu} \xi- \xi \partial_{\mu} \xi^{\dagger})/2 = i \xi^{\dagger} \partial_{\mu} \Sigma \xi^{\dagger}/2$ and is transforming under $G$ as $\mathcal A_{\mu} \to U \mathcal A_{\mu} U^{\dagger}$. Its role will become apparent later.
\par
The Lagrangian of the SM is not chiral\index{chiral invariance} invariant. The chiral\index{chiral symmetry breaking} symmetry of the strong interactions is broken by the electroweak interactions generating in particular non-zero quark masses. The basic assumption of $\chi$PT\index{$\chi$PT!assumptions} is that the chiral\index{chiral limit} limit constitutes a realistic starting point for a systematic expansion in chiral\index{chiral symmetry breaking} symmetry breaking interactions. Namely we extend the chiral\index{chiral invariance} invariant QCD Lagrangian\index{QCD!Lagrangian}~(\ref{eq_2_1}) by coupling  the quarks to external hermitian matrix fields $v_{\mu} = r_{\mu} + l_{\mu}$, $a_{\mu} = r_{\mu} - l_{\mu}$, $s$, $p$\cite{Ecker:1994gg}:\index{chiral expansion}
\begin{equation}
\mathcal L = \mathcal L^0_{QCD} + \sum_{m,n=1}^{N_f} \left[\overline q^{(m)} (\slashed{v}_{(m,n)} + \slashed{a}_{(m,n)}\gamma_5) q^{(n)} - \overline q^{(m)} (s_{(m,n)} - i p_{(m,n)}\gamma_5) q^{(n)} \right].
\end{equation}
Here $v_{\mu}$ and $a_{\mu}$ will contain external photons and weak gauge\index{gauge boson} bosons so that Green's functions\index{Green's function} for electromagnetic and semileptonic weak currents can be obtained as functional derivatives of the generating functional $Z[v,a,s,p]$ with respect to external photon and weak boson fields. The scalar and pseudoscalar fields $s$, $p$ on the other hand give rise to Green's functions\index{Green's function} of (pseudo)scalar quark currents, as well as providing a very convenient way of incorporating explicit chiral\index{chiral symmetry breaking} symmetry breaking through the quark masses. To preserve the manifest chiral\index{chiral symmetry} symmetry of the effective Lagrangian\index{effective Lagrangian}, we promote it to a local symmetry and treat the external fields as spurions\index{spurion} with the transformation properties $r_{\mu} \to R r_{\mu} R^{\dagger} + i R \partial_{\mu} R^{\dagger}$, $l_{\mu} \to L l_{\mu} L^{\dagger} + i L \partial_{\mu} L^{\dagger}$ and $s+ i p \to R(s + i p) L^{\dagger}$. Accordingly we have to introduce the covariant derivatives for pion fields $D_{\mu}\Sigma = \partial_{\mu} \Sigma - i r_{\mu} \Sigma + i \Sigma l_{\mu}$ (as well as the appropriate external field stress-energy tensors). The physically interesting Green's functions\index{Green's function} are then functional derivatives of the generating functional $Z[v,a,s,p]$ at chosen values of the spurion\index{spurion} fields. In particular for the quark masses we use $s=m_q \equiv \mathrm{diag}(m_u,m_d,\ldots)$.
Even more generally, any effective quark operator (e.g. from the OPE\index{OPE} of the effective weak Hamiltonian \index{effective weak Hamiltonian}) can be incorporated into the effective chiral Lagrangian by coupling it to the appropriate external chiral\index{spurion} spurion field and then constructing the corresponding source terms out of the Goldstone fields\index{pseudo-Goldstone boson}.
\par
The effective chiral\index{chiral Lagrangian} Lagrangian is usually organized in a derivative expansion based on the chiral\index{chiral power counting} power counting rules. One prescribes chiral\index{chiral power counting} powers $p$ to all the constituent field operators and then builds the Lagrangian out of them by constructing all the terms adherent to the symmetries up to a given chiral\index{chiral power counting} power. In general, this procedure counts the powers of derivatives on the Goldstone\index{pseudo-Goldstone boson} fields as well as the number of external field insertions which at the level of Green's functions\index{Green's function} translates to counting the powers of pseudo-Goldstone\index{pseudo-Goldstone boson} masses and exchanged momenta. Using the simplest choice for the quark mass chiral\index{chiral power counting} counting $s \sim p^2$ one arrives at the lowest $\mathcal O(p^2)$ order Lagrangian\index{chiral Lagrangian} describing the low energy strong interactions of light pseudoscalar mesons~\cite{Donoghue:1992dd,Gasser:1984gg}
\begin{equation}
\mathcal L^{(2)}_{\chi} = \frac{f^2}{8} \partial_{\mu} \Sigma_{ab} \partial^{\mu} \Sigma^{\dagger}_{ba} + \lambda_0 \left[(m_q)_{ab} \Sigma_{ba} + (m_q)_{ab} \Sigma^{\dagger}_{ba}\right].
\end{equation}
A trace is taken over the repeated light quark flavor indices. For the sake of clarity we have omitted all external currents ($l=r=p=0$) except $s$ in the second term, which induces masses of the pseudo-Goldstone\index{pseudo-Goldstone boson} bosons $m^2_{ab}= 4 \lambda_0 (m_a + m_b)/f^2$. While the normalization of the first term is canonical, the second term contains an unknown constant $\lambda_0$, which we can fit to the light pseudoscalar meson masses. Conversely, lattice QCD\index{lattice QCD} simulations often work in the exact $SU(2)$ flavor isospin symmetry limit. There, one can parameterize the pseudo-Goldstone\index{pseudo-Goldstone boson!parameterization of masses} masses according to the Gell-Mann formulae as~\cite{Becirevic:2004uv}
\begin{equation}
\begin{array}{rclrclrcl}
m^2_{\pi} & = & \frac{8 \lambda_0 m_s}{f^2} r, & m_K^2 & = & \frac{8 \lambda_0 m_s}{f^2} \frac{r + 1}{2}, &
m^2_{\eta_8} & = & \frac{8 \lambda_0 m_s}{f^2} \frac{r+2}{3},
\end{array}
\label{eq_2.6}
\end{equation}
where $r=m_{u,d}/m_s$ and $8 \lambda_0 m_s / f^2 = 2m_K^2-m_{\pi}^2$.
\par
Higher order terms in the chiral\index{chiral power counting} power counting can be constructed in this manner as well as terms involving any general external fields. The higher order terms in this expansion also serve a double role as the counterterms\index{counterterms} absorbing loop\index{loop divergencies} divergences from diagrams with insertions of lower order terms in the Lagrangian, thus keeping the theory renormalizable (in the general sense of the word).

\subsection{Light flavor singlet mixing and the $\eta'$}
\index{meson!$\eta'$}
\index{mixing!of $\eta$ and $\eta'$}
\index{meson!$\eta$}
\index{meson!$\eta$!mixing}

The eight $SU(3)$ pseudo-Goldstone\index{pseudo-Goldstone boson} bosons: $\pi^+$, $\pi^-$, $\pi^0$, $K^+$, $K^-$, $K^0$, $\overline {K^0}$ and $\eta_8$ have the same quantum numbers as the following quark-antiquark pairs: $u\bar d$, $d\bar u$, $u\bar u - d\bar d$, $u\bar s$, $s\bar u$, $d\bar s$, $s\bar d$ and $u\bar u + d\bar d - 2 s\bar s$. This suggests the existence of a ninth meson $\eta_0$ that would correspond to the $u\bar u + d \bar d + s \bar s$ singlet as the pseudo-Goldstone\index{pseudo-Goldstone boson} boson of the $U(1)_A$ axial\index{axial symmetry} symmetry of QCD\index{QCD!axial symmetry}. However, due to the axial anomaly,\index{axial anomaly} the mass of $\eta_0$ is not protected and can be much larger than those of the other eight states.

\par

Nevertheless, for practical reasons $\eta_0$ still has to be incorporated into the theory. Namely, its existence would entail\index{mixing!of $\eta$ and $\eta'$} mixing with the $\eta_8$ state to form two distinct physical states ($\eta$ and $\eta'$) and this scenario needs to be taken into account. One of the common approaches is to pretend that there is no axial\index{meson!$\eta$!mixing} anomaly\index{axial anomaly} and add $\eta_0$ to the matrix $\Pi$~(\ref{eq_2.3}) as an $SU(3)$ singlet:
\begin{equation}
\Pi = \begin{pmatrix}
    \frac{1}{\sqrt 6}\eta_8 + \frac{1}{\sqrt 3}\eta_0 + \frac{1}{\sqrt 2} \pi^0 & \pi^+ & K^+ \\
   \pi^- & \frac{1}{\sqrt 6}\eta_8 + \frac{1}{\sqrt 3}\eta_0 - \frac{1}{\sqrt 2} \pi^0 & K^0 \\
   K^- & \overline K^0 & -\sqrt{\frac{2}{3}}\eta_8 + \frac{1}{\sqrt 3}\eta_0
    \end{pmatrix}.
\end{equation}
The mixing\index{mixing!of $\eta$\index{meson!$\eta$!mixing} and $\eta'$} of $\eta_q$ and $\eta_0$ to form physical states can in principle involve other states (e.g. $\eta(1279)$ or even $\eta_c$), can depend on the energy of the state or can be influenced by the axial anomaly\index{axial anomaly}. Consequently the mixing scheme can be very complicated. In this thesis, we use the approach developed by Feldman et al.~\cite{Feldmann:1998vh}. There, the physical states $\eta$ and $\eta'$ can be written as linear combinations of $\eta_q$ and $\eta_s$: $\eta=\eta_q \cos\phi - \eta_s \sin_\phi$, $\eta' = \eta_q \sin \phi + \eta_s \cos \phi$, where $\eta_q$ has a $(u \bar u + d \bar d)/\sqrt 2$ flavor structure and $\eta_s$ is an $s \bar s$ state, while $\phi$ is the mixing\index{mixing!of $\eta$ and $\eta'$}\index{meson!$\eta$!mixing} angle.

\par

The decay\index{decay constant} constants of $\eta$ and $\eta'$ follow the same pattern of state mixing\index{mixing!of $\eta$ and $\eta'$}\index{meson!$\eta$!mixing}:
\begin{eqnarray}
f_{\eta}^q = f_q \cos \phi, && f_{\eta}^s = - f_q \sin \phi, \nonumber\\
f_{\eta'}^q = f_q \sin \phi, && f_{\eta'}^s = f_q \cos \phi,
\end{eqnarray}
with the decay\index{decay constant} constants defined as
\begin{eqnarray}
\bra{\eta(p)}\bar q \gamma_{\mu} \gamma_5 q \ket{0} = i f_{\eta}^q p_{\mu}, && \bra{\eta(p)}\bar s \gamma_{\mu} \gamma_5 s \ket{0} = i f_{\eta}^s p_{\mu},\nonumber\\
\bra{\eta'(p)}\bar q \gamma_{\mu} \gamma_5 q \ket{0} = i f_{\eta'}^q p_{\mu}, && \bra{\eta'(p)}\bar s \gamma_{\mu} \gamma_5 s \ket{0} = i f_{\eta'}^s p_{\mu}.
\end{eqnarray}
Due to the $SU(3)$ flavor symmetry breaking effects and the axial anomaly\index{axial anomaly} $f_q/f_s\neq 1$. In the first order of flavor symmetry breaking, it can be deduced that $f_q = f_\pi$, $f_s = \sqrt{2 f_K^2-f_{\pi}^2}$. Therefore, if the $\eta_0$ -- $\eta_{8}$ basis is used instead, two mixing angles\index{mixing!of $\eta$ and $\eta'$} rather than one are needed\index{meson!$\eta$!mixing}
$\eta = \eta_8 \cos \theta_8 - \eta_0 \sin \theta_0$, $\eta' = \eta_8 \sin \theta_8 + \theta_0 \cos \theta_0$.
The angles $\theta_0$ and $\theta_8$ are connected with $\phi$, $f_s$ and $f_d$ as
$\theta_8 = \phi - \arctan (\sqrt2 f_s/f_q)$, $\theta_0 = \phi - \arctan (\sqrt 2 f_q / f_s)$.

\par

The value of $\phi$ can be obtained phenomenologically from the various measured processes involving $\eta$ and $\eta'$ states. The value that fits the data best is $\phi=39.3^{\circ}$.

\section{Symmetries of heavy quarks\label{sec:2.2}}
\index{HQET}
\index{Heavy Quark Effective Theory|see{HQET}}
\index{QCD!with heavy quarks|see{HQET}}
\index{symmetries!of heavy quarks}

Since their early applications, symmetries of heavy quarks have been one of the key ingredients in the theoretical investigations of processes involving heavy quarks. They have been successfully applied to the heavy hadron spectroscopy, to the inclusive as well as a number of exclusive decays (for reviews of the heavy quark effective theory and related issues see~\cite{Neubert:1993mb, Manohar:2000dt}).
\par
The important observation here is that for heavy enough quarks, the effective strong coupling of QCD\index{QCD coupling|see{$\alpha_s$}}, due to its renormalization group running and asymptotic freedom, will be small at the mass scale of the heavy quark. This implies that on length scales, comparable to the compton wavelength $\lambda_Q \sim 1/m_Q$, the strong interactions are perturbative and much like the electromagnetic interactions. Furthermore, heavy quark spin participates in the strong interactions only through relativistic chromomagnetic effects. Since these vanish in the limit of the infinite quark mass, the spin of the heavy quarks decouples as well. The resulting theory therefore contains an approximate $SU(2)$ spin symmetry with the spin states of the heavy quark transforming in the fundamental representation. To arrive at this result formally, we start with the QCD Lagrangian\index{QCD!Lagrangian} for a single flavor $Q$ of heavy quarks
\begin{equation}
\mathcal L^{Q}_{\mathrm{heavy~quarks}} = \overline Q (i \slashed{D} - m_Q) Q\,.
\end{equation}
We separate the quark fields into their positive and negative frequency parts -- i.e. into "quark" and "anti-quark" fields $Q = Q^{(+)} + Q^{(-)}$, where $Q^{(+)}$ annihilates the $Q$ quark while $Q^{(-)}$ creates the corresponding anti-quark. For an infinitely heavy (anti)quark field travelling with velocity $v$, it is useful to rescale the ground state energy of the effective theory Fock space relative to the mass of the heavy (anti)quark in the frame of reference. This is done by factoring out the dominant kinetic phase factor $\exp (\pm i m_Q v\cdot x)$ from the (anti)quark fields $Q^{(\pm)}$. Then we further project out the large components of the heavy (anti)quark spinors using a velocity dependent projection operators\index{HQET projection operator} $P_{\pm} = (1\pm\slashed{v})/2$ to obtain the effective heavy (anti)quark fields $h^{\pm}_v(x) = P_{\pm} \exp (\pm i m_Q v\cdot x) Q^{(\pm)}(x)$. They satisfy $\slashed{v} h^{\pm}_v = \pm h^{\pm}_v$. We construct the HQET\index{HQET!Lagrangian} Lagrangian from QCD\index{QCD} by using the combined field $h_v = h_v^{(+)} + h_v^{(-)}$ while its orthogonal small (anti)quark spinor components $\widetilde h^{\pm}_v(x) = P_{\mp} \exp (\pm i m_Q v\cdot x) Q^{(\pm)}(x)$ can be integrated\index{integrating out!small components of heavy quark fields} out using their equations of motions~\cite{Wu:2005fd} or more elegantly via direct Gaussian path integration of the generating functional $Z[\rho_v]$, where external sources $\rho_v$ only couple to the $h_v$ fields and none to $\widetilde h_v$. Consequently $\widetilde h_v$ contribute only spin symmetry breaking corrections to the interactions among $h_v$ fields. They are proportional to the inverse powers of the heavy quark mass, yielding for the HQET\index{HQET!Lagrangian} Lagrangian\index{heavy quark expansion}
\begin{equation}
\mathcal L^{Q}_{\mathrm{HQET}} = \overline h_v (i v \cdot D) h_v + \mathcal O\left( 1/{m_Q} \right) + \mathcal L_{\mathrm{gauge}} + \mathcal L_{\mathrm{light~quarks}}\,.
\label{eq_2_8}
\end{equation}
The decoupling of heavy quark spin contributions in the leading term of eq. (\ref{eq_2_8}) is now intuitively manifest due to the absence of Dirac gamma matrices. Alternatively one can show, that it is invariant under the the generators of the $SU(2)$ transformations $S^i = \gamma_5 \slashed v \slashed \epsilon^i/2$, where $i=1,2,3$, $v\cdot \epsilon = 0$ and the heavy quark fields transform in the spinor representation $D(S)$ ($h_v \to D(S) h_v$, $D(S)^{-1} = \gamma_0 D(S)^{\dagger} \gamma_0$). Also at leading order in the expansion, there are no quark-antiquark couplings as it would take an infinite amount of energy (twice) to pair-produce infinitely heavy quarks. We can generalize the above arguments to $N_h$ flavors of heavy quarks ($c,b,\ldots$). Since in QCD\index{QCD} different quark flavors are only distinguished by their \index{QCD!Lagrangian}Lagrangian masses, for infinitely heavy quarks, QCD interactions become blind to the flavor of heavy quarks, exhibiting in total a $U(2N_h)$ spin-flavor symmetry. The HQET\index{HQET!Lagrangian} Lagrangian then becomes
\begin{equation}
\mathcal L_{\mathrm{HQET}} = \sum_{n=1}^{N_h}\overline h^{(n)}_v (i v \cdot D) h^{(n)}_v + \mathcal O\left( 1/{m_Q} \right) + \mathcal L_{\mathrm{gauge}} + \mathcal L_{\mathrm{light~quarks}}\,.
\label{eq_2_9}
\end{equation}
There are two important issues related to such HQET\index{HQET} formulation. Firstly, the choice of the heavy quark velocity to be factored out of the fields is arbitrary and we can formally get a separate independent set of quark fields for each choice. The result is sometimes called velocity superselection rule, and related to it is the heavy quark velocity reparametrization invariance. It simply states, that any shift in the velocity of the heavy quark by $v \to v + \epsilon/m_Q$, where $\epsilon$ satisfies $v \cdot \epsilon = 0$, can be accommodated by a corresponding redefinition of the heavy quark field $h_v \to \exp(i \epsilon \cdot x) (1 + \slashed{\epsilon}/2m_Q) h_v$.
Secondly, apart from the kinetic term in eq.~(\ref{eq_2_9}) whose normalization is fixed via velocity reparametrization invariance, any effective operators involving heavy quark fields in HQET have to be properly matched\index{matching!HQET to QCD} to the corresponding operators in full QCD\index{QCD}. Fortunately, due to the heavy mass scale, this matching can be performed perturbatively. As an example let us consider the heavy-light left-handed current operator (the case for the right-handed current proceeds identically)\index{heavy quark expansion}
\begin{equation}
J^{\mu}_{(V-A)\mathrm{QCD}} = \overline q_L \gamma^{\mu} Q.
\end{equation}
At the tree level the HQET\index{HQET!heavy-to-light current} current can be written as
\begin{equation}
J^{\mu}_{(V-A)\mathrm{HQET,~Tree}} = \overline q_L \gamma^{\mu} h_v + \mathcal O\left(1/{m_Q}\right).
\label{eq_2_11}
\end{equation}
Radiative corrections modify this result. The effective current operators present at the tree level are renormalized and additional operators are induced. Since in HQET\index{HQET} the heavy quark velocity $v$ is not a dynamical degree of freedom, the effective current operators can explicitly depend on it. The most general short-distance expansion of the vector current in the effective theory contains two operators of lowest dimension (three):\index{heavy quark expansion}
\begin{equation}
J^{\mu}_{(V-A)\mathrm{HQET}} = C_1(\nu) \overline q_L \gamma^{\mu} h_v + C_2(\nu) \overline q_L v^{\mu} h_v + \mathcal O\left(1/{m_Q}\right),
\label{eq_2_12}
\end{equation}
where $\nu$ is the regularization scale. After we have integrated out\index{integrating out!degrees of freedom} degrees of freedom when going from QCD\index{QCD} to HQET\index{HQET!matching to QCD}, the $C_{1,2}$ scale dependence reflects the non-trivial RG running of the effective theory operators (see section~\ref{sec:ope} for details). At the tree level the coefficients are $C_1= 1$ and $C_2= 0$, and one recovers eq.~(\ref{eq_2_11}). Explicit expressions for $C_i(\mu)$ at higher orders in $\alpha_s$\index{$\alpha_s$!corrections} are obtained from the comparison of the loop matrix elements of the currents in the full and in the effective theory. In addition to this, higher order power corrections in the $1/m_Q$ expansion may be considered where operators of higher dimensions in HQET\index{HQET!matching to QCD} are taken into account in the matching\index{matching} procedure.

\section{Combining heavy quark and chiral symmetries\label{sec:hmcpt}}
\index{HM$\chi$PT}
\index{Heavy Meson Chiral Perturbation Theory|see{HM$\chi$PT}}
\index{$\chi$PT!for heavy mesons|see{HM$\chi$PT}}
\index{symmetries!of heavy quarks}
\index{symmetries!of light quarks}
\index{chiral symmetry}

Heavy hadrons contain a heavy quark as well as light quarks and/or antiquarks and gluons (the heavy quark -- antiquark pairs being suppressed in the $m_Q\to\infty$ limit of HQET\index{HQET!static limit}). All the degrees of freedom other than the heavy quark are referred to as the light degrees of freedom $\ell$. The total angular momentum $\bf J$ is a conserved operator with eigenvalues ${\bf J}^2 = j(j+1)$. We have also seen that the spin of the heavy quark ${\bf S}_{Q}$ is conserved in the $m_Q\to\infty$ limit (we define its eigenvalues $s_{Q}$ through ${\bf S}_Q^2 = s_Q (s_Q+1)$). Therefore, the spin of the light degrees of freedom $\bf S_{\ell}$ defined by ${\bf S}_{\ell} \equiv {\bf J} - {\bf S}_Q$ is also conserved in the heavy quark limit (eigenvalues ${\bf S}_{\ell}^2 = s_{\ell} (s_{\ell}+1)$). Heavy hadrons come in doublets (unless $s_{\ell}=0$) containing states with the total spin $j_{\pm} = s_{\ell} \pm 1/2$ obtained by combining the spin of the light degrees of freedom with the spin of the heavy quark $s_Q=1/2$. These doublets are degenerate in the $m_Q\to\infty$ limit. Mesons containing a heavy quark $Q$ are made up of a heavy quark and a light antiquark $\bar q$ (plus gluons and $q \bar q$ pairs). The ground state mesons are composed of a heavy quark with $s_Q=1/2$ and light degrees of freedom with $s_{\ell}=1/2$ forming a multiplet of hadrons with spin $j=1/2 \otimes 1/2 = 0 \oplus 1$ and negative parity, since quarks and antiquarks have opposite intrinsic parity. These states are the $D$\index{meson!$D$} and $D^*$ mesons if $Q$ is a charm quark, and the $\overline B$ and $\overline B^*$ mesons if $Q$ is a $b$ quark. The field operators which annihilate these heavy quark mesons with velocity $v$ are denoted by $P_v^{(Q)}$ and $P_{v\mu}^{*(Q)}$ respectively, with $P_v^{*(Q)} \cdot v = 0$. Since these operators will mix\index{mixing!of operators} under the heavy quark spin transformations, it is convenient to collect them into a single tensor field operator transforming accordingly under the heavy quark spin and flavor symmetries\index{heavy quark symmetry} of the HQET\index{HQET!Lagrangian} Lagrangian~(\ref{eq_2_9}). Lorentz contractions with $\gamma_{\mu}$ and $\gamma_5$ convert vectors and pseudoscalars into bi-spinors so we can immediately identify the field $H_v^{(Q)}$ annihilating the ground state $Q\bar q$ (or better $h_v^{+}\bar q$) mesons as $H_v^{+} = P_{+} [ \slashed P_v^{*+} - P_v^{+} \gamma_5 ]$, where $P_{+}$ serves to project out the large particle components of the heavy quark $Q$ ($h_v^+$) while the factoring out of the large momentum phase factor is implicit. The expression for the creation field $\overline{H_v^{+}}= \gamma_0 H_v^{+\dagger} \gamma_0$ follows from its bi-spinor transformation properties under $CPT$\index{CPT}.  An analogous procedure can be performed for the mesons containing a heavy antiquark $\overline Q$ ($h_v^-$) leading to $H_v^{-} = P_{-} [ \slashed P_v^{*-} + P_v^{-} \gamma_5 ]$, which {\it creates} the corresponding particles. The apparent difference in the relative sign between the pseudoscalar and vector components is conventional and defined so that the particle creation and anti-particle annihilation fields (and vice-versa) appear with the same relative sign between vector and pseudoscalar operators. The only further fields needed for the remainder of the thesis will be for the lowest lying mesons of positive parity (scalar $P_0$ and axial-vector $P_{1\mu}^*$), which can be represented by $S_v^{\pm} = P_{\pm} [ \slashed P_{1v}^{*\pm}\gamma_5 \mp P_{0v}^{\pm} ]$\footnote{For a general treatment of hadronic states with higher spins see e.g.~\cite{Falk:1993iu,Falk:1991nq}}.
\par
Due to the peculiar construction of the hadron fields, the normalization of states in HQET\index{HQET!normalization of states} is different from that of full QCD\index{QCD}. Namely the standard relativistic normalization of hadronic states of mass dimension $-1$ (possible spin labels are suppressed)
\begin{equation}
\braket{H(p')|H(p)}_{QCD} = 2 E_{\bf p} (2\pi)^3 \delta^3({\bf p - p'})
\end{equation}
is modified to factor out any dependence on the mass of the heavy quark, while the states are labelled by their four-velocity
\begin{equation}
\braket{H(v')|H(v)}_{HQET} = 2 v^0 (2\pi)^3 \delta_{v v'}.
\end{equation}
States normalized by using this HQET\index{HQET!normalization of states} convention have mass dimension $-3/2$ and the two normalizations differ by a factor $\sqrt{m_H}$ as well as possible power corrections\index{heavy quark expansion}
\begin{equation}
\ket{H(p)}_{QCD} = \sqrt{m_H} \left [ \ket{H(v)}_{HQET} + \mathcal O\left( 1/{m_Q} \right) \right ].
\label{eq_2.15}
\end{equation}
\par
To take into considerations also the interactions with the pseudo-Goldstone\index{pseudo-Goldstone boson}\index{Goldstone boson|see{pseudo-Goldstone boson}} bosons due to the chiral\index{chiral dynamics} dynamics of the light antiquark ($u,~d,\ldots$) inside the heavy meson, these are factored out of the quark (and consequently hadron) fields. Under the chiral\index{chiral symmetry group} group $G$, $H_v$ and $S_v$ therefore transform as $H(S)^{+}_v \to H(S)^{+}_v U^{\dagger}$ and $H(S)^{-}_v \to U H(S)^{-}_v$. The most general effective \index{effective Lagrangian}Lagrangian containing positive and negative parity heavy mesons containing a heavy quark (we will drop the super-  and subscripts '$+$' and $v$ respectively and keep in mind, that an analogous Lagrangian can be written down for the heavy mesons containing a heavy antiquark, and that velocity reparametrization invariance connects different heavy quark velocity representations) to order $\mathcal O(p)$ in the chiral\index{chiral expansion} expansion and at leading order in the heavy quark mass expansion, that is invariant under heavy quark and chiral\index{chiral symmetry} symmetries, and is a Lorentz scalar is~\cite{Casalbuoni:1996pg,Manohar:2000dt}\index{heavy quark expansion}\index{HM$\chi$PT!parameters}
\begin{eqnarray}
    \mathcal L^{(1)}_{\mathrm{HM}\chi\mathrm{PT}} &=& \mathcal L^{(1)}_{\frac{1}{2}^-} + \mathcal L^{(1)}_{\frac{1}{2}^+} + \mathcal L^{(1)}_{\mathrm{mix}}, \nonumber\\
    \mathcal L^{(1)}_{\frac{1}{2}^-} &=& - \mathrm{Tr}\left[ \overline H_a (i v \cdot \mathcal{D}_{ab} - \delta_{ab} \Delta_H ) H_b\right] + g \mathrm{Tr} \left[ \overline H_b H_a \slashed{\mathcal A}_{ab} \gamma_{5} \right], \nonumber\\
    \mathcal L^{(1)}_{\frac{1}{2}^+} &=& \mathrm{Tr} \left[\overline S_a ( i v \cdot \mathcal{D}_{ab} - \delta_{ab} \Delta_S) S_b \right] + \widetilde g \mathrm{Tr} \left[ \overline S_b S_a \slashed{\mathcal A}_{ab} \gamma_{5}  \right], \nonumber\\
    \mathcal L^{(1)}_{\mathrm{mix}} &=& h \mathrm{Tr} \left[ \overline H_b S_a \slashed{\mathcal A}_{ab} \gamma_{5}  \right] + \mathrm{h.c.}.
    \label{eq_2_13}
\end{eqnarray}
h.c. denotes an additional Hermitian conjugate term, $\mathcal{D}^{\mu}_{ab} = \delta_{ab}\partial^{\mu} -\mathcal{V}^{\mu}_{ab}$ is the chiral\index{chiral covariance} covariant heavy meson derivative\index{covariant derivative!in $HM\chi$PT}, while the trace Tr runs over Dirac indices. Chiral vector ($\mathcal V$) and axial-vector ($\mathcal A$) pseudo-Goldstone\index{pseudo-Goldstone boson} current operators have been defined in section~\ref{sec_2_2}, while $g$, $h$ and $\widetilde g$\index{HM$\chi$PT!parameters} are three unknown effective couplings between heavy and light mesons.
The $\Delta_H$ and $\Delta_S$ are the so-called residual masses of the $H$ and $S$ fields respectively. In a theory with only $H$ {\it or} $S$ fields, one is free to set $\Delta_H=0$ ($\Delta_S=0$) since all loop\index{loop divergencies} divergences are cancelled by $\mathcal O(p^2)$ (counter)terms at zero order in heavy quark expansion. However, once both fields are added to the theory, another dimensionful quantity $\Delta_{SH} = \Delta_S - \Delta_H$  enters calculations and does not vanish in the chiral\index{chiral limit} and heavy quark limits~\cite{Mehen:2005hc}. It is of the order $\mathcal O(p^0)$ in the standard chiral\index{chiral power counting} power counting and is usually it is accounted for via the appropriate pole offset in the heavy meson propagators. Although one is free to offset both positive and negative parity heavy meson poles, the end results of any calculation with well defines mass-shell conditions will only depend on the difference of both quantities. Alternatively, one could also boost the heavy mesons of different parities to different velocities so as to factor out this additional scale e.g.
$H(v') = \exp{(i \Delta_{H} v \cdot x)}H(v)$ with $v'=v - \Delta_{H}/m_Q$, and $S(v'') = \exp{(i \Delta_{S} v \cdot x)}S(v)$ with $v''=v - \Delta_{S}/m_Q$. In this case however the splitting reappears in the form of a phase factor difference in $\mathcal L^{(1)}_{\mathrm{mix}}$ in eq.~(\ref{eq_2_13}). In terms of the momentum space Feynman rules this corresponds to modification of the momentum conservation at the $H S \pi$ vertex inducing at leading order in $1/m_H$ a new $\Delta_{SH}/f$  coupling between positive and negative parity mesons and pseudo-Goldstone\index{pseudo-Goldstone boson} bosons (in addition to new $1/m_H$ corrections). We see immediately that if $\Delta_{SH}$  is comparable or larger than $f$, this leads to a strongly coupled theory. In addition it is not suppressed\index{chiral suppression} by powers of pseudo-Goldstone\index{pseudo-Goldstone boson} momenta (it has chiral\index{chiral power counting} power counting zero) and therefore spoils the chiral\index{chiral limit} limit of the theory. However, the physical content of both theory representations is the same, meaning that at any perturbation order in both heavy quark expansion, chiral\index{chiral power counting} counting {\it and} $\Delta_{SH}$ , both Feynman rule sets yield the same results for any Green's function\index{Green's function}. The main difference is that in the first case we are actually able to re-sum all $\Delta_{SH}$\index{HM$\chi$PT!parameters}  contributions to all orders by solving the free field theory (including the extra $\Delta_H$ and $\Delta_S$ terms) exactly and obtaining the free heavy field propagators. This will prove to be of major importance in our calculations, there we adopt this approach throughout this work (for the list of derived Feynman rules used, see Appendix~A)\index{HM$\chi$PT!Feynman rules}.
\par
In the same way as sketched in the previous paragraph we must consider bosonization\index{bosonization!of weak current} of HQET\index{HQET!heavy-to-light current} currents and more general operators that appear in electro-weak processes. Again we chose eq.~(\ref{eq_2_12}) as an example. At the effective hadronic level of HM$\chi$PT\index{HM$\chi$PT} we must construct all operators consistent with transformation properties (with respect to chiral\index{chiral symmetry}, heavy quark\index{heavy quark symmetry} and Lorentz symmetries) of the two effective HQET current operators up to the given order in the chiral\index{chiral expansion} and heavy quark expansions (formally this is done by inserting the same external spurion\index{spurion} currents into generating functionals of both theories). At the $\mathcal O(p^0)$ order in the chiral\index{chiral power counting} counting and at the leading order in the heavy quark expansion the effective current containing a single positive or negative parity heavy mesons simply reads\index{heavy quark expansion}\index{HM$\chi$PT!matching to HQET}
\begin{equation}
J^{(0)\mu}_{(V-A)\mathrm{HM}\chi\mathrm{PT}} = \frac{i \alpha}{2} \mathrm{Tr} [ \gamma^{\mu}(1-\gamma_5) H_b] \xi_{ba}^{\dagger} - \frac{i \alpha'}{2} \mathrm{Tr} [ \gamma^{\mu}(1-\gamma_5) S_b] \xi_{ba}^{\dagger} + \mathcal O \left( 1/{m_Q} \right),
\label{eq_2_17}
\end{equation}
where $\alpha$ and $\alpha'$\index{HM$\chi$PT!parameters} are unknown parameters which can be matched\index{matching} to the heavy meson decay\index{decay constant} constants. Note that at this order there exists only one distinct operator for each parity because any insertions of the heavy meson velocity $v$ can be reduced to this form by the use of the heavy meson velocity projection identities $\slashed v H(S) = H(S)$ and $H(S) \slashed v = - H(S)$. In fact any general structure heavy-to-light current $\bar q \Gamma Q$, where $\Gamma=1,\gamma_5,\gamma_{\mu} \gamma_5 \gamma_{\mu}, \sigma_{\mu\nu}$ can be translated into the effective bosonized form\index{bosonization!of weak current} in the same manner~\cite{Manohar:2000dt, Wise:1992hn}.

\chapter[Hadronic amplitudes]{Hadronic amplitudes -- effective approaches and resonances}

In this chapter we will briefly review some standard methods used in the phenomenology of weak interactions of hadronic systems. Here we can consider as ''weak'' all possible interactions apart from QCD\index{QCD}, which may contribute significantly to quark dynamics at high enough energies but are almost completely swapped by strong interactions which confine quarks into hadrons at energies well below the electroweak scale. In addition to the prototype weak interactions of the electroweak SM, these may include contributions from possible new physics beyond the SM\index{new physics}. The methods of OPE\index{OPE} allow us to integrate out\index{integrating out!degrees of freedom} all degrees of freedom not directly associated with the external hadronic states and split the problem into a perturbative calculation of all short distance contributions using asymptotically free quarks on one hand, and an essentially nonperturbative calculation of hadronic matrix elements of operators, which however now contain only light degrees of freedom of QCD\index{QCD}. We will also briefly touch upon some of the general properties, approximations and relations among these hadronic amplitudes.

\section{Operator product expansion\label{sec:ope}}
\index{OPE}
\index{Wilsonian expansion|see{OPE}}

In this section we will briefly review the ideas behind OPE\index{OPE} and its application to weak interactions. The original idea dates back to Wilson~\cite{Wilson:1969zs}, who conjectured that the singular part (as $x\to y$) of the product $A(x)B(y)$ of two operators is given by a sum over other local operators
\begin{equation}
A(x) B(y) \xrightarrow[x\to y]{} \sum_{n} C_{n}^{AB}(x-y) \mathcal O_n (y),
\label{eq_3_1}
\end{equation}
where $C_n^{AB}(x-y)$ are singular c-number functions. Dimensional analysis\index{dimensional analysis} suggests that $C_n^{AB}(x-y)$ behaves fox $x\to y$ like the power $d_{\mathcal O_n} - d_A - d_B$ of $x-y$, where $d_O$ is the dimensionality of the operator $O$ in powers of mass or momentum. Since $d_O$ increases as we add more fields or derivatives to an operator $O$, the strength of the singularity of $F^{AB}_n$ decreases for operators $\mathcal O_n$ of increasing complexity, making their contributions to the sum~(\ref{eq_3_1}) less and less relevant. The simple power counting argument is modified slightly by quantum effects in the renormalization group treatment, where anomalous dimensions\index{anomalous dimension} of operators come into play. Another remarkable property of eq.~(\ref{eq_3_1}) is that it is an operator\index{operator relation} relation: it holds regardless of what the states it acts on are. The OPE\index{OPE} in general reads
\begin{equation}
T\{A_1(x_1) A_2(x_2)\ldots A_k(x_k) \xrightarrow[x_i\to x]{} \sum_{n} C_{n}^{A_1\ldots A_k}(x-x_1,\ldots, x-x_k) \mathcal O_n (x),
\end{equation}
with $T$ being the time ordering operator\index{operator ordering}.

\par

The application of the OPE\index{OPE} to weak interactions comes from the observation that the distances at which weak interactions occur are set by the mass of the intermediate $W$ and $Z$ bosons, i.e., $x-y \sim 1/m_W$. If one is interested in the processes at energy scales $\mu$ much smaller than the weak scale ($\mu\ll m_W$), or in other words, in the processes effectively occurring at typical distances $1/\mu$ that are much larger than $x-y\sim 1/m_W$, we can take the limit $x\to y$ (or equivalently $m_W\to \infty$) and use the OPE\index{OPE}.

\par

Formally we consider the generating functional of correlation functions and we focus on the relevant integration over the $W$ and $Z$ degrees of freedom. The charged current part of the action then contributes e.g.~\cite{Buras:1998ra}
\begin{equation}
Z_W \sim \int [\mathrm{d} W^+] [\mathrm{d} W^-] \mathrm{e}^{i \int d^4 x \mathcal L_W},
\end{equation}
where
\begin{equation}
\mathcal L_W = -\frac{1}{2} \left( \partial_{\mu} W^+_{\nu} - \partial_{\nu} W^+_{\mu} \right)\left( \partial^{\mu} W^{-\nu} - \partial^{\nu} W^{-\mu} \right) + m^2_W W^+_{\mu} W^{-\mu} + \frac{g_2}{2\sqrt2}\left( J_{\mu}^+ W^{+\mu} + J_{\mu}^- W^{-\mu} \right),
\end{equation}
with $J^+_{\mu} = (\bar u d')_{V-A} + (\bar c s')_{V-A} + (\bar t b')_{V-A} + (\bar \nu_e e)_{V-A} + (\bar \nu_{\mu} \mu)_{V-A} + (\bar \nu_{\tau} \tau)_{V-A}$,
and $J^-_{\mu} = (J^+_{\mu})^{\dagger}$. $q'_i = V_{ij}^{CKM} q_j$\index{CKM} are the rotated weak states and $g_2$ is the weak isospin coupling constant. We use the unitary gauge\index{unitary gauge} for the $W$ field and introduce $K_{\mu\nu}(x,y) = \delta^{(4)}(x-y) [ g_{\mu\nu} (\partial^2 + m_W^2) -\partial_{\mu} \partial_{\nu} ]$. After discarding a total derivative in the $W$ kinetic term we have
\begin{equation}
Z_W \sim \int [\mathrm{d} W^+] [\mathrm{d} W^-] \mathrm{e}^{i \int d^4 x d^4 y W^+_{\mu}(x) K^{\mu\nu}(x,y)W^-_{\nu}(y) + i\frac{g_2}{2\sqrt2} \int \mathrm{d}^4 x \left( J_{\mu}^+ W^{+\mu} + J_{\mu}^- W^{-\mu} \right)}.
\end{equation}
Performing a Gaussian functional integration over $W^{\pm}(x)$ explicitly, we arrive at
\begin{equation}
Z_W \sim \mathrm{e}^{- i \frac{g^2_2}{8} \int d^4 x d^4 y \left( J_{\mu}^-(x) \Delta^{\mu\nu}(x-y) J_{\mu}^+(y) \right)},
\label{eq_3.6}
\end{equation}
where $\Delta_{\mu\nu}(x)$ is the $W$ propagator in the unitary gauge\index{unitary gauge}. This result implies a nonlocal action functional for the quarks which we can expand in powers of $1/m_W^2$ to obtain a series of local interaction operators of dimensions that increase with the order in $1/m_W^2$. To lowest order $\Delta_{\mu\nu}(x) \approx \delta^{(4)}(x) g^{\mu\nu}/m_W^2$ and the effective action in eq.~(\ref{eq_3.6}) becomes
\begin{equation}
-\frac{g^2_2}{8m_W^2}\int d^4 x J^-_{\mu} J^{+\mu},
\end{equation}
corresponding to the usual effective charged current interaction Hamiltonian of the Fermi\index{Fermi theory} theory\index{G$_F$}
\begin{equation}
\mathcal H_{\mathrm{eff}} = -\frac{G_F}{\sqrt 2} C_2 \mathcal O_2,
\label{eq_3.9}
\end{equation}
where the definition of the Fermi\index{Fermi constant|see{G$_F$}}\index{G$_F$} constant $G_F/\sqrt 2 = g_2^2/8m_W^2$ has been used and a local four-quark operator\index{four quark operator} $\mathcal O_2=J^-_{\mu} J^{+\mu}$ with a Wilson coefficient $C_2=1$ has been defined. Actually, the effective Hamiltonian~(\ref{eq_3.9}) is valid only in the absence QCD\index{QCD corrections} interactions. Once these are taken into account, another four-quark operator\index{four quark operator} appears in the OPE\index{OPE} $\mathcal O_1 = J^-_{\alpha\beta\mu} J^{+\mu}_{\beta\alpha}$, where summation over the color indices $\alpha$, $\beta$ of current quarks is understood. Formally we are integrating out\index{integrating out!degrees of freedom} all degrees of freedom at scales equal or larger than $m_W$, including hard (energetic) gluons, while practically, due to the asymptotic freedom of QCD\index{asymptotic freedom, in QCD}, the corrections can be computed perturbatively by considering all the relevant correlation functions with free quarks in the asymptotic states. The effective weak Hamiltonian\index{effective weak Hamiltonian} is then\index{G$_F$}
\begin{equation}
\mathcal H_{\mathrm{eff}} = -\frac{G_F}{\sqrt 2} \left( C_1 \mathcal O_1 + C_2 \mathcal O_2 \right),
\label{eq_3.10}
\end{equation}
where the coefficient $C_1$ is proportional to $\alpha_s$\index{$\alpha_s$}, while $C_2\sim1$ is nonzero already at tree level in the perturbative QCD\index{QCD corrections} expansion as discussed above.
A similar procedure can be employed to integrate out\index{integrating out!heavy quarks} any possible heavy quarks in which case the current operators $\mathcal O_{1,2}$ in eq.~(\ref{eq_3.10}) appear in the properly reduced form containing only the light quark fields. Finally, we may also consider neutral current weak processes including flavor changing neutral current (FCNC)\index{FCNC} processes in which case a number of new operators appears in the OPE\index{OPE} corresponding to tree-level, penguin\index{penguin diagrams} and box\index{box diagrams} diagram contributions in the original theory.

\par

In the calculation of the Wilson coefficients\index{Wilson coefficient} typically expressions of the form $\alpha_s \mathrm{ln}(\mu/m_W)$\index{$\alpha_s$!corrections} appear, where $\mu$ is a typical scale at which we want to study the processes. In the case of $c$\index{decay of $c$ quark} and $b$ quark decays\index{decay of $b$ quark}, these are typically of the order of a few GeV. Thus, the ratio of scales in the argument of the logarithm can be very large, of order 100, and consequently the factor $\alpha_s \mathrm{ln}(\mu/m_W)$\index{$\alpha_s$!corrections} of the order $\mathcal O(1)$. Even though the QCD\index{$\alpha_s$} coupling $\alpha_s$\index{$\alpha_s$} is not terribly large at the heavy quark scales and could be used as a perturbative expansion parameter, the appearance of large logarithms prevents the straightforward application of perturbation theory. All large logarithms of the form $[\alpha_s \mathrm{ln} (\mu/m_W)]^n$\index{$\alpha_s$!corrections} have to be summed up using renormalization group equations. This is done by again considering correlation functions $\braket{\mathcal O_i}$ of operators appearing in the OPE\index{OPE} both in the effective and in full theory. $\braket{\mathcal O_i}$ do not depend on the renormalization, whereas even after accounting for the renormalization of the quark fields in the original theory, due to the different UV structure of the effective theory, the OPE\index{OPE} expressions have to be multiplicatively renormalized. In terms of the unrenormalized Wilson coefficients\index{Wilson coefficient} $C_i^{(0)}$ and operators $\mathcal O_i^{(0)}$, of which neither depend on $\mu$, the renormalization condition can be written as
\begin{equation}
C_i^{(0)}\mathcal O_i^{(0)} = C_i(\mu)Z_{ij}^{-1}(\mu) Z_{ji}(\mu)\mathcal O_i(\mu),
\end{equation}
where the scale dependence of both the renormalized Wilson coefficients\index{Wilson coefficient} and the operators is fully determined by the renormalization matrix $Z_{ij}(\mu)$. In a compact form we can write
\begin{equation}
\frac{dC_i}{d\mathrm{ln}\mu} = \gamma_{ij} C_j, \quad \gamma = Z^{-1} \frac{dZ}{d\mathrm{ln}\mu},
\label{eq_3.11}
\end{equation}
where we have introduced the anomalous dimension\index{anomalous dimension} matrix $\gamma$, which we determine be identifying the leftover singularities (or equivalently the logarithmic $\mu$ dependence as in eq.~(\ref{eq_3.11}), but for the operators $\mathcal O_i$) in the process of matching\index{matching!OPE to QCD} Green's\index{Green's function} functions in both the original and the effective (OPE\index{OPE}) theory. Using the evolution of the QCD coupling constant $g_S$ (in the $\overline{\mathrm{MS}}$\index{renormalization scheme, $\overline{\mathrm{MS}}$} renormalization scheme) and their expansion
\begin{eqnarray}
\frac{d g_S(\mu)}{d\ln \mu} = \beta(g_S) &=& -\beta_0 \frac{g_S^3}{16\pi^2} +\ldots,\\
\gamma(\alpha_s) &=&  \gamma^{(0)} \frac{\alpha_s}{4\pi} +\ldots,
\end{eqnarray}
where $\alpha_s=g_S^2/4\pi$\index{$\alpha_s$} and $\beta_0 = (11 N_c- 2 N_f)/3$ for $N_f$ active flavors and $N_c$ colors, the evolution equation~(\ref{eq_3.11}) can be solved to any given order.  As an example we consider the leading order renormalization group (RG\index{RGE}\index{renormalization group equations|see{RGE}}) evolution of a single Wilson coefficient\index{Wilson coefficient}
\begin{equation}
C(\mu) = \left[\frac{\alpha_s(m_W)}{\alpha_s(\mu)}  \right]^{\gamma^{0}/2\beta_0} C(m_W).
\label{eq_3.12}
\end{equation}
The strong coupling constant $\alpha_s(\mu)$\index{$\alpha_s$} appearing in~(\ref{eq_3.12}) is at the one-loop order
\begin{equation}
\alpha_s(\mu) = \frac{4\pi}{\beta_0 \ln(\mu^2/\Lambda^2_{QCD})},
\end{equation}
where $\Lambda_{QCD}$ is the QCD\index{QCD scale} scale (the value of which depends on the number of active flavors). Using the precisely measured value of the strong coupling constant at the $Z$\index{gauge boson!$Z$} boson mass, $\alpha_s(m_Z)=0.1172\pm0.002$\index{$\alpha_s$!value} one arrives at $\Lambda_{QCD}^{(5)}=216\pm25~\mathrm{MeV}$, while for $\mu<m_b$ with four active flavors, the matching\index{matching} at $m_b=4.25~\mathrm{GeV}$ gives $\Lambda_{QCD}^{(4)}=311\pm33~\mathrm{MeV}$. The important observation about eq.~(\ref{eq_3.12}) is that it contains all the terms of the form $[\alpha_s\ln(\mu/m_W)]^n$\index{$\alpha_s$} as has been announced at the beginning of the paragraph.

\section{Vacuum saturation and resonance dominance approximations}
\index{VSA}
\index{Vacuum Saturation Approximation|see{VSA}}
\index{resonance dominance approximation}

As discussed in the previous section, the weak interactions can be described at low energies by means of an effective Lagrangian\index{effective weak Hamiltonian} obtained through the OPE\index{OPE} and the RGE\index{RGE}. The Wilson coefficients\index{Wilson coefficient} $C_i$ contain contributions from hard gluon exchanges and can be calculated perturbatively as described in the previous section. They are scale and at NLO also renormalization scheme dependent. This dependence is canceled by the scale and renormalization scheme dependence of the local four-quark operators\index{four quark operator} $\mathcal O_i$. The matrix elements in the hadronic weak transitions\index{G$_F$}
\begin{equation}
\mathcal M_{fi} = \frac{G_F}{\sqrt 2} \sum_i C_i \bra{f}\mathcal O_i \ket{i},
\end{equation}
are thus scale and scheme independent. The nonperturbative nature of these transitions is hidden in the matrix elements $\bra{f}\mathcal O_i \ket{i}$ between hadronic final and initial states. Evaluation of these elements is a very hard problem and lies at the core of all the difficulties connected with the weak transitions between hadronic states. Currently the best way to estimate them is to calculate them on the lattice\index{lattice QCD}. However the problem is so involved, especially for the heavy-to-light\index{transition!heavy-to-light} hadron transitions, that even the ''exact'' calculations on the lattice have to resort to a number of approximations. One of such phenomenologically and theoretically motivated approximations is the very simple but extremely useful vacuum saturation (or complete factorization)\index{factorization approximation|see{VSA}} approximation (VSA). It comes in when the currents appearing in the operators $\mathcal O_i$, which are proportional to interpolating stable or quasistable hadronic fields, are approximated by asymptotically free hadronic fields in the ''in'' and ''out'' states. Whence the currents are assumed to factor completely. Formally this is achieved by rewriting the time-ordered products of the interpolating fields and OPE operators in terms of commutators, then inserting a full set of states in-between the commutators and finally discarding all but the vacuum. I.e. taking the operator $\mathcal O_i = J^a_i \otimes J^b_i$ one obtains
\begin{equation}
\bra{f}\mathcal O_i \ket{i} \propto \sum_{\mathrm{perm.}\{a,b\}}\sum_{n} \bra{f} J^a_i \ket{n}\otimes\bra{n} J^b_i \ket{i} \to \sum_{\mathrm{perm.}\{a,b\}}\bra{f} J^a_i \ket{0}\otimes\bra{0} J^b_i \ket{i},
\end{equation}
where the sum over $\mathrm{perm.}\{a,b\}$ denotes all the possible (distinct) ways of inserting the full set of states (see e.g. section 3.3.2 in ref.~\cite{Prelovsek:2000rj} for details). Due to the rather ad-hoc nature of this procedure, the information on the sizes of the individual factorized current contributions is in principle lost and the effective Wilson coefficients\index{Wilson coefficient} multiplying these contributions have to be estimated from experimental data or alternatively inferred from complementary theoretical approaches such as $1/N_c$ expansion. Such breakdown of productivity of this approach is already signalled by the lack of $\mu$ dependence of the factorized current matrix elements, which therefore by themselves cannot cancel the renormalization dependence of the original Wilson coefficients\index{Wilson coefficient}.

\par

The relevance of VSA can actually be extended beyond the association of factorized currents with single asymptotically free hadronic fields in the final state by the use of resonance\index{resonance dominance approximation} dominance approximation. This approximation has considerable phenomenological vindication in hadron physics at energies less then about $1~\mathrm{GeV}$~\cite{Donoghue:1992dd} where it was first proposed in the form of {\it vector dominance}. It states that all the main dynamical effects of hadronic long-distance (final state) strong interactions are associated with exchange of intermediate resonances. The idea is based on a ``polology'' theorem (See chapter 10.2 of~\cite{Weinberg:1995mt}) stating that the poles and cuts in the configuration manifold of any correlation function can be associated with propagation of virtual intermediate composite quasi-stable single and multi-particle states coupling to asymptotic states. The resonance\index{resonance dominance approximation} dominance approximation saturates the correlation function with contributions from a few of the relevant phenomenologically determined resonances. In connection with the VSA, the asymptotic final state field configurations are coupled to intermediate resonances using effective models or phenomenological Lagrangians. These resonances then saturate the operator matrix elements as asymptotically free hadronic fields. Formally we write
\begin{equation}
\bra{f}\mathcal O_i \ket{i} \to \sum_{n} \bra{f} \mathcal L_{\mathrm{eff}} \ket{n} \otimes G_n \otimes  \bra{n} \mathcal O_i  \ket{i},
\label{eq_3.18}
\end{equation}
where we have denoted the propagation of intermediate resonance states with $G_n$ and $\mathcal L_{\mathrm{eff}}$ contains the effective vertex coupling resonant and final states. We see that the accuracy of this approach relies on the number of phenomenological resonances we consider before truncating the sum in eq.~(\ref{eq_3.18}) as well as on the calculation of the effective vertexes $\bra{f} \mathcal L_{\mathrm{eff}} \ket{n}$ coupling these resonances to final states. A preferred approach here is to employ effective theories based on symmetries  of QCD\index{symmetries!of QCD} such as (HM)$\chi$PT\index{HM$\chi$PT}\index{$\chi$PT}. We can then map the quantum numbers of the resonances onto dynamical fields in the effective theory or conversely introduce the appropriate external resonance currents into the effective theory.

\section{Parameterization of hadronic amplitudes}
\index{amplitude parameterization}
\index{amplitude factorization|see{VSA}}

In this section we will briefly review some general properties of
(hadronic) matrix elements of operators, which we encounter in the
OPE\index{OPE} as well as in other approaches describing processes of
hadrons. We will focus on the matrix elements entering the two-body
leptonic, three-body semileptonic, two- and three-body nonleptonic
decays of pseudoscalar mesons as well as mixing of neutral
pseudoscalar mesons with their anti-particles. We will mostly
consider pseudoscalar mesons in the initial state as they are always
the lowest lying states with a given single quark and anti-quark
flavor quantum numbers\footnote{This is due to the opposite
intrinsic parities of particles and anti-particles, complemented by
parity conservation of QCD\index{QCD!parity conservation}.}, and in case of open flavors, where the
quark and the anti-quark are of different flavor, cannot decay\index{strong decay}
strongly or electromagnetically due to flavor conservation of QCD\index{QCD!flavor conservation}
and QED\index{QED flavor conservation}. Thus they open a window to the underlying weak dynamics.

\par

General matrix elements of operators between initial and final particle states are generalized functions of the particle degrees of freedom and can always be decomposed into generalized scalar functions of Lorentz invariants multiplying available Lorentz structures. A simple example is the matrix element of the unit operator between two pseudoscalar and a vector state. The pseudoscalar states can be uniquely labeled by their four-momenta (and any additional internal quantum numbers) and we denote them by $P_1(p_1)$ and $P_2(p_2)$. The vector state can in term be labeled by its momentum and polarization\index{polarization of vector meson} $V(P_V,\epsilon)$. Due to Lorentz invariance and in presence of parity conservation the matrix element can be reduced to a scalar c-number parameter multiplying a Lorentz invariant label\index{amplitude parameterization}
\begin{equation}
\braket{P_1(p_1) P_2(p_2) | V(p_V,\epsilon)}= G_{P_1 P_2 V} \epsilon\cdot p_1,
\label{eq_strong_v}
\end{equation}
where $g_{P_1 P_2 V}$ is an effective on-shell vertex. Note that there is an ambiguity in labeling the momentum $p_1$ contributing to the amplitude via $\epsilon\cdot p_1$ since both pseudoscalar states are equivalent. We shall therefore impose the convention of always taking the momentum of the lighter or the two pseudoscalars. Also since $\epsilon \cdot p_V=0$ on-shell as well as due to Lorentz momentum conservation $P_V=p_1+p_2$, actually the $\epsilon\cdot (p_1-p_2)$ structure contributes to the decay  rate. However such different definitions of the vertex are all simply related by fixed normalization factors due to the decomposition $p_1 = 1/2 (p_1+p_2) + 1/2 (p_1-p_2)$ and the vector state transversality condition. The choice only becomes relevant when considering effective approaches such as vector resonance\index{resonance dominance approximation} dominance approximation~(\ref{eq_3.18}) where the vector state $V(P_V,\epsilon)$ may be intermediate and off-shell. In such cases our choice in eq.~(\ref{eq_strong_v}) turns out to be beneficial.

\par

In weak leptonic decays\index{amplitude parameterization} of pseudoscalar $P(p)$ and vector
$V(p,\epsilon)$ mesons, the hadronic
amplitudes comprise of matrix elements of weak currents between the
mesonic states and the vacuum and can be parameterized in terms of
these Lorentz covariants multiplying c-number parameters  -- decay
\index{decay constant}constants. We will define them as
\begin{subequations}
\begin{eqnarray}
\label{eq_3_19}
\bra{0}J_{\mu} \ket{P(p)} &=& i f_P p_{\mu}, \\
\bra{0}J_{\mu} \ket{V(p,\epsilon)} &=& f_{V} m_{V} \epsilon_{\mu}.
\label{eq_3_19b}
\end{eqnarray}
\end{subequations}
First note that Lorentz invariance of the amplitudes projects out
the vector, or axial current components, depending on the parity of
the initial states, and secondly that transversality condition of
the on-shell vector states ($\epsilon \cdot p=0$) together with
Lorentz invariance prevents a term proportional to $p_{\mu}$ in the
second line of eq.~(\ref{eq_3_19}). This can be most easily seen in
the rest frame of the vector meson. In this frame the components of
the four-vector current factorize into its three-vector
(proportional to $\epsilon$) and three-scalar (proportional to the
time component of $p$) parts (referring here to the three spatial
dimensions).
We see immediately that only the three-vector part of the current
can couple to the on-shell vector state, thus projecting out the
term proportional to the meson momentum.

\par

In weak semileptonic decays\index{amplitude parameterization}, the hadronic part of the transition
amplitude is described by the weak quark current matrix element
between initial and final hadronic states. Again if these states
comprise of single pseudoscalar mesons (e.g. $P_i$ and $P_f$), the
$P_i\to P_f$ current matrix element can then be
parameterized in terms of the appropriate Lorentz covariants made
from momenta $p_i$ and $p_f$ reproducing the Lorentz structure of the
current, multiplied by form factors\index{form factor} -- scalar functions of the
Lorentz invariant (Mandelstam variable) $s=(p_i-p_f)^2$ -- the
exchanged momentum squared. Parity of the external states also
projects out the axial component of the current, so only the vector
part contributes and we can write~\cite{Marshak:1969tw}\index{form factor!decomposition}
\begin{equation}
\bra{ P_f (p_f) } J_{V}^{\mu} \ket{ P_i (p_i)} = F_+(s) (p_i+p_f)^{\mu} + F_{-}(s) (p_i-p_f)^{\mu},
\label{def-ff_0}
\end{equation}
where $F_{\pm}(s)$ are the two form factors. The physical region for $s$ is defined by $m_{\ell}^2\leq s \leq (m_{P_i}-m_{P_f})^2$, where $m_{\ell}$ is the invariant mass of the final state leptons. From the previous section we recall that analytic structure of the matrix element can be identified with the propagation of virtual intermediate single and multiparticle states. In our case, these states when on-shell will contribute poles and cuts in the complex $s$ Riemann sheet of both form factors. Let us try to analyze these contributions in more detail and consider the matrix element obtained from the above by crossing the final state $P_f$
\begin{equation}
\bra{ 0 } J_{V}^{\mu} \ket{ P_i (p_i) P_f (p_f)} = F_+(t) (p_i-p_f)^{\mu} + F_{-}(t) (p_i+p_f)^{\mu},
\label{def-ff_1}
\end{equation}
where $t=(p_i+p_f)^2$. One convenient way of classifying the various contributions to the crossection is via their spin or angular momentum properties. For this purpose we go to the
center of mass (c.m.) frame of the $P_i P_f$ system. There we have
${\bf p_i} + {\bf p_f} = 0$ and we now expand the state into spherical
harmonics
\begin{equation}
\ket{ P_i ({\bf p_i}) P_f (-{\bf p_i})} = \sum_{L,m} Y^m_L({\bf p/|\bf p|}) \ket{ P_i P_f ({\bf p},L,m)},
\label{eq_3.22}
\end{equation}
where $Y^m_L(\bf p/|\bf p|)$ and $\ket{ P_i P_f ({ p},L,m)}$ represent a state of $P_i$ and  $P_f$ with c.m. three-momentum $\bf p$, total angular momentum $L$ and its third component $m$. Since the current $J_V$ contains a 3-dimensional vector and a 3-dimensional scalar, the only intermediate states contributing in the sum on the r.h.s of eq.~(\ref{eq_3.22}) are necessarily states with $J^P=1^-$ or $0^+$. Thus
\begin{equation}
\bra{ 0 } J_{V}^{\mu} \ket{ P_i (p_i) P_f (p_f)} = \bra{ 0 } J_{V}^{\mu} \ket{ P_i P_f ({\bf p},0,0)} + \sum_m \bra{ 0 } J_{V}^{\mu} \ket{ P_i P_f ({\bf p},1,m)} Y^m_1(\bf p/|\bf p|).
\label{eq_3.23}
\end{equation}
But the first term on the r.h.s. of eq.~(\ref{eq_3.23}) only gets contributions from the time component of the current and so vanishes except when $\mu = 0$ while the second term is analogously nonzero only for $\mu=1,~2,~3$. Thus, noting that in the c.m. frame $t=(p_i^0+p_f^0)^2$ and $p_i^0-p_f^0 = (m_{P_i}^2-m_{P_f}^2)/\sqrt t$ we obtain from eq.~(\ref{def-ff_1}) for $\mu=0$
\begin{equation}
\bra{ 0 } J_{V}^{0} \ket{ P_i P_f (p,0,0)} = \sqrt t \left[  \frac{m_{P_i}^2-m_{P_f}^2}{t} F_+(t) + F_-(t)\right] = \sqrt t F_0(t),
\label{eq_3.24}
\end{equation}
and by taking $\mu=1,~2,~3$ we obtain
\begin{equation}
\sum_m \bra{ 0 } {{\bf J}_{V}} \ket{ P_i P_f (p,1,m)} Y^m_1({\bf p/|\bf p|}) = 2 {\bf p_i} F_+(t).
\label{eq_3.25}
\end{equation}
From eqs.~(\ref{eq_3.24}) and~(\ref{eq_3.25}), it is clear that
$F_+$ receives pole contributions from the intermediate on-shell
$J^P=1^-$ states (it is therefore called the vector form factor) and
$F_0$ receives contributions from states with $J^P=0^+$ (and is thus
termed the scalar form factor). Note that this decomposition only
refers to intermediate propagation of on-shell states which
satisfy the usual scalar invariance and vector transversality
conditions. Nonetheless this allows us to identify and distinguish
pole and cut contributions to both of the form factors\index{form factor!$P\to P$}. Also note
that in order for this new matrix element decomposition\index{form factor!decomposition}
\begin{eqnarray}
\bra{ P_f (p_f) } J_{V-A}^{\mu} \ket{ P_i (p_i)} &=& F_+(s) \left( (p_i+p_f)^{\mu} -
    \frac{m_{P_i}^2-m_{P_f}^2}{s}(p_i-p_f)^{\mu} \right) \nonumber\\*
    &&+ F_0(s) \frac{m_{P_i}^2-m_{P_f}^2}{s}(p_i-p_f)^{\mu},
\label{def-ff}
\end{eqnarray}
be finite at $s=0$, the form factors\index{form factor!kinematic relations} must satisfy the kinematic relation
\begin{equation}
F_+(0)=F_0(0)~.
\label{PP_form_factor_relations}
\end{equation}

\par

Another advantage of using form factors as defined in
eq.~(\ref{def-ff}) is that if one neglects the charged lepton mass
(reasonable approximation in case of electrons and arguably also
muons), the scalar current component does not couple to the two
chiral leptons in the final state. Thus only $F_+$ contributes to
the total decay width which may in this case be simply written
as~\cite{Bajc:1995km}
\begin{equation}
\Gamma = \frac{|C|^2 m_H^2 }{24 \pi^3} \int_0^{y_m^P} \mathrm{d}y  | F_+(m_{P_i}^2 y) |^2
|{\bf p}_f (y)|^3,
\label{eq_gamma_HP}
\end{equation}
where $C$ is the appropriate Wilson coefficient\index{Wilson coefficient} from the effective weak Hamiltonian\index{effective weak Hamiltonian}, $y=s/m_{P_i}^2$, $y_m^P = (1-m_{P_f}/m_{P_i})^2$ and the
three-momentum of the final state meson is given by
\begin{equation}
|{\bf p}_f (y)|^2 = \frac{[m_{P_i}^2 (1-y) + m_{P_f}^2]^2}{4
m_{P_i}^2} -m_{P_f}^2.
\end{equation}

\par

A similar decomposition can be done for the current matrix elements relevant to semileptonic decays between a pseudoscalar meson state $\ket{P(p_P)}$ with momentum $p_P^{\nu}$ and a vector meson state $\ket{V(p_V,\epsilon)}$ with momentum $p_V^{\nu}$ and polarization\index{polarization of vector meson} vector $\epsilon^{\nu}$ arriving at\index{form factor!$P\to V$}
\begin{eqnarray}
    \bra{ V (\epsilon,p_V) } J^{\mu}_{V} \ket{P (p_P)} &=& \frac{2 V(s)}{m_P + m_V} \varepsilon^{\mu\nu\alpha\beta} \epsilon_{\nu}^* p_{P\alpha} p_{V\beta}, \nonumber\\*
    \bra{ V (\epsilon,p_V) } J^{\mu}_{A} \ket{P (p_P)}  &=&  - i \epsilon^* \cdot (p_P-p_V) \frac{2m_V}{s} (p_P-p_V)^{\mu} A_0(s) \nonumber\\*
    &&- i(m_P + m_V) \left[\epsilon^{*\mu} - \frac{\epsilon^{*} \cdot (p_P-p_V)}{s} (p_P-p_V)^{\mu}\right] A_1(s) \nonumber\\*
    &&+ i \frac{\epsilon^{*}\cdot (p_P-p_V)}{(m_P + m_V)}\left[(p_P+p_V)^{\mu} - \frac{m_P^2-m_V^2}{s} (p_P-p_V)^{\mu}\right] A_2(s).\nonumber\\
\label{eq:PtoV}
\end{eqnarray}
First note that $\epsilon \cdot p_V=0$ and thus only the projection $\epsilon \cdot p_P$ contributes in the expressions when the final state vector meson is on shell. However, when using effective approaches such as vector meson dominance approximation, the $V(\epsilon,p_V)$ may denote an off-shell intermediate state and keeping the full $\epsilon^{*}\cdot (p_P-p_V)$ dependence in the form factor definition turns out to be beneficial. Then $V$ denotes the vector form factor and receives pole and cut contributions from intermediate vector states, the axial $A_1$ and $A_2$ form factors contain axial state contributions, while $A_0$ denotes the pseudoscalar form factor and is populated by pseudoscalar states~\cite{Wirbel:1985ji}.
In order that these matrix elements are finite at $s=0$, the form factors\index{form factor!kinematic relations} must also satisfy the well known relation
\begin{equation}
A_0(0)-\frac{m_P+m_V}{2m_V}A_1(0)+\frac{m_P-m_V}{2m_V}A_2(0)=0\,.
\label{PV_form_factor_relations}
\end{equation}

\par

In $P\to V \ell \nu$ decays  it is sometimes convenient to introduce helicity amplitudes\index{helicity amplitudes}~\cite{Ball:1991bs}\footnote{In refs.~\cite{Fajfer:2005ug, Fajfer:2006uy} there is a typo in the last term of the second line, where an additional factor of $|{\bf p}_V(y)|$ is missing. I am grateful to Damir Be\'cirevi\'c for bringing it to my attention.}:
\begin{eqnarray}
H_{\pm}(y) &=&  (m_P+m_V) A_1(m_P^2 y) \mp \frac{2 m_P |{\bf p}_V(y)|}{m_P+m_V} V(m_P^2 y),\nonumber\\*
H_0(y) &=& \frac{m_P+m_V}{2 m_P m_V \sqrt y} [ m_P^2 (1-y) -m_V^2] A_1(m_P^2 y) - \frac{2 m_P |{\bf p}_V(y)|^2}{m_V(m_P+m_V) \sqrt y } A_2(m_P^2 y),\nonumber\\
\label{eq_helicity}
\end{eqnarray}
where as before $y=s/m_P^2$ and the three-momentum of the final
state vector meson is given by:
\begin{equation}
|{\bf p}_V (y)|^2 = \frac{[m_P^2 (1-y) + m_V^2]^2}{4 m_P^2} -m_V^2.
\end{equation}
In the approximation where one neglects the lepton masses, the decay
rates for the polarized\index{polarization of vector meson} final vector mesons are then simply
proportional to~\cite{Bajc:1995km}:
\begin{equation}
\Gamma_a = \frac{|C|^2 m_H^2 }{96 \pi^3} \int_0^{y_m^V} y \mathrm{d}y  | H_a( y) |^2 |{\bf p}_V (y)|,
\label{eq_gamma_HV}
\end{equation}
where $a=+,-,0$ and $y_m^V = \left(1-{m_V}/{m_P} \right)^2$.

\chapter{\label{ch_strong}Strong decays of heavy mesons}
\index{strong decay!of charmed mesons} \index{decay of $D$ meson}
\index{meson!$D$!decay} \index{chiral corrections!in strong decays}
\index{HM$\chi$PT!in strong decays}

\index{extrapolation!chiral|see{chiral extrapolation}}

Discoveries of resonances\index{resonance!of charmed meson} in the
spectrum of open charm hadrons have stimulated many studies. The
measured properties of the $D_{0}^*$ and $D_1'$ states support their
interpretation as belonging to the $(0^+,1^+)$ heavy quark
spin-parity multiplet of $c \bar u$ and $c \bar d$ mesons.
Conversely, the $D_{sJ}$ states have been proposed as members of the
$(0^+,1^+)$ spin-parity doublet of $c \bar s$
mesons~\cite{Bardeen:2003kt,Nowak:2004uv}. The strong and
electromagnetic transitions of these new states have been studied
within a variety of
approaches~\cite{Bardeen:2003kt,Colangelo:1995ph,Colangelo:1997rp,Colangelo:2003vg,Colangelo:2005gb,Mehen:2004uj,Nielsen:2005zr,
Lu:2006ry, Wang:2006bs,Wang:2006fg, Wang:2006id}. In these
investigations HM$\chi$PT\index{HM$\chi$PT!in strong decays} at
leading order was used as well in attempts to explain the observed
strong and electromagnetic decay
rates~\cite{Colangelo:2003vg,Colangelo:2005gb,Mehen:2004uj}.

\par

In ref.~\cite{Stewart:1998ke} the chiral \index{loop corrections!in
strong decays} loop corrections to the $D^* \to D \pi$ and $D* \to D
\gamma$ decays\index{decay!$D^* \to D \pi$}\index{decay!$D^* \to D
\gamma$} were calculated and a numerical extraction of the one-loop
bare couplings was first performed. Since this calculation preceded
the discovery of even-parity meson states, it did not involve loop
contributions containing the even-parity meson states. The ratios of
the radiative and strong decay widths, and the isospin violating
decay\index{decay!$D_s^* \to D_s \pi^0$} $D_s^* \to D_s \pi^0$ were
used to extract the relevant couplings. However, since that time,
the experimental situation has improved and therefore we consider
the chiral  loop contributions to the strong decays\index{strong
decay} of both the even and odd parity charmed meson states using
HM$\chi$PT\index{HM$\chi$PT!in strong decays}. In our calculation we
consider the strong decay modes given in table~\ref{table_input}.
The existing data on the decay widths enable us to constrain the
leading order parameters: the $D^* D\pi$ coupling $g$, $D^{*}_0 D
\pi$ coupling $h$, and the coupling $\widetilde g$ which enters in
the interaction of even parity charmed mesons and the light
pseudo-Goldstone\index{pseudo-Goldstone boson} bosons in the
HM$\chi$PT\index{HM$\chi$PT!Lagrangian} Lagrangian~(\ref{eq_2_13}).
Although the coupling $\widetilde g$ is not yet experimentally
constrained, we will see, that it moderatelly affects the decay
amplitudes which we consider.

\par

In the work presented in~\cite{Mehen:2005hc}, the next to leading
terms ($1/m_H$) were included in the study of charm meson mass
spectrum. Due to the very large number of unknown couplings the
combination of $1/m_H$ and chiral\index{chiral corrections!in strong
decays} corrections does not seem to be possible for the decay modes
we consider here. Also, recent lattice QCD studies\index{lattice
QCD!in strong decays}~\cite{Abada:2003un,McNeile:2004rf} of the
strong couplings of heavy mesons have noticed that $1/m_H$
corrections seem not to be very significant but pointed out the
importance of controlling chiral loop corrections.

\par

The precise knowledge of the effective strong couplings in the
leading order HM$\chi$PT\index{HM$\chi$PT!Lagrangian}
Lagrangian~(\ref{eq_2_13}) is essential for theoretical calculations
of heavy meson weak processes within HM$\chi$PT\index{HM$\chi$PT!in
strong decays} as they enter in all the chiral loop corrections to
any HM$\chi$PT effective operator. Currently the most reliable
method of estimating hadronic matrix elements are the numerical
lattice QCD\index{lattice QCD!in strong decays} simulations. Due to
the increase of simulation time, when approaching the
chiral\index{chiral limit!in strong decays} limit,
lattice\index{lattice QCD!in strong decays} studies use large values
of the light quark masses. To make their results physically
relevant, they need to extrapolate\index{extrapolation!of lattice
QCD data} them to the physical (basically chiral) \index{chiral
limit!in strong decays}limit. This extrapolation induces systematic
uncertainties that are hard to control as the spontaneous
chiral\index{chiral symmetry breaking} symmetry breaking effects are
expected to become increasingly pronounced as one lowers the light
quark mass~\cite{Yamada:2001xp,Kronfeld:2002ab,Becirevic:2002mh}.
HM$\chi$PT allows us to gain some control over these uncertainties
because it predicts the chiral behavior of the hadronic quantities
relevant to the heavy-light quark phenomenology which then can be
implemented to guide the extrapolation of the lattice results. In
HM$\chi$PT\index{HM$\chi$PT!in strong decays} one computes the
chiral logarithmic corrections (the so-called non-analytic terms)
which are expected to be relevant to the very low energy region,
i.e., $m_q \ll \Lambda_{\rm QCD}$. While this condition is satisfied
for  $u$- and $d$-quarks, the situation with the $s$-quark is still
unclear~\cite{Descotes-Genon:2003cg,Descotes-Genon:2002yv}.  Also
ambiguous is the size of the chiral\index{chiral symmetry breaking
scale} symmetry breaking scale, $\Lambda_\chi$. Some authors
consider it to be around $4\pi f_\pi \simeq
1$~GeV~\cite{Manohar:1983md}, while the others prefer identifying it
with the mass of the first vector resonance,
$m_\rho=0.77$~GeV~\cite{Pich:1995bw,Gasser:1984gg}, and sometimes
even lower. In the heavy-light quark systems the situation becomes
more complicated because the first orbital excitations
($j_\ell^P=1/2^+$) are not far away from the lowest lying states
($j_\ell^P=1/2^-$).  The recent experimental evidence for the scalar
$D_{0s}^{\ast}$ and axial $D_{1s}$ mesons indicate that this
splitting is only $\Delta_{S_s} \equiv
m_{D_{0s}^{\ast}}-m_{D_s}=350$~MeV~\cite{Aubert:2003fg,Vaandering:2004ix,Besson:2003jp,Abe:2003jk},
and somewhat larger for the non-strange states
$\Delta_{S_q}=430(30)$~MeV~\cite{Abe:2003zm,
Link:2003bd}.~\footnote{We note, in passing, that the experimentally
established fact that $\Delta_{Ss} < \Delta_{Sq}$ is not yet
understood~\cite{Colangelo:2004vu,Swanson:2006st,Godfrey:2005ww,Becirevic:2004uv}
although a recent lattice\index{lattice QCD!in strong decays} study
with the domain wall quarks indicates a qualitative agreement with
experiment~\cite{Lin:2006vc}.  }  This and the result of the
lattice\index{lattice QCD!in strong decays} QCD study in the
static\index{HQET!static limit} heavy quark
limit~\cite{Green:2003zz} suggest that the size of this mass
difference remains as such in the $b$ quark sector as well. One
immediately observes that both  $\Delta_{S_s}$ and  $\Delta_{S_q}$
are  smaller than $\Lambda_\chi$, $m_\eta$, and even $m_K$, which
requires revisiting the predictions based on
HM$\chi$PT\index{HM$\chi$PT!in strong decays}. In this chapter we
therefore study systematically the effects of the positive parity
states' contributions on the chiral extrapolation of strong decay
amplitudes of the ground state heavy mesons.

\section{Heavy quark and chiral expansion}
\index{chiral expansion!in strong decays} \index{heavy quark
expansion!in strong decays}

In our calculation we employ the leading order HM$\chi$PT
Lagrangian~(\ref{eq_2_13})\index{HM$\chi$PT!Lagrangian} containing
both positive and negative parity heavy meson doublets. From it we
derive the Feynman rules, which can be found in
Appendix~A\index{HM$\chi$PT!Feynman rules}. Due to the divergences
coming from the chiral loops one needs to include the appropriate
counterterms\index{counterterms!in strong decays} -- interaction
terms of higher order in the chiral\index{chiral expansion!in strong
decays} expansion, which will only contribute tree level terms in
our calculations. Therefore we construct a full operator
 basis of the relevant counterterms\index{counterterms!in strong decays} and include it into our effective theory Lagrangian\index{HM$\chi$PT!Lagrangian}: following refs.~\cite{Stewart:1998ke,Boyd:1994pa}, we absorb the infinite and scale dependent pieces from one loop amplitudes into the appropriate counterterms\index{counterterms!in strong decays} at order $\mathcal O (m_q)$
\begin{eqnarray}
    \mathcal L^{\mathrm{ct}} &=& \mathcal L_{\frac{1}{2}^-}^{\mathrm{ct}} + \mathcal L_{\frac{1}{2}^+}^{\mathrm{ct}} + \mathcal L_{\mathrm{mix}}^{\mathrm{ct}}, \nonumber\\
    \mathcal{L}_{\frac{1}{2}^-}^{\mathrm{ct}} &=& \lambda_1 \mathrm{Tr}\left[ \overline H_b H_a   (m_q^{\xi})_{ba}\right]  + \lambda'_1 \mathrm{Tr}\left[ \overline H_a H_a (m_q^{\xi})_{bb}\right] + \frac{g \kappa_1}{\Lambda^2_{\chi}} \mathrm{Tr}\left[ (\overline H H \slashed{\mathcal A} \gamma_5)_{ab}(m_q^{\xi})_{ba} \right]  \nonumber\\
    &&\hspace{-35pt} + \frac{g \kappa_3}{\Lambda^2_{\chi}} \mathrm{Tr}\left[ (\overline H H \slashed{\mathcal A} \gamma_5)_{aa}(m_q^{\xi})_{bb} \right] + \frac{g \kappa_5}{\Lambda^2_{\chi}} \mathrm{Tr}\left[ \overline H_a H_a \slashed{\mathcal A}_{bc} \gamma_5 (m_q^{\xi})_{cb} \right] + \frac{g \kappa_9}{\Lambda^2_{\chi}} \mathrm{Tr}\left[ \overline H_c H_a (m_q^{\xi})_{ab} \slashed{\mathcal A}_{bc} \gamma_5 \right] \nonumber\\
    &&\hspace{-35pt} + \frac{\delta_2}{\Lambda_{\chi}} \mathrm{Tr}\left[ \overline H_a H_b i v \cdot \mathcal D_{bc} \slashed{\mathcal A}_{ca} \gamma_5 \right] + \frac{\delta_3}{\Lambda_{\chi}} \mathrm{Tr}\left[ \overline H_a H_b i \slashed{ \mathcal D}_{bc} v \cdot \mathcal A_{ca} \gamma_5 \right] + \ldots,\nonumber\\
    \mathcal{L}_{\frac{1}{2}^+}^{\mathrm{ct}} &=& - \widetilde \lambda_1 \mathrm{Tr}\left[  S_a \overline S_b (m_q^{\xi})_{ba}\right]  - \widetilde \lambda'_1 \mathrm{Tr}\left[ S_a \overline S_a (m_q^{\xi})_{bb} \right]  + \frac{\widetilde g \widetilde \kappa_1}{\Lambda^2_{\chi}} \mathrm{Tr}\left[ (\overline S S \slashed{\mathcal A} \gamma_5)_{ab}(m_q^{\xi})_{ba} \right] \nonumber\\
    &&\hspace{-35pt} + \frac{\widetilde g \widetilde \kappa_3}{\Lambda^2_{\chi}} \mathrm{Tr}\left[ (\overline S S \slashed{\mathcal A} \gamma_5)_{aa}(m_q^{\xi})_{bb} \right]  + \frac{\widetilde g \widetilde \kappa_5}{\Lambda^2_{\chi}} \mathrm{Tr}\left[ \overline S_a S_a \slashed{\mathcal A}_{bc} \gamma_5 (m_q^{\xi})_{cb} \right] + \frac{\widetilde g \widetilde \kappa_9}{\Lambda^2_{\chi}} \mathrm{Tr}\left[ \overline S_c S_a (m_q^{\xi})_{ab} \slashed{\mathcal A}_{bc} \gamma_5 \right] \nonumber\\
    &&\hspace{-35pt} + \frac{\widetilde \delta_2}{\Lambda_{\chi}} \mathrm{Tr}\left[ \overline S_a S_b i v \cdot \mathcal D_{bc} \slashed{\mathcal A}_{ca} \gamma_5 \right] + \frac{\widetilde \delta_3}{\Lambda_{\chi}} \mathrm{Tr}\left[ \overline S_a S_b i \slashed{ \mathcal D}_{bc} v \cdot \mathcal A_{ca} \gamma_5 \right]  + \ldots,\nonumber
\end{eqnarray}
\begin{eqnarray}
    \mathcal L_{\mathrm{mix}}^{\mathrm{ct}} &=& \frac{h \kappa'_1}{\Lambda^2_{\chi}} \mathrm{Tr}\left[ (\overline H S \slashed{\mathcal A} \gamma_5)_{ab}(m_q^{\xi})_{ba} \right]  + \frac{h \kappa'_3}{\Lambda^2_{\chi}} \mathrm{Tr}\left[ (\overline H S \slashed{\mathcal A} \gamma_5)_{aa}(m_q^{\xi})_{bb} \right] \nonumber\\
    &&\hspace{-35pt} + \frac{h \kappa'_5}{\Lambda^2_{\chi}} \mathrm{Tr}\left[ \overline H_a S_a \slashed{\mathcal A}_{bc} \gamma_5 (m_q^{\xi})_{cb} \right] + \frac{h \kappa'_9}{\Lambda^2_{\chi}} \mathrm{Tr}\left[ \overline H_c S_a (m_q^{\xi})_{ab} \slashed{\mathcal A}_{bc} \gamma_5 \right] \nonumber\\
    &&\hspace{-35pt} + \frac{ \delta'_2}{\Lambda_{\chi}} \mathrm{Tr}\left[ \overline H_a S_b i v \cdot \mathcal D_{bc} \slashed{\mathcal A}_{ca} \gamma_5 \right] + \frac{ \delta'_3}{\Lambda_{\chi}} \mathrm{Tr}\left[ \overline H_a S_b i \slashed{ \mathcal D}_{bc} v \cdot \mathcal A_{ca} \gamma_5 \right]  + \mathrm{~h.c.~} + \ldots.\nonumber\\
    \label{eq_L_1}
\end{eqnarray}
Here $m_q^{\xi} = (\xi m_q \xi + \xi^{\dagger} m_q \xi^{\dagger})$,
$\mathcal{D}^{\mu}_{ab} \mathcal A^{\nu}_{bc} = \partial^{\mu}
\mathcal A^{\nu}_{ac} + [ \mathcal{V}^{\mu},\mathcal A^{\nu}]_{ab}$
is the covariant derivative acting on the pseudo-Goldstone
boson\index{pseudo-Goldstone boson} fields and $\Lambda_{\chi}\simeq
4\pi f$ is the effective chiral\index{chiral symmetry breaking
scale} symmetry breaking scale and the cutoff of the effective
theory. Ellipses denote terms contributing only to processes with
more then one pseudo-Goldstone\index{pseudo-Goldstone boson} boson
as well as terms with $(iv\cdot \mathcal D)$ acting on $H$ or $S$,
which do not contribute at this order: since they only enter at tree
level in our calculations, they can be integrated\index{integrating
out} out using their equations of motion~\cite{Stewart:1998ke}. In
the chiral \index{chiral power counting}power counting ($m_q \sim
p^2$) all the $\lambda$ terms in $\mathcal L^{\mathrm{ct}}$ are of
the order $\mathcal O(p^2)$ while the $\delta$ and $\kappa$ terms
are of the order $\mathcal O(p^3)$. Parameters $\lambda'_1$ and
$\widetilde \lambda'_1$ can be absorbed into the definition of heavy
meson masses by a phase redefinition of $H$ and $S$, while
$\lambda_1$ and $\widetilde \lambda_1$ split the masses of $SU(3)$
flavor triplets of $H_a$ and $S_a$, inducing residual mass terms in
heavy meson propagators: $\Delta_a = 2\lambda_1 m_a$ and $\widetilde
\Delta_a = 2\widetilde \lambda_1 m_a$
respectively~\cite{Stewart:1998ke}. As with $\Delta_{H(S)}$, only
differences between these  $\mathcal O(p^2)$ residual mass terms
enter our expressions. We denote them as $\Delta_{ba}=\Delta_b -
\Delta_a$, $\Delta_{\widetilde b a}=\widetilde \Delta_b - \Delta_a$
and $\widetilde \Delta_{ba}=\widetilde \Delta_b - \widetilde
\Delta_a$.
Note that their contributions to the non-analytic (logarithmic)
pseudo-Goldstone\index{pseudo-Goldstone boson} mass dependencies of
the amplitudes will be of higher order in the chiral\index{chiral
power counting} counting and can there be neglected. For the
$\kappa_1$ and $\kappa_9$\index{HM$\chi$PT!parameters} terms only
the combination $\kappa_{19} = \kappa_1 + \kappa_9$ will enter in an
isospin conserving manner here~\cite{Stewart:1998ke} (the $\kappa_1
- \kappa_9$ combination contributes to isospin violating $D_s^*\to
D_s\pi^0$ decay\index{decay!$D_s^* \to D_s \pi^0$}, which we do not
consider). In the same manner we only consider contributions of
$\kappa'_{19} = \kappa'_1 + \kappa'_9$ and $\widetilde \kappa_{19} =
\widetilde \kappa_1 + \widetilde \kappa_9$. At any fixed value of
$m_q$, the finite parts of $\kappa_3$, $\widetilde \kappa_3$ and
$\kappa'_3$ can be absorbed into the definitions of $g$, $\widetilde
g$ and $h$ respectively~\cite{Stewart:1998ke}. However, one needs to
keep in mind that these terms introduce a non-trivial mass
dependence on the couplings when chiral extrapolation is considered.
The $\delta_2$ and $\delta_3$ enter in a fixed linear combination,
introducing momentum dependence into the definition of $g$ of the
form $g-(\delta_2+\delta_3) v\cdot k / \Lambda_{\chi}$. For decays
with comparable outgoing pseudo-Goldstone \index{pseudo-Goldstone
boson}energy, this contribution cannot be disentangled from that of
$g$~\cite{Stewart:1998ke}. On the other hand, these contributions
have to be considered when combining processes with different
outgoing pseudo-Goldstone\index{pseudo-Goldstone boson} momenta. The
same holds for contributions of $\widetilde \delta_2$ and
$\widetilde \delta_3$ with respect to $\widetilde g$, as well as
$\delta'_2$ \index{HM$\chi$PT!parameters}and $\delta'_3$ with
respect to $h$. At order $\mathcal O(m_q)$ we are thus left with
explicit analytic contributions from $\kappa_5$, $\kappa_9$,
$\widetilde \kappa_5$, $\widetilde \kappa_{19}$, $\kappa'_5$,
$\kappa'_{19}$, $\delta_2+\delta_3$,$\widetilde \delta_2 +
\widetilde \delta'_3$ and $\delta'_2+\delta'_3$.

\section{Chiral corrections including excited states}
\index{chiral corrections!in strong decays}

\subsection{Wave-function renormalization}
\index{heavy meson wave-function renormalization}

We first calculate the wave-function renormalization $Z_{2H}$ of the
heavy $H=P,~P^*$ and $P_0$, $P^*_1$ fields. This is done by
calculating the heavy meson self-energy $\Pi(v\cdot p)$, where $p$
is the residual heavy meson momentum, and using the prescription
\begin{equation}
Z_{2H} =  1 - \frac{1}{2}\frac{\partial \Pi(v\cdot p)}{\partial
v\cdot p}\Big |_{\mathrm{on~mass-shell}},
\end{equation}
where the mass-shell condition is different for the $H$ and $S$
fields due to their residual mass terms $\Delta_{H(S)}$ as well as
for the different light quark flavors due to the $\Delta_a$ terms.
In general it evaluates to $v\cdot p - \Delta_{H(S)}-\Delta_a=0$.

\par

\begin{figure}
\begin{center}
\epsfxsize5cm\epsffile{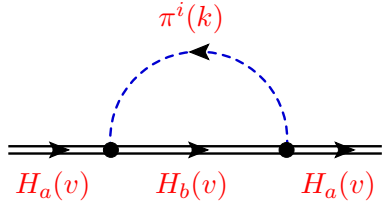}
\end{center}
\caption{\small\it \label{diagram_sunrise}"Sunrise topology" diagram
contributing to heavy meson wave-function renormalization. The
double line indicates the heavy-light meson and the dashed one the
pseudo-Goldstone\index{pseudo-Goldstone boson} boson propagator. The
full dot is proportional to the effective strong coupling.}
\end{figure}

At the $\mathcal O(p^2)$ power counting order we get non-zero
contributions to the heavy meson wave-function renormalization from
the self energy ("sunrise") topology diagrams in
fig.~\ref{diagram_sunrise} with leading order couplings in the loop.
\psfrag{pijl}[bc]{$\Red{ \pi^i(q)}$}
\psfrag{Ha}[cc]{${\Red{H_a(v)}}$} \psfrag{Hb}[cc]{$\Red{H_b(v)}$}
\psfrag{pi}[bc]{$\Red {\pi^i(k)}$}
In the case of the $P$ mesons both vector $P^*$ and scalar $P_0$
mesons can contribute in the loop yielding for the wave-function
renormalization coefficient
\begin{equation}
Z_{2P_a} = 1 - \frac{\lambda^i_{ab}\lambda^i_{ba}}{16\pi^2 f^2}
\left [ 3 g^2 C'_1\left(\frac{\Delta_{ba}}{m_i},m_i\right) - h^2
C'\left(\frac{\Delta_{\widetilde b a} + \Delta_{SH}}{m_i}
,m_i\right)\right]. \label{eq_Z_2_P}
\end{equation}
As in ref.~\cite{Stewart:1998ke}, a trace is assumed over the inner
repeated index(es) (here $b$) throughout the text, while the
loop\index{loop functions} functions $C_i$ and their analytic
properties are defined in the Appendix~B. In the chiral\index{chiral
power counting} power counting scheme, their non-analytic
$\Delta_{ba}$ dependence is of the order $\mathcal O(p^4\log p)$ or
higher and can be neglected at this order. However $\Delta_{ba}$
also enter analytically and we have to check for sensitivity of our
results to these parameters. On the other hand the
chiral\index{chiral power counting} power counting of
$\Delta_{SH}\sim p^0$\index{chiral power counting} leads anomalous
contributions to $C_i$ and will cause problems when employing these
formulae for chiral extrapolation. We shall deal with this problem
in the next sections. At leading order in heavy quark expansion, due
to heavy quark spin symmetry\index{heavy quark symmetry!in strong
decays}, the wave-function renormalization coefficient for the $P^*$
field is identical to that of $P$ although it gets contributions
from three different sunrise diagrams with states $P$, $P^*$ and
$P_1^*$ in the loops \index{loop corrections!in strong
decays}~\cite{Cheng:1993kp}.

\par

The positive parity $P_0$ and $P^*_1$ obtain wave-function
renormalization contributions from self energy diagrams
(fig.~\ref{diagram_sunrise}) with $P^*_1$, $P$ and $P_0$, $P^*_1$,
$P^*$ mesons in the loops respectively, which yield identically
\begin{equation}
Z_{2P_{0a}} = 1 - \frac{\lambda^i_{ab}\lambda^i_{ba}}{16\pi^2 f^2}
\left[ 3 \widetilde g^2 C'_1\left(\frac{\widetilde
\Delta_{ba}}{m_i},m_i\right) - h^2 C'\left(\frac{\Delta_{b\widetilde
a} - \Delta_{SH}}{m_i},m_i\right) \right]. \label{eq_Z_S}
\end{equation}

\subsection{Vertex corrections}
\index{heavy meson vertex corrections} \index{resonance
contribution!in chiral loops} \index{resonance contribution!in
strong decays} \index{chiral corrections!in strong decays}
\index{heavy quark expansion!in heavy-to-light decays}

Next we calculate loop corrections for the $PP^*\pi$, $P_0 P^*_1
\pi$ and $P_0 P\pi$ vertices. At zeroth order in $1/m_Q$ expansion
these are identical to the $P^*P^*\pi$, $P^*_1 P^*_1 \pi$ and
$P^*_1P^*\pi$ couplings respectively due to heavy quark spin
symmetry\index{heavy quark symmetry!in strong decays}. Again we
define vertex renormalization factors for on-shell initial and final
heavy and light meson fields. Specifically, for the vertex
correction amplitude $\Gamma(v\cdot p_i, v\cdot p_f, k_{\pi}^2)$
with heavy meson residual momentum conservation condition $p_f = p_i
+ k_{\pi}$ one can write the renormalization coefficient $Z_{1 H_i
H_f \pi}$ schematically
\begin{equation}
Z_{1 H_i H_f \pi} = 1 - \frac{\Gamma(v\cdot p_i, v\cdot p_f,
k_{\pi}^2)}{\Gamma_{l.o.}(v \cdot p_i,v\cdot
p_f,k_{\pi}^2)}\Big|_{\mathrm{on~mass-shell}}.
\end{equation}
Here $\Gamma_{l.o.}$ is the tree level vertex amplitude, $p_{i(f)}$
is the residual momentum of the initial (final) state heavy meson
$H_{i(f)}$, while $k_{\pi}$ is the pseudo-Goldstone
momentum\index{pseudo-Goldstone boson}. This implies that in the
heavy quark limit, one must evaluate the above expression at
$k_{\pi}^2 = m_{\pi}^2$, $v\cdot p_{i(f)} = \Delta_{i(f)}$ and
consequently $v\cdot k_{\pi} = \Delta_{fi} = \Delta_f - \Delta_i$,
where $\Delta_{i(f)}$ is the residual mass of the initial (final)
heavy meson fields. Such prescription
ensures that in all expressions one only encounters the physical,
re-parametrization invariant quantities $\Delta_{SH}$ ,
$\Delta_{ab}$, $\Delta_{\widetilde a b}$ and $\widetilde
\Delta_{ab}$~\footnote{This is different from the prescription in
ref.~\cite{Stewart:1998ke}, where $v\cdot k_{\pi}$ entering loop
calculations was evaluated as the physical pion energy. However, for
the processes considered there, the discrepancy between the two
prescriptions is only of the order of a few percent due to the small
hyperfine splitting between the relevant initial and final state
mesons.}. Satisfying both conditions should systematically be
possible at any order in the $1/m_Q$ expansion by properly
accounting for the effective coupling of heavy meson fields at
different velocities as in $B \to D$ meson transitions (see
e.g.~\cite{Manohar:2000dt}).

\par

\begin{figure}
\begin{center}
\epsfxsize9cm\epsffile{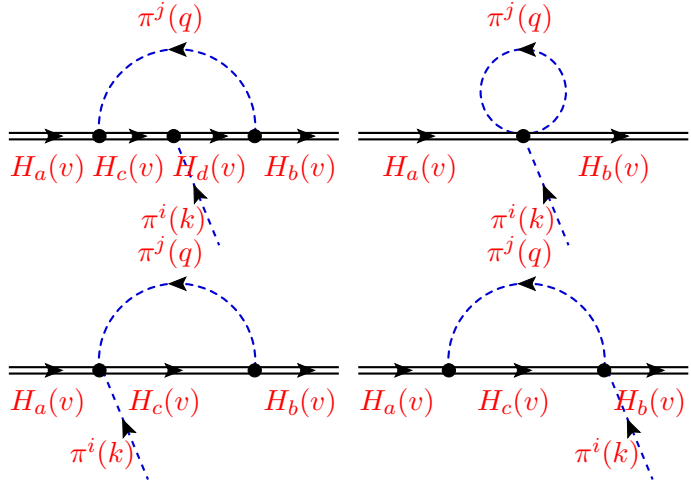}
\end{center}
\caption{\small\it \label{diagram_sunrise_road}"Sunrise road"
topology diagram contributing to effective strong vertex
correction.}
\end{figure}

At the $\mathcal O(p^3)$ order, the relevant contributions to the
vertices between the heavy and light mesons come from the one-loop
topology diagrams in fig.~\ref{diagram_sunrise_road} with leading
order couplings in the loop. \psfrag{pijl}[cc]{$\Red{ \pi^i(q)}$}
\psfrag{Ha}[cc]{${\Red{H_a(v)}}$} \psfrag{Hb}[cc]{$\Red{H_b(v)}$}
\psfrag{pi}[cc]{$\Red {\pi^i(k)}$} \psfrag{Hc}[cc]{$\Red{H_c(v)}$}
\psfrag{Hd}[cc]{$\Red{H_d(v)}$} \psfrag{pij}[bc]{$\Red{\pi^j(q)}$}
Contributions from two lower diagrams vanish because the heavy meson
covariant derivative emitting two
pseudo-Goldstone\index{pseudo-Goldstone boson} bosons either
annihilates ((axial)vector) external heavy meson states, or the two
diagrams (with (pseudo)scalar states in the loop) cancel each-other.
The upper right diagram on the other hand is subtracted completely
by the pseudo-Goldstone\index{pseudo-Goldstone boson!wave-function
renormalization} wave-function renormalization
$Z_{2\pi^i}$~\cite{Cheng:1993kp} at the given order. Thus the only
one-loop topology contributing, are the ''sunrise road'' diagrams as
the one on the upper left. For the case of the $PP^*\pi$ vertex,
only $(P^*,P)$, $(P^*,P^*)$ and $(P_0,P^*_1)$ contribute pairwise in
such loops. Adding the relevant $\mathcal O(p^3)$ counterterm
contributions we thus obtain \index{chiral corrections!in strong
decays}
\begin{eqnarray}
\hspace{-10pt}Z_{1P^*_a P_b\pi^i} &=& 1 + \frac{
\lambda^j_{ac}\lambda^i_{cd}\lambda^j_{db}}{\lambda^i_{ab} 16\pi^2
f^2} \times \Bigg\{ g^2
            C'_1\left(\frac{\Delta_{ca}}{m_j},\frac{\Delta_{db}}{m_j},m_j\right)
\nonumber\\&&\hskip3.2cm + \frac{h^2 \widetilde g }{g }
        C'\left(\frac{\Delta_{\widetilde c a } + \Delta_{SH}}{m_j},\frac{\Delta_{\widetilde d b } + \Delta_{SH}}{m_j},m_j\right)
\Bigg\}
    \nonumber\\
&& \hskip0.3cm +
\frac{\lambda^{i}_{ac}(m_q)_{cb}}{\Lambda_{\chi}\lambda^i_{ab}}(\kappa_{19}
+ \delta^{ab} \kappa_5) - \frac{\Delta_{ba}}{\Lambda_{\chi}}
\frac{\delta_2 + \delta_3}{g}.
\end{eqnarray}
The same expression is obtained for the $P^*P^*\pi$ vertex
renormalization from pairs of $(P,P^*)$, $(P^*,P^*)$, $(P^*,P)$ and
$(P^*_1,P^*_1)$ running in the loops \index{loop corrections!in
strong decays}.

\par

Similarly for the $P_0 P^*_1 \pi$ and $P^*_1 P^*_1 \pi$ vertices we
get contributions from pairs of $(P_1^*,P_0)$, $(P^*_1,P^*_1)$,
$(P,P^*)$ and $(P_0,P^*_1)$, $(P^*_1,P_0)$, $(P^*_1,P^*_1)$,
$(P^*,P^*)$ respectively running in the loops which yield
identically \index{chiral corrections!in strong decays}
\begin{eqnarray}
Z_{1P^*_{1a} P_{0b}\pi^i} &=& 1 + \frac{
\lambda^j_{ac}\lambda^i_{cd}\lambda^j_{db}}{\lambda^i_{ab} 16\pi^2
f^2} \times \Bigg\{ \widetilde g^2
             C'_1\left(\frac{\widetilde \Delta_{ca}}{m_j},\frac{\widetilde \Delta_{db}}{m_j},m_j\right)
\nonumber\\&&\hskip3.2cm + \frac{h^2 g}{\widetilde g}
        C'\left(\frac{\Delta_{c\widetilde a} - \Delta_{SH}}{m_j},\frac{\Delta_{d\widetilde b} - \Delta_{SH}}{m_j},m_j\right)
\Bigg\}
    \nonumber\\
&&\hskip0.3cm+
\frac{\lambda^{i}_{ac}(m_q)_{cb}}{\Lambda_{\chi}\lambda^i_{ab}}(\widetilde
\kappa_{19} + \delta^{ab} \widetilde\kappa_5)- \frac{\widetilde
\Delta_{ba}}{\Lambda_{\chi}} \frac{\widetilde \delta_2 + \widetilde
\delta_3}{\widetilde g}.
\end{eqnarray}

\par

Finally for the $PP_0\pi$ and $P^*P^*_1\pi$ vertices the pairs of
$(P_0,P)$, $(P^*,P^*_1)$ and $(P^*_1,P^*)$, $(P^*,P^*_1)$, $(P,P_0)$
contribute in the loops respectively, yielding for the
renormalization identically \index{chiral corrections!in strong
decays}
\begin{eqnarray}
Z_{1P_{0a} P_{b}\pi^i} &=& 1 + \frac{
\lambda^j_{ac}\lambda^i_{cd}\lambda^j_{db}}{\lambda^i_{ab} 16\pi^2
f^2} \times \Bigg\{ 3 g \widetilde g
        C'_1\left(\frac{\widetilde \Delta_{ca}}{m_j},\frac{ \Delta_{db}}{m_j},m_j\right)
\nonumber\\&&\hskip3.2cm - h^2
            C'\left(\frac{\Delta_{c\widetilde a} - \Delta_{SH}}{m_j},\frac{\Delta_{\widetilde d b} + \Delta_{SH}}{m_j},m_j\right)
\Bigg\}
    \nonumber\\
&&\hskip0.3cm+
\frac{\lambda^{i}_{ac}(m_q)_{cb}}{\Lambda_{\chi}\lambda^i_{ab}}(\kappa'_{19}
+ \delta^{ab} \kappa'_5) - \frac{\Delta_{b\widetilde a} -
\Delta_{SH}}{\Lambda_{\chi}} \frac{\delta'_2 + \delta'_3}{h}.
\end{eqnarray}

\section[Extraction of phenomenological couplings]{Extraction of phenomenological couplings from charmed meson decays}
\index{decay of $D$ meson} \index{strong decay}

Using known experimental values for the decay widths of $D^{+*}$,
$D^{+*}_0$, $D^{0*}_0$ and $D^{'}_1$, and the upper bound on the
width of $D^{0*}$ one can extract the values for the bare couplings
$g$, $h$ and $\widetilde g$ from a fit to the data. The decay rates
are namely given by\index{amplitude!in strong decays}
\begin{subequations}
\begin{eqnarray}
\Gamma(P_a^{*}\to \pi^i P_b) &=& \frac{|g^{\mathrm{eff.}}_{P^{*}_a P_b\pi^i}|^2}{6 \pi f^2} |{\bf k}_{\pi^i}|^3,\\
\Gamma(P_{0a}\to \pi^i P_b)&=& \frac{|h^{\mathrm{eff.}}_{P_{0a}
P_b\pi^i}|^2}{2 \pi f^2} E_{\pi^i}^2 |{\bf k}_{\pi^i}|.
\end{eqnarray}
\end{subequations}
Here ${\bf k}_{\pi}$ is the three-momentum vector of the outgoing
pion and $E_{\pi}$ its energy. The renormalization condition for the
couplings can be written as
\begin{equation}
g^{\mathrm{eff.}}_{P_a^*P_b\pi^i} = g
\frac{\sqrt{Z_{2P_a}}\sqrt{Z_{2P_b^*}}\sqrt{Z_{2\pi^i}}}{Z_{1P_aP_b^*\pi^i}}
= g Z^g_{P_a^*P_b\pi^i} \label{eq_g_renorm}
\end{equation}
with similar expressions for the $h$ and $\widetilde g$ couplings.

\par

Due to the large number of unknown
counterterms\index{counterterms!in strong decays} entering our
expressions ($\kappa_5$, $\kappa_{19}$, $\widetilde \kappa_5$,
$\widetilde \kappa_{19}$, $\kappa'_5$,  $\kappa'_{19}$,
$\delta_2+\delta_3$, $\widetilde\delta_2+\widetilde\delta_3$ and
$\delta'_2+\delta'_3$) we cannot fix all of their values. Therefore
we first perform a fit with a renormalization scale set to
$\mu\simeq1~\mathrm{GeV}$~\cite{Stewart:1998ke} and we choose to
neglect counterterm contributions altogether. Our choice of the
renormalization scale in dimensional
regularization\index{dimensional regularization} is arbitrary and
depends on the renormalization scheme. Therefore any quantitative
estimate made with such a procedure cannot be considered meaningful
without also thoroughly investigating
counterterm\index{counterterms!in strong decays}, quark mass and
scale dependencies.

\par

We constrain the range of the fitted bare couplings by using
existing knowledge of their dressed values and assuming the first
order loop corrections to be moderate and thus also maintaining
convergence of the perturbation series:
\begin{itemize}
\item $g$ - following quark model predictions for the positive sign of this coupling~\cite{Becirevic:1999fr} as well as previous determinations~\cite{Stewart:1998ke, Abada:2003un, McNeile:2004rf,Colangelo:1995ph, Colangelo:1997rp, Wang:2006id} we constrain its bare value to the range $g \in [0,1]$.
\item $h$ - this coupling only enters squared in our expressions for the decay rates and was recently found to be quite large~\cite{Colangelo:1995ph, Colangelo:1997rp, Wang:2006bs,Wang:2006id}. We constrain its bare value to the region $|h| \in [0,1]$.
\item $\widetilde g$ - non-relativistic quark models predict this coupling to be smaller in absolute value than $g$~\cite{Falk:1992cx}. Similar results were also obtained using light-cone sum rules~\cite{Colangelo:1997rp}, while the chiral\index{chiral partners model} partners HM$\chi$PT\index{HM$\chi$PT!chiral partners model} model predicts $|\widetilde g| = |g|$~\cite{Bardeen:2003kt,Nowak:2004uv,Mehen:2005hc}. A recent lattice\index{lattice QCD!in strong decays} QCD study~\cite{Becirevic:2005zu} found this coupling to be smaller and of opposite sign than $g$. We combine these different predictions and constrain the bare $\widetilde g$ to the region $\widetilde g \in [-1,1]$.
\end{itemize}
We perform a Monte-Carlo randomized least-squares fit for all the
three couplings in the prescribed regions using the experimental
values for the decay rates from table~\ref{table_input} to compute
$\chi^2$ and using values from PDG~\cite{Eidelman:2004wy} for the
masses of final state heavy and light mesons. In the case of excited
$D^*_0$ and $D'_1$ mesons, we also assume saturation of the measured
decay widths with the strong decay channels to ground state charmed
mesons and pions ($D^*_0 \to D \pi$ and $D'_1\to D^*
\pi$)~\cite{Mehen:2004uj}.
We obtain the best-fitted values for the couplings $g=0.66$, $|h|=0.47$ and $\widetilde g=-0.06$ at $\chi^2/\mathrm{d.o.f.}=3.9/3$. The major contribution to the value of $\chi^2$ comes from the discrepancy between decay rates of $D'_1$ and $D_0^*$ mesons. While the former favor a smaller value for $|h|$, the later, due to different kinematics of the decay, prefer a larger $|h|$ with small changes also for the bare $g$ and $\widetilde g$ couplings. Similarly, as noted in ref.~\cite{Becirevic:2004uv}, such differences are due to the uncertainties in the measured masses of the broad excited meson resonances\index{resonance!of charmed meson}. We expect these uncertainties to dominate our error estimates and have checked that they can shift our fitted values of the bare couplings by up to $5\%$ in the case of $g$ and $h$  and as much as $70\%$ for $\widetilde g$ depending on which experimental mass values are considered. In a similar fashion our results are sensitive to the values of the residual mass splittings between heavy fields (which we fix to their phenomenological values) as $30\%$ variations in $\Delta_{SH}$ , $\Delta_{ab}$, $\Delta_{\widetilde a b}$ and $\widetilde \Delta_{ab}$ shift our fitted values for the bare couplings a few percent in the case of $g$ and $h$, while $\widetilde g$ can recieve much larger corrections. A large uncertainty in the determination of this coupling was to be expected since it only features in our fit indirectly via its loop \index{loop corrections!in strong decays} contributions. 

\par

If we do not include positive parity states' contributions in the
loops (and naturally fix $\Delta_H=0$), we obtain a best fit for
this coupling $g=0.53$. We see that chiral loop corrections
including positive parity heavy meson fields tend to increase the
bare $g$ value compared to its phenomenological (tree level) value
of $g_{l.o.}=0.61$~\cite{Anastassov:2001cw}, while in a theory
without these fields, the bare value would decrease. The fitted
value of $|h|$, is close to its tree level phenomenological value
obtained from the decay widths of $D_0^{*}$ and $D'_1$ mesons (and
using the tree level value for $f=130~\mathrm{MeV}$) $h_{l.o}=0.52$.
Our determined magnitude for $\widetilde g$ is close to the
QCD\index{QCD sum rules} sum rules determination of its dressed
value~\cite{Colangelo:1997rp,Dai:1997df}, but somewhat smaller
compared to parity doubling model
predictions~\cite{Bardeen:2003kt,Nowak:2004uv}. Its sign is also
consistent with the lattice\index{lattice QCD!in strong decays} QCD
result of ref.~\cite{Becirevic:2005zu}. Based on this calculation we
can derive a prediction for the phenomenological coupling between
the heavy axial and scalar mesons and light pseudo-Goldstone
bosons\index{pseudo-Goldstone boson!coupling to heavy mesons}
$G_{P_{1}^* P_0 \pi}$, which we defined for the case of two
pseudoscalar and a vector state in eq.~(\ref{eq_strong_v}) with the
identification $P_1=\pi$, $P_2=P_0$ and $V=P_1^*$, and is related to
the bare $\widetilde g$ coupling as (see e.g.
ref.~\cite{Casalbuoni:1996pg})
\begin{equation}
G_{P_{1}^* P_0 \pi} = \frac{2 \sqrt{m_{P_1^*}m_{P_0}}}{f} \widetilde
g^{\mathrm{eff.}}_{P_1^* P_0 \pi}. \label{eq_4.11}
\end{equation}
Using our best fitted value for $\widetilde g = -0.06$ and excited
meson masses from table~\ref{table_input}, we predict the absolute
value of this phenomenological coupling for the case of
$P_1^*=D_1^{'0}$, $P_0=D_0^{*+}$ and $\pi =\pi^-$: $|G_{D_1^{'0}
D_0^{*+} \pi^-}| =  6.0$ corresponding to an effective tree level
coupling value of $|\widetilde g^{\mathrm{eff.}}_{l.o.}|=0.15$,
which is marginally consistent with other estimates of $\widetilde
g^2/g^2 \approx 1/9$~\cite{Colangelo:1997rp, Lu:2006ry,
Becirevic:2005zu, Dai:1997df}.

\par

We can summarize the best-fitted values for the bare couplings in
table~\ref{table_summary}.
\begin{table}[!t]
\begin{center}
\begin{tabular}{|l|c|c|c|}
\hline
 Calculation scheme & $g$ & $|h|$ & $\widetilde g$ \\
 \hline
 \hline
 Leading order & $0.61$~\cite{Anastassov:2001cw} & $0.52$ & $-0.15$\footnotemark[3]\\
 \hline
 One-loop without positive parity states & $0.53$ & & \\
 \hline
 One-loop with positive parity states  & $0.66$ & $0.47$ & $-0.06$ \\
 \hline
\end{tabular}
\end{center}
\caption{\small\it \label{table_summary}Summary of our results for
the effective couplings as explained in the text. The listed
best-fit values for the one-loop calculated bare couplings were
obtained by neglecting counterterms' contributions at the
regularization scale $\mu\simeq 1~\mathrm{GeV}$.}
\end{table}
\footnotetext[3]{Effective tree level coupling value derived from
one loop calculation for the case $D^{'0}_1 \to D_0^{*+} \pi^-$.}
\setcounter{footnote}{3} One should remember that the quantitatively
different results of ref.~\cite{Stewart:1998ke} appeared before the
observation of the even parity heavy meson states and in that study
a combination of strong and radiative decay modes was considered in
constraining $g$.

\subsection{Renormalization scale dependence, counterterm contributions and $1/m_H$ corrections}
\index{renormalization scale dependence} \index{counterterm
contributions!in strong decays} \index{heavy quark expansion!in
strong decays}

\begin{figure}[!t]
\psfrag{muScale}[cc]{\Blue{$~~~~~~~~~\mu~[\mathrm{GeV}]$}}
\psfrag{cpl}[tc][tc][1][90]{\Blue{$ g,h,\widetilde g$}}
\psfrag{gCoupl}[tc][tc][1][0]{$g$}
\psfrag{hCoupl}[tc][tc][1][0]{$|h|$}
\psfrag{gtCoupl}[tc][tc][1][0]{\hskip-6pt $\widetilde g$}
\begin{center}
\hspace{-1cm} \includegraphics{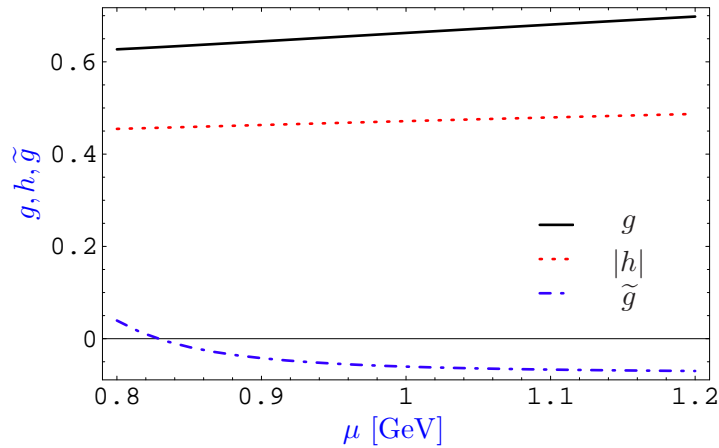}
\end{center}
\caption{\small\it \label{fig_4.3}Renormalization scale dependence
of the fitted bare couplings as explained in the text.}
\end{figure}

In a full NLO HM$\chi$PT analysis, the renormalization scale
dependence of the non-analytic (log) terms cancels completely
against the one in the relevant counterterms for any physical
quantity. However, in our coupling extraction we neglect the
contributions of the unknown counterterms, thus spoiling such
cancelation. If we probe our results to the sensitivity to the
renormalization scale $\mu$ we obtain a moderate dependence (see
also fig.~\ref{fig_4.3}),
namely a $20\%$ variation of scale around $1~\mathrm{GeV}$ results
in roughly $10\%$ variation in $g$, $6\%$ variation in $h$ whereas
the value of $\widetilde g$ is more volatile and can even change
sign for small values of $\mu$. This behavior could be expected
since $\widetilde g$ only features in logarithmic corrections which
diminish at small scales comparable to pseudo-Goldstone masses.
Therefore in order to compensate for this in the absence of any
counterterm contributions, the value of $\widetilde g$ has to change
drastically, while the values of the other two couplings are held
fixed close to their tree level estimates.

\par

\begin{figure}[!t]
\psfrag{xk}[cc]{\Blue{$\overline{|\kappa|}$}}
\psfrag{yg}[tc][tc][1][90]{\Blue{$g$}}
\psfrag{yh}[tc][tc][1][90]{\Blue{$|h|$}}
\psfrag{ygt}[tc][tc][1][90]{\Blue{$\widetilde g$}}
\hspace*{-0.6cm}\scalebox{0.83}{\includegraphics{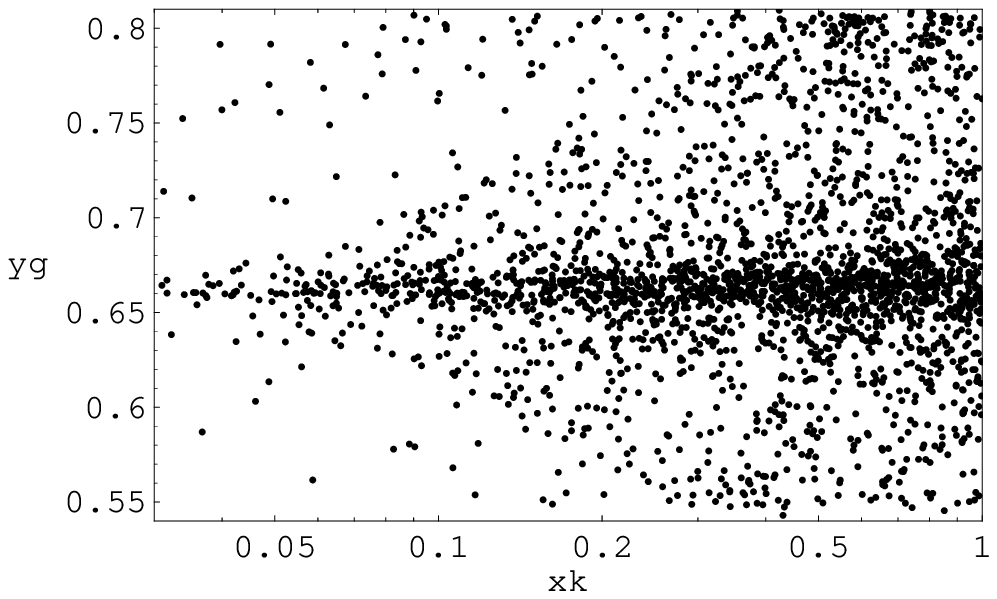}}
\hspace*{-0.4cm}\scalebox{0.83}{\includegraphics{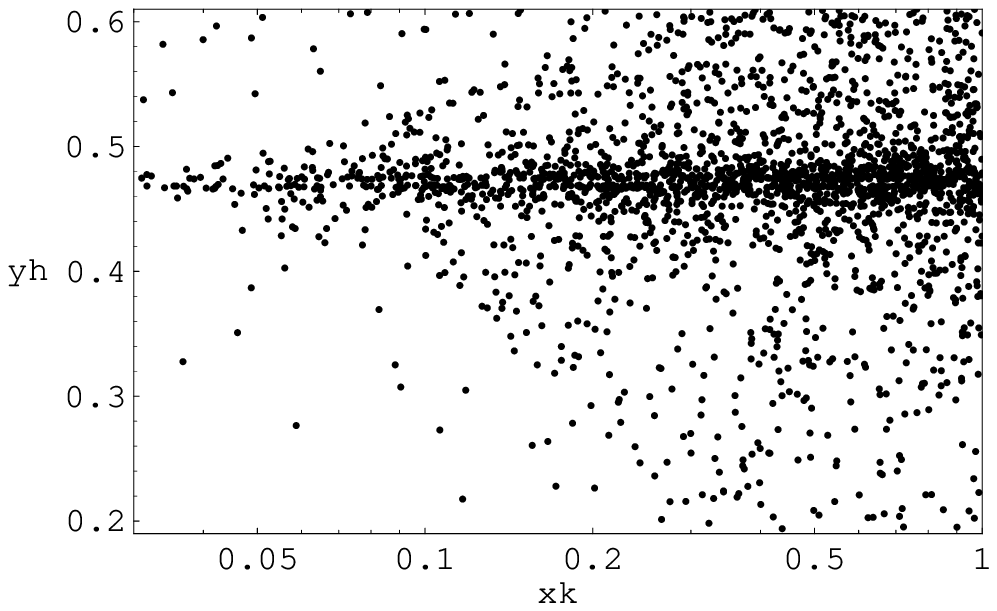}}
\hspace*{3cm}\scalebox{0.83}{\includegraphics{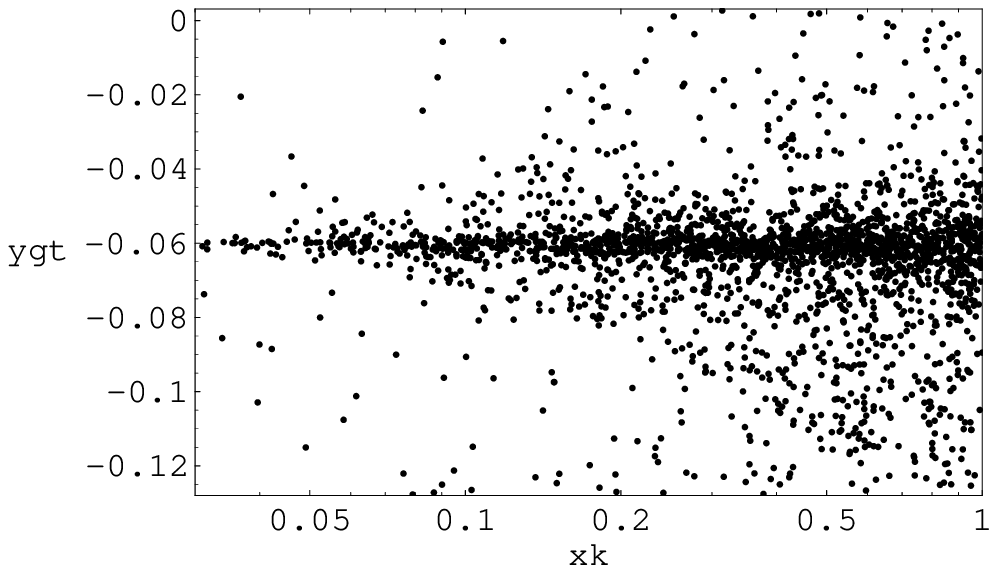}}
\caption{\small\it \label{figure_MC}Effect of the $m_q$ and
$E_{\pi}$ counterterms of the size order $\overline{|\kappa|}$ on
the solutions for the couplings $g$ (top left ), $|h|$ (top right)
and $\tilde g$ (bottom) as explained in the text.}
\end{figure}

Since we consider decay modes with the  pion in the final state, one
should not expect sizable contribution of the counterterms. Namely,
the counterterms which appear in our study are proportional to the
$u$ and $d$ quark masses, and not to the strange quark
mass~\cite{Stewart:1998ke}. Nonetheless we study the effects of the
counterterms on our couplings fit. Following the approach of
ref.~\cite{Stewart:1998ke} we take the values of $\kappa_5$,
$\kappa'_5$, $\kappa_{19}$, $\kappa'_{19}$, $\delta_2+\delta_3$ and
$\delta'_2+\delta'_3$ entering our decay modes to be randomly
distributed at $\mu\simeq 1~\mathrm{GeV}$ in the interval $[-1,1]$.
Near our original fitted solution, we generate 5000 values of $g$,
$|h|$ and $\widetilde g$ by minimizing $\chi^2$ at each counterterm
sample. For each solution also the average absolute value of the
randomized counterterms ($\overline{|\kappa|}$) is computed. We plot
the individual coupling solution distributions against this
counterterm size measure in fig.~\ref{figure_MC}.
We see that the inclusion of counterterms\index{counterterm
contributions!in strong decays} spreads the fitted values of the
three couplings. From this we can estimate roughly the uncertainty
of the solutions due to the counterterms to be at the one sigma
level $g  =  0.66^{ +0.08}_{-0.06}$, $|h|  =  0.47^{ +0.07}_{-0.04}$
and $\widetilde g  =  -0.06^{+0.03}_{-0.04}$ if we assume the
counterterms do not exceed values of the order $\mathcal O (1)$.
This result is in a way complementary to the study of
renormalization scale dependence of our couplings' fit. Both are
important since although it is always possible in principle to trade
the counterterms contributions for a specific choice of the
renormalization scale, the latter will be different for different
amplitudes where the combination of
counterterms\index{counterterms!in strong decays} will be different.

\par

A full calculation of the strong decay\index{strong decay} couplings
should contain, in addition to the corrections we determine, also
the relevant $1/m_H$ corrections as discussed in
ref.~\cite{Mehen:2005hc}. However, the number of unknown couplings
is yet too high to be determined from the existing data. In
addition, the studies  of the lattice groups~\cite{Hein:2000qu,
Dougall:2003hv, Abada:2003un} indicate that the $1/m_H$ corrections
do not contribute significantly to their determined values of the
couplings, and we therefore assume the same to be true in our
calculations of chiral  corrections.

\par

To summarize, the counterterm contributions of order $\mathcal O
(1)$ can spread the best fitted values of $ g$, $|h|$ by roughly
$15\%$ and $\widetilde g$ by as much as $60\%$. Similarly, up to
$20\%$ shifts in the renormalization scale modify the fitted values
for the $g$ and $|h|$ by less than $10\%$ while $\widetilde g$ may
even change sign at high renormalization scales. Combined with the
estimated uncertainty due to discrepancies in the measured excited
heavy meson masses and resulting mass splittings, we consider these
are the dominant sources of error in our determination of the
couplings as summarized also in table~\ref{table:strong_errors}.
\begin{table}[!t]
\begin{center}
\begin{tabular}{|l|c|c|c|}
\hline
 Input (variation) & $\delta g~[\%]$ & $\delta |h|~[\%]$ & $\delta \widetilde g~[\%]$ \\
 \hline
 \hline
 $\Delta_{SH}$ (30\%) & 7 & 5 & 70\\
 \hline
 $\Delta_{a3}$ (30\%) & 5 & $<1$ & 16\\
 \hline
 $m_{D^*_0}$, $m_{D'_1}$ (\cite{Abe:2003zm},~\cite{Link:2003bd},~\cite{Anderson:1999wn}) & 2 & 5 & 70\\
 \hline
 $\mu$ (20\%) & 10 & 6 & $>100$\\
 \hline
 c.t. (-1,1) & 15 & 15 & 60\\
 \hline
\end{tabular}
\end{center}
\caption{\small\it \label{table:strong_errors}Summary of probed
input parameter ranges and coresponding fitted couplings' variations
as explained in the text.}
\end{table}
One should keep in mind however that without better experimental
data and/or lattice QCD inputs, the phenomenology of strong decays
of charmed mesons presented above ultimately cannot be considered
reliable at this stage.

\section{Chiral extrapolation of lattice QCD simulations}
\index{chiral extrapolation!in strong decays}

\begin{figure}
\vspace*{-0.3cm}
\begin{center}
\begin{tabular}{@{\hspace{-0.25cm}}c}
\epsfxsize6cm\epsffile{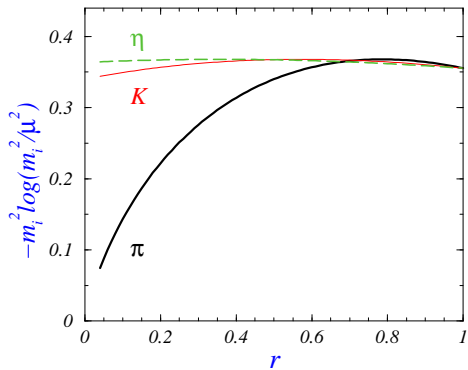}    \\
\end{tabular}
\caption{\small\it \label{fig5}{Typical chiral logarithmic
contributions $-m_i^2\log(m_i^2/\mu^2)$ are shown for pion, kaon and
$\eta$ as a function of $r=m_{d}/m_s$, with $m_s$ fixed to its
physical value, and $\mu=1$~GeV.} }
\end{center}
\end{figure}

Next we study the contributions of the additional
resonances\index{resonance contribution!in strong decays} in the
chiral loops to the chiral\index{chiral extrapolation!in strong
decays} extrapolations employed by lattice\index{lattice QCD!in
strong decays} QCD studies to run the light meson masses from the
large values used in the simulations to the chiral\index{chiral
limit!in strong decays} limit~\cite{Abada:2003un, McNeile:2004rf}.
In the extrapolation of the lattice data the kaon\index{meson!$K$!in
chiral loops} and the $\eta$-meson\index{meson!$\eta$!in chiral
loops} loops essentially do not alter the quark mass dependence,
whereas the important nonlinearity comes from the pion chiral loops.
As an illustration, in fig.~\ref{fig5} we plot the  typical chiral
logarithm, $-m_i^2\log(m_i^2/\mu^2)$, as a function of $r=m_{d}/m_s$
which appear in the Gell-Mann--Oakes--Renner formulae~(\ref{eq_2.6})
with  $2 B_0 m_s = 2 m_K^2-m_\pi^2=0.468\ \gev^2$.

\par

The results of the previous section suggest that the inclusion of
heavy excited mesons in the chiral loops introduces relatively large
corrections into the renormalization of the coupling constants.
Formally, the problem was already mentioned in
section~\ref{sec:hmcpt} when the two equivalent realizations of the
theory involving the new $\Delta_{SH}$  scale were considered. There
we encountered a possibility of a strongly coupled mixed sector of
the theory in the case the $\Delta_{SH}/f$  ratio grew large. Within
our chosen parametrization, when the $\Delta_{SH}$  contributions
are re-summed into the heavy meson propagators the problem can be
explored by analyzing the dimensionally
regularized\index{dimensional regularization} loop
integrals\index{loop integral} involving the off-set propagators.
The large splitting between the ground state and excited heavy
mesons in the loops causes the pseudo-Goldstone
bosons\index{pseudo-Goldstone boson} in the loops to carry large
momenta. They can be highly virtual or, in the cases of $P_0 P \pi$,
$P_1^* P^* \pi$ and $P_1^* P_0 \pi$ couplings' renormalization, real
in a considerable portion of the phase space. Such behavior casts
doubts on the validity of this extended perturbation scheme, as
contributions from higher lying excited heavy meson states seem to
dominate the loop amplitudes. As an example we consider
$I_1^{\mu\nu}(m,\Delta)$, which can be found in the Appendix~B,
while all the other loop\index{loop integral} integrals relevant for
this chapter can be obtained from this one via algebraic
manipulation. The integral is characterized by two dimensionful
scales ($m$ and $\Delta$). In addition $\chi$PT\index{$\chi$PT}
requires pion momenta (also those integrated over in the loops) to
be much smaller than the chiral\index{chiral symmetry breaking
scale} symmetry breaking scale $\Lambda_{\chi}$. The first integral
scale $m$ is the mass of the pseudo-Goldstone\index{pseudo-Goldstone
boson} bosons running in the loop \index{loop corrections!in strong
decays}. In lattice\index{lattice QCD!in strong decays} studies, its
value is varied and can be taken as large as $m\sim 1~\mathrm{GeV}$.
Within $\chi$PT\index{$\chi$PT} however, it is protected by chiral
symmetry to be small. On the other hand, once $\Delta$ contains the
splitting between heavy meson states of different parity, it is not
protected by either heavy quark or chiral symmetries and can be
arbitrarily large (its size should mainly be determined by $\mathcal
O(\Lambda_{\chi})$ effects up to chiral\index{chiral corrections!in
strong decays}, $m_q$ and $1/m_Q$ corrections). Once we attempt to
integrate over loop momenta probing also this scale, we are
effectively including harder and harder momentum scales in the
dimensionally regularized\index{dimensional regularization}
expression as this splitting grows. Finally, as these approach
$\Lambda_{\chi}$, the perturbativity and predictability of such
scheme break down.

\par

While the phenomenological couplings' fit seems mainly unaffected by
such problems (e.g. the results depend only mildly on the actual
value of the mass splitting in the range probed), they play a much
more profound role in the chiral\index{chiral extrapolation!in
strong decays} extrapolation. As customary we expect the
non-analytic chiral log terms to dominate the extrapolation, while
any analytic dependence on the
pseudo-Goldstone\index{pseudo-Goldstone boson} masses can be
absorbed into the appropriate counterterms\index{counterterms!in
strong decays}. As an example we write down the dominating
contributions to the chiral log extrapolation of the $g$ coupling
\begin{eqnarray}
\frac{1}{m_j^2} \frac{\mathrm{d} g^{\mathrm{eff.}}_{P_a^* P_b \pi^i}}{\mathrm{d} \log m_j^2} &=& \frac{g}{(4\pi f)^2}  \Bigg\{ \frac{\lambda^j_{ac} \lambda^j_{ca} + \lambda^j_{bc} \lambda^j_{cb}}{2} \left[ -3 g^2 - h^2 \left(1-\frac{6 \Delta_{SH}^2}{m_j^2}\right) \right] \nonumber\\
&& -
\frac{\lambda^j_{ac}\lambda^i_{cd}\lambda^j_{db}}{\lambda^i_{ab}}
\left[ g^2 - h^2 \frac{\widetilde g}{g} \left(1-\frac{6
\Delta_{SH}^2}{m_j^2}\right) \right] \Bigg\}. \label{eq_g_extra2}
\end{eqnarray}
In the above expression we have for the sake of simplicity neglected
the light flavor splittings between the heavy mesons which are
always small compared to  $\Delta_{SH}$, are of higher order in the
power counting and vanish in the chiral limit. On the other hand one
can immediately see, that the $\Delta_{SH}$ contributions due to
excited heavy mesons in the loops seemingly dominate the
chiral\index{chiral limit!in strong decays} limit, where they
diverge. The issue seems therefore to be especially severe in the
case of pions, which due to their small masses can also develop
sizable imaginary parts in their analytic contributions introducing
uncontrollable final state interactions\index{final state
interactions}. This seems to fly in the face of one of the basic
principles of QFT\index{QFT}, namely that the description of low
energy processes should not be sensitive to the UV completion of the
theory. The whole idea of effective field theories is based upon
this foundation, that for low enough energies, the contributions
from heavier states should decouple. Eq.~(\ref{eq_g_extra2}) however
suggests, that chiral corrections to strong decays of heavy mesons
are dominated by higher resonance states and we need to understand
the mechanisms that could restore the proper decoupling limit in
this case.

\subsection{\label{sec_dsh}Taming resonance contributions - the decoupling limit}
\index{resonance contribution!in strong decays} \index{resonance
decoupling} \index{$1/\Delta_{SH}$ expansion} \index{integrating
out!resonances} \index{resonance!integrating out}

We explore the issue by focusing on the chiral\index{chiral limit!in
strong decays} limit of the theory and attempt an expansion of the
relevant loop integral expressions (coming from opposite parity
heavy mesons propagating in the loops) in powers of the
pseudo-Goldstone\index{pseudo-Goldstone boson} mass. First we notice
that in the case of $g$ and $\widetilde g$ vertex corrections both
heavy meson states propagating in the loop \index{loop
corrections!in strong decays} always belong to the same spin-parity
multiplet. Due to the identity between the loop\index{loop
functions} integral functions $C'(x,x,m)=C'(x,m)$ (see Appendix~B)
in this limit, the expressions entering the leading order chiral
corrections of heavy meson wave-function and vertex renormalization
are the same and we must only evaluate the limit of
\begin{equation}
\lim_{m\to0} \left[\left. \frac{1}{m} \frac{d}{dx} C\left(x ,
m\right)\right|_{x=\Delta/m}\right] = 6\Delta^2
\log\frac{4\Delta^2}{\mu^2} - 2\Delta^2 - m^2
\log\frac{4\Delta^2}{\mu^2} - 3 m^2 + \ldots, \label{eq_4.13}
\end{equation}
where the dots stand for higher powers in $m^2$. We see immediately
that actually the diverging analytic and logarithmic parts cancel
exactly in the chiral limit washing out any leading order
contributions to the chiral running from such loops. In other words,
below  $\Delta\equiv\Delta_{SH}$ the presence of the nearby opposite
parity states does not affect the leading order pionic logarithmic
behavior of the $g$ and $\widetilde g$ couplings at all.
\par
In order to generalize this result also to the $h$ coupling, we need
to consider a slightly different and more general route. Namely, we
attempt on an perturbatively approximative solution. We expand the
integrand of  $I_1^{\mu\nu}(m,\Delta)$ over powers of $\Delta$. We
may do this, assuming the relevant loop momentum integration region
lies away from the ($v\cdot q - \Delta$) pole, which is true for
$\chi$PT\index{$\chi$PT} involving soft
pseudo-Goldstone\index{pseudo-Goldstone boson} bosons and for a
large enough $\Delta\sim\Lambda_{\chi}$. We obtain a sum of
integrals of the form
\begin{equation}
I_1^{\mu\nu}(m,\Delta)|_{\Delta = \mathrm{large}} =
\frac{\mu^{4-D}}{(2\pi)^D} \int \mathrm{d}^D q \frac{q^{\mu}
q^{\nu}}{(q^2-m^2)} \frac{-1}{\Delta} (1+\frac{q\cdot v}{\Delta} +
\ldots), \label{eq_int1}
\end{equation}
where the ellipses denote terms of higher order in the $1/\Delta$
expansion. This greatly simplified integral has a characteristic,
that all terms with odd powers of loop momenta in the numerator
vanish exactly. Thus, the first correction to the leading $\mathcal
O (1/\Delta)$ order truncation appears only at $\mathcal O
(1/\Delta^3)$.
\par
The above described procedure is similar to what is done in the
"method of regions" (see e.g.~\cite{Beneke:1997zp}) when one
separates out the different momentum scales, appearing in problems
involving collinear degrees of freedom. However, here we are only
interested in the low momentum part of the whole integral and assume
the high momentum contributions are properly accounted for in the
counterterms\index{counterterms!in strong decays}. The leading order
term in~(\ref{eq_int1}) then yields for the loop\index{loop
functions} functions
\begin{eqnarray}
C_1(x,m) &=& -\frac{m}{4 x} \left[ \frac{m^2}{2} - m^2 \log \left(\frac{m^2}{\mu^2}\right) \right] + \mathcal O \left( \frac{m^3}{x^3} \right), \nonumber\\
C_2(x,m) &=& \mathcal O \left( \frac{m^3}{x^3} \right).
\end{eqnarray}
When compared to the special limiting case of eq.~(\ref{eq_4.13}),
we are here effectively throwing out of the loop integrals all
contributions involving positive powers of $x(\Delta)$ (those
written out in eq.~(\ref{eq_4.13})), since they originate from hard
pseudo-Goldstone exchange and shifting them into the
counterterms\index{counterterms!in strong decays} (which now appear
appropriately rescaled). It is important to stress that the relevant
ratio for the validity of this approach is $\Delta/E_{\pi}\gg 1$ as
we are expanding in powers of loop momentum, not
pseudo-Goldstone\index{pseudo-Goldstone boson} masses.
\par
This approach can alternatively be understood as the expansion
around the decoupling limit of the positive parity states with the
corresponding contributions being just a series of local operators
with $\Delta$ dependent prefactors - effective
counterterms\index{counterterms!in strong decays} of a theory with
no positive parity mesons. For example in counting the
chiral\index{chiral power counting} powers in the first term of the
expansion~(\ref{eq_int1}) we see that the obtained structure
corresponds to a $\mathcal O(p^3 \log p)$ contribution coming from a
counterterm\index{counterterms!in strong decays} loop insertion. For
the case considered in eq.~(\ref{eq_4.13}), we were able to show
such decoupling explicitly because the associated loop integral
effectively factorizes\index{loop integral!factorization} in that
case into soft and hard contributions. In general this is not always
possible and we have to rely on the expansion of eq.~(\ref{eq_int1})
instead. Any large deviations of this approach from the predictions
of a theory without positive parity states and with the couplings
properly refitted would signal the breaking of such expansion and
the fact that the contributions from "dynamical" positive parity
states cannot be neglected. We expect such an expansion to hold well
for the $SU(2)$ chiral theory involving only pions as
pseudo-Goldstone\index{pseudo-Goldstone boson} bosons, as their
masses are much lighter then the phenomenological value of
$\Delta_{SH}$. For an illustration we can sketch the relevant energy
scales of the effective theory as follows
\begin{equation}
m_{u,d} \sim \frac{m_{\pi}^2}{\Lambda_{\chi}} < \Delta_{SH} \lesssim
m_s \sim \frac{m_{K,\eta}^2}{\Lambda_{\chi}} < \Lambda_{\chi} \ll
m_Q.
\end{equation}
Within a full $SU(3)$ chiral theory involving positive and negative
parity heavy states we are expanding in the powers of
$\{m_{\pi,K,\eta},\Delta_{SH}\}/\Lambda_{\chi}$ and
$\{m_{\pi,K,\eta},\Delta_{SH},\Lambda_{\chi}\}/m_{Q}$, whereas in a
$SU(2)$ chiral theory with a $1/\Delta_{SH}$\index{$1/\Delta_{SH}$
expansion} loop \index{loop momentum expansion}momentum expansion,
we are instead considering $m_{\pi}/\{\Lambda_{\chi},\Delta_{SH}\}$
and $\{m_{\pi},\Delta_{SH},\Lambda_{\chi}\}/m_{Q}$.

\subsection{Chiral extrapolation of the effective meson couplings}
\index{chiral extrapolation!in strong decays}

We apply the two above described approaches to the one-loop chiral
extrapolation of the effective strong couplings $g$, $h$ and
$\widetilde g$ and first write down the leading chiral log
contributions of the $SU(2)$ theory together with the leading
corrections due to the opposite parity states contributions:
\index{$1/\Delta_{SH}$ corrections}\index{chiral corrections!in
strong decays}
\begin{subequations}
\begin{eqnarray}
\label{eq_g_extra1}
g^{\mathrm{eff.}}_{P^{*}_a P_b \pi} &=& g \left\{1 + \frac{1}{(4\pi f)^2} m_{\pi}^2 \log\frac{m_{\pi}^2}{\mu^2} \left[ - 4 g^2 - \frac{m_{\pi}^2}{8\Delta_{SH}^2} h^2 \left( 3 + \frac{\widetilde g}{g}\right) \right]\right\}, \\
h^{\mathrm{eff.}}_{P^{'}_{a0} P_b \pi} &=& h \left\{1 + \frac{1}{(4\pi f)^2} m_{\pi}^2 \log\frac{m_{\pi}^2}{\mu^2} \left[ \frac{3}{4} (2g\widetilde g - 3 g^2 - 3\widetilde g^2) - \frac{m^2_{\pi}}{2\Delta_{SH}^2} h^2 \right]\right\},\\
\widetilde g^{\mathrm{eff.}}_{P^{*}_{a1} P^{'}_{b0} \pi} &=&
\widetilde g \left\{1 + \frac{1}{(4\pi f)^2} m_{\pi}^2
\log\frac{m_{\pi}^2}{\mu^2} \left[ - 4 \widetilde g^2 +
\frac{m_{\pi}^2}{8\Delta_{SH}^2} h^2 \left( 3 + \frac{ g}{\widetilde
g}\right) \right]\right\},
\label{eq_g_extra}
\end{eqnarray}
\end{subequations}\index{$1/\Delta_{SH}$ corrections}
where we do not distinguish between decay modes with either a
charged or neutral pion in the final state. We shall compare these
expressions with the full $SU(3)$ chiral log contributions including
those from opposite parity states (eqs.~(\ref{eq_g_extra2})).

\par

In the following quantitative analysis we take the fitted values of
the couplings from the previous section. We compare:
\begin{description}
\item[(I)] Loop integral expansion in the $SU(2)$ limit. The leading order contribution is given by the chiral loop contributions in a theory with only a single heavy parity multiplet, while we also probe the next-to-leading contributions due to the $1/\Delta_{SH}^2$ terms.
\item[(II)] Complete $SU(3)$ leading log extrapolation with heavy multiplets of both parities contributing.
\item[(III)] The same as (II) but in the degenerate limit $\Delta_{SH}=0$,
\item[(0)] Chiral $SU(3)$ extrapolation without the $1/\Delta_{SH}^2$ dependent contributions in eqs.~(\ref{eq_g_extra1}-\ref{eq_g_extra}).
\end{description}
We assume exact $SU(2)$ isospin symmetry\index{isospin symmetry!in
chiral extrapolation} and parameterize the
pseudo-Goldstone\index{pseudo-Goldstone boson} masses according to
the formulae~(\ref{eq_2.6}). Consequently in the chiral
\index{chiral extrapolation!in strong decays}extrapolation we only
vary the ratio $r$ -- the light quark mass with respect to the
strange quark mass which is kept fixed to its physical value. Since
we are only interested in the nonanalytic $r$-dependence of our
amplitudes, the common renormalization scale dependence can be
subtracted together with counterterm\index{counterterm
contributions!in strong decays} contributions which are analytic in
$r$. The leading slope of these can in principle be inferred from
lattice QCD\index{lattice QCD!in strong decays}. In
ref.~\cite{Becirevic:2005zu} the $g$ and $\widetilde g$ couplings
were calculated on the lattice at different $r$ values. However that
study used large values of pion masses ($r\sim 1$) where the
predispositions for our extrapolation expansion in scenario (I) are
not justified. Also in order to use lattice data to infer on the
validity of our approach from such a chiral \index{chiral
extrapolation!in strong decays}extrapolation one would in addition
need lattice\index{lattice QCD!results} results for the $h$
coupling, since it enters in the new potentially large
next-to-leading chiral logs of the other two couplings considered in
eqs.~(\ref{eq_g_extra1}-\ref{eq_g_extra}). Instead we normalize our
results for the $g$ coupling renormalization in all scenarios at $8
r_{\mathrm{ab}} \lambda_0 m_s / f^2= \Delta_{SH}^2 $  (corresponding
in our case to $r_{\mathrm{ab}}=0.34$) to a common albeit arbitrary
value of $Z^g_{P_a^*P_b\pi^i}(r_{\mathrm{ab}})=1$ and zero slope. In
order to fit our results to lattice data, one would instead need to
add the (counter) terms constant and linear in $r$ to the
chiral\index{chiral extrapolation!in strong decays} extrapolation
formulae, representing contributions from $s$ and $u,d$ quark
masses. Their values could then be inferred together with the values
for the bare couplings from the combined fit for all the three
couplings to the lattice results\index{lattice QCD!results}.
\par
\begin{figure}
\begin{center}
\hspace*{-2cm}{\includegraphics{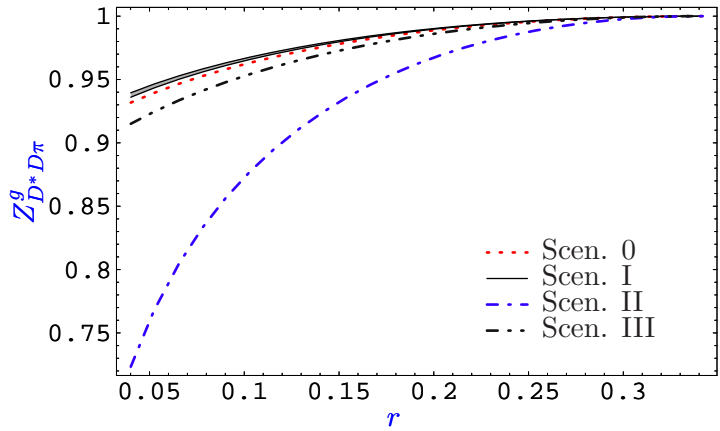}}
\end{center}
\caption{\small\it \label{coupling_plot_1}The $g$ coupling
renormalization in $D^{*}\to D\pi$. Comparison of chiral
extrapolation in (I) $SU(2)$ limit and loop integral expansion
(black, solid), (II) complete $SU(3)$ log contribution of both
parity heavy multiplets (blue, dash-dotted), (III) its degenerate
limit (gray, dash-double-dotted), and (0) $SU(3)$ log contributions
of negative parity states (red, dashed line) as explained in the
text.}
\end{figure}

As an example we again consider the strangeless process $D^{*}\to
D\pi$ in fig.~\ref{coupling_plot_1}.\index{chiral behavior!of
HM$\chi$PT couplings} \psfrag{xk1}[bc]{\Blue{$r$}}
\psfrag{xm1}[tc][tc][1][90]{\Blue{$Z^g_{D^{*}D\pi}$}}
\psfrag{s0}{Scen. 0} \psfrag{s1}{Scen. I} \psfrag{s2}{Scen. II}
\psfrag{s3}{Scen. III}
\index{chiral extrapolation!in strong decays}
\psfrag{xk1}[bc]{\Blue{$r$}}
\psfrag{xm1}[tc][tc][1][90]{\Blue{$|Z^h_{D^{*}_0 D\pi}|$}}
\begin{figure}
\begin{center}
\hspace*{-2cm}{\includegraphics{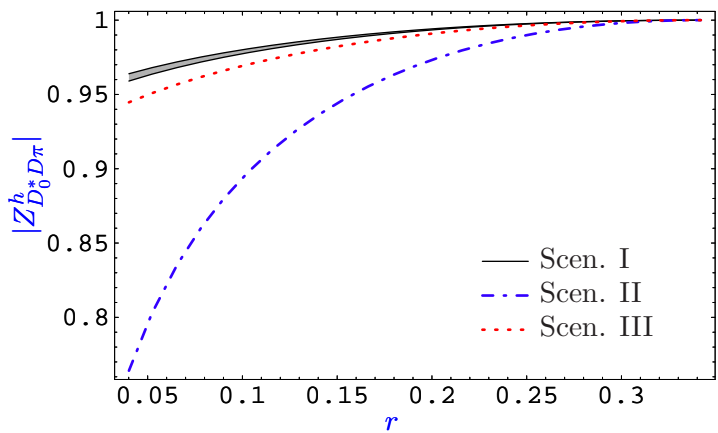}}
\end{center}
\caption{\small\it \label{coupling_plot_2} Chiral extrapolation of
the $h$ coupling renormalization in $D^{*}_0\to D\pi$.Comparison of
chiral extrapolation with (I) loop integral expansion in the $SU(2)$
limit (black, solid), (II) complete $SU(3)$ log contribution (blue,
dash-dotted), and (III) its degenerate limit (red, dotted)  as
explained in the text.}
\end{figure}
We can see that including the complete chiral log contributions from
excited states in the loops \index{loop corrections!in strong
decays}, introduces large ($\gtrsim 30\%$) deviations from the
extrapolation without these states due to the
$\Delta_{SH}^2/m^2_{\pi}$ divergence of the log terms in the
chiral\index{chiral limit!in strong decays} limit and are obviously
flawed. If one instead applies the decoupling expansion discussed
above, the deviations diminish considerably. The somewhat
non-physical case of degenerate multiplets is better in this respect
producing extrapolation closer to those in the $SU(2)$ or $SU(3)$
theories without dynamical positive parity states. Such corrections
to the running due solely to the $h^2$ terms are less then 5\%. The
$SU(2)$ and $SU(3)$ scenarios are almost identical, since it is
always the pions which contribute mostly to the chiral\index{chiral
limit!in strong decays} running near the chiral limit, while
kaon\index{meson!$K$!in chiral loops} and eta\index{meson!$\eta$!in
chiral loops} contributions are almost constant. We can estimate the
leading effects of the integrated out positive parity
resonances\index{resonance contribution!in strong decays} on the
chiral log running of the $g$ via the broadening of the gray shaded
area between the black curves of scenario (I). These represent the
leading order contributions and the dominating next-to-leading order
contributions due to factored out positive parity states. Their
difference amounts to the order of $0.5\%$, signaling a well
converging perturbative expansion. It is important to stress, that
this expansion is only applicable in the mass extrapolation region
we are considering here namely below $\Delta_{SH}$  and in order to
obtain accurate results via such extrapolation,
lattice\index{lattice QCD!in strong decays} studies must approach
below this value with their simulated pion masses. Above this value,
we are unable to disentangle the effects of positive parity
resonances with hard pions from those of leading order soft kaon and
eta loops unless we discern all the counterterms\index{counterterm
contributions!in strong decays} as well and fix them e.g. from the
lattice\index{lattice QCD!in strong decays}.
\par
For completeness we also plot the chiral\index{chiral
extrapolation!in strong decays} extrapolation diagram for the
$Z^h_{D^*_0 D \pi}$ in fig.~\ref{coupling_plot_2} (while the
extrapolation for the $\widetilde g$ goes along the same lines as
the one for $g$ discussed above, only with the couplings $g$ and
$\widetilde g$ interchanged -- see
eqs.~(\ref{eq_g_extra1}-\ref{eq_g_extra})).
Here scenario (0) is meaningless, as is the $h$ coupling in a theory
without dynamical heavy multiplets of both parities. On the other
hand, the $1/\Delta_{SH}$\index{$1/\Delta_{SH}$ expansion} expansion
of loop\index{loop integral!expansion} integrals  in scenario (I)
still makes sense, although its physical meaning here is more
clouded. Namely we are not integrating\index{integrating
out!resonances}\index{resonance!integrating out} out heavy fields of
either parity, but rather truncating a class of diagrams, where the
kinematics of the intermediate heavy states cause the exchanged
pions to be hard. Therefore contributions involving a single $h$
coupling sandwiched between $g$ and $\widetilde g$ vertices are not
suppressed\index{chiral suppression}, whereas contributions of
opposite parity intermediate states in heavy meson wave-function
renormalization as well as triple-parity changing loop \index{loop
corrections!in strong decays} contributions to vertex
renormalization (all involving three powers of $h$) all contribute
only at the next-to-leading order in our expansion. The results for
the chiral extrapolation of the $h$ coupling renormalization are
thus very similar to the $g$ coupling case. Here, in all
approximations, the main contributions to the extrapolation come
from the $g^2$, $\widetilde g^2$ and mixed $g\widetilde g$ terms,
with smaller corrections due to the $\Delta_{SH}$  dependent $h^2$
terms (except for the complete scenario II, where these terms give
large deviations). Next-to-leading contributions of
$1/\Delta_{SH}$\index{$1/\Delta_{SH}$ expansion} expansion again
give an effect of the order of $0.5\%$.

\par

To summarize, our analysis of chiral extrapolation of the coupling
$g$ shows that the full loop contributions of excited charmed mesons
give sizable effects in modifying the slope and curvature in the
limit $m_\pi \to 0$. We argue that this is due to the inclusion of
hard pion momentum scales inside chiral\index{loop integral} loop
integrals containing the large mass splitting between charmed mesons
of opposite parity $\Delta_{SH}$  which does not vanish in the
chiral\index{chiral limit!in strong decays} limit. If we instead
impose physically motivated approximations for these contributions
-- we expand them in terms of $1/\Delta_{SH}$\index{$1/\Delta_{SH}$
expansion} -- the effects reduce dramatically, with explicit $h$
contributions appearing at the next-to-leading order in the
expansion and contributing of the order of $0.5\%$ to the running.
Consequently one can infer on the good convergence of the
$1/\Delta_{SH}$ expansion. We conclude\index{conclusions} that
chiral loop corrections in strong charm meson decays\index{strong
decay!of charmed mesons} can be kept under control provided the
extrapolation is performed below the $\Delta_{SH}$  scale, give
important contributions and are relevant for the precise extraction
of the strong coupling constants $g$, $h$ and $\widetilde g$.

\chapter{\label{chapter_semileptonic}Semileptonic decays of heavy mesons}
\index{semileptonic decay}
\index{meson!$D$!decay}

Presently, one of the most important issues in hadronic physics is the extraction of the CKM\index{CKM!parameters} parameters from exclusive decays. An essential ingredient in this approach is the knowledge of the form factors' shapes and sizes in {\it heavy to heavy} and {\it heavy to light} weak transitions. Most of the attention has traditionally been devoted to $B$\index{meson!$B$!decay} decays\index{decay of $B$ meson} and the determination of the CKM\index{CKM!phase} phase and of the $V_{ub}$\index{CKM!matrix elements!$V_{ub}$} and $V_{cb}$\index{CKM!matrix elements!$V_{cb}$} CKM\index{CKM!matrix elements} matrix element moduli. At the same time in the charm sector, the most accurate determination of the size of $V_{cs}$\index{CKM!matrix elements!$V_{cs}$} and $V_{cd}$\index{CKM!matrix elements!$V_{cd}$} matrix elements is not from a direct measurement, mainly due to theoretical uncertainties in the calculations of the relevant form factors\index{form factor}. In both sectors, the presence of nearby excited heavy meson resonances\index{resonance!of heavy meson} might affect the present picture substantially. In this chapter we therefore explore the leading effects due to possible excited heavy meson states on the determination of the relevant form factors\index{form factor} in semileptonic transitions involving heavy mesons.
\section{Heavy to light transitions}
\index{transition!heavy-to-light}
\index{semileptonic decay}

Semileptonic decays\index{decay of $D$ meson} of charmed mesons are necessary for extracting moduli of CKM elements $V_{cs}$\index{CKM!matrix elements!$V_{cs}$} and $V_{cd}$ directly and thus checking the values fixed by imposing CKM\index{CKM!unitarity} unitarity. The knowledge of the form factors which describe the weak $heavy \to light$ semileptonic transitions is very important in such an endeavor. Namely, one needs to know an accurate value of the relevant form factors\index{form factor!heavy-to-light} obtained in QCD at (at least) one value of $s$ at which both theory and experiment can reach an accuracy comparable to the error on $|V_{cs}|$\index{CKM!matrix elements!$V_{cd}$} ($|V_{cd}|$) actually fixed from CKM\index{CKM!unitarity} unitarity. To do so, lattice QCD\index{lattice QCD!in heavy-to-light transitions} will hopefully help us in the near future.

\par

However, the actual shape of the form factors has been a subject of many discussions in the literature and at least its qualitative understanding can possibly help us solve the hadrodynamics in situations that are far more complicated (notably in the decays\index{decay of baryon} of baryons etc.).  A lack of precise information about the shapes of various form factors is still the main source of uncertainties in many of these processes.

\par

On the experimental side there exist a number of interesting results on $D$\index{meson!$D$!decay} meson semileptonic decays \index{semileptonic decay}\index{decay of $D$ meson}.  The CLEO\index{CLEO} and FOCUS\index{FOCUS} collaborations have studied semileptonic decays \index{semileptonic decay}\index{decay!$D^0\to \pi^- \ell \nu$}\index{decay!$D^0\to K^- \ell \nu$} $D^0\to \pi^- \ell^+ \nu$ and $D^0\to K^- \ell^+ \nu$~\cite{Huang:2004fr,Link:2004dh}. Their data provides relevant information on the $D^0\to \pi^- \ell^+ \nu$ and $D^0\to K^- \ell^+ \nu$ form factors'  shapes. Usually in $D$\index{meson!$D$!decay} semileptonic decays \index{semileptonic decay}\index{decay of $D$ meson} a simple pole parametrization has been used in the past. The results of refs.~\cite{Huang:2004fr,Link:2004dh} for the single pole parameters required by the fit of their data, however, suggest pole masses, which are inconsistent with the physical masses of the lowest lying charm meson resonances. In their analysis they also utilized a modified pole fit as suggested in~\cite{Becirevic:1999kt} and their results indeed suggest the existence of contributions beyond the lowest lying charm meson resonances\index{resonance!of charmed meson}~\cite{Huang:2004fr}. The experimental situation in $D \to V \ell \nu_{\ell}$ at the same time has also been gaining pace~\cite{Link:2002wg,Link:2004qt,Link:2004gp,Link:2005ge,Coan:2005iu,Huang:2005iv}.

\par

There exist many theoretical calculations describing semileptonic decays \index{semileptonic decay}\index{transition!heavy-to-light} of heavy to light mesons: quark models
(QM)~\cite{Wirbel:1985ji,Isgur:1988gb,Scora:1995ty,Choi:1999nu,Melikhov:2000yu,Wang:2002zb}, QCD\index{QCD sum rules} sum rules
(SR)~\cite{Ball:1991bs,Ball:1993tp,Yang:1997ar,Khodjamirian:2000ds,Du:2003ja,Aliev:2004qd}, lattice QCD\index{lattice QCD!in heavy-to-light transitions}~\cite{Flynn:1997ca,Aubin:2004ej} and a few attempts to use combined heavy meson and chiral Lagrangian theory
(HM$\chi$T)~\cite{Casalbuoni:1996pg,Wise:1992hn}. Most of the above methods have limited range of applicability. For example, QCD\index{QCD sum rules} sum rules are suitable only for describing the low $s$ region while
lattice QCD\index{lattice QCD!in heavy-to-light transitions} and HM$\chi$T\index{HM$\chi$PT!in heavy-to-light transitions} give good results only for the high $s$ region.  However, the quark models, which do provide the full $s$ range of the form factors , cannot easily be related to the QCD Lagrangian\index{QCD!Lagrangian} and require input parameters, which may not be of fundamental significance~\cite{Melikhov:2000yu}. In addition to studies of heavy to light pseudoscalar meson weak transitions ($H\to P$), transitions of heavy pseudoscalar mesons to light vector mesons ($H\to V$) such as $D_s\to \phi \ell \nu_{\ell}$ an $D_s\to K^* \ell \nu_{\ell}$ offer an opportunity to extract the size of the relevant CKM\index{CKM!matrix elements} matrix elements or probe different approaches to form factor  calculations. The $H\to V$ transitions were also already carefully investigated within many different frameworks such as perturbative QCD~\cite{Kurimoto:2001zj,Mahajan:2004dx}, QCD\index{QCD sum rules} sum rules~\cite{Ball:1993tp, Du:2003ja, Aliev:2004qd, Ball:2004rg, Bakulev:2000fb, Wang:2001bh}, lattice QCD\index{lattice QCD!in heavy-to-light transitions}~\cite{Flynn:1995dc, DelDebbio:1997kr, Demchuk:1997uz, Abada:2002ie}, a few attempts to use combined heavy meson and chiral Lagrangians (HM$\chi$T)~\cite{Casalbuoni:1996pg,Bajc:1995km}, quark models~\cite{Wirbel:1985ji, Scora:1995ty, Melikhov:2000yu, Faustov:1995bf}, large energy effective theory (LEET\index{LEET})\index{large energy effective theory|see{LEET}}~\cite{Charles:1998dr} and soft collinear effective theory (SCET)\index{SCET}~\cite{Beneke:2000wa,Bauer:2000yr, Burdman:2000ku, Ebert:2001pc, Hill:2004if, Hill:2004rx}.

\par

The purpose of this section is to investigate whether a theoretically and phenomenologically consistent form factor\index{form factor!parameterization} parameterization can be conceived by saturating the dispersion relations for the form factors\index{form factor!dispersion relations} by one or more effective poles in $s$ and at the same time encompassing all the relevant symmetry constraints. Therefore we will review the general BK parameterization of the $H\to P$ form factors\index{form factor!parameterization} due to Be\'cirevi\'c and Kaidalov~\cite{Becirevic:1999kt}, and devise a similar parameterization also for all the form factors relevant to $H\to V$ weak transitions which would take into account known experimental results on heavy meson resonances\index{resonance!of heavy meson} as well as known theoretical limits of HQET\index{HQET!in heavy-to-light transitions} and LEET\index{LEET} relevant to $H\to V$ weak  transitions. Furthermore we would like to investigate contributions of the newly discovered charm mesons discussed in the Introduction to $D\to P$ and $D\to V$ semileptonic decays\index{transition!$D\to P$}\index{transition!$D\to V$} within an effective model based on HM$\chi$PT\index{HM$\chi$PT inspired model} by incorporating the newly discovered heavy meson fields into the HM$\chi$PT\index{HM$\chi$PT!Lagrangian} Lagrangian and utilizing the general form factor parameterization. We restrain our discussion to the leading chiral and $1/m_H$ terms in the expansion, but we hope to capture the main physical features about the impact of
the nearest poles in the $t$-channel to the $s$-dependence of the form factors .

\par

\subsection{Semileptonic heavy to light meson form factors}
\index{form factor!heavy-to-light}
\index{form factor!$H\to P$}
\index{form factor!$H\to V$}

We will work in the static limit of HQET\index{HQET!in heavy-to-light transitions}\index{HQET!static limit} where the eigenstates of QCD\index{QCD} and HQET\index{HQET!Lagrangian} Lagrangians are related via eq.~(\ref{eq_2.15}). In ref.~\cite{Becirevic:2002sc} it was pointed out that in the limit of a static heavy meson one can use the following decomposition:
\begin{eqnarray}
\langle P (p_P) | J^{\mu}_{V} | H (v) \rangle_{\mathrm{HQET}} &=& \left[ p_P^{\mu} - (v\cdot p_P) v^{\mu} \right] f_p(v\cdot p_P) + v^{\mu} f_v(v\cdot p_P),
\label{HQET_ff}
\end{eqnarray}
where the form factors  $f_{p,v}$ are functions of the variable
$v\cdot p_P = (m_H^2 +m_P^2-s)/2m_H$, which in the heavy meson rest frame is the energy of the light meson $E_P$. The important thing to note is that $f_{p,v}$ as defined in eq.~(\ref{HQET_ff}) are independent of the heavy quark mass and thus do not scale with it. The form factors  $F_{+,0}$ given in (\ref{def-ff}) and the form factors  $ f_{p,v}(v\cdot p_P)$ are related to each other by matching\index{matching!HQET to QCD} QCD to HQET\index{HQET!in heavy-to-light transitions}\index{HQET!matching to QCD} at the scale $\mu \simeq m_Q$~\cite{Eichten:1989zv,Broadhurst:1994se}. We compare compare the time and space components of eqs.~(\ref{def-ff}) and~(\ref{HQET_ff}) in the static frame of the heavy initial state meson ($v^0=1$, ${\bf v} = 0$) to obtain:\index{heavy quark expansion!in heavy-to-light decays}
  \begin{eqnarray}
  F_+(s) + \frac{m_H^2-m_P^2}{s} \left[ F_+(s) - F_0(s) \right]|_{s\approx s_{max}} &=& C_{\gamma_1}(m_Q) \sqrt m_H \left[ f_p(v\cdot p) + \mathcal O (1/m_Q) \right],\nonumber\\
  &&\hspace{-7.5cm} (m_H+E_P) F_+(s) + (m_H-E_P)\frac{m_H^2-m_P^2}{s} \left[ F_+(s) - F_0(s) \right]|_{s\approx s_{max}} \nonumber\\*
  &=& C_{\gamma_0}(m_Q) \sqrt m_H \left[ f_v(E_p) + \mathcal O (1/m_Q) \right].\nonumber\\
  \end{eqnarray}
We fix the matching  constants $C_{\gamma_i}$ to their tree level values $C_{\gamma_i}=1$. This approach immediately accounts for the $F_{+,0}$ behavior at $s_{\mathrm{max}}$.
At the leading order in heavy quark expansion, the two definitions are then related near zero recoil momentum ($s\simeq s_{\mathrm{max}}=(m_H-m_P)^2$ or equivalently $|{\bf p}_P|\simeq0$) as
\begin{subequations}
\begin{eqnarray}
 F_0(s)|_{s\approx s_{\mathrm{max}}} &=& \frac{1}{\sqrt{m_H}} f_v(v\cdot p_P)
 \label{ff_rel1}\\*
 F_+(s)|_{s\approx s_{\mathrm{max}}} &=& \frac{\sqrt{m_H}}{2} f_p(v\cdot p_P).
 \label{ff_rel2}
\end{eqnarray}
\end{subequations}

\par

Similarly in this limit it is also more convenient to use parametrization of the $H\to V$ transitions in which the form factors\index{form factor!$H\to V$} are independent of the heavy meson mass, namely we propose
\begin{subequations}
\begin{eqnarray}
    \langle V (\epsilon,p_V) | J_V^{\mu} | H (v) \rangle_{\mathrm{HQET}} &=& g_v \epsilon^{\mu\nu\alpha\beta} \epsilon_{\nu}^* v_{\alpha} p_{V\beta}, \\
    \langle V (\epsilon,p_V) | J_A^{\mu} | H (v) \rangle_{\mathrm{HQET}} &=& - i a_2 (\epsilon^*_V \cdot v) \left[p^{\mu}_V - (v\cdot p_V) v^{\mu}\right]\nonumber\\*
    && - i a_1 \left[ \epsilon^{*\mu}_V - (v \cdot \epsilon_V^*) v^{\mu} \right]\nonumber\\*
    && - i a_0 (v\cdot \epsilon^{*}_V) v^{\mu},
\end{eqnarray}
\end{subequations}
The form factors $g_v$, $a_1$, $a_2$ and $a_0$ are functions of the variable $ v \cdot p_V = (m_H^2 + m_V^2 - s)/2m_H$. In such decomposition, again all the form factors ($g_v$, $a_1$, $a_2$ and $a_0$) scale as constants with the heavy meson mass. The relation between the two form factor\index{form factor!decomposition} decompositions is obtained by correctly matching\index{matching!HQET to QCD} QCD\index{QCD} and HQET\index{HQET!in heavy-to-light transitions}\index{HQET!matching to QCD} at the scale $\mu \sim m_Q$~\cite{Eichten:1989zv,Broadhurst:1994se}:\index{heavy quark expansion!in heavy-to-light decays}
\begin{eqnarray}
    \frac{C_{\gamma_1}(m_Q)}{\sqrt m_H} \left[ g_v(v\cdot p_V) + \mathcal O(1/m_H) \right] &=&  \frac{2 V(s)}{m_H+m_V}|_{s\approx s_{\mathrm{max}}},\nonumber\\*
    \frac{C_{\gamma_0\gamma_5}(m_Q)}{ \sqrt{m_H}} \left[ a_0(v\cdot p_V) +\mathcal O(1/m_H)\right] &=& \nonumber\\*
     &&\hspace{-3.5cm} \Bigg\{ \frac{(m_H-E_V)}{s} \left[ 2 m_V A_0(s) + (m_H+m_V) A_1(s) - (m_H-m_V) A_2(s) \right] \nonumber\\*
    &&\hspace{-3.cm} + \frac{(m_H+E_V)}{m_H+m_V} A_2(s) - \frac{(m_H+m_V)}{m_H} A_1(s) \Bigg\} \Bigg|_{s\approx s_{\mathrm{max}}},\nonumber\\*
    C_{\gamma_1\gamma_5}(m_Q) \sqrt{m_H} \left[a_1(v\cdot p_V) + \mathcal O(1/m_H) \right] &=&  (m_H+m_V)  A_1(s) |_{s\approx s_{\mathrm{max}}},\nonumber\\*
    \frac{C_{\gamma_1\gamma_5}(m_Q)}{\sqrt m_H} \left[a_2(v\cdot p_V) + \mathcal O(1/m_H) \right] &=& \nonumber\\*
     &&\hspace{-3.5cm} \Bigg\{ \frac{m_H+m_V}{s} \left[A_1(s) + A_0(s)\right] - \frac{m_H-m_V}{s} \left[ A_2(s) + A_0(s)\right] \nonumber\\*
    && \hspace{-3.cm}- \frac{A_2(s)}{m_H+m_V} \Bigg\} \Bigg|_{s\approx s_{\mathrm{max}}}.
\end{eqnarray}
In the following we set the matching  constants $C_{\Gamma}$ to their tree level values $(C_{\gamma_i}=1)$. At leading order in $1/m_Q$ we thus get
\begin{subequations}
\begin{eqnarray}
\label{eq_HQET_scaling1}
V(s)|_{s\approx s_{\mathrm{max}}} &=& \frac{\sqrt m_H}{2 } g_v (v \cdot p_V), \\*
A_1(s)|_{s\approx s_{\mathrm{max}}} &=& \frac{1}{\sqrt m_H} a_1 (v \cdot p_V), \\*
A_2(s)|_{s\approx s_{\mathrm{max}}} &=& \frac{\sqrt m_H}{2} a_2 (v \cdot p_V), \\*
A_0(s)|_{s\approx s_{\mathrm{max}}} &=& \frac{\sqrt m_H}{2 m_V } a_0 (v \cdot p_V),
\label{eq_HQET_scaling}
\end{eqnarray}
\end{subequations}
which exhibit the usual heavy meson mass scaling laws for the semileptonic form factors ~\cite{Isgur:1990kf}. This parametrization is especially useful when calculating the form factors within HM$\chi$PT\index{HM$\chi$PT!in heavy-to-light transitions}\index{form factor!heavy-to-light}. The individual contributions of different terms in the HM$\chi$PT\index{HM$\chi$PT!Lagrangian} Lagrangian to various form factors can be easily projected out.

\subsection{Relations in HQET and SCET and Form Factor Parameterization\label{sec_5.1.2}}
\index{form factor!relations}
\index{SCET}

Now we turn to the discussion of the form factor  $s$ distribution, where we follow the analysis of ref.~\cite{Becirevic:1999kt}. As already evident from eqs.~(\ref{ff_rel1},\ref{ff_rel2}) and~(\ref{eq_HQET_scaling1}-\ref{eq_HQET_scaling}), due to the heavy mass invariance of the HQET\index{HQET!static limit} form factors, there exist the well known HQET scaling laws in the limit of zero recoil~\cite{Isgur:1990kf}.
On the other hand in the large energy limit $s\to 0$, one obtains the following expressions for the form factors ~\cite{Charles:1998dr}
\begin{subequations}
\begin{eqnarray}
\label{eq_LEET_scaling1}
F_{+}(s)|_{s\approx 0} &=& \xi(E_P), \\*
\label{eq_LEET_scaling2}
F_{0}(s)|_{s\approx 0} &=& \frac{2 E_P}{m_H} \xi(E_P), \\*
V(s)|_{s\approx 0} &=& \frac{m_H + m_V}{m_H} \xi_{\perp} (E_V), \\*
A_1(s)|_{s\approx 0} &=& \frac{2 E_V} {{m_H + m_V}} \xi_{\perp} (E_V), \\*
A_2(s)|_{s\approx 0} &=& \frac{m_H + m_V}{m_H} \left[ \xi_{\perp} (E_V) - \frac{m}{E} \xi_{\parallel} (E_V) \right],\\*
A_0(s)|_{s\approx 0} &=& \left( 1 - \frac{m_V^2}{2 E_V m_H} \right) \xi_{\parallel} (E_V) \approx \xi_{\parallel} (E_V),
\label{eq_LEET_scaling}
\end{eqnarray}
\end{subequations}
where in the limit of large energy of the light meson (large recoils), in the rest frame of the heavy meson
\begin{equation}
E_{P,V} = \frac{m_H}{2}\left( 1-\frac{s}{m_H^2} + \frac{m_{P,V}^2}{m_H^2}\right)
\end{equation}
These scaling laws were subsequently confirmed by means of SCET~\cite{Beneke:2000wa, Ebert:2001pc}\index{SCET}. This is important since the LEET\index{LEET} description breaks down beyond the tree level due to missing soft gluonic degrees of freedom which are however systematically taken into account within SCET\index{SCET}~\cite{Burdman:2000ku, Hill:2004rx}. Still one needs to keep in mind that these scaling laws are also subject to $1/m_H$\index{heavy quark expansion!in heavy-to-light decays} power corrections and sizable deviations might occur for finite heavy meson masses, especially in the case of charmed mesons or conversely for large enough $m_{P,V}$ (e.g. for final state kaons\index{meson!$K$} etas or light vector mesons).

\par

\begin{figure}[!t]
\begin{center}
\hspace*{-2cm}\resizebox{!}{5.5cm}{\includegraphics{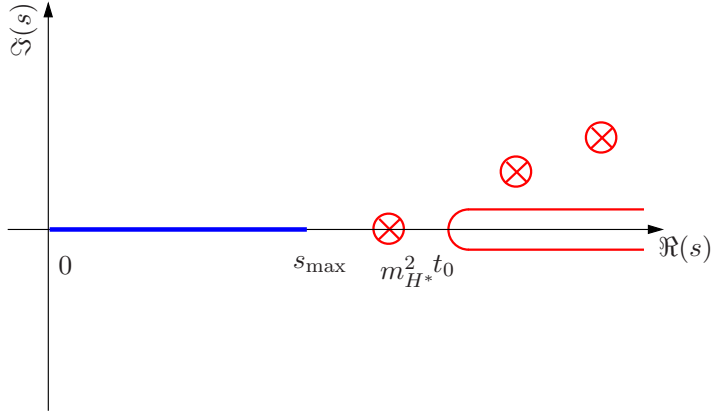}}
\end{center}
\caption{\small\it \label{fig_5_1}A schematic view of the $s$ worldsheet in heavy to light ($H\to P$) semileptonic decays, with imaginary contributions to $F_+$ form factor  marked in red. Crossed circles indicate quasi-stable particle (resonance) poles, while the cut along the real axis represents the $t$-channel $HP$ pair emission above threshold $t_0$. The physical kinematical region in the $s$-channel is marked with blue.}
\end{figure}

The starting point are the vector form factors $V$ and $F_+$, which
in the part of the phase space that is close to zero recoil are
dominated by the first known pole due to the heavy vector meson
resonance\index{resonance!of heavy meson} at $t=m_{H^*}^2$. The important thing is that this pole
lies below the $H P$ pair production threshold for both charm and
bottom meson sectors $t_0 = (m_H + m_{P,V})^2$. Therefore the analytic
structure of these form factors  dictates the following dispersion
relations~\cite{Peskin:1995ev, Donoghue:1992dd}\index{form factor!dispersion relation}\index{form factor!resonance contributions}
\begin{equation}
F_+(s) = \frac{\mathrm{Res}_{s=m_{H^*}^2}F_+(s)}{m_{H^*}^2-s} + \frac{1}{\pi} \int_{t_0}^{\infty} dt \frac{\mathrm{\Im} [F_+(t)]}{t-s-i\epsilon},
\end{equation}
with an analogous expression for $V$. The imaginary parts are
consisting of all single and multiparticle states with $J^P=1^-$,
thus a multitude of poles and cuts above $t_0$ (see
fig.~\ref{fig_5_1}).
\psfrag{s}[cl]{$\Re(s)$}
\psfrag{ImF}[cc][cc][1][90]{$\Im(s)$}
\psfrag{t0}[tl]{$t_0$}
\psfrag{mH}[tl]{$m_{H^*}^2$}
\psfrag{s0}[tl]{$0$}
\psfrag{smax}[tl]{$s_{\mathrm{max}}$}
However the $H^*$ contribution can be singled out and the residues at
the pole at $s=m_{H^*}^2$ can be identified using the resonance\index{resonance dominance approximation} dominance approximation, which is certainly valid in the vicinity of the isolated resonance. We write the $H\to P$ current matrix element in the $P$ crossed channel as $\bra{0} J_{\mu} \ket{H(p_H) P(p_P)} \simeq \bra{0} J_{\mu} \ket{ H^*} \otimes G_{H^*} \otimes \braket{ H^*|H(p_H) P(p_P)}$, where $G_{H^*}$ is the $H^*$ propagator. We use the standard definitions for $\braket{H(p_H) P(p_P) | H^*(p_H+p_P,\epsilon)}$ (eq.~(\ref{eq_strong_v})) and  $\bra{0} J_{\mu} \ket{H^*(p_{H^*},\epsilon)}$ (eq.~(\ref{eq_3_19})).
Inserting the vector particle propagator for $H^*$ we obtain
\begin{equation}
\bra{0} J_{\mu} \ket{H(p_H) P(p_P)} \simeq -\frac{G_{H P H^*} m_{H^*} f_{H^*}}{2(t-m_{H^*}^2)} \left[(p_{H}-p_{P})_{\mu} - \frac{m_H^2-m_P^2}{m_{H^*}^2} (p_{H}+p_{P})_{\mu}\right],
\end{equation}
from which we can immediately read off the form factor  residual\index{form factor!resonance contributions}
\begin{eqnarray}
\mathrm{Res}_{s=m_{H^*}^2} F_+(s) &=& \frac{1}{2}G_{H^* H P} f_{H^*} m_{H^*}
\end{eqnarray}
It scales as $\sim m_H^{3/2}$ with the heavy meson
mass~\cite{Becirevic:1999kt} as can be easily inferred from the HQET\index{HQET!static limit} scaling of $G_{H^* H P}$~(\ref{eq_4.11}) and $f_{H^*}$\footnote{In the case of $f_{H^*}$ we have to take into account the appropriate HQET\index{HQET!static limit} scaling of external states in~(\ref{eq_3_19}), leading to $f_{H^*}\sim m_{H^*}^{-1/2}$ up to logarithmic corrections due to the perturbative matching to QCD\index{QCD}.}. For the heavy to light transitions this situation is expected to be realized near the zero recoil where
the HQET\index{HQET!static limit} scaling~(\ref{ff_rel1},\ref{ff_rel2},\ref{eq_HQET_scaling1}-\ref{eq_HQET_scaling}) applies. However,
since the kinematically accessible region $s\in(0,s_{\mathrm{max}}]$
is large, the pole dominance can be used only on a small fraction of
the phase space, $i.e.$ for $|{\bf p}_i - {\bf p}_f|\approx 0$. Even
in this region the situation for $H\to V$ form factors\index{form factor!$H\to V$} is more
complex than in the case of $H\to P$ transitions, where
$s_{\mathrm{max}}$ is indeed very close to the vector pole due to
low mass of the light pseudoscalar mesons. Here, due to larger
masses of the light vector mesons, $s_{\mathrm{max}}$ is pushed away
from the resonance pole and $V(s)$ may not be completely saturated
by it.
For the sake of clarity we shall, however, at present
neglect such possible discrepancies and assume
complete saturation of both vector form factors ($F_+$ in $H\to P$
and $V$ in $H\to V$ transitions) in this region by the first
physical resonance\index{resonance}. On the other hand, in the region of large
recoils, LEET\index{LEET} dictates the scaling~(\ref{eq_LEET_scaling1}-\ref{eq_LEET_scaling}). We see immediately, that a simple pole ansatz for the vector form factors, which assumes that all the states above $t_0$ which couple to the vector components of the current would eventually cancel, would produce the wrong scaling at $s=0$ of $F_+(0)\sim V(0) \sim m_{H}^{-1/2}$. Instead, we can try to take into account possible non-vanishing contributions to the form factors  from all states above $t_0$ by adding an additional effective pole term to the form factor expression
\begin{equation}
F_+(s) = c_H \left ( \frac{1}{1-x} - \frac{a}{1- \gamma x} \right),
\label{eq_5_12}
\end{equation}
where $c_H=-g_{H^* H P} f_{H^*}/m_{H^*}$, $x=s/m_{H^*}^2$  ensures, that the form factor  is dominated by the physical $H^*$ pole, while $a$ and $\gamma$ are positive constants. The form factor scaling laws in the LEET\index{LEET} and HQET\index{HQET!static limit} limits give their scaling with the heavy meson mass as $(1-a) \sim (1-\gamma) \sim 1/m_H$.

\par

Next we use the form factor relations at $s=0$ and construct the scalar form factor ($F_0$) in the same way, such that it also satisfies all scaling limits
\begin{equation}
F_0(s) = c_H  \frac{1-a}{1-b x},
\label{eq_5_f0}
\end{equation}
where $b$ now parameterizes the effective pole due to all states coupling to the scalar component of the current. Finally, we note that the LEET\index{LEET} limit is even more constraining on the two form factors, as it imposes the following relation~(\ref{eq_LEET_scaling1},\ref{eq_LEET_scaling2}): $F_0(s) = 2 E_P F_{+}(s)/m_H $, which when translated in terms of our parameters reads $(1-a) = (1-\gamma)[1 + \mathcal O(1/m_H)]$. The constraint can be satisfied for $a = \gamma$, leading to a much simplified expression for the $F_+$ form factor \index{form factor!parameterization}
\begin{equation}
F_+(s) = c_H  \frac{1-a}{(1-x)(1- a x)}.
\label{eq_5_f+}
\end{equation}

\par

In the full analogy with the discussion made above\cite{Becirevic:1999kt, Hill:2005ju}, the vector form factor $V$
also receives contributions from two poles\index{two poles parameterization of heavy-to-light form factors} and can be
written as\index{form factor!parameterization}
\begin{equation}
V(s) = c'_H \frac{1-a'}{(1-x)(1-a' x)},
\label{eq_v_ff}
\end{equation}
where again $a'$ measures the contribution of higher states which are parametrized by the second effective pole at $m_{\mathrm{eff}}^2=m_{H^*}^2/a'$.  Note that although similar in parameterization, the $a'$ and $c_H'$ are not the same as $a$ and $c_H$, since neither the $s=m_{H^*}^2$ pole residual nor the threshold region above $t_0$ are the same. Still, the parameters $c'_H$ and $a'$ scale with the heavy meson mass as before $c'_H\sim m_H^{-1/2}$ and $(1-a')\sim 1/m_{H}$ to ensure the correct form factor scaling in both small and large recoil regions. Again using the large energy limit relation between $V$ and $A_1$~\cite{Charles:1998dr}
\begin{equation}
\left[ V(s)/A_1(s) \right]|_{s\approx 0} = \frac{(m_H+m_V)^2}{2 E_V m_H},
\label{eq_VA_LEET}
\end{equation}
(valid up to terms $\propto 1/m_H^2$~\cite{Ebert:2001pc}) we can impose a single pole structure on $A_1$. We thus continue in the same line of argument as before and write\index{form factor!parameterization}
\begin{equation}
A_1(s) = c'_H \xi \frac{1-a'}{1-b' x}.
\label{eq_a1_ff}
\end{equation}
Here $\xi=m_H^2/(m_H+m_V)^2$ is the proportionality factor between $A_1$ and $V$ from~(\ref{eq_VA_LEET}), while $b'$ measures the contribution of higher states with spin-parity assignment $1^+$ which are parametrized by the effective pole at $m_{H'^*_{\mathrm{eff}}}^2=m_{H^*}^2/b'$. It can be readily checked that also $A_1$, when parametrized in this way, satisfies all the scaling constraints.

\par

Next we parametrize the $A_0$ form factor, which is completely independent of all the others so far as it is dominated by the pseudoscalar pole and is proportional to a different universal function in LEET\index{LEET}. To satisfy both HQET\index{HQET!static limit} and LEET\index{LEET} scaling laws we parametrize it as\index{form factor!parameterization}
\begin{equation}
A_0(s) = c''_H \frac{1-a''}{(1-\widetilde y)(1-a'' \widetilde y)},
\label{eq_a0_ff}
\end{equation}
where $\widetilde y = s/m_{\widetilde H}^2$ ensures the physical $0^-$ pole dominance at small recoils. Imposing $c''_H\sim m_H^{-1/2}$ and $(a'- 1)\sim 1/m_H$ preserves all scaling laws, while $a''$ again parametrizes the contribution of higher pseudoscalar states by an effective pole at $m_{\mathrm{eff}}^2=m_{\widetilde H}^2/a''$. Note that $m_{\widetilde H}$ mass appearing in $\widetilde y$ is due to the intermediate $0^-$ heavy meson state with the flavor quantum numbers of the quark current, and not the initial state. The resemblance to $V$ and $F_+$ is obvious and due to the same kind of analysis~\cite{Becirevic:1999kt} although the parameters appearing in the two form factors are again completely unrelated.

\par

Finally for the $A_2$ form factor, due to the pole behavior of the $A_1$ form factor on one hand and different HQET\index{HQET!static limit} scaling at $s_{\mathrm{max}}$~(\ref{eq_HQET_scaling1}-\ref{eq_HQET_scaling}) on the other hand, we have to go beyond a simple pole formulation. Thus we impose\index{form factor!parameterization}
\begin{equation}
A_2(s) = \frac{c'''_H}{(1-b' x)(1-b'' x)},
\label{eq_a2_ff}
\end{equation}
where $c'''_H = [(m_H+m_V) \xi c'_H (1-a) + 2 m_V c''_H (1-a')]/(m_H-m_V)$ is determined from form factor relations at $s=0$ and kinematic constraint~(\ref{PV_form_factor_relations}) so that we only gain one new parameter in this formulation, $b''$. This however causes the contribution of the $1^+$ resonances\index{resonance contribution!in heavy-to-light transitions} to be shared between the two effective poles in this form factor. At the end we have parametrized the two $H\to P$ form factors and four $H\to V$ form factors in terms of three ($c_H$, $a$, $b$) and  six parameters ($c'_H$, $a'$, $b'$, $a''$, $c''_H$, $b''$) respectively.

\par

We can now shortly comment on the LEET\index{LEET} and HQET\index{HQET!static limit} limits of the $H\to V$ transitions in our parameterization. As shown in ref.~\cite{Ebert:2001pc} the helicity amplitudes\index{helicity amplitudes} of eq.~(\ref{eq_helicity}) can be related to individual form factors near $s=0$. Using relations~(\ref{eq_LEET_scaling1}-\ref{eq_LEET_scaling}), valid in the large energy limit, one can write
\begin{eqnarray}
H_{-}(y)|_{y\approx 0} &\approx& 2(m_H+m_V) A_1(m_H^2 y),\nonumber\\
H_{+}(y)|_{y\approx 0} &\simeq& 0.
\end{eqnarray}
Thus in this region we can probe directly for the parameter $c'_H(1-a')$. 

\par

On the other hand in the region of small recoil ($|{\bf p}_V| \simeq 0$ or $y\approx y_{\mathrm{max}}$) the helicity amplitudes\index{helicity amplitudes} are saturated by the $A_1$ form factor
\begin{eqnarray}
H_{\pm}(y)|_{y\approx y_{\mathrm{max}}} &\approx& (m_H+m_V) A_1(m_H^2 y), \nonumber\\
H_{0}(y)|_{y\approx y_{\mathrm{max}}} &\approx& -2 (m_H+m_V) \frac{m_V}{m_H} A_1(m_H^2 y).
\end{eqnarray}
Consequently we can also directly probe for the value of the $b'$ parameter determining the position of the first effective axial resonance pole by taking a ratio of $H_-$ helicity amplitude\index{helicity amplitudes} values at small and large recoils
\begin{equation}
\frac{H_{-}(y)|_{y\approx 0}}{H_{-}(y)|_{y\approx y_{\mathrm{max}}}} \approx 2 \left[ 1 - b' (m_H-m_V)^2/m_H^2 \right].
\end{equation}

\index{excited state|see{resonance}}

\subsection{HM$\chi$PT description including excited states\label{sec_hmchpt_HL}}
\index{HM$\chi$PT!in heavy-to-light transitions}
\index{resonance contribution!in heavy-to-light transitions}

Going beyond the general discussion of the previous subsections, we now attempt to determine the parameters of the general form factor parameterizations within an effective theory description based on heavy quark and chiral\index{chiral symmetry} symmetries and by including phenomenologically motivated dynamical heavy meson states into the model. These states will represent the lowest lying excited heavy meson resonances\index{resonance!of heavy meson}\index{resonance contribution!in heavy-to-light transitions} which we will then assume to saturate the pole structure of the form factor parameterizations.

\par

\begin{figure}[!t]
\begin{center}
\epsfxsize8cm\epsffile{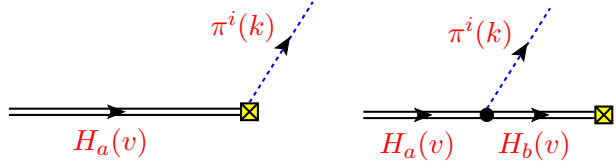}
\end{center}
\caption{\small\it \label{diagram_HP}Diagrams contributing to $H\to P$ form factors. The square stands for the weak current vertex. }
\end{figure}

For the semileptonic decays \index{semileptonic decay} the weak Lagrangian can be given by the
effective current-current Fermi\index{Fermi interaction} interaction\index{G$_F$}
\begin{equation}
\mathcal L_{\mathrm{eff}} = - \frac{G_F}{\sqrt 2} \left[ \bar \ell \gamma^{\mu} (1-\gamma^5) \nu_{\ell} \mathcal J_{\mu}  \right],
\end{equation}
where $G_F$ is the Fermi\index{G$_F$} constant and $\mathcal J_{\mu}$ is the effective
hadronic current. In heavy to light meson decays\index{transition!heavy-to-light} it can be written as
$\mathcal J_{\mu} = K_a J^a_{\mu}$, where constants $K^a$ parametrize the {\it heavy-light}
flavor mixing\index{mixing!of quark flavors}. In the HM$\chi$PT\index{HM$\chi$PT!in heavy-to-light transitions}\index{form factor!heavy-to-light} description,the $H\to P$ leading order weak current $J^a_{\mu}$ in $1/m_H$ and chiral expansion is given by eq.~(\ref{eq_2_17}). In HM$\chi$PT\index{HM$\chi$PT!Feynman rules} derived Feynman rules are valid near zero recoil ($|{\bf p}_V|\simeq0$) where we get leading order contributions to the effective current matrix elements from Feynman diagrams in fig.~\ref{diagram_HP}.
\psfrag{Ha}[cc]{${\Red{H_a(v)}}$}
\psfrag{Hb}[cc]{$\Red{H_b(v)}$}
\psfrag{pi}[bc]{$\Red {\pi^i(k)}$}
We see that the left diagram structure already mimics resonant contributions\index{resonance contribution!in heavy-to-light transitions}. However, when examining the pole structure of the form factor parameterization, we see, that only a single pole of $F_+$ can be identified with the single $1^-$ state in a HM$\chi$PT\index{HM$\chi$PT!in heavy-to-light transitions}\index{form factor!heavy-to-light} construction when taking into account a single $1/2^-$ and $1/2^+$ heavy meson multiplet. We attempt to cope with the problem by introducing a second $1/2^-$ heavy meson multiplet $\widetilde H$ representing radially excited pseudoscalar and vector states. Experimentally, hints have been given in the past of the existence of such relatively long lived radially excited states in the charmed sector~\cite{Evdokimov:2004iy, Abreu:1998vk}, however, the initial observations have not been confirmed~\cite{Rodriguez:1998ng,Abbiendi:2001qp}. The required modifications of the HM$\chi$PT\index{HM$\chi$PT!Lagrangian} strong and weak Lagrangians are straightforward\index{chiral expansion!in heavy-to-light transitions}\index{heavy quark expansion!in heavy-to-light decays}
\begin{eqnarray}
    \mathcal L^{(1)}_{\mathrm{HM}\chi\mathrm{PT}} &+=& \widetilde{\mathcal L}^{(1)}_{\frac{1}{2}^-} + \widetilde{\mathcal L}^{(1)}_{\mathrm{mix}}, \nonumber\\
    \widetilde{\mathcal L}^{(1)}_{\frac{1}{2}^-} &=& - \mathrm{Tr}\left[ \overline {\widetilde H}_a (i v \cdot \mathcal{D}_{ab} - \delta_{ab} \Delta_{\widetilde H} ) \widetilde H_b\right], \nonumber\\
    \widetilde{\mathcal L}^{(1)}_{\mathrm{mix}} &=& \widetilde h \mathrm{Tr} \left[ \overline H_b \widetilde H_a \slashed{\mathcal A}_{ab} \gamma_{5}  \right] + \mathrm{h.c.},
    \label{eq_5.25}
\end{eqnarray}
and
\begin{equation}
J^{(0)\mu}_{a(V-A)\mathrm{HM}\chi\mathrm{PT}} += \frac{i \widetilde \alpha}{2} \mathrm{Tr} [ \gamma^{\mu}(1-\gamma_5) \widetilde H_b] \xi_{ba}^{\dagger},
\end{equation}
where we have introduced three new parameters: the $\Delta_{\widetilde H}$ residual mass term of the second  $1/2^-$ multiplet, $\widetilde h$ coupling between the two $1/2^-$ heavy meson multiplets and pseudo-Goldstones\index{pseudo-Goldstone boson} and $\widetilde \alpha$\index{HM$\chi$PT!parameters} as the effective weak coupling of the new states, which is related to their decay\index{decay constant!of heavy meson} constants. Together with contributions from the ground state $1/2^-$ and lowest lying $1/2^+$ mesons we get for the weak $H\to P$ hadronic current matrix element
\begin{eqnarray}
\langle P_{ba}(p_P) | J^{\mu}_a | H_b(v) \rangle &=& \frac{\alpha}{f} v^{\mu}  + \frac{\alpha}{f} g \frac{p_P^{\mu}- v^{\mu} v\cdot p_P}{v\cdot p_P+\Delta_{ba}} \nonumber\\*
&& + \frac{\widetilde\alpha}{f} \widetilde h \frac{p_P^{\mu}-v^{\mu} v\cdot p_P}{v\cdot p_P + \Delta_{\widetilde H_b H_a}} + \frac{\alpha'}{f} h \frac{v^{\mu} v\cdot p_P}{v \cdot p_P + \Delta_{S_b H_a}},
\label{J_HMCT}
\end{eqnarray}
where $a,b$ as before denote light flavor indexes and we have introduced the mass splitting parameter between the two $1/2^-$ multiplets $\Delta_{\widetilde H_b H_a} = \Delta_{\widetilde H_b} - \Delta_{H_a}$, between $1/2^+$ and $1/2^-$ states $\Delta_{S_b H_a} = \Delta_{S_b} - \Delta_{H_a}$ as well as the mass splitting between the initial state and the intermediate $1/2^-$ ground state (vector) resonance $\Delta_{ba}=\Delta_{H_b}-\Delta_{H_a}$. While in the exact heavy quark and chiral symmetry\index{chiral limit} limits, the first two would be flavor and spin independent and the last would vanish, we are here keeping the flavor dependence as we expect large flavor $SU(3)$ breaking and will work only in the exact $SU(2)$ limit where $\Delta_{3a}=-\Delta_{a3}\approx 100~\mathrm{MeV}$ are nonvanishing for $a=1,2$. In this approximation $\Delta_{\widetilde H_b H_a} =\Delta_{\widetilde H_b H_b} + \Delta_{ba}$. In the remainder of the section we shall use these definitions while suppressing the flavor indices, which should from here on be considered implicit, and writing just $\Delta_{\widetilde HH}$, $\Delta_{SH}$  and $\Delta$.
We now apply the projectors $v_{\mu}$ and $p_{P\mu} - v_{\mu} v\cdot p_P$ on eq. (\ref{J_HMCT})  and extract the form factors $F_{+}(s)$, $F_0(s)$ using eqs.~(\ref{HQET_ff}) and~(\ref{ff_rel1}-\ref{ff_rel2}) \index{form factor!resonance contributions}
\begin{subequations}
\begin{eqnarray}
\label{F+FFq}
F_+(s)|_{s\approx s_{\mathrm{max}}} &=& \frac{\alpha}{2\sqrt m_H f} g \frac{m_H}{v\cdot p_P + \Delta} + \frac{\widetilde\alpha}{2\sqrt m_H f} \widetilde h \frac{m_H}{v\cdot p_P +
\Delta_{H\widetilde H}},\\
F_0(s)|_{s\approx s_{\mathrm{max}}} &=& \frac{\alpha}{\sqrt m_H f} + \frac{\alpha'}{\sqrt m_H f} h \frac{v\cdot p_P}{v\cdot p_P +
\Delta_{SH}}.
\label{F0FFq}
\end{eqnarray}
\end{subequations}
If one uses directly relation~(\ref{def-ff}) instead of this extraction of form factors at large $s_{\mathrm{max}}$~\cite{Becirevic:2002sc} one ends up with the mixed\index{mixing!of form factor resonance contributions} leading $\sqrt m_H$ terms and the subleading $1/\sqrt m_H$ terms in (\ref{F0FFq}). Furthermore, the scalar meson contribution appears in the $F_+$ form factor. The extraction of form factors we follow here~\cite{Becirevic:2002sc} gives a correct $1/m_H$ behavior of the form factors and the contributions of $1^-$ resonances\index{resonance contribution!in heavy-to-light transitions} enter in $F_+$, while $0^+$ resonances contribute to $F_0$ as they must~\cite{Marshak:1969tw}.

\par

One can attempt a similar procedure in the case of $H\to V$ transitions by using an effective $SU(3)$ model description of the light vector mesons and append it to the HM$\chi$PT\index{HM$\chi$PT!Lagrangian}\index{HM$\chi$PT!inspired model!of light vector interactions} Lagrangian\footnote{Note the important difference in the treatment so far, namely HM$\chi$PT is an effective theory based on the approximate symmetries of QCD\index{symmetries!of QCD}, spontaneous symmetry breaking and the Goldstone\index{Goldstone theorem} theorem, and its corrections can be computed perturbatively. On the other hand the $SU(3)$ description of vector (as well as scalar and other) light resonances\index{resonance!of light meson} employed here (and later) can only be cast into an effective model, about whose corrections we can only speculate.}. A common procedure of achieving this is the hidden symmetry\index{hidden symmetry}\index{gauge symmetry!hidden} approach (c.f.~\cite{Casalbuoni:1996pg}). In this approach, the light vector mesons are introduced in the HM$\chi$PT\index{HM$\chi$PT!Lagrangian} Lagrangian as gauge fields of an extended\index{HM$\chi$PT!extended $SU(3)_V$ symmetry} $SU(3)_V$ symmetry. The $\Sigma$ field belongs to its singlet representation -- hence the origin of the term hidden symmetry\index{hidden symmetry}. Light vector meson fields are introduced as gauge fields of $SU(3)_V$ and are described by $\hat\rho_{\mu} = i \frac{g_V}{\sqrt{2}} \rho_{\mu}$, where $\rho_{\mu}$ is the light vector meson field matrix in the adjoint octet representation of $SU(3)_V$
\begin{equation}
\rho_{\mu} =
   \begin{pmatrix}
    \frac{1}{\sqrt 2} (\omega_{\mu} + \rho^0_{\mu}) & \rho^+_{\mu} & K^{*+}_{\mu} \\
    \rho^-_{\mu} & \frac{1}{\sqrt 2} (\omega_{\mu} - \rho^0_{\mu}) & K^{*0}_{\mu} \\
    K^{*-}_{\mu} & \overline K^{*0}_{\mu} & \phi_{\mu}
   \end{pmatrix}.
\end{equation}
The kinetic and mass Lagrangian\index{HM$\chi$PT!Lagrangian} terms for the $\hat \rho$ fields are then simply
\begin{equation}
\mathcal L_V = \frac{1}{2g_V^2}[ F_{\mu\nu}(\hat\rho)_{ab} F^{\mu\nu}(\hat\rho)_{ba}] - \frac{a f^2}{2} (\mathcal V^{\mu}_{ab} - \hat \rho^{\mu}_{ab})(\mathcal V_{\mu,ba} - \hat \rho_{\mu,ba}),
\end{equation}
where the gauge field tensor is defined as $F_{\mu\nu}(\hat\rho) = \partial_{\mu} \hat\rho_{\nu} - \partial_{\nu} \hat\rho_{\mu} + [\hat \rho_{\mu} , \hat \rho_{\nu}]$, while $m_{\rho}^2 = a g_V^2 f^2/2$ is the degenerate mass squared of the vector octet. If we also examine the gauged vector current $J^{\mu}_V = i a f^2 \xi (\mathcal V^{\mu} - \hat \rho^{\mu}) \xi^{\dagger}$, we obtain $f_{\rho} = f^2 g_V a /\sqrt 2 m_{\rho}$ as the common vector decay\index{decay constant} constant fixing the model parameter $g_V=5.9$ phenomenologically\index{HM$\chi$PT!parameters}.

\par

At leading order in $1/m_H$ expansion, strong interactions between lowest lying  heavy meson fields, and light vector meson fields are described by the interaction Lagrangians\index{HM$\chi$PT!Lagrangian}~\cite{Bajc:1995km,Casalbuoni:1996pg}\index{chiral expansion!in heavy-to-light transitions}\index{heavy quark expansion!in heavy-to-light decays}
\begin{eqnarray}
\mathcal L^{\mathrm{int}}_{1/2^-} &=& - i \beta \mathrm{Tr} [H_b v_{\mu} \hat\rho^{\mu}_{ba} \overline H_a ] + i \lambda \mathrm{Tr} [ H_b \sigma^{\mu\nu} F_{\mu\nu}(\hat\rho)_{ba} \overline H_a ],\\
\mathcal L^{\mathrm{int}}_{\mathrm{mix}} &=&  - i \zeta \mathrm{Tr} [ H_b v_{\mu} \hat\rho^{\mu}_{ba} \overline S_a ] + \mathrm{h.c.} \nonumber\\*
&& + i \mu \mathrm{Tr} [ H_b \sigma^{\mu\nu} F_{\mu\nu}(\hat\rho)_{ba} \overline S_a ] + \mathrm{h.c.},\\
\widetilde{\mathcal L}^{\mathrm{int}}_{\mathrm{mix}} &=&  - i \widetilde \zeta \mathrm{Tr} [ H_b v_{\mu} \hat\rho^{\mu}_{ba} \overline {\widetilde{H}}_a ] + \mathrm{h.c.} \nonumber\\*
&& + i \widetilde \mu \mathrm{Tr} [ H_b \sigma^{\mu\nu} F_{\mu\nu}(\hat\rho)_{ba} \overline {\widetilde H}_a ] + \mathrm{h.c.},
\label{L_even_odd}
\end{eqnarray}
where the first terms in each row of the above expression are even under na\"ive parity transformation ($H(x) \to H(\mathcal P x)$, $\xi(x) \to \xi (\mathcal P x)$ with $\mathcal P (t,{\bf x} ) =  (t, -{\bf x})$), and the second terms are na\"ive parity odd (all are invariant under the real parity transformations $H(x) \to \gamma_0 H(\mathcal P x) \gamma_0$ and $\xi(x) \to \xi^{\dagger} (\mathcal P x)$). Similarly one can construct the corresponding weak current operator containing light vector fields and append it to the effective weak Lagrangian\index{HM$\chi$PT!Lagrangian}
\begin{eqnarray}
J^{(0)\mu}_{a(V-A)\mathrm{HM}\chi\mathrm{PT}} &+=& \alpha_1 \mathrm{Tr} [\gamma^5 H_b \hat\rho^{\mu}_{ba} ] + \alpha_2 \mathrm{Tr} [ \gamma^{\mu} \gamma^{5} H_b v_{\alpha} \hat\rho^{\alpha}_{ba} ].
\end{eqnarray}
Using these building blocks, one obtains exactly the same topology diagrams contributing to the $H\to V$ form factors, as the ones displayed in fig.~\ref{diagram_HP} with the replacement of the external pseudo-Goldstone\index{pseudo-Goldstone boson} lines with light vector boson lines. For the corresponding transition matrix element we get (we are again suppressing flavor indices)
\begin{eqnarray}
\langle V(p_V) | J^{\mu} | H(v) \rangle &=& - i \sqrt 2 g_V \left( \alpha_1 \epsilon_{V}^{\mu} - \alpha_2 v \cdot \epsilon_{V} v^{\mu} \right) \nonumber\\*
&& -\sqrt 2 g_V \alpha  \frac{\lambda \epsilon^{\mu\nu\alpha\beta} v_{\nu} p_{V\alpha} \epsilon_{V\beta}}{v\cdot p_V + \Delta} -\sqrt 2 g_V \widetilde \alpha  \frac{\widetilde\mu \epsilon^{\mu\nu\alpha\beta} v_{\nu} p_{V\alpha} \epsilon_{V\beta}}{v\cdot p_V + \Delta_{H\widetilde H}} \nonumber\\*
&& - i \frac{g_V}{\sqrt 2} \alpha \frac{\beta v \cdot \epsilon_V v^{\mu} }{v\cdot p_V + \Delta} - i \frac{g_V}{\sqrt 2} \widetilde\alpha  \frac{\widetilde\zeta v \cdot \epsilon_V v^{\mu}}{v\cdot p_V + \Delta_{H \widetilde H}} \nonumber\\*
&& - i \frac{g_V}{\sqrt 2} \alpha' \frac{\epsilon_V^{\mu} \left( \zeta - 2\mu v\cdot p_V \right) + \left( 2\mu p_V^{\mu} - \zeta v^{\mu} \right) v\cdot \epsilon_V }{v\cdot p_V + \Delta_{SH}},
\label{J_HMCPT}
\end{eqnarray}
where we the mass splitting $\Delta $ now refers to the initial and intermediate pseudoscalar heavy meson ground states. We apply the projectors $v_{\mu}$ and $v_{\mu} v\cdot p_P - p_{P\mu}$ on eq. (\ref{J_HMCPT})  and extract the form factors $V(s)$, $A_1(s)$, $A_2(s)$ and $A_0(s)$ at $s_{\mathrm{max}}$ using eqs.~(\ref{eq_HQET_scaling1}-\ref{eq_HQET_scaling}): \index{form factor!resonance contributions}
\begin{subequations}
\begin{eqnarray}
\label{eq_ff_HMcT1}
V(s)|_{s\approx s_{\mathrm{max}}} &=& -\frac{g_V}{\sqrt 2} \alpha m_H \sqrt m_H \frac{\lambda }{v\cdot p_V + \Delta} \\*
&& - \frac{g_V}{\sqrt 2} \widetilde\alpha m_H \sqrt m_H \frac{\widetilde \mu }{v\cdot p_V + \Delta_{H\widetilde H}} \\
A_1(s)|_{s\approx s_{\mathrm{max}}} &=& \frac{g_V}{\sqrt 2} \alpha' \frac{\sqrt m_H}{m_H + m_V} \frac{ \zeta - 2\mu (v\cdot p_V) }{v\cdot p_V + \Delta_{SH}} \\*
&& - \sqrt 2 g_V \alpha_1 \frac{\sqrt m_H}{m_H +m_V} \\
A_2(s)|_{s\approx s_{\mathrm{max}}} &=& \frac{g_V}{\sqrt 2} \alpha' \frac{m_H +m_V}{\sqrt m_H} \frac{\mu}{v\cdot p_V + \Delta_{SH}} \\*
A_0(s)|_{s\approx s_{\mathrm{max}}} &=& \frac{g_V}{2\sqrt 2} \frac{\sqrt m_H}{m_V} \Big( 2 \alpha_1 - 2 \alpha_2 \\*
&& + \alpha \frac{\beta}{v\cdot p_V + \Delta} + \widetilde \alpha \frac{\widetilde\zeta}{v\cdot p_V + \Delta_{H\widetilde H}} \Big).
\label{eq_ff_HMcT}
\end{eqnarray}
\end{subequations}
Again we see that the heavy meson resonance\index{resonance!of heavy meson}\index{resonance contribution!in heavy-to-light transitions} pole structure of the form factors in this model setup nicely reproduces the one of the general parameterization from the previous section for all the form factors except $A_2$, which receives only a single resonant pole contribution while the general parameterization would require two effective poles.

\subsection{Determination of model parameters -- comparison with experiment}

In order to test our approach against experimental data, we need to extrapolate\index{extrapolation!of heavy-to-light form factors} the HM$\chi$PT\index{HM$\chi$PT!inspired model} model calculations of the form factors from the previous subsection, which are valid near zero recoil over the whole physical phase space region. We use the general HQET\index{HQET!in heavy-to-light transitions} and LEET\index{LEET} compatible extrapolation\index{extrapolation!of heavy-to-light form factors} formulae as guidance in such an extrapolation, but want to use as much as possible known experimentally measured phenomenological parameters.

\subsubsection{Resonance pole saturation}
\index{resonance saturation approximation}

In order to trim down the number of undetermined parameters we first fix the effective pole parameters $a$, $a'$, $a''$, $b$, $b'$ and $b''$ in eqs.~(\ref{eq_5_f0}),~(\ref{eq_5_f+}),~(\ref{eq_v_ff}),~(\ref{eq_a0_ff}) and~(\ref{eq_a1_ff}) by the next-to-nearest resonances\index{resonance!of heavy meson} in the heavy meson spectrum as already hinted by the HM$\chi$PT\index{HM$\chi$PT!in heavy-to-light transitions} results from the previous subsection. Although in the original idea~\cite{Becirevic:1999kt} the extra pole in $F_+$ parametrized all the neglected higher resonances, we are here saturating each pole by a single nearest resonance\index{resonance contribution!in heavy-to-light transitions} in all the form factors. The recent numerous discoveries of excited charmed meson states enable as to use physical pole masses in this procedure. In our numerical analysis we therefore make use of available experimental information in addition to theoretical predictions on charm meson resonances\index{resonance!of charmed meson}. Particularly, we use the spectroscopic data in TABLE~\ref{table_input} for the first scalar, vector and axial resonances. On the other hand for the radially excited pseudoscalar and vector states, no reliable experimental results exist, while recent theoretical studies~\cite{DiPierro:2001uu,Vijande:2003uk} indicate that radially excited states of $D$\index{meson!$D$!resonances} as well as $D_s$\index{meson!$D_s$!resonances} should have masses of $m_{D^{'*}}\simeq2.7~\mathrm{GeV}$ and $m_{D_{s}^{'*}}\simeq2.8~\mathrm{GeV}$~\cite{DiPierro:2001uu}. We use these values in our analysis.

\subsubsection{$D\to P$ transitions}
\index{transition!$D\to P$}

In our  calculations we use for the heavy meson weak current coupling $\alpha = f_H \sqrt{m_H}$ from the tree level matching\index{matching!HQET to QCD} of HQET\index{HQET!matching to QCD} to QCD~\cite{Wise:1992hn,Becirevic:2002sc}, which we calculate from the lattice QCD\index{lattice QCD!in heavy-to-light transitions} value of $f_{D} = 0.235(8)(14)~\mathrm{GeV}$~\cite{Chiu:2005ue} and experimental $D$ meson mass $m_{D}=1.87~\mathrm{GeV}$~\cite{Eidelman:2004wy} yielding $\alpha=0.32~\mathrm{GeV^{3/2}}$. Since we expect large $SU(3)$ light flavor symmetry\index{chiral symmetry breaking} corrections, we use a different value of $\alpha_{(3)}= f_{D_s} \sqrt{m_{D_s}} = 0.37~\mathrm{GeV^{3/2}}$ from the lattice QCD value of $f_{D_s} = 0.266(10)(18)~\mathrm{GeV}$~\cite{Chiu:2005ue} and experimental $D_s$ meson mass $m_{D_s}=1.97~\mathrm{GeV}$~\cite{Eidelman:2004wy}.
For light pseudoscalar mesons we use $f = 130~\mathrm{MeV}$, while for the $g$ and $h$ couplings we use the tree level values from the first row of table~\ref{table_summary}.


\par

Eqs.~(\ref{eq_5_12}) and~(\ref{F+FFq}) can be combined to obtain a theoretical estimate for the value of $\widetilde h \widetilde \alpha$ in the limit of infinite heavy meson mass. By equating both
terms in eqs.~(\ref{eq_5_12}) and~(\ref{F+FFq}) at $s_{\mathrm{max}}$ and then imposing $a=\gamma$ one obtains
in the exact chiral and heavy quark limits \index{HM$\chi$PT!parameters}
\begin{equation}
\widetilde\alpha\widetilde h = -\alpha g = - 0.2~\mathrm{GeV}^{3/2}.
\label{agtilde}
\end{equation}
On the other hand the $1/m_D$  and chiral corrections might still modify this result significantly. Such corrections were explicitly written out in ref.~\cite{Boyd:1994pa} but they include additional parameters which cannot be fixed within our context.

\par

Similarly,  in this limit we can infer on the value of $\alpha'$. By applying equalities~(\ref{PP_form_factor_relations}) and~(\ref{agtilde}) to eqs.~(\ref{F0FFq}) and~(\ref{eq_5_f0}) one finds that the chiral\index{chiral limit} limit is ill defined. Namely, by first taking the exact heavy quark limit, one obtains a relation \index{HM$\chi$PT!parameters}
$\frac{\alpha'}{\alpha} h = -\frac{\Delta_{SH}}{m_P}$.
The expression blows up in the limit $m_P\to0$, while when applied to the $D\to\pi$ transitions gives a large value of $\alpha' \sim 1.5~\mathrm{GeV}^{3/2}$. On the other hand $D\to K$ transitions give $\alpha'_{(3)} \sim 0.4~\mathrm{GeV}^{3/2}$. Recently the decay\index{decay constant} constant of the $1/2^{+}$ charmed-strange meson has been estimated on the lattice\index{lattice QCD!in heavy-to-light transitions} $f_{D'_s}=340(110)~\mathrm{MeV}$~\cite{Herdoiza:2006qv,McNeile:2004rf} yielding for $\alpha'_{(3)}\approx 0.5$ which is in good agreement with our model's prediction. The same cannot be claimed at present for the value of $\alpha'$ involved in pion transitions. The situation is reminiscent of the discussions in the last chapter in the sense that the off-shellness of the intermidiate scalar resonance, large compared to the pion mass, invalidates affected HM$\chi$PT\index{HM$\chi$PT!in heavy-to-light transitions} results. Importantly however, the weak current coupling of $1/2^+$ mesons is in no case suppressed with respect to the $1/2^-$ ones.  If one instead first imposes the chiral\index{chiral limit} limit on eq.~(\ref{F0FFq}), the $h$ contributions in~(\ref{eq_5_f0}) decouple and we instead obtain a nontrivial relation $g=\Delta_{SH}/\Delta_{\widetilde H H}$  , which is roughly satisfied by current experimental values and theoretical estimates for the three quantities in the charmed sector. Again one should expect possibly large $1/m_H$ corrections to these relations.

\par

Alternatively the values of the new model parameters can be determined by fitting the model predictions to known experimental values of branching ratios  $\mathcal B (D^0\to K^- \ell^+ \nu)$, $\mathcal B (D^+\to \overline K^0 \ell^+ \nu)$, $\mathcal B (D^0\to \pi^- \ell^+ \nu)$, $\mathcal B (D^+\to \pi^0
\ell^+ \nu)$, $\mathcal B (D^+_s\to \eta \ell^+ \nu)$ and $\mathcal B (D^+_s\to \eta' \ell^+
\nu)$~\cite{Eidelman:2004wy}. In our decay  width calculations we shall neglect the lepton mass, so the decay  width is given by eq.~(\ref{eq_gamma_HP}), with the Wilson\index{Wilson coefficient} coefficient $C = G_F K_{HP}$\index{G$_F$}, where the constants $K_{HP}$ parametrize the flavor mixing\index{mixing!of quark flavors} relevant to a particular transition, and are given in table~\ref{PP_mixing_table} together with the pole mesons.\index{CKM!matrix elements!$V_{cs}$}\index{CKM!matrix elements!$V_{cd}$}
\begin{table}[!t]
\begin{center}
\begin{tabular}{|c|c|c|c|c|c|}
\hline
$H$ & $P$ & $H^*$ & $\widetilde H^*$ & $S$ & $K_{HP}$\\
\hline
\hline
$D^0$ & $K^-$ & $D^{*+}_s$ & $D'^{*+}_{s}$ & $D_{sJ}(2317)^+$ & $V_{cs}$\\
    $D^+$ & $\overline K^0$ & $D^{*+}_s$ & $D'^{*+}_{s}$ & $D_{sJ}(2317)^+$ & $V_{cs}$\\
    $D^+_s$ & $\eta$ & $D^{*+}_s$ & $D'^{*+}_{s}$ & $D_{sJ}(2317)^+$ & $ V_{cs} \sin\phi$\\
    $D^+_s$ & $\eta'$ & $D^{*+}_s$ & $D'^{*+}_{s}$ & $D_{sJ}(2317)^+$ & $V_{cs} \cos\phi$\\
    $D^0$ & $\pi^-$ & $D^{*+}$ & $D'^{*+}$ & $D'^+$ & $V_{cd}$\\
    $D^+$ & $\pi^0$ & $D^{*+}$ & $D'^{*+}$ & $D'^+$ & $V_{cd}/\sqrt 2$\\
    $D^+$ & $\eta$ & $D^{*+}$ & $D'^{*+}$ & $D'^+$ & $V_{cd}\cos\phi/\sqrt 2$\\
    $D^+$ & $\eta'$ & $D^{*+}$ & $D'^{*+}$ & $D'^+$ & $V_{cd}\sin\phi/\sqrt 2$\\
    $D^+_s$ & $K^0$ & $D^{*+}$ & $D'^{*+}$ & $D'^+$ & $V_{cd}$\\
    \hline
\end{tabular}
\end{center}
\caption{\small\it \label{PP_mixing_table} The pole mesons and the flavor mixing constants $K_{HP}$ for the $D\to P$ semileptonic decays.}
\end{table}

\par

We calculate the result for $\widetilde h \widetilde \alpha$ \index{HM$\chi$PT!parameters}by fitting to the most precisely measured decay \index{decay!$D^0\to \pi^- \ell \nu$} rate of $D^0\to \pi^-\ell\nu_{\ell}$. We also expect chiral corrections to be smallest in this case. The calculation yields $\widetilde\alpha\widetilde h = -0.04\mathrm{~GeV^{3/2}}$, which is rather small in absolute terms compared to estimation given by (\ref{agtilde}). This discrepancy can be attributed  to the presence of $1/m_D$ and chiral corrections which are not included systematically into consideration here due too many new parameters~\cite{Boyd:1994pa} which cannot be fixed within this approach. However, we estimate the influence of such corrections on the fitted value of $\widetilde\alpha\widetilde g$ by varying the value of the input parameters $\alpha g/f$ in eq.~(\ref{F+FFq}) by $20\%$~\cite{Stewart:1998ke} and inspecting the fit results. We obtain a range of $\widetilde\alpha\widetilde h \in [-0.3,+0.2]~\mathrm{GeV}^{3/2}$\footnote{An alternative method would be to fit the parameter to $D^0\to K^-\ell^+\nu_{\ell}$, but the variation of $\widetilde\alpha\widetilde h$ obtained in this way is very small.}.

\par

In the same way we also estimate the value of $\alpha'$\index{HM$\chi$PT!parameters} by fitting relation~(\ref{PP_form_factor_relations}) for $D^0\to\pi^-$ transition and using the $\widetilde \alpha \widetilde h$ value from the previous paragraph in the fit. We obtain a value of $\alpha'=1.5~\mathrm{GeV}^{3/2}$ which agrees very well with the theoretical estimate in the case of pions. If we instead apply the same procedure on $D\to K$ channels, we obtain a value of $\alpha'_{(3)} = 0.6~\mathrm{GeV}^{3/2}$ indicating indeed large $SU(3)$ light flavor symmetry corrections\index{chiral symmetry breaking}.

\par

\begin{figure}[!t]
\begin{center}
\scalebox{1}{\includegraphics{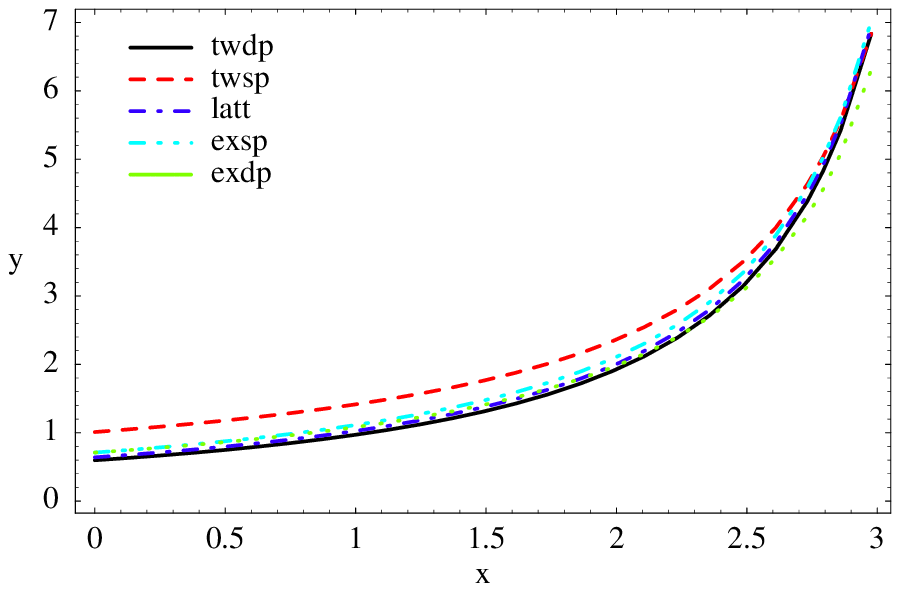}}
\end{center}
\caption{\small\it \label{FplotDPi}Comparison of $D^0 \to \pi^-$ transition $F_+$ form factor
 $s$ dependence of our model two poles extrapolation (solid (black) line), single pole extrapolation (dashed (red) line), lattice QCD fitted to two poles (dot-dashed (blue) line) and experimental two poles fits ((green) dotted and dash-double dotted lines).}
\end{figure}

\begin{figure}[!t]
\begin{center}
\scalebox{1}{\includegraphics{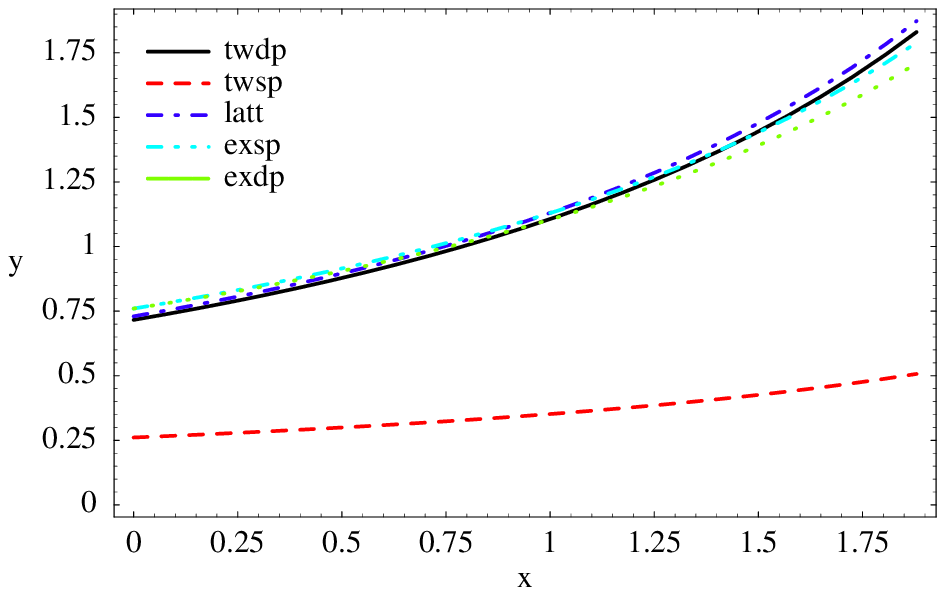}}
\end{center}
\caption{\small\it \label{FplotDK}Comparison of the  $D^0 \to K^-$ transition $F_+$ form factor
 $s$ dependence of our model two poles extrapolation (solid (black) line), single pole extrapolation (dashed (red) line), lattice QCD fitted to two poles (dot-dashed (blue) line) and experimental two poles fits ((green) dotted and dash-double dotted lines).}
\end{figure}

We next draw  the $s$ dependence of the  $F_+$ form factors for the $D^0\to K^-$ and $D^0\to \pi^-$ transitions  and compare it with results of   lattice QCD\index{lattice QCD!form factor fit} two poles\index{two poles parameterization of heavy-to-light form factors} fit analysis~\cite{Aubin:2004ej}, as well as  the experimental results of a two poles\index{two poles parameterization of heavy-to-light form factors} fit from CLEO\index{CLEO}~\cite{Huang:2004fr} and FOCUS\index{FOCUS}~\cite{Link:2004dh} with $F_+(0)$ values taken from ref.~\cite{Ablikim:2004ej}. The results are depicted in fig.~\ref{FplotDK} and fig.~\ref{FplotDPi}.
\psfrag{x}[tc]{\Blue{$s~[\mathrm{GeV}^2]$}}
\psfrag{twdp}[cl]{This work (two poles)}
\psfrag{twsp}[cl]{This work (single pole)}
\psfrag{latt}[cl]{Latt.~\cite{Aubin:2004ej} (two poles)}
\psfrag{exsp}[cl]{Expt.~\cite{Huang:2004fr} (two poles)}
\psfrag{exdp}[cl]{Expt.~\cite{Link:2004dh} (two poles)}
\psfrag{y}[bc][bc][1][90]{\Blue{$F_+^{D^0\to \pi^-}$}}
\psfrag{y}[bc][bc][1][90]{\Blue{$F_+^{D^0\to K^-}$}}
For comparison we also plot results when single pole fit is used. Also in this case we calculate $F_{+} (s_{\mathrm{max}})$ within HM$\chi$T, take into account both resonances (eq.~(\ref{F+FFq})) and fit the free parameter $\widetilde \alpha \widetilde h$ to the $D^0\to \pi^-$ semileptonic decay  rate. From the plots it becomes apparent, that our model's predictions for both $D^0\to K^-$ and $D^0\to \pi^-$ transition $F_+$ form factors   are in good agreement with experimental and lattice data\index{lattice QCD!results} when extrapolated with two poles\index{two poles parameterization of heavy-to-light form factors}, while single pole extrapolations\index{extrapolation!of heavy-to-light form factors} are not in good agreement with experimental results, especially for $D\to K$ transitions indicating that in this extrapolation a single parameter cannot fit both pionic and kaonic decays\index{decay of $K$ meson}\index{meson!$K$!decay}. This discrepancy further increases if only the first resonance\index{resonance contribution!in heavy-to-light transitions} contribution is kept in the $F_{+} (s_{\mathrm{max}})$ calculation for the single pole extrapolation\index{extrapolation!of heavy-to-light form factors} within HM$\chi$T as was done in previous studies~\cite{Casalbuoni:1996pg}. In such calculations only the $D^*$ resonance\index{resonance contribution!in heavy-to-light transitions} contributed, and a lower value of the $g$ strong coupling was used. At that time only few decay  rates were measured. In comparison with the present experimental data the predicted branching ratios  were too large. Note also that the experimental fits on the single pole parametrization of the $F_+$ form factor\index{form factor!$D^0\to\pi^-$}\index{form factor!$D^0\to K^-$}
 in $D^0\to\pi^- (D^0\to K^-)$ transitions done in refs.~\cite{Huang:2004fr,Link:2004dh} yielded effective pole masses which are somewhat lower than the physical masses of the $D^* (D^{*}_s)$ meson resonances\index{resonance!of charmed meson} used in this analysis. The approach of ref.~\cite{Bajc:1995km} was developed to treat $D$\index{meson!$D$!decay} meson semileptonic decay \index{semileptonic decay}\index{decay of $D$ meson} within heavy light meson symmetries in the allowed kinematic region by using the full propagators.  We find that this approach cannot reproduce the observed  $s$ shape of the $F_+$ form factors .

\par

\begin{figure}[!t]
\begin{center}
\scalebox{1}{\includegraphics{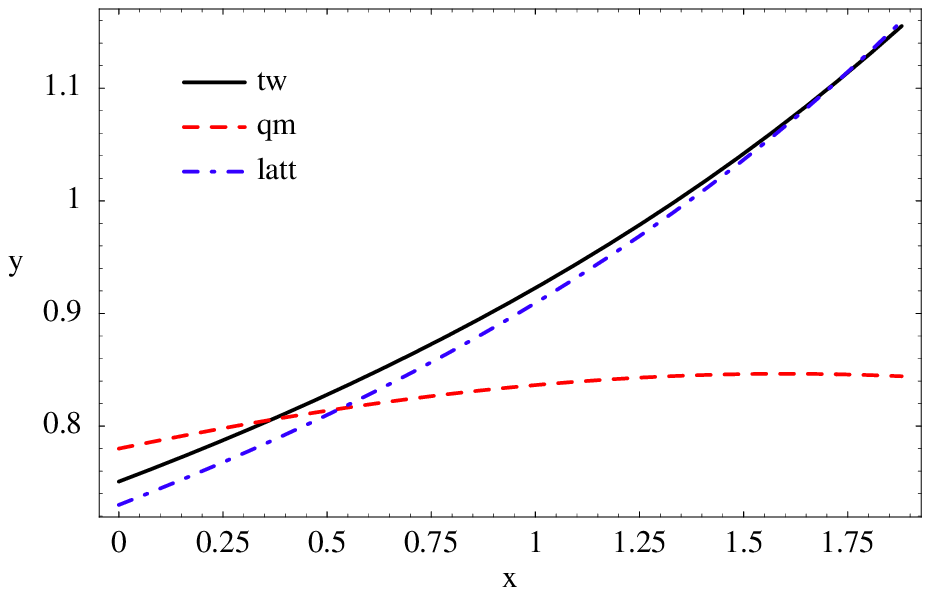}}
\end{center}
\caption{\small\it \label{F0plotDK}Comparison of the  $D^0 \to K^-$ transition $F_0$ form factor  $s$ dependence of our model (solid (black) line),
quark model of Melikhov \& Stech (dashed (red) line) and lattice QCD fitted to a pole (dot-dashed (blue) line).}
\end{figure}

\begin{figure}[!t]
\begin{center}
\scalebox{1}{\includegraphics{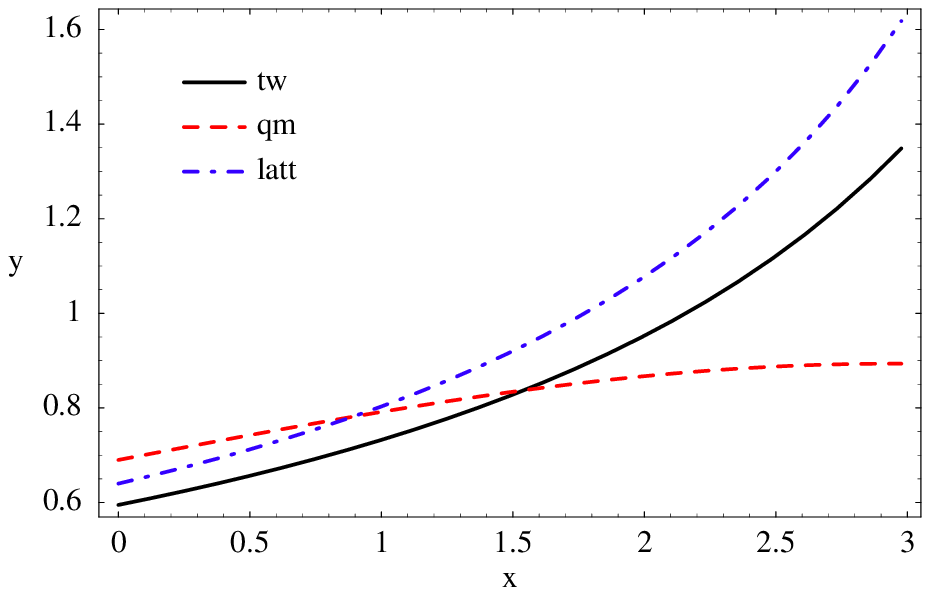}}
\end{center}
\caption{\small\it \label{F0plotDPi}Comparison of the  $D^0 \to \pi^-$ transition $F_0$ form factor $s$ dependence of our model (solid (black) line),
quark model of Melikhov \& Stech (dashed (red) line) and lattice QCD fitted to a pole (dot-dashed (blue) line).}
\end{figure}

We also compare our predictions for the $F_0$ scalar  form factor  $s$ dependence for the $D^0\to K^-$ and $D^0\to \pi^-$ transitions with those of a successful quark model in ref.~\cite{Melikhov:2000yu} and with lattice QCD\index{lattice QCD!form factor fit} pole fit analysis of ref.~\cite{Aubin:2004ej}. The results are depicted in fig.~\ref{F0plotDK} and fig.~\ref{F0plotDPi}.
\psfrag{x}[tc]{\Blue{$s~[\mathrm{GeV}^2]$}}
\psfrag{y}[bc][bc][1][90]{\Blue{$F_0^{D^0\to K^-}$}}
\psfrag{tw}[cl]{This work (effective pole)}
\psfrag{qm}[cl]{Quark model~\cite{Melikhov:2000yu}}
\psfrag{latt}[cl]{Latt.~\cite{Aubin:2004ej} (effective pole)}
\psfrag{y}[bc][bc][1][90]{\Blue{$F_0^{D^0\to \pi^-}$}}
Note that without the scalar resonance, one only gets a contribution from the $\alpha / \sqrt m_H f$ term from eq.~(\ref{F0FFq}). This gives for the $s$ dependence of $F_0$ a constant value $F_0(s)=1.81(2.05)$ for $D\to\pi$ ($D\to K$) transitions, which largely disagrees with lattice QCD\index{lattice QCD!results} results as well as heavily violates relation~(\ref{PP_form_factor_relations}).

\subsubsection{$D\to V$ transitions}
\index{transition!$D\to V$}

We next apply the strategy from the previous section to the extrapolation\index{extrapolation!of heavy-to-light form factors} and parameter extraction in $D\to V$ transitions. We use the information on the contributions of different resonances\index{resonance contribution!in heavy-to-light transitions} to the form factors  as suggested by our model. For the vector form factor $V$ we thus propose $a'=a=m_{H^*}^2/m_{\widetilde H^*}^2$ which saturates the effective second pole by the first vector radial excitation $\widetilde H^*$. Similarly we set $a''=m_{H}^2/m_{\widetilde H}^2$ with $a''=a'=a$ holding in the exact heavy quark limit, and $b'=m_{H^*}^2/m_{S^*}^2\simeq b$ saturating the poles of the $A_0$ and $A_1$ form factors  and the first pole of the $A_2$ form factor  with the $\widetilde H$ pseudoscalar radial excitation and the $S^*$ orbital axial excitation respectively. Since our model does not contain a second resonance contribution to the $A_2$ form factor , we impose $b''=0$, effectively sending the second pole mass of this form factor to infinity. At the end we have fixed all the pole parameters appearing in the general form factor\index{form factor!parameterization} parameterization formulas of sec.~\ref{sec_5.1.2} using physical information and model predictions on the resonances\index{resonance contribution!in heavy-to-light transitions} contributing to the various form factors . The remaining parameters ($c'_H$ and $c''_H$) are on the other hand related to the parameters of HM$\chi$T via the model matching conditions at zero recoil.

\par

We again restrict our present study to $D$\index{meson!$D$!decay} decays\index{decay of $D$ meson}, in order to use the available experimental information in the charm sector, although our calculations can readily be applied to semileptonic decays\index{decay of $B$ meson} of $B$\index{meson!$B$!decay} mesons once more experimental information becomes available on excited $B$\index{meson!$B$!resonances}\index{resonance!of $B$ meson} meson resonances. In our numerical analysis we use available experimental information and theoretical predictions on charm meson resonances as in the previous section.

\par

Our HM$\chi$PT\index{HM$\chi$PT!inspired model} model calculations of section~\ref{sec_hmchpt_HL} contain several parameters. The $\lambda$ coupling was usually~\cite{Casalbuoni:1996pg, Cheng:2004ru} determined from the value of $V(0)$. However, this derivation employed a single pole ansatz for the shape of $V(s)$. One can instead use data on $D^* \to D \gamma$ radiative decays\index{radiative decay}\index{decay!$D^* \to D \gamma$}. Following discussion in refs.~\cite{Prelovsek:2000rj,Fajfer:1997bh},  using the  most recent data on $D^*$ radiative and strong decays\index{strong decay}~\cite{Eidelman:2004wy}, and accounting for the $SU(3)$ flavor symmetry\index{chiral symmetry breaking} breaking effects, we calculate $\lambda =- 0.526$ GeV$^{-1}$.
The coupling $\beta \simeq 0.9$ has been estimated in ref.~\cite{Isola:2003fh} relying on the assumption that the electromagnetic interactions of the light quark within heavy meson are dominated by the exchange of $\rho^0$, $\omega$, $\phi$ vector mesons. We fix the other free parameters ($\alpha_1, \alpha_2, \alpha', \widetilde\alpha, \zeta, \mu, \widetilde\zeta, \widetilde\mu$) appearing in the HM$\chi$T Lagrangian\index{HM$\chi$PT!Lagrangian} and weak currents by comparing our model predictions to known experimental values of branching ratios  $\mathcal B (D^0\to K^{*-}\ell^+\nu)$, $\mathcal B (D_s \to \phi\ell^+\nu)$, $\mathcal B (D^+\to \rho^0\ell^+\nu)$, $\mathcal B (D^+\to K^{0*}\ell^+\nu)$, as well as partial decay  width ratios $\Gamma_L/\Gamma_T (D^+\to K^{0*}\ell^+\nu)$ and $\Gamma_+/\Gamma_- (D^+\to K^{0*}\ell^+\nu)$~\cite{Eidelman:2004wy}. In order to compare the results of our approach with experimental values, we calculate the decay  rates for polarized\index{polarization of vector meson} final light vector mesons in eq.~(\ref{eq_gamma_HV}) again with the Wilson coefficient\index{Wilson coefficient} $C=G_F K_{HV}$\index{G$_F$}. The constants $K_{HV}$ parametrize the flavor mixing\index{mixing!of quark flavors} relevant to a
particular transition, and are given in table~\ref{PV_mixing_table}
together with the pole mesons.
\begin{table}[!t]
\begin{center}
\begin{tabular}{|c|c|c|c|c|c|}
\hline
$H$ & $V$ & $H^*$ & $H'$ & $S^*$ & $K_{HV}$\\
\hline
\hline
$D^0$ & $K^{*-}$ & $D^{*+}_s$, $D^{'*+}_{s}$ & $D_s^+$, $D_s^{'+}$ & $D_{sJ}(2463)^+$ & $V_{cs}$\\
$D^+$ & $\overline K^{*0}$ & $D^{*+}_s$, $D^{'*+}_{s}$ & $D_s^+$, $D_s^{'+}$ & $D_{sJ}(2463)^+$ & $V_{cs}$\\
$D^+_s$ & $\phi$ & $D^{*+}_s$, $D^{'*+}_{s}$ & $D_s^+$, $D_s^{'+}$ & $D_{sJ}(2463)^+$ & $V_{cs}$\\
$D^0$ & $\rho^-$ & $D^{*+}$, $D^{'*+}$ & $D^+$, $D^{'+}$ & $D_1(2420)$ & $V_{cd}$\\
$D^+$ & $\rho^0$ & $D^{*+}$, $D^{'*+}$  & $D^+$, $D^{'+}$ & $D_1(2420)$ & $-\frac{1}{\sqrt 2} V_{cd}$\\
$D^+$ & $\omega$ & $D^{*+}$, $D^{'*+}$  & $D^+$, $D^{'+}$ & $D_1(2420)$ & $\frac{1}{\sqrt 2} V_{cd}$\\
$D^+_s$ & $K^{*0}$ & $D^{*+}$, $D^{'*+}$  & $D^+$, $D^{'+}$ & $D_1(2420)$ & $V_{cd}$\\
\hline
\end{tabular}
\end{center}
\caption{\small\it \label{PV_mixing_table} The pole mesons and the flavor mixing constants $K_{HV}$ for the $D\to V$ semileptonic decays.}
\end{table}
The $A_0$ form factor  does not contribute to any decay  rate in this approximation and we can not fix the parameters $\alpha_2$ and $\widetilde\zeta$ solely from comparison with experiment. Although $A_0$ actually does contribute indirectly through the relation~(\ref{PV_form_factor_relations}) at $s=0$, this constraint is not automatically satisfied by our model. On the other hand, we can still enforce it "by hand" after the extrapolation\index{extrapolation!of heavy-to-light form factors} to $s=0$ to obtain some information on these parameters. Due to the specific combinations in which the parameters appear in eqs.~(\ref{eq_ff_HMcT1}-\ref{eq_ff_HMcT}) we are further restrained to determining only the products $\widetilde\alpha\widetilde\mu$, $\alpha'\zeta$ and $\alpha'\mu$ using this kind of analysis. Lastly, since the only relevant contribution of $\alpha_1$ is to the $A_1$ form factor , we cannot disentangle it from the influence of $\alpha'\zeta$. Yet again we can impose the large energy limit relation~(\ref{eq_VA_LEET}) to extract both values independently.

\par

We calculate the result for $\widetilde\alpha\widetilde\mu$, $\alpha'\zeta$, $\alpha'\mu$ and $\alpha_1$ by a weighted average of values obtained from all the measured decay  rates and their ratios taking into account for the experimental uncertainties. Furthermore, the values of $\alpha_1$ and $\alpha'\zeta$ are extracted separately by minimizing the fit function $(V(0)\xi - A_1(0))^2/(V(0)\xi + A_1(0))^2$. Both minimizations are performed in parallel and the global minimum is sought on the hypercube of dimensions $[-1,1]^4$ in the hyperspace of the fitted parameters. At the end we obtain the following values of parameters:
\begin{eqnarray}
\widetilde\alpha\widetilde\mu &=& 0.090~\mathrm{GeV}^{1/2} \nonumber\\*
\alpha'\zeta &=& 0.038~\mathrm{GeV}^{3/2} \nonumber\\*
\alpha'\mu &=& -0.066~\mathrm{GeV}^{1/2} \nonumber\\*
\alpha_1 &=& -0.128~\mathrm{GeV}^{1/2}
\end{eqnarray}
These values qualitatively agree with the analysis done in ref.~\cite{Casalbuoni:1996pg} using a combination of quark model predictions and single pole experimental fits for all the form factors .

\par

\begin{figure}[!t]
\begin{center}
\scalebox{1}{\includegraphics{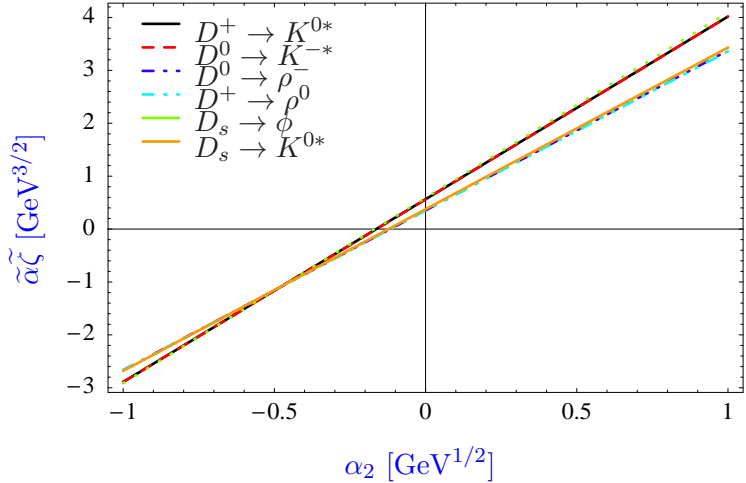}}
\end{center}
\caption{\small\it \label{fig_a2_tz} Solutions of eq.~(\ref{PV_form_factor_relations}) in the $\alpha_2 \times \widetilde\alpha\widetilde\zeta$ parameter plane for the various decay channels considered.}
\end{figure}

We next use these values in relation~(\ref{PV_form_factor_relations}) to extract information on the the parameters $\alpha_2$ and $\widetilde\zeta$. From eqs.~(\ref{eq_ff_HMcT1}-\ref{eq_ff_HMcT}) it is easy to see that the solutions lie on a straight line in the $\alpha_2 \times \widetilde\alpha\widetilde\zeta$ plane. We draw these for the various decay  channels used in our analysis in fig.~\ref{fig_a2_tz}.
\psfrag{x}[tc]{\Blue{$\alpha_2~[\mathrm{GeV}^{1/2}]$}}
\psfrag{y}[bc][bc][1][90]{\Blue{$\widetilde \alpha \widetilde \zeta~[\mathrm{GeV}^{3/2}]$}}
\psfrag{dks0}[cl]{$D^+\to K^{0*}$}
\psfrag{d0ks}[cl]{$D^0\to K^{-*}$}
\psfrag{d0r}[cl]{$D^0\to \rho^-$}
\psfrag{dr0}[cl]{$D^+\to \rho^{0}$}
\psfrag{dsf}[cl]{$D_s\to \phi$}
\psfrag{dsk0}[cl]{$D_s\to K^{0*}$}
We can see that all the decay  channels considered fit approximately the same solution in the plane. Consequently, we can use any point on the approximate solution line to obtain the same prediction for the $s$ dependence of the $A_0$ form factor .

\par

It is important to note at this point that due to a high degree of interplay of the various HM$\chi$T parameters in the model predictions used in the fit, the values of the new model parameters obtained in such a way are very volatile to changes in the other inputs to the fit. Furthermore these are tree level leading order parameter values and may in addition be very sensitive to chiral and $1/m_H$ corrections. Therefore their stated values should be taken {\it cum grano salis}. However more importantly, the form factor , branching ratio  and polarization\index{polarization width ratio} width ratio predictions based on this approach are more robust since they are insensitive to particular combinations of parameter values used, as long as they fit the experimental data. We estimate that chiral\index{chiral symmetry breaking} and heavy quark symmetry\index{heavy quark symmetry breaking} breaking corrections could still modify these predictions by as much as $30\%$.

\par

\begin{figure}[!t]
\begin{center}
\scalebox{1}{\includegraphics{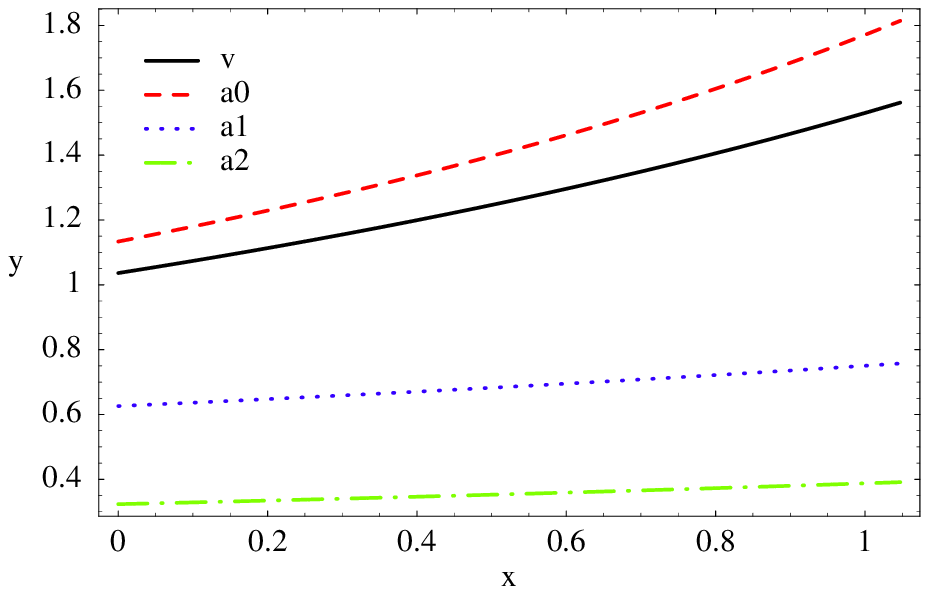}}
\end{center}
\caption{\small\it \label{fig_ff_d0k} Predictions of our model for the $s$ dependence of the form factors $V(s)$ (black solid line), $A_0(s)$ (red dashed line), $A_1(s)$ (blue dotted line) and $A_2(s)$ (green dash-dotted line) in $D^0\to K^{-*}$ transition.}
\end{figure}

\begin{figure}[!t]
\begin{center}
\scalebox{1}{\includegraphics{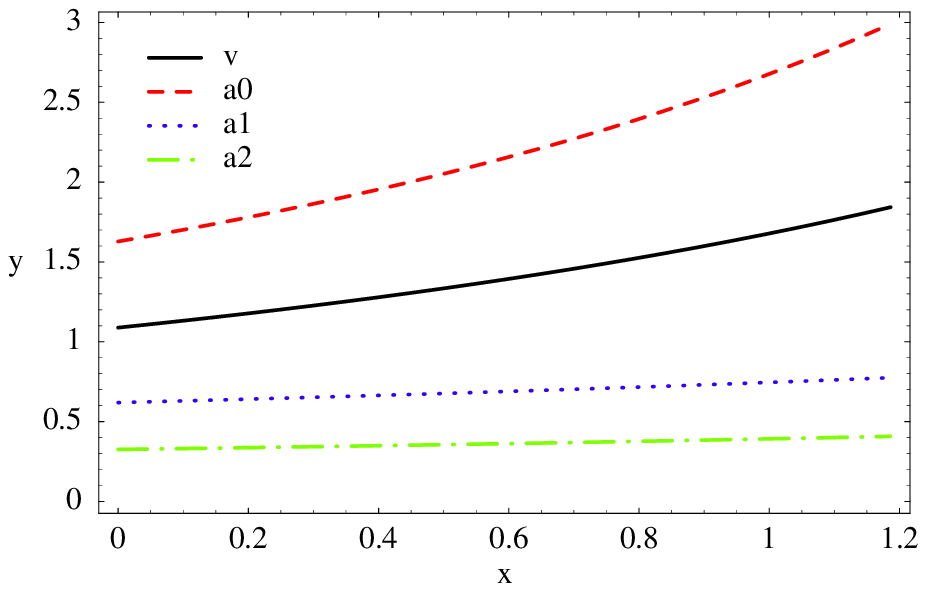}}
\end{center}
\caption{\small\it \label{fig_ff_d0rho} Predictions of our model for the $s$ dependence of the form factors $V(s)$ (black solid line), $A_0(s)$ (red dashed line), $A_1(s)$ (blue dotted line) and $A_2(s)$ (green dash-dotted line) in $D^0\to \rho^-$ transition.}
\end{figure}

\begin{figure}[!t]
\begin{center}
\scalebox{1}{\includegraphics{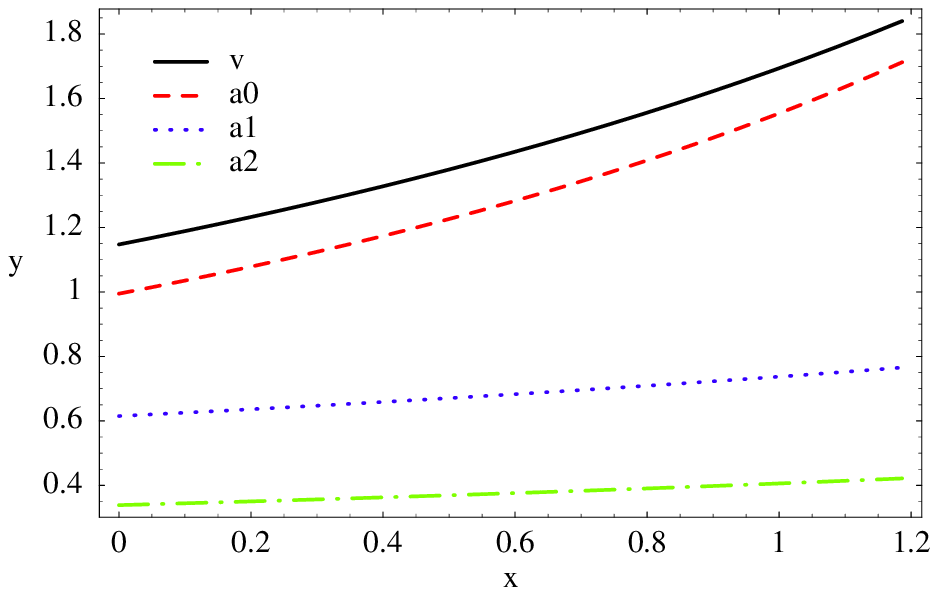}}
\end{center}
\caption{\small\it \label{fig_ff_dsphi} Predictions of our model for the $s$ dependence of the form factors $V(s)$ (black solid line), $A_0(s)$ (red dashed line), $A_1(s)$ (blue dotted line) and $A_2(s)$ (green dash-dotted line) in $D_s\to \phi$ transition.}
\end{figure}

We are now ready to draw the $s$ dependence of all the form factors\index{form factor!$D^0\to K^{-*}$}\index{form factor!$D^0\to \rho^-$}\index{form factor!$D_s\to \phi$} for the $D^0\to K^{-*}$, $D^0\to \rho^-$ and $D_s\to \phi$ transitions. The results are depicted in figs.~\ref{fig_ff_d0k},~\ref{fig_ff_d0rho}, and~\ref{fig_ff_dsphi}.
\psfrag{x}[tc]{\Blue{$s~[\mathrm{GeV}^2]$}}
\psfrag{y}[bc][bc][1][90]{\Blue{$F^{D^0\to K^{-*}}(s)$}}
\psfrag{v}[cl]{$V$}
\psfrag{a0}[cl]{$A_0$}
\psfrag{a1}[cl]{$A_1$}
\psfrag{a2}[cl]{$A_2$}
\psfrag{y}[bc][bc][1][90]{\Blue{$F^{D^0\to \rho^-}(s)$}}
\psfrag{y}[bc][bc][1][90]{\Blue{$F^{D_s\to \phi}(s)$}}
We also compare our model predictions with recent experimental analysis of helicity amplitudes\index{helicity amplitudes} $H_{+,-,0}$ performed by the FOCUS\index{FOCUS} collaboration. Because of the arbitrary normalization of the form factors\index{form factor!normalization} in~\cite{Link:2005dp}, we fit our model predictions for a common overall scale in order to compare the results. We plot the $s$ dependence of the predicted helicity amplitudes\index{helicity amplitudes} and compare them with the experimental results of FOCUS\index{FOCUS}, scaled by an overall factor determined by the least square fit of our model predictions, in figures~\ref{1},~\ref{2} and~\ref{3}.
The scale factor is common to all form factors\index{form factor!normalization}.
\psfrag{hp}[bc][bc][1][90]{\Blue{$\left|H_+^{(D^+ \to \overline K^{*0})}\right|^2~[\mathrm{GeV}^2]$}}
\psfrag{dp}[cl]{This work}
\psfrag{sp}[cl]{This work (single pole)}
\psfrag{exp}[cl]{FOCUS~\cite{Link:2005dp}}
\begin{figure}[!t]
\begin{center}
\scalebox{1}{\includegraphics{H1plotDK0.eps}}
\end{center}
\caption{\small\it \label{1} Predictions of our model (two poles in black solid line and single pole in red dashed line) for the $s$ dependence of the
helicity amplitude $H_+^2(s)$\index{helicity amplitudes} in comparison with scaled FOCUS data on $D^+ \to \overline K^{*0}$ semileptonic decay.}
\end{figure}
\psfrag{hm}[bc][bc][1][90]{\Blue{$\left|H_-^{(D^+ \to \overline K^{*0})}\right|^2~[\mathrm{GeV}^2]$}}
\psfrag{sp}[cl]{This work (single p.)}
\begin{figure}[!t]
\begin{center}
\scalebox{1}{\includegraphics{H2plotDK0.eps}}
\end{center}
\caption{\small\it \label{2} Predictions of our model (two poles in black solid line and single pole in red dashed line) for the $s$ dependence of the
 helicity amplitude $H_-^2(s)$\index{helicity amplitudes} in comparison with scaled FOCUS data on $D^+ \to \overline K^{*0}$ semileptonic decay.}
\end{figure}
\psfrag{h0}[bc][bc][1][90]{\Blue{$\left|H_0^{(D^+ \to \overline K^{*0})}\right|^2~[\mathrm{GeV}^2]$}}
\psfrag{sp}[cl]{This work (single pole)}
\begin{figure}[!t]
\begin{center}
\scalebox{1}{\includegraphics{H3plotDK0.eps}}
\end{center}
\caption{\small\it \label{3} Predictions of our model (two poles in black solid line and single pole in red dashed line) for the $s$ dependence of the
helicity amplitude $H_0^2(s)$\index{helicity amplitudes} in comparison with scaled FOCUS data on $D^+ \to \overline K^{*0}$ semileptonic decay.}
\end{figure}
In addition to the two pole contributions we calculate helicity amplitudes\index{helicity amplitudes}
in the case when all the form factors  exhibit single pole behavior.
Putting contributions of higher charm resonances\index{resonance!of charmed meson} to zero, we fit the remaining model parameters to existing branching ratios  and partial decay  width ratios. We obtain the values for the following parameter combinations:
\begin{eqnarray}
\widetilde\alpha\widetilde\mu &=& 0 \nonumber\\*
\alpha'\zeta &=& -0.180~\mathrm{GeV}^{3/2} \nonumber\\*
\alpha'\mu &=& -0.00273~\mathrm{GeV}^{1/2} \nonumber\\*
\alpha_1 &=& -0.203~\mathrm{GeV}^{1/2}
\end{eqnarray}
As shown in figures.~\ref{1} and~\ref{2} the experimental data for $H_{\pm}$ do not favor such a parametrization, while in the case of $H_0$ helicity amplitude \index{helicity amplitudes}there is almost no difference since the $H_0$ helicity amplitude\index{helicity amplitudes} is defined via the $A_{1,2}$ form factors , which are in our approach both effectively dominated by a single pole. The agreement between the FOCUS\index{FOCUS} results and our model predictions for the $s$ dependence of the helicity amplitudes\index{helicity amplitudes} is good, although as noted already
in~\cite{Link:2005dp}, the uncertainties of the data points are still rather large.
In figures.~\ref{4} and~\ref{5} we present helicity amplitudes\index{helicity amplitudes} for the $D^+ \to \rho^0 \ell \nu$ and $D^+_s \to \phi  \ell \nu$ decays\index{decay!$D^+ \to \rho^0 \ell \nu$}\index{decay!$D^+_s \to \phi  \ell \nu$}. Both decay modes are most promissing for the future experimental studies. We make predictions for the shapes of helicity amplitudes\index{helicity amplitudes} for both cases: where two poles\index{two poles parameterization of heavy-to-light form factors} contribute to the vector form factor and a single pole to the axial form factors, and the second case where all form factors  exhibit single pole behavior.
\psfrag{y}[bc][bc][1][90]{\Blue{$\left|H_i^{(D^+ \to \rho^0)}\right|^2~[\mathrm{GeV}^2]$}}
\psfrag{hp}[cl]{$H^+$}
\psfrag{hps}[cl]{$H^+$ (single pole)}
\psfrag{hm}[cl]{$H^-$}
\psfrag{hms}[cl]{$H^-$ (single pole)}
\psfrag{h0}[cl]{$H^0$}
\psfrag{h0s}[cl]{$H^0$ (single pole)}
\begin{figure}[!t]
\begin{center}
\scalebox{1}{\includegraphics{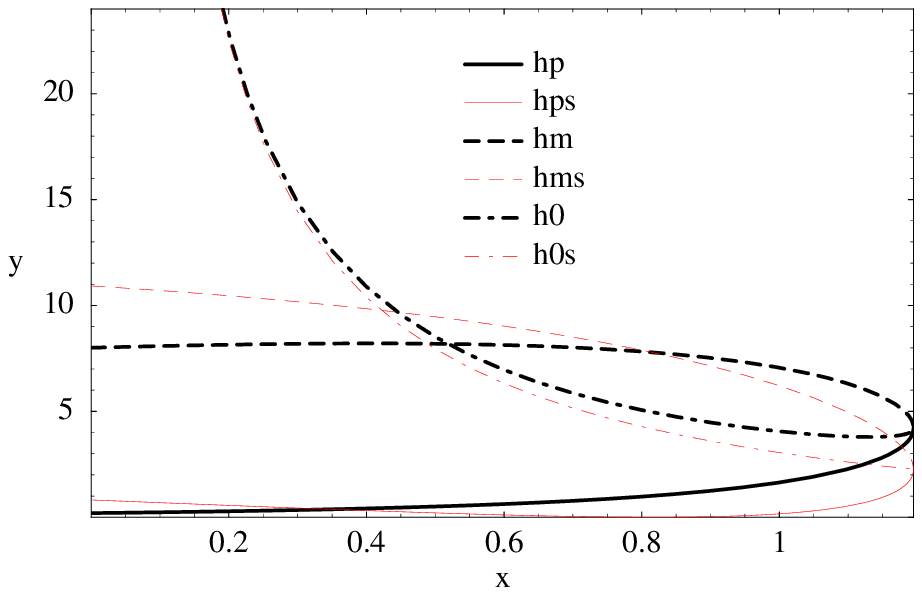}}
\end{center}
\caption{\small\it \label{4} Predictions of our model for the $s$ dependence of the helicity amplitudes\index{helicity amplitudes} $H_i^2(s)$ for the $D^+\to \rho^0$ semileptonic decay. Two poles'\index{two poles parameterization of heavy-to-light form factors} predictions are rendered in thick (black) lines while single pole predictions are rendered in thin (red) lines: $H_+$ (solid lines), $H_-$ (dashed lines) and $H_0$ (dot-dashed lines).}
\end{figure}
\psfrag{y}[bc][bc][1][90]{\Blue{$\left|H_i^{(D_s^+ \to \phi)}\right|^2~[\mathrm{GeV}^2]$}}
\begin{figure}[!t]
\begin{center}
\scalebox{1}{\includegraphics{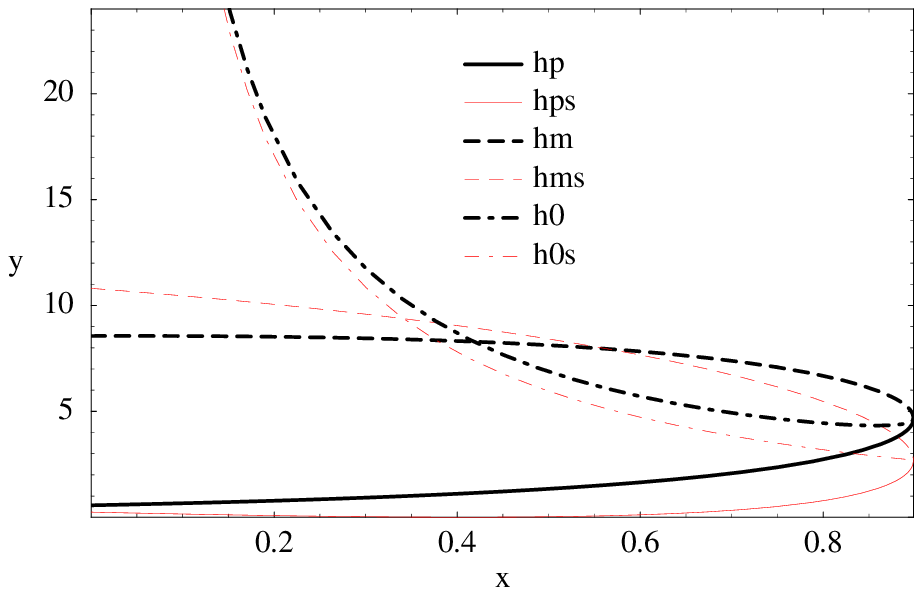}}
\end{center}
\caption{\small\it \label{5} Predictions of our model for the $s$ dependence of the helicity amplitudes\index{helicity amplitudes} $H_i^2(s)$ for the $D_s^+\to \phi$ semileptonic decay. Two poles' predictions are rendered in thick (black) lines while single pole predictions are rendered in thin (red) lines: $H_+$ (solid lines), $H_-$ (dashed lines) and $H_0$ (dot-dashed lines).}
\end{figure}

\subsection{Summary}
\index{summary}
\index{conclusions}

Our predictions for the shapes of the various form factors  can also be summarized using the general formulas\index{form factor!parameterization}
\begin{eqnarray}
F_+(s) &=& \frac{F_+(0)}{(1-x)(1-ax)}, \nonumber\\
F_0(s) &=& \frac{F_0(0)}{(1-bx)}, \nonumber\\
V(s) &=& \frac{V(0)}{(1-x)(1-a'x)}, \nonumber\\
A_0(s) &=& \frac{A_0(0)}{(1-y)(1-a''y)}, \nonumber\\
A_1(s) &=& \frac{A_1(0)}{1-b'x}, \nonumber\\
A_2(s) &=& \frac{A_2(0)}{(1-b'x)(1-b''x)},
\label{eq_ff_general}
\end{eqnarray}
where as before $x=s/m_{H^*}^2$ and $y=s/m_{H'}^2$. These expressions are actually simplifications of the form factor\index{form factor!parameterization} parameterizations~(\ref{eq_v_ff}), (\ref{eq_a0_ff}), (\ref{eq_a1_ff}) and (\ref{eq_a2_ff}) respectively. The parameters $F_+(0)=F_0(0)$, $V(0)$, $A_0(0)$, $A_1(0)$, $A_2(0)$, $a$, $a'$, $b'$ and $b''$, which we fix by nearest resonance \index{resonance saturation approximation}saturation approximation and HM$\chi$T calculation at $s_{\mathrm{max}}$, are listed in tables~\ref{table_params} and~\ref{table_params_V} for the various decay  channels considered.
\begin{table*}[!t]
\begin{center}
\begin{tabular}{|l|cc|cc|}
\hline
Decay  & $F_+(0)$ & $F_0(0)$ & $a$ & $b$ \\
\hline\hline
    ${D_0\to \pi^-}^{\dagger}$ & $0.60$ & $0.60$ & $0.55$ & $0.76$\\
    ${D^0\to K^{-}}$ & $0.72$ & $0.72$ & $0.57$ & $0.83$\\
    $D^+\to\pi_0$ & $0.60$ & $0.62$ & $0.55$ & $0.76$\\
    $D^+\to \overline K_0$ & $0.72$ & $0.71$ & $0.57$ & $0.83$\\
    $D_s\to\eta$ & $0.73$ & $0.81$ & $0.57$ & $0.83$\\
    $D_s\to\eta'$ & $0.87$  & $0.66$ & $0.57$ & $0.83$\\
    $D^+\to\eta$ & $0.60$ & $0.62$ & $0.55$ & $0.76$\\
    $D^+\to\eta'$ & $0.60$ & $0.62$ & $0.55$ & $0.76$\\
    $D_s\to\overline K_0$ & $0.60$ & $0.62$ & $0.55$ & $0.76$\\
\hline
\end{tabular}
\end{center}
\caption{\small\it \label{table_params} Predictions of our model for the parameter values appearing in the general form factor formulas~(\ref{eq_ff_general}) for the various $D\to P\ell\nu_{\ell}$ decay channels considered. The $D^0\to \pi^-$ decay channel marked with a dagger ${\dagger}$ has been used to fit the model parameters.}
\end{table*}
\begin{table*}[!t]
\begin{center}
\begin{tabular}{|l|cccc|cc|}
\hline
Decay  & $V(0)$ & $A_0(0)$ & $A_1(0)$ & $A_2(0)$ & $a''=a'$ &  $b'$ \\
\hline
\hline
    ${D^0\to \rho^-}^{\dagger}$ & $1.05$ & $1.32$ & $0.61$ & $0.31$ & $0.55$ & $0.76$ \\
    ${D^0\to K^{-*}}^{\dagger}$ & $0.99$ & $1.12$ & $0.62$ & $0.31$ & $0.57$ & $0.83$  \\
    ${D^+\to \rho^0}^{\dagger}$ & $1.05$ & $1.32$ & $0.61$ & $0.31$ & $0.55$ & $0.76$ \\
    ${D^+\to K^{0*}}^{\dagger}$ & $0.99$ & $1.12$ & $0.62$ & $0.31$ & $0.57$ & $0.83$ \\
    $D^+\to \omega$ & $1.05$ & $1.32$ & $0.61$ & $0.31$ & $0.55$ & $0.76$ \\
    ${D_s\to \phi}^{\dagger}$ & $1.10$ & $1.02$ & $0.61$ & $0.32$ & $0.57$ & $0.83$ \\
    $D_s\to K^{0*}$ & $1.16$ & $1.19$ & $0.60$ & $0.33$ & $0.55$ & $0.76$ \\
\hline
\end{tabular}
\end{center}
\caption{\small\it \label{table_params_V} Predictions of our model for the parameter values appearing in the general form factor formulas~(\ref{eq_ff_general}) for the various $D\to V\ell\nu_{\ell}$ decay channels considered ($b''=0$ for all decay modes as explained in the text).The decay channels marked with a dagger ${\dagger}$ have been used to fit the model parameters.}
\end{table*}

\par

Finally, using numerical values as explained above, we calculate the branching ratios  for all the relevant $D\to P$ and $D\to V$ semileptonic decays and compare the predictions of our model with experimental data from PDG~\cite{Eidelman:2004wy}. The results are summarized in tables~\ref{PP_results_table} and \ref{table_results}.

\begin{table}[!t]
\begin{center}
\begin{tabular}{|c|c|c|c|}
   \hline
    Decay  & $\mathcal{B}\mathrm{(two~poles)}~[\%]$ & $\mathcal{B}\mathrm{(single~pole)}~[\%]$ & $\mathcal{B}\mathrm{(Exp.)}~[\%]$\\
    \hline
    \hline
   ${D^0\to \pi^-}^{\dagger}$ & $0.36$ & $0.36$ & $0.36\pm0.06$\\
   $D^0\to K^-$ & $3.8$ & $0.43$ & $3.43\pm0.14$ \\
   $D^+\to \pi^0$ & $0.46$ & $0.51$ & $0.31\pm0.15$\\
   $D^+\to \overline K^0$ & $9.7$ & $1.1$ & $6.8\pm0.8$ \\
   $D^+_s\to \eta$ & $2.6$ & $0.38$ & $2.5\pm0.7$ \\
   $D^+_s\to \eta'$ & $0.86$ & $0.03$ & $0.89\pm0.33$ \\
   $D^+\to \eta$ & $0.11$ & $0.006$ & $<0.5$ \\
   $D^+\to \eta'$ & $0.016$ & $0.0003$ & $<1.1$ \\
   $D^+_s\to K^0$ & $0.33$ & $0.06$ & \\
    \hline
\end{tabular}
\end{center}
\caption{\small\it \label{PP_results_table} The branching ratios  for the $D\to P$ semileptonic decays. Comparison of model predictions with experiment as explained in the text. The $D^0\to \pi^-$ decay channel marked with a dagger ${\dagger}$ has been used to fit the model parameters.}
\end{table}

\begin{table}[!t]
\begin{center}
\begin{tabular}{|l|cc|cc|cc|}
\hline
Decay  & $\mathcal{B}$ [\%] & $\mathcal{B}$ (Exp.) [\%] & $\Gamma_L/\Gamma_T$ &  $\Gamma_L/\Gamma_T$ (Exp.) &  $\Gamma_+/\Gamma_-$ &  $\Gamma_+/\Gamma_-$ (Exp.) \\\hline
    \hline
     ${D^0\to \rho^-}^{\dagger}$ & $0.20$ & $0.194(41)$~\cite{Blusk:2005fq}\footnotemark[3] & $1.10$ & & $0.13$ & \\
    ${D^0\to K^{-*}}^{\dagger}$ & $2.2$ & $2.15(35)$~\cite{Eidelman:2004wy}\footnotemark[3] & $1.14$ & & $0.22$ & \\
    ${D^+\to \rho^0}^{\dagger}$ & $0.25$ & $0.25(8)$~\cite{Eidelman:2004wy}\footnotemark[3] & $1.10$ & & $0.13$ & \\
    ${D^+\to K^{0*}}^{\dagger}$ & $5.6$ & $5.73(35)$~\cite{Eidelman:2004wy}\footnotemark[3] & $1.13$ & $1.13(8)$~\cite{Eidelman:2004wy}\footnotemark[3] & $0.22$ & $0.22(6)$~\cite{Eidelman:2004wy}\footnotemark[3] \\
    ${D_s\to \phi}^{\dagger}$ & $2.4$ & $2.0(5)$~\cite{Eidelman:2004wy}\footnotemark[3] & $1.08$ & & $0.21$ & \\
    $D^+\to \omega$ & $0.25$ & $0.17(6)$~\cite{Blusk:2005fq} & $1.10$ & & $0.13$ & \\
    $D_s\to K^{0*}$ & $0.22$ &  & $1.03$ & & $0.13$ &\\
    \hline
\end{tabular}
\end{center}
\caption{\small\it \label{table_results} The branching ratios  and partial decay width ratios for the $D\to V$ semileptonic decays. Predictions of our model and experimental results as explained in the text. The decay channels marked with a dagger ${\dagger}$ have been used to fit the model parameters.}
\end{table}
For comparison we also include in table~\ref{PP_results_table} the results for the rates obtained with our approach for $F_+(q_{\mathrm{max}}^2)$ (eq.~\ref{F+FFq}) but using a single pole fit. It is very interesting that our model extrapolated\index{extrapolation!of heavy-to-light form factors} with two poles\index{two poles parameterization of heavy-to-light form factors} gives branching ratios  for $D \to P(V) \ell \nu_{\ell}$ in rather good agreement with experimental results for the already measured (partial) decay  rates. It is also obvious that the single pole fit gives rates largely incompatible with the experimental results.

\par

We expect $1/m_D$ as well as chiral corrections in the case of $D \to K^{(*)} \ell \nu_{\ell}$ and $D_s \to \eta (\eta',\phi,\omega) \ell \nu_{\ell}$ might further improve the agreement with the experimental data, but due to the presence of a large number of new couplings it is impossible to include them into the calculation within present framework.
\footnotetext[3]{Values used in the fit of our model parameters.}
\setcounter{footnote}{3}
We expect that the errors in the predicted decays  rates stemming from the uncertainties in the input parameters we used can be $~ 20\%$.
In addition, the semileptonic $D\to V$ decay rates in our model fit are numerically
dominated by the longitudinal helicity amplitude $H_0$\index{helicity amplitudes} which has a broad
$1/\sqrt{s}$ pole~\footnote{Naive HQET\index{HQET!in heavy-to-light transitions} scaling predicts that the $H_-$
helicity amplitude\index{helicity amplitudes}, which scales as $\sqrt{m_H}$ should dominate the decay
rate.}. This is true especially for $D\to V$ but to minor extent also for
$B\to V$ transitions. Since our model parameters are determined at
$s_{\mathrm{max}}$, this gives a poor handle on the dominating effects in
the overall decay  rate. Thus, accurate determination of the magnitude and
shape of the $H_0$ helicity amplitude\index{helicity amplitudes} near $s=0$ would contribute much to
clarifying this issue.


\par

In principle one can apply the above procedure to the $B \to P(V) \ell \nu_{\ell}$
decays\index{decay!$B \to P(V) \ell \nu_{\ell}$}. However, due to the much broader leptons invariant mass dependence in
this case, our procedure is much more sensitive to the values of the form factors  at $s \approx  0$ and additional contributions beyond the nearest resonances\index{resonance contribution!in heavy-to-light transitions} considered here.

\par

To summarize, we have devised a general parametrization of all the $H\to P$ and $H\to V$ by saturating the form factor\index{form factor!resonance contributions} dispersion relations with effective poles and encompassing relevant symmetry relations. We conclude that in order to satisfy all the constraints and to be compatible with the available experimental data, one needs at least 1-2 resonant excitations contributing and saturating the form factors'  shapes in the entire physical $s$-region. Our results show, that a single pole parametrization of all the form factors  cannot be considered meaningful. Quantitatively, though, we cannot give reliable predictions about the form factors. Our model approach of saturating the effective form factor poles with experimentally measured charmed meson resonances\index{resonance!of charmed meson} is to be considered as an illustration of the general principles behind the form factor\index{form factor!parameterization} parameterizations.

\par

\vskip1cm

\section{Heavy to heavy transitions}
\index{transition!heavy-to-heavy}
\index{transition!$B\to D^{(*)}$}
\index{semileptonic decay}

In our quest for the precise determination of the $V_{cb}$\index{CKM!matrix elements!$V_{cd}$} CKM matrix element the studies of $B$\index{meson!$B$!decay} meson decays into charm resonances\index{resonance!of charmed meson} have been playing a prominent role. In experiments aimed to determine $V_{cb}$\index{CKM!matrix elements!$V_{cd}$},
actually the product $|V_{cb} {\cal F } (1) |$ is extracted, where  ${\cal F } (1)$ is the $B \to D$ or $B \to D^*$ hadronic form factor \index{form factor!$B \to D^{(*)}$} at zero recoil. A lack of precise information about the shapes of various form factors is thus still the main source of uncertainties. In theoretical studies, heavy quark symmetry\index{heavy quark symmetry!in heavy-to-heavy transitions} has been particularly appealling due to the reduction of six form factors\index{form factor!relations} in the case of $B \to D(D*) l \nu_l$ transitions to only one~\cite{Isgur:1989ed,Isgur:1990jf}. In addition, at zero recoil, when the final state meson is at rest in the $B$ rest frame, the normalization of the form factors\index{form factor!normalization} is fixed by symmetry. However, the results obtained within heavy meson effective theories obtain important corrections coming from operators which are suppressed\index{HQET supressed operators} as $1/M_{B,D}$~\cite{Boyd:1995pq} as well as of higher order in the chiral\index{chiral expansion!in $B\to D$ transitions} expansion~\cite{Casalbuoni:1996pg,Falk:1993iu,Cho:1992cf,Jenkins:1992qv}. The knowledge of both kinds of corrections has improved during the last few years.  The $B \to D^* l \nu_l$ decay\index{decay!$B \to D^* l \nu_l$} amplitude is corrected by  $1/M_{B,D}$ only at the second order in this expansion making it more appropriate for the experimental studies~\cite{Casalbuoni:1996pg,Luke:1990eg}. In addition to heavy meson effective theory, other approaches have been used in the study of the $B \to D (D^*)$ form factors\index{form factor!$B \to D^{(*)}$}, such as quark models~\cite{DeVito:2006sq} and
QCD\index{QCD sum rules} sum rules~\cite{Dai:1998ca}, while the most reliable results should be expected from lattice QCD\index{lattice QCD!in heavy-to-heavy transitions}~\cite{Kronfeld:2003sd}. In the treatment of hadronic properties using lattice QCD the main problems arise due to the small masses of the light quarks. Namely, lattice\index{lattice QCD!in heavy-to-heavy transitions} studies have to consider light quarks with larger masses and then extrapolate\index{extrapolation!of lattice QCD data} results to their physical values. In these studies the chiral behavior of the amplitudes is particularly important. HM$\chi$PT\index{HM$\chi$PT!in heavy-to-heavy transitions} is very useful in giving us some control over the uncertainties appearing when the chiral\index{chiral limit!in heavy-to-heavy transitions} limit is approached. Most recently in ref.~\cite{Laiho:2005ue}, the authors have discussed $B \to D l \nu_l$ and $B \to D l \nu_l$  form factors\index{form factor!$B \to D^{(*)}$} in staggered $\chi$PT\index{$\chi$PT!staggered} by including the chiral\index{loop corrections!to Isgur-Wise function} loop corrections.

\par


In this section we investigate chiral loop corrections within HM$\chi$PT\index{HM$\chi$PT!in heavy-to-heavy transitions} in the semileptonic transitions of $B$ mesons into charm mesons of positive and negative parities to determine their impact on the chiral\index{chiral extrapolation!of Isgur-Wise function} extrapolation used by lattice QCD studies of the relevant form factors.

\subsection[$\overline B\to D^{(*)}$ form factors]{$\bf \overline B\to D^{(*)}$ form factors}
\index{form factor!$B \to D^{(*)}$}
The weak vector current\index{amplitude!in $B\to D^{(*)}$ decay} matrix element between heavy $B$ and $D$ mesons with velocities $v = p/m_B$ and $v' = p'/m_D$ respectively can be parametrized in terms of two velocity dependent form factors\index{form factor!$B \to D^{(*)}$}
~\cite{Manohar:2000dt}
\begin{equation}
\frac{\bra{D(p')}J^{\mu}_V\ket{\overline B(p)}}{\sqrt{m_B m_D}} = h_+( w) (v + v')^{\mu} + h_-( w) (v - v')^{\mu},
\end{equation}
where $ w = v \cdot v' = (m_B^2+m_D^2-s)/2 m_B m_D$ and with similar formulae for the vector and axial current matrix elements between $B$ and the vector or scalar charmed states. The differential decay rate\index{differential decay rate, in $B\to D$ decay} in terms of these form factors
 is\index{CKM!matrix elements!$V_{cb}$}\index{G$_F$}
\begin{equation}
\frac{d\Gamma}{d w}(\overline B \to D \ell e\nu_e) = \frac{G_F^2 |V_{cb}|^2 m_B^5}{48 \pi^3} ( w^2-1)^{3/2} r^3 (1+r)^2 \mathcal F( w)^2,
\end{equation}
where $r=m_D/m_B$ and we have assumed the form factor
 is real throughout the kinematically allowed region $0\leq  w-1\leq (m_B-m_D)^2/2m_B m_D$. It can be related to $h_{\pm}$ via
\begin{equation}
\mathcal F( w)^2 = \left[h_+( w) + \left( \frac{1-r}{1+r} \right)h_-( w) \right]^2.
\end{equation}
Again similar formulae are valid for $B\to D^*$ and $B\to D^*_0$ transitions.

\subsection{Framework and Calculation of Chiral Loop Corrections}
\index{HM$\chi$PT!in heavy-to-heavy transitions}
\index{loop corrections!to Isgur-Wise function}
\index{loop corrections!to Isgur-Wise function}
\index{Isgur-Wise function}
\index{Isgur-Wise function!chiral corrections}

We use the formalism of heavy meson chiral\index{HM$\chi$PT!Lagrangian} Lagrangians of the previous sections. The weak part of the Lagrangian describing transitions among heavy quarks can be matched\index{matching!HM$\chi$PT to HQET} upon weak heavy quark currents\index{HQET heavy-to-heavy current}\index{HM$\chi$PT!heavy-to-heavy current} in HQET~\cite{Falk:1993iu,Wise:1992hn}\index{bosonization!of weak current}\index{HM$\chi$PT!matching to HQET}
\begin{eqnarray}
\overline c_{v'} \Gamma b_{v} &\to& C_{cb} \Huge\{ \,-\, \xi(w) \mathrm{Tr} \left[ \overline{H}_a(v') \Gamma H_a(v) \right] \nonumber\\
&&\phantom{C_{cb} \{}-\widetilde \xi(w) \mathrm{Tr} \left[ \overline{S}_a(v') \Gamma S_a (v) \right] \nonumber\\
&&\phantom{C_{cb} \{}-\tau_{1/2}(w) \mathrm{Tr} \left[ \overline{H}_a(v') \Gamma S_a (v) \right] + \mathrm{h.c.} \Huge\}\,,
\end{eqnarray}
at leading order in chiral\index{chiral expansion!in $B\to D$ transitions} and heavy quark expansion\index{heavy quark expansion!in heavy to heavy transitions} and where $\Gamma=\gamma_{\mu}(1-\gamma_5)$. Evaluating the traces in the first term on the r.h.s we can identify for example
\begin{equation}
\bra{D(v')} \overline c_{v'} \gamma_{\mu} b_{v} \ket{\overline B(v)} = \xi( w) (v + v')_{\mu}\,,
\end{equation}
resulting in the HQET\index{HQET form factor relation}\index{form factor!relations} predictions $\mathcal F( w) = h_+( w) = \xi ( w)$ and $h_-( w)=0$. Note that heavy quark symmetry\index{heavy quark symmetry!in heavy-to-heavy transitions} dictates the values of $\xi(1)=\widetilde \xi(1) =1$, which should not receive any chiral corrections\index{chiral corrections!to Isgur-Wise functions}. On the other hand $\tau_{1/2}(w)$ is not constrained and we use the recently determined value of~\cite{Becirevic:2004ta} $\tau_{1/2}(1)=0.38$.

\par

\begin{figure}[!t]
\begin{center}
\epsfxsize5cm\epsffile{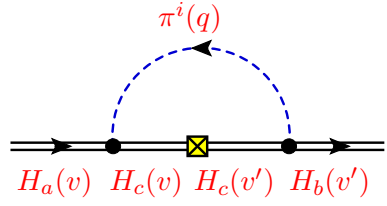}
\end{center}
\caption{\small\it\label{diagram_xi}Weak vertex correction diagram.}
\end{figure}

We first calculate the wave function renormalization $Z_{2H}$ of the heavy $H(v)=P(v),~P^*(v)$ and $P_0(v)$, $P^*_1(v)$ fields. This has been done in the chapter~\ref{ch_strong}. We get non-zero contributions to the heavy meson wavefunction renormalization from the self energy ("sunrise" topology) diagrams in fig.~\ref{diagram_sunrise} with leading order couplings in the loop \index{loop corrections!to Isgur-Wise function}. In the case of the $P(v)$ mesons both vector $P^*(v)$ and scalar $P_0(v)$ mesons can contribute in the loop. The positive parity $P_0(v)$ and $P^*_1(v)$ similarly obtain wavefunction renormalization contributions from self energy diagrams (fig.~\ref{diagram_sunrise}) with $P^*_1(v)$, $P(v)$ and $P_0(v)$, $P^*_1(v)$, $P^*(v)$ mesons in the loops respectively. Then we calculate loop corrections to the Isgur -Wise functions $\xi( w)$, $\widetilde \xi ( w)$ and $\tau_{1/2}( w)$. These come from the diagram topologies as the one shown fig.~(\ref{diagram_xi}).
\psfrag{pi}[bc]{$\Red{\pi^i(q)}$}
\psfrag{Ha}[cc]{$\Red{H_a(v)}$}
\psfrag{Hb}[cc]{$\Red{H_b(v')}$}
\psfrag{Hc}[cc]{$\Red{H_c(v)}$}
\psfrag{Hd}[cc]{$\Red{H_c(v')}$}
Namely the initial and final heavy states may exchange a pseudo-Goldstone\index{pseudo-Goldstone boson}, while pairs of positive and negative parity heavy mesons may propagate in the loop. Again not all heavy states contribute due to parity conservation in effective strong interaction vertices. Thus, when initial and final states are pseudoscalars we get contributions from pairs of $P^*(v')P^*(v)$, $P_0(v')P^*(v)$, $P^*(v')P_0(v)$ and $P_0(v')P_0(v)$ propagating in the loop, while for pseudoscalar initial and scalar final state we get contributions from pairs of $P^*(v')P(v)$, $P^*(v')P_1^*(v)$, $P_0(v')P(v)$ and $P_0(v')P_1^*(v)$ in the loop (due to heavy quark symmetry\index{heavy quark symmetry!in heavy-to-heavy transitions}, the same results are obtained for (axial)vector external states, although different intermediate states contribute). The complete expressions for the loop\index{loop corrections!to Isgur-Wise function}\index{loop corrections!to Isgur-Wise function} corrected $\xi( w)$, $\widetilde \xi ( w)$ and $\tau_{1/2}( w)$ we obtain, are listed below. For the $\xi(w)$ we get\index{chiral corrections!to Isgur-Wise functions}\index{Isgur-Wise function!chiral corrections}
\begin{eqnarray}
\xi_{ab}(w) &=&  \xi(w) \Bigg\{ \delta_{ab} + \frac{1}{2} \delta Z_{2P_a(v')} + \frac{1}{2} \delta Z_{2P_b(v)}  + \frac{\lambda^i_{ac}\lambda^i_{cb}}{16\pi^2 f^2}   \nonumber\\
&& \hskip-2.cm \times \Bigg[ g^2 \left((w+2) C_1(w,m,0,0) + (w^2-1) C_2(w,m,0,0)\right) \nonumber\\
&&\hskip-2cm  - h^2 \frac{\widetilde \xi (w)}{\xi(w)} \left( \sum_{i=1}^4 C_i (w,m,\Delta_{SH},\Delta_{SH}) + (w^2-w+1) C_2(w,m,\Delta_{SH},\Delta_{SH})  \right) \nonumber\\
&&\hskip-2.cm  - 2 h g \frac{\tau_{1/2} (w)}{\xi(w)} (w-1) \left(
C_1(w,m,\Delta_{SH},0) + w C_2(w,m,\Delta_{SH},0) + C_4(w,m,\Delta_{SH},0) \right) \Bigg]\Bigg\},\nonumber\\
\end{eqnarray}
where the same formulae can be applied to $\widetilde \xi (w)$ with the
substitution $g\leftrightarrow \widetilde g$ and $\Delta_{SH}
\leftrightarrow -\Delta_{SH}$.  For the $\tau_{1/2}(w)$ on the other
hand we obtain\index{chiral corrections!to Isgur-Wise functions}\index{Isgur-Wise function!chiral corrections}
\begin{eqnarray}
\tau_{1/2 ab}(w) &=&  \tau_{1/2} (w) \Bigg\{ \delta_{ab} + \frac{1}{2} \delta Z_{2P_a(v')} + \frac{1}{2} \delta Z_{2P_{0b}(v)}  + \frac{\lambda^i_{ac}\lambda^i_{cb}}{16\pi^2 f^2}  \nonumber\\
&&\hskip-1cm  \hskip-1.5cm \times \Bigg[ g \widetilde g \left((w-2) C_1(w,m,0,0) + (w^2-1) C_2(w,m,0,0)\right) \nonumber\\
&& \hskip-1cm \hskip-1.5cm - h^2 \left(w \sum_{i=1}^4 C_i (w,m,\Delta_{SH},-\Delta_{SH}) + (w^2-w+1) C_2(w,m,\Delta_{SH},-\Delta_{SH})  \right) \nonumber\\
&&\hskip-1cm  \hskip-1.5cm + h g (w+1) \frac{\xi(w)}{\tau_{1/2} (w)} \left( C_1(w,m,0,-\Delta_{SH}) + w C_2(w,m,0,-\Delta_{SH}) + C_3(w,m,0,-\Delta_{SH}) \right) \nonumber\\
&&\hskip-1cm  \hskip-1.5cm - h \widetilde g (w+1) \frac{\widetilde\xi(w)}{\tau_{1/2}
(w)} \left( C_1(w,m,\Delta_{SH},0) + w C_2(w,m,\Delta_{SH},0) +
C_4(w,m,\Delta_{SH},0) \right) \Bigg]\Bigg\}.\nonumber\\
\end{eqnarray}
In the above expressions $\delta Z_{2P} = (Z_{2P}-1)$ are the chiral loop\index{loop corrections!to Isgur-Wise function} corrections to the heavy meson wavefunction renormalization
for the negative and positive parity doublets is given in eqs.~(\ref{eq_Z_2_P}) and~(\ref{eq_Z_S}). As in ref.~\cite{Stewart:1998ke} and in previous sections, a trace is assumed over the inner repeated index(es) (here $b$), while the complete expressions for the loop\index{loop functions} integral functions $C_i$ can be found in Appendix~\ref{ch_loop}.

\subsection{Chiral Extrapolation}
\index{chiral extrapolation!of Isgur-Wise function}
\index{Isgur-Wise function!chiral extrapolation}

We study the contributions of the additional resonances\index{resonance contribution!in heavy-to-heavy transitions} in the chiral\index{resonance contribution!in chiral loops} loops to the chiral extrapolations employed by lattice QCD\index{lattice QCD!in heavy-to-heavy transitions} studies to run the light meson masses from the large values used in the simulations to the chiral\index{chiral limit!in heavy-to-heavy transitions} limit~\cite{Abada:2003un, McNeile:2004rf}. In order to tame the chiral\index{chiral behavior!of Isgur-Wise function} behavior of the amplites containing the mass gap between the ground state and excited heavy meson states $\Delta_{SH}$  we again use the $1/\Delta_{SH}$\index{$1/\Delta_{SH}$ expansion} expansion of the chiral\index{loop integral!expansion} loop integrals from section~\ref{sec_dsh} so that we obtain for the non-analytic terms\index{$1/\Delta_{SH}$ corrections}\index{chiral corrections!to Isgur-Wise functions}
\begin{eqnarray}
\xi_{aa}(w) &=&  \xi(w) \Bigg\{ 1 + \frac{\lambda^i_{ab}\lambda^i_{ba}}{16\pi^2 f^2} m^2 \log \frac{m^2}{\mu^2} \Bigg[ g^2 2 (r(w)-1) \nonumber\\
&&\hskip1cm- h^2  \frac{m^2}{4\Delta_{SH}^2} \left(1-w\frac{\widetilde \xi(w)}{\xi(w)}\right) - h g \frac{m^2}{\Delta_{SH}^2} w(w-1)\frac{\tau_{1/2}(w)}{\xi(w)}\Bigg] \Bigg\},
\label{eq:5}
\end{eqnarray}
and \index{chiral corrections!to Isgur-Wise functions}
\begin{eqnarray}
\tau_{1/2 aa}(w) &=&  \tau_{1/2}(w) \Bigg\{ 1 + \frac{\lambda^i_{ab}\lambda^i_{ba}}{16\pi^2 f^2} m^2 \log \frac{m^2}{\mu^2} \Bigg[ - g\widetilde g(2r(w)-1) - \frac{3}{2} (g^2+\widetilde g^2)  \nonumber\\
&& + h^2  \frac{m^2}{4\Delta_{SH}^2} \left(w-1\right)- h g \frac{m^2}{2\Delta^2} \frac{\xi(w)}{\tau_{1/2}(w)} w(1+w) + h\widetilde g \frac{m^2}{2\Delta_{SH}^2} \frac{\widetilde \xi(w)}{\tau_{1/2}(w)} w(1+w) \Bigg] \Bigg\},\nonumber\\
\label{eq:6}
\end{eqnarray}
where
\begin{equation}
r(x) = \frac{\log(x+\sqrt{x^2-1})}{\sqrt{x^2-1}},
\end{equation}
so that $r(1)=1$ and $r'(1)=-1/3$. The first lines of eqs.~(\ref{eq:5}) and~(\ref{eq:6}) contain the leading contributions while the calculated  $1/\Delta_{SH}$\index{$1/\Delta_{SH}$ corrections} corrections are contained in the second lines. Note that the positive parity heavy mesons contribute only at the $1/\Delta_{SH}^2$\index{$1/\Delta_{SH}$ expansion} order in this expansion since all the possible $1/\Delta_{SH}$ contributions vanish in dimensional regularization\index{dimensional regularization} and the affected loop integral expressions have to be expanded up to the second order in $1/\Delta_{SH}$.
\par
As argued in section~\ref{sec_dsh} the $1/\Delta_{SH}$ expansion works well in an $SU(2)$ theory where kaons\index{meson!$K$!in chiral loops} and etas\index{meson!$\eta$!in chiral loops}, whose masses would compete with the $\Delta_{SH}$ splitting, do not propagate in the loops. Therefore we write down explicit expressions for the chiral loop corrected Isgur-Wise functions specifically for the strangeless states ($a=u,d$) in the $SU(2)$\index{chiral symmetry} theory:\index{loop corrections!to Isgur-Wise function}\index{chiral corrections!to Isgur-Wise functions}
\begin{eqnarray}
\xi_{aa}(w) &=&  \xi(w) \Bigg\{ 1 + \frac{3}{32\pi^2 f^2} m^2_{\pi} \log \frac{m^2_{\pi}}{\mu^2} \Bigg[ g^2 2 (r(w)-1) \nonumber\\
&&\hskip1cm - h^2  \frac{m^2_{\pi}}{4\Delta^2} \left(1-w\frac{\widetilde \xi(w)}{\xi(w)}\right) - h g \frac{m^2_{\pi}}{\Delta^2} w(w-1)\frac{\tau_{1/2}(w)}{\xi(w)}\Bigg] \Bigg\},
\end{eqnarray}
and
\begin{eqnarray}
\tau_{1/2 aa}(w) &=&  \tau_{1/2}(w) \Bigg\{ 1 + \frac{3}{32\pi^2 f^2} m^2_{\pi} \log \frac{m^2_{\pi}}{\mu^2} \Bigg[ - g\widetilde g(2r(w)-1) - \frac{3}{2} (g^2+\widetilde g^2) \nonumber\\
&&\hskip1cm + h^2  \frac{m^2_{\pi}}{4\Delta^2} \left(w-1\right) - h g \frac{m^2_{\pi}}{2\Delta^2} \frac{\xi(w)}{\tau_{1/2}(w)} w(1+w) + h\widetilde g \frac{m^2_{\pi}}{2\Delta^2} \frac{\widetilde \xi(w)}{\tau_{1/2}(w)} w(1+w) \Bigg] \Bigg\}.\nonumber\\
\end{eqnarray}

\par

\begin{figure}
\begin{center}
\hspace*{-0.4cm}\scalebox{1}{\includegraphics{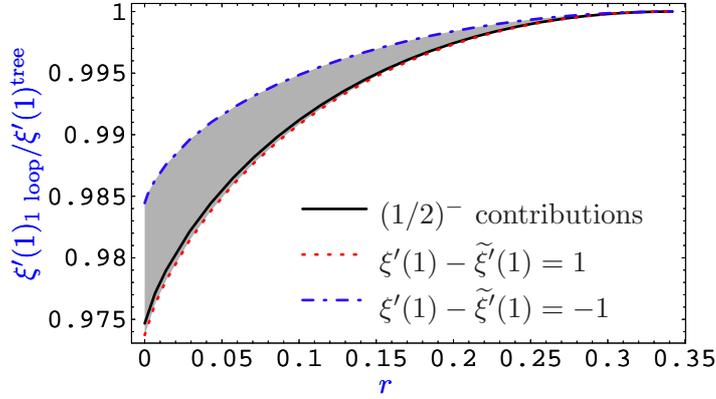}}
\end{center}
\caption{\small\it\label{plot_1} Chiral extrapolation of the slope of the IW
function at $w=1$ ($\xi'(1)$). Negative parity heavy states'
contributions (black line) and a range of possible positive parity
heavy states' contribution effects when the difference of slopes of
$\xi(1)$ and $\widetilde \xi(1)$ is varied between $1$ (red dashed line)
and $-1$ (blue dash-dotted line).}
\end{figure}

\begin{figure}
\begin{center}
\hspace*{-0.4cm}\scalebox{1}{\includegraphics{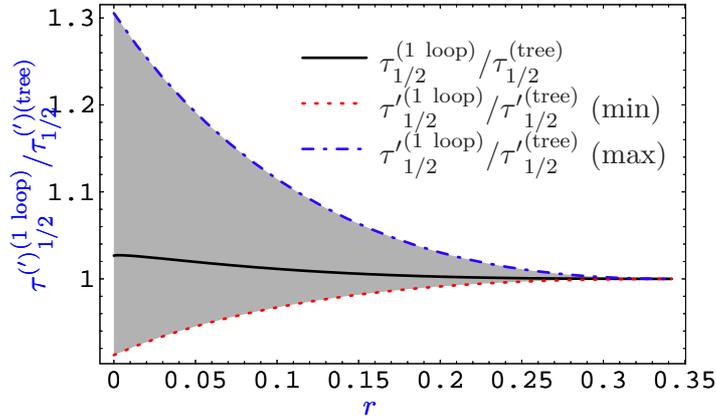}}
\end{center}
\caption{\small\it\label{plot_2} Chiral extrapolation of the $\tau_{1/2}$
function and its slope at $w=1$. $\tau_{1/2}(1)$ extrapolation
including $1/\Delta_{SH}$ contributions (black solid line), and a range of possible extrapolation effects of its slope -- $\tau_{1/2}'(1)$ -- (gray shaded region) when the difference of slopes $\xi'(1)$, $\widetilde \xi'(1)$ and $\tau_{1/2}'(1)$ is varied between $1$ (red dashed line) and $-1$ (blue dash-dotted line).}
\end{figure}

We then plot the chiral behavior of the Isgur-Wise function renormalization in the chiral limit below the $\Delta_{SH}$  scale in figs.~\ref{plot_1} and~\ref{plot_2}\index{chiral behavior!of Isgur-Wise function}. \psfrag{xk1}[bc]{\Blue{$r$}}
\psfrag{xm1}[tc][tc][1][90]{\Blue{$\xi'(1)_{\mathrm{1~loop}}/\xi'(1)^{\mathrm{tree}}$}}
\psfrag{s1}[cl]{$(1/2)^-$ contributions}
\psfrag{s3}[cl]{$\xi'(1)-\widetilde \xi'(1)=1$}
\psfrag{s4}[cl]{$\xi'(1)-\widetilde \xi'(1)=-1$}
\index{Isgur-Wise function!chiral extrapolation}
\index{chiral extrapolation!of Isgur-Wise function}
\psfrag{xk1}[bc]{\Blue{$r$}}
\psfrag{xm1}[tc][tc][1][90]{\Blue{${\tau^{(')}}^{\mathrm{(1~loop)}}_{1/2}/\tau^{(')\mathrm{(tree)}}_{1/2}$}}
\psfrag{s1}[cl]{$\tau^{\mathrm{(1~loop)}}_{1/2}/\tau^{\mathrm{(tree)}}_{1/2}$}
\psfrag{s3}[cl]{${\tau'}^{\mathrm{(1~loop)}}_{1/2}/{\tau'}^{\mathrm{(tree)}}_{1/2}$ (min)}
\psfrag{s4}[cl]{${\tau'}^{\mathrm{(1~loop)}}_{1/2}/{\tau'}^{\mathrm{(tree)}}_{1/2}$ (max)}
We again normalize the values of the extrapolated quantities at $m_{\pi}\sim \Delta_{SH}$  to $1$ and perform the chiral\index{chiral extrapolation!of Isgur-Wise function} extrapolation using the Gell-Mann formulae (eq.~(\ref{eq_2.6})). Presently no reliable estimates exist for the values of $\widetilde \xi'(1)$ and $\tau_{1/2}'(1)$, which feature in chiral extrapolation involving opposite parity heavy states. Therefore we estimate their possible effects by varying their relative values in respect to $\xi'(1)$ between $1$ and $-1$ in our extrapolations. We see that the effects of positive parity states' in the chiral loops on the chiral extrapolation of $\xi'(1)$ appear to be mild (around one percent in our estimate) below the $\Delta_{SH}$  scale (the gray shaded region around the leading order result in black solid line). Actually if $\xi'(1)-\widetilde \xi'(1)$ is positive as reasoned in~\cite{Falk:1993iu}
and around $1$, these leading $1/\Delta_{SH}$\index{$1/\Delta_{SH}$ corrections} corrections almost vanish.
The same general chiral behavior can be attributed to $\widetilde \xi'(1)$
with the substitutions $g \leftrightarrow \widetilde g$, $\Delta_{SH}
\leftrightarrow -\Delta_{SH}$ and $\xi'(1) \leftrightarrow \widetilde \xi'(1)$.
Also, the chiral extrapolation
(including small leading $1/\Delta_{SH}$ contributions) of the
$\tau_{1/2}(1)$ normalization appears fairly flat, indicating a
linear extrapolation as a good approximation, whereas the effects of
chiral loops on the extrapolation of its slope $\tau'_{1/2}(1)$
appear to be sizable, up to $30\%$ in our crude estimate.

\section{Discussion and Conclusion}
\index{conclusions}
\index{summary}

In this chapter we have calculated chiral loop corrections to $\xi$ and $\tau_{1/2}$ functions within a HM$\chi$PT framework, which includes even and odd parity heavy meson interactions with light pseudoscalars as pseudo-Goldstone bosons. Our analysis confirms that the form of the leading pionic logarithmic corrections to the Isgur-Wise functions is not changed by the inclusion opposite parity heavy mesons; they only contribute at the $m^4\log m^2$ order as can be inferred by comparing Eq.~(\ref{eq:5}) with Eq.~(8) of Ref.~\cite{Jenkins:1992qv}.
\par
These results are particularly important for the lattice QCD\index{lattice QCD} extraction of the form factors. The present errors on the $V_{cb}$\index{CKM!matrix elements!$V_{cb}$} parameters in the exclusive channels
are of the order few percent. This calls for careful control over
theoretical uncertainties in its extraction. The understanding of chiral corrections is
crucial in assuring validity of the form factor\index{form factor!$B \to D^{(*)}$} extraction and error estimation coming from the lattice\index{lattice QCD!error estimation}. Our estimates for the leading $1/\Delta_{SH}$ corrections also constrain the accuracy of such extrapolations. From these results one can deduce also the chiral corrections in the $B_s \to D_s\ell\nu$ decays\index{decay!$B_s \to D_s\ell\nu$}\index{meson!$B_s$!decay} which are not approached by experiment. Due to the strange quark flavor of final and intital heavy meson states, there is no leading pion logarithmic corrections making the lattice extraction below the heavy meson parity splitting gap $\Delta_{SH}$ much simpler.

\par

\begin{figure}[!t]
\begin{center}
\epsfxsize9.5cm\epsffile{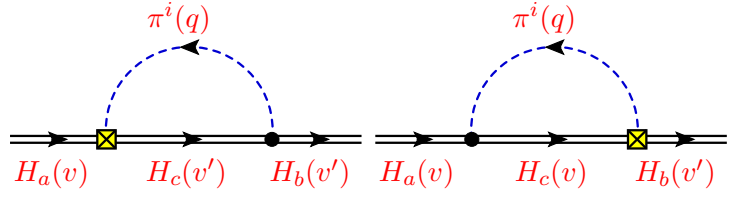}
\end{center}
\caption{\small\it\label{diagram_counter_xi}Example counterterm loop contributions yielding possible $1/\Lambda_{\chi}$ and $1/\Lambda_{\chi}\Delta_{SH}$ chiral corrections. The pseudo-Goldstone in the loop is emitted from a weak vertex counterterm.}
\end{figure}

In the $1/\Delta_{SH}$\index{$1/\Delta_{SH}$ expansion} expansion the opposite parity contributions yield formally next-to-leading chiral log order corrections in a theory with dynamical heavy meson fields of only single parity. Therefore they compete with $1/\Lambda_{\chi}$ corrections due to counterterm\index{counterterms!in heavy-to-heavy transitions} operators of higher chiral\index{chiral power counting} powers within chiral loops (see e.g. fig.~\ref{diagram_counter_xi}), where $\Lambda_{\chi}$ is the chiral symmetry breaking\index{chiral symmetry breaking scale} cut-off scale of the effective theory.
In a theory containing propagating heavy meson states of both parities, the inclusion of such terms would in addition also yield $1/\Lambda_{\chi}\Delta_{SH}$ terms. Our present approach to the estimation of the positive parity effects on the chiral extrapolation is therefore valid with the assumption $\Delta_{SH}\ll \Lambda_{\chi}$ where these additional contributions are further suppressed\index{chiral suppression}.

\chapter{Heavy neutral meson mixing}
\index{mixing!of heavy neutral mesons}
\index{oscillations!of heavy neutral mesons}
\index{SUSY!SM|see{MSSM}}

The oscillations\index{oscillations!$B^0_d-\overline B^0_d$}\index{oscillations!$B^0_s-\overline B^0_s$} in the $B^0_{d,s}-\overline B^0_{d,s}$ systems  are mediated by FCNCs\index{FCNC}\index{Flavor Changing Neutral Currents|see{FCNC}} which  are forbidden at tree level of the SM and therefore their detection gives access to the particle content in the corresponding loop diagrams.  First experimental measurement of a {\it large} value of $\Delta m_{B_d}$ indicated that the top quark mass was very heavy~\cite{Albajar:1986it}, which was confirmed almost a decade later in the direct measurements through the $p\bar p$-collisions, $M_t=172.5(1.3)(1.9)\ \gev$~\cite{Whiteson:2006zp}.
Nowadays, the accurately measured  $\Delta m_{B_d}=0.509(5)(3)\ {\rm ps}^{-1}$~\cite{Barberio:2006bi}, and $\Delta m_{B_s}=17.31(^{33}_{17})(7)\ {\rm ps}^{-1}$~\cite{Abulencia:2006mq}\index{meson!$B_s$!oscillations}\index{meson!$B$!oscillations}, are used to constrain the shape of the CKM unitarity\index{CKM!unitarity} triangle and thereby determine the amount
of the CP-violation\index{CP violation} in SM~\cite{Bona:2006ah,Charles:2004jd}.  This goal is somewhat hampered  by the theoretical
uncertainties in computing the values for the two decay\index{decay constant!of heavy meson} constants,
$f_{B_{s,d}}$, and the corresponding ``bag" parameters, $B_{B_{s,d}}$. These quantities can, in principle, be  computed on the lattice\index{lattice QCD!in heavy meson mixing}.~\footnote{Recent reviews on the current status of the lattice QCD computations of $\bbarq$ mixing amplitudes can be found in ref.~\cite{Onogi:2006km,Wingate:2006ie,Hashimoto:2004hn,Becirevic:2003hf}.} However, a major obstacle in the current lattice\index{lattice QCD!in heavy meson mixing} studies is that the $d$-quark cannot be reached directly but through an extrapolation of the results obtained by working with larger light quark masses down to the physical $d$-quark mass. In this chapter we investigate the effects of including positive parity heavy mesons into chiral\index{chiral extrapolation!of heavy meson bag parameter} extrapolation calculations on the specific examples of the decay constants $f_{B_{d,s}}$ and the
bag parameters which enter the investigation of the SM and SUSY
effects in the $\bbard$ and $\bbars$ mixing amplitudes~\cite{Becirevic:2001jj,Ball:2003se,Buras:2002vd,Barger:2004qc,Ball:2006xx}.

\section[$\Delta B = 2$ operator basis and mixing]{$\bf \Delta B = 2$ operator basis and mixing}
\index{$\Delta B = 2$ operator basis}

The SUSY contributions to the $\bbarq$ mixing amplitude, where $q$ stands for either $d$- or $s$-quark,
are usually discussed in the so called SUSY\index{SUSY!basis of $\Delta B = 2$ operators} basis of $\Delta B=2$ operators~\cite{Gabbiani:1996hi}:
\begin{eqnarray}
 \label{baseS}
{\phantom{{l}}}\raisebox{-.16cm}{\phantom{{j}}}
O_1 &=& \ \bar b^i \gamma_\mu (1- \gamma_{5} )  q^i \,
 \bar b^j  \gamma^\mu (1- \gamma_{5} ) q^j \,  ,
  \nonumber \\
{\phantom{{l}}}\raisebox{-.16cm}{\phantom{{j}}}
O_2&=& \ \bar b^i  (1- \gamma_{5} ) q^i \,
\bar b^j  (1 - \gamma_{5} )  q^j \, ,  \nonumber  \\
{\phantom{{l}}}\raisebox{-.16cm}{\phantom{{j}}}
O_3&=& \ \bar b^i  (1- \gamma_{5} ) q^j \,
 \bar b^j (1 -  \gamma_{5} ) q^i \, ,  \\
{\phantom{{l}}}\raisebox{-.16cm}{\phantom{{j}}}
O_4 &=& \ \bar b^i  (1- \gamma_{5} ) q^i \,
 \bar b^j   (1+ \gamma_{5} ) q^j \,  ,  \nonumber \\
{\phantom{{l}}}\raisebox{-.16cm}{\phantom{{j}}}
O_5 &=& \ \bar b^i  (1- \gamma_{5} ) q^j \,
 \bar b^j   (1+ \gamma_{5} ) q^i \,  ,  \nonumber
 \end{eqnarray}
where $i$ and $j$ are the color indices.  Although the operators in the above bases are written with both parity even and parity odd parts, only the
parity even ones survive in the matrix elements. In SM,   only  $O_1$ (left--left) operator is relevant in
describing  the $\bbarq$ mixing amplitude.
The matrix elements of the above operators are conventionally parameterized
in terms of bag parameters\index{bag parameter of heavy meson}, $B_{{1}{\mathrm -}{5}}$, as a measure of the discrepancy with respect to
the estimate obtained by using the VSA,
\begin{eqnarray}
{\langle \overline B^0_a\vert O_{{1}{\mathrm -}{5}}(\nu)\vert B^0_a\rangle
\over
\langle \overline B^0_a\vert O_{{1}{\mathrm -}{5}}(\nu)\vert B^0_a\rangle_{\rm VSA} } = B_{{1}{\mathrm -}{5}}(\nu)\,,
\end{eqnarray}
where $\nu$ is the renormalization scale of the logarithmically divergent operators, $O_i$,
 at which the separation between the long-distance (matrix elements) and short-distance
 (Wilson coefficients\index{Wilson coefficient}) physics is made.  Explicit calculation yields\index{bag parameter of heavy meson}
\begin{subequations}
\begin{eqnarray}
 \label{params}
\langle \overline B^0_a \vert  O_1 \vert  B^0_a \rangle_{\rm VSA}    &=& 2 \left( 1 + {1\over 3}\right)  \langle \overline B^0_a \vert A_\mu\vert 0\rangle\   \langle  0 \vert A^\mu \vert B^0_a \rangle \,,  \\
\langle \overline B^0_a \vert  O_2 \vert  B^0_a \rangle_{\rm VSA}  &=& -2 \left( 1 - \frac{1}{6} \right)  \, \left|
 \langle 0 \vert P \vert B^0_a \rangle \right|^2\,,\\
\langle \overline B^0_a \vert  O_3 \vert  B^0_a \rangle_{\rm VSA} &=&  \left( 1 - \frac{2}{3} \right)  \, \left|
 \langle 0 \vert P \vert B^0_a \rangle  \right|^2\,,  \\
\langle \overline B^0_a \vert  O_4 \vert  B^0_a \rangle_{\rm VSA}  &=&  {1\over 3}  \langle \overline B^0_a \vert A_\mu\vert 0\rangle\   \langle 0 \vert A^\mu\vert B^0_a \rangle + 2 \left|
 \langle 0 \vert P \vert B^0_a \rangle \right|^2\,,\\
\langle \overline B^0_a \vert  O_5 \vert  B^0_a \rangle_{\rm VSA}  &=&    \langle \overline B^0_a \vert A_\mu\vert 0\rangle\   \langle 0 \vert A^\mu\vert B^0_a \rangle + {2\over 3} \left|
 \langle 0 \vert P \vert B^0_a \rangle  \right|^2\,,
 \end{eqnarray}
\end{subequations}
with  $A_\mu = \bar b \gamma_\mu\gamma_5 q$ and $P = \bar b \gamma_5 q$ being the axial current and the pseudoscalar density, respectively. Next we switch to HQET\index{HQET!in heavy meson mixing}, by replacing the the field $\bar b$ with the static\index{static quark|see{HQET!static limit}} one introduced in section~\ref{sec:2.2} -- $h_v^\dagger$, satisfying  $h_v^\dagger \gamma_0=h^\dagger$. This equation and the fact that that the amplitude is invariant under the Fierz transformation in Dirac indices, eliminate the operator $O_3$ from further discussion, i.e.,
$\langle  \overline B^0_a\vert  \widetilde O_3 + \widetilde O_2 +\frac{1}{2} \widetilde O_1 \vert   B^0_a \rangle = 0$, where the tilde is used to stress that the operators are now being considered in the static limit of HQET\index{HQET!static limit} ($\vert {\bf v}\vert =0$). Furthermore, in the same limit
\begin{eqnarray}
\lim_{m_b\to \infty}{  \langle 0 \vert A_\mu \vert B^0_a(p) \rangle_{\rm QCD} \over \sqrt{2m_B}} =\lim_{m_b\to \infty}{ \langle 0 \vert P \vert B^0_a(p) \rangle_{\rm QCD} \over \sqrt{2m_B}}  =    \langle 0 \vert \widetilde A_0 \vert B^0_a(v) \rangle_{\rm HQET} = i \hat f_a v_\mu\,,
\end{eqnarray}
where $\hat f_a$ is the decay constant of the static $1/2^-$ heavy-light meson\index{HQET!static limit}, and the HQET\index{HQET!normalization of states} states are normalized as $\langle B_a^0(v)\vert B_a^0(v^\prime)\rangle=\delta(v-v^\prime)$,
so that we finally have
\begin{subequations}
\begin{eqnarray}
\label{hqet-basis}
\langle \overline B^0_a \vert  \widetilde O_1(\nu) \vert  B^0_a \rangle   &=&  {8\over 3} \hat f_a(\nu)^2  \widetilde B_{1q}(\nu)\,,  \\
\langle \overline B^0_a \vert  \widetilde O_2(\nu) \vert  B^0_a \rangle   &=&  -{5\over 3} \hat f_a(\nu)^2  \widetilde B_{2q}(\nu)\,,  \, \\
\label{hqet-basis_middle}
\langle \overline B^0_a \vert  \widetilde O_4(\nu) \vert  B^0_a \rangle &=& {7\over 3} \hat f_a(\nu)^2  \widetilde B_{4q}(\nu)\,, \\
\langle \overline B^0_a \vert \widetilde  O_5(\nu) \vert  B^0_a \rangle &=&    {5\over 3} \hat f_a(\nu)^2  \widetilde B_{5q}(\nu)\,.
\label{hqet-basis_end}
 \end{eqnarray}
\end{subequations}
One of the reasons why lattice QCD\index{lattice QCD!in heavy meson mixing} is the best currently available method for computing these matrix elements is the fact that it enables a control over the $\nu$-dependence by verifying the corresponding
renormalization group equations, which is essential for the cancellation against the $\nu$-dependence in the corresponding perturbatively computed Wilson coefficients\index{Wilson coefficient}~\cite{Sommer:2006sj,Dimopoulos:2006es,Becirevic:2004ny}.  From now on we will assume that the UV divergences are being taken care of and the scale $\nu$ will be implicit.

\section{Chiral logarithmic corrections}
\index{chiral corrections!to heavy meson bag parameters}

In this section we use HM$\chi$PT\index{HM$\chi$PT!in heavy meson mixing} to describe the low energy behavior of the matrix elements~(\ref{hqet-basis}-\ref{hqet-basis_end}).
Before entering the details, we notice that the operators $\widetilde O_4$ and $\widetilde O_5$ differ only in the color indices, i.e., by a gluon exchange, which is a local effect that cannot influence the long distance behavior\index{long distance behavior} described by $\chi$PT\index{$\chi$PT}. In other words, from the point of view of $\chi$PT\index{$\chi$PT}, the entire difference of the chiral\index{chiral behavior!of heavy meson bag parameters} behavior of the bag parameters $\widetilde B_{4q}$ and $\widetilde B_{5q}$ is
encoded in the local counterterms\index{counterterms!in heavy meson mixing}, whereas their chiral logarithmic behavior is the same.  Similar
observation has been made for the operators entering the SUSY\index{SUSY} analysis of the $\overline K^0$-$K^0$\index{meson!$K$!oscillations} mixing amplitude, as well as for the electromagnetic penguin\index{penguin operators} operators in $K\to \pi\pi$ decay\index{decay!$K\to \pi\pi$}~\cite{Becirevic:2004qd}.
Thus, in the static\index{HQET!static limit} heavy quark limit ($m_Q\to \infty$),  we are left with the first three operators~(\ref{hqet-basis}-\ref{hqet-basis_middle}) which, in their bosonized version\index{bosonization!of mixing amplitudes}, can be  written as~\cite{Detmold:2006gh}\index{chiral expansion!in heavy meson mixing}\index{HM$\chi$PT!matching to HQET}
\begin{subequations}
\begin{eqnarray}\label{bosbase1}
\widetilde  O_1 &=& \sum_X \beta_{1X} {\rm Tr} \left[ (\xi^{\dagger} H)_a \gamma_{\mu}
(1-\gamma_5) X \right] {\rm Tr} \left[ (\xi^{\dagger} H)_a \gamma^{\mu} (1-\gamma_5) X \right] + {\rm c.t.}\,,\\
 &&\hfill \nn \\
\label{bosbase2}
\widetilde O_2 &=&\sum_X \beta_{2X} {\rm Tr} \left[ (\xi^{\dagger} H)_a (1-\gamma_5) X \right] {\rm Tr} \left[
(\xi^{\dagger} H)_a (1-\gamma_5) X \right]+ {\rm c.t.}\,,\\
 &&\hfill \nn \\
\widetilde O_4 &=& \sum_X \beta_{4X} {\rm Tr} \left[ (\xi^{\dagger} H)_a (1-\gamma_5) X \right] {\rm Tr} \left[
(\xi H)_a (1+\gamma_5) X \right] + {\rm c.t.}\,,
\label{bosbase3}
\end{eqnarray}
\end{subequations}
where $X\in\{ 1, \gamma_5, \gamma_{\nu},
\gamma_{\nu}\gamma_5, \sigma_{\nu\rho}\}$~\footnote{Contraction of Lorentz
indices and HQET\index{HQET!parity conservation} parity conservation requires the same $X$ to appear in
both traces of a summation term. Any insertions of $\vdir $ can be
absorbed via $\vdir  H = H$, while any non-factorisable
contribution with a single trace over Dirac matrices can be reduced to
this form by using the $4\times 4$ matrix identity
\begin{eqnarray} 4 {\rm Tr}(AB) &=&
{\rm Tr}(A){\rm Tr}(B) +
{\rm Tr}(\gamma_5 A){\rm Tr}(\gamma_5 B) +
{\rm Tr}(A\gamma_{\mu}){\rm Tr}(\gamma^{\mu}B) \nn \\
&&+{\rm Tr}(A\gamma_{\mu}\gamma_5){\rm Tr}(\gamma_5\gamma^{\mu}B) +
1/2{\rm Tr}(A\sigma_{\mu\nu}){\rm Tr}(\sigma^{\mu\nu}B).\nn \end{eqnarray}}. As before
the index ``$a$" denotes the light quark flavor, and ``c.t." stands for the local counterterms\index{counterterms!in heavy meson mixing}. To relate  $\beta_i$'s to the bag parameters in eq.~(\ref{hqet-basis}-\ref{hqet-basis_end}) we evaluate the traces in eq.~(\ref{bosbase1}-\ref{bosbase3}) to obtain\footnote{Our convention differs by a factor of two, compared to those of~\cite{Detmold:2006gh}. This is due to our use of combined positive and negative  frequency $H^++H^-$ fields which yield this additional factor in mixed $H^+ H^-$ terms which always appear twice.} \index{bag parameter of heavy meson}
\begin{eqnarray}
\widetilde B_1 ={1\over 3\hat f^2}\widehat \beta_1\,,\quad
\widetilde B_2 ={24\over 5\hat f^2}\widehat\beta_2\,,\quad
\widetilde B_4 ={24\over 7\hat f^2}\widehat\beta_4\,,\quad
\widetilde B_5 ={24\over 5\hat f^2}\widehat\beta_4\, ,
\end{eqnarray}
where
\begin{subequations}
\begin{eqnarray}\label{betas}
\widehat \beta_1 &=& \beta_1 + \beta_{1\gamma_5} + 4 (\beta_{1\gamma_{\nu}} +
\beta_{1\gamma_{\nu}\gamma_5}) - 12 \beta_{1\sigma_{\nu\rho}},\\
\widehat \beta_2&=& \beta_2 + \beta_{2\gamma_5} + \beta_{2\gamma_{\nu}} +
\beta_{2\gamma_{\nu}\gamma_5},\\
\widehat \beta_4 &=& \beta_4 - \beta_{4\gamma_5} + \beta_{4\gamma_{\nu}} -
\beta_{4\gamma_{\nu}\gamma_5}.
\end{eqnarray}
\end{subequations}
We will use the known form of the HM$\chi$PT\index{HM$\chi$PT!Lagrangian} Lagrangian~(\ref{eq_2_13}). To get the chiral\index{chiral corrections!to heavy meson bag parameters} logarithmic corrections to $\widetilde B_{iq}$-parameters, we should subtract twice the chiral corrections to the decay constant $\hat f_a$ from the chiral corrections to the four-quark operators\index{four quark operator!in heavy meson mixing}~(\ref{bosbase1}-\ref{bosbase3}).  The former is obtained from the study of the bosonized\index{bosonization!of weak current} left-handed weak current~(\ref{eq_2_17}) for the negative parity heavy mesons. Here we shall also consider the next-to-leading order chiral counterterm contributions and we write it as\index{counterterm contributions!to heavy meson decay constants}
\begin{equation}\label{eqVA}
J^{(1)\mu}_{(V-A)\mathrm{HM}\chi\mathrm{PT}} = J^{(0)\mu}_{(V-A)\mathrm{HM}\chi\mathrm{PT}} + \varkappa _2 \mathrm{Tr} [ \gamma^{\mu}(1-\gamma_5) (\xi^{\dagger} H)_a] (m_q)_{cc} +
 \varkappa _1 {\rm Tr}\left[\gamma^\mu(1-\gamma_5) (\xi^{\dagger} H)_b \right] (m_q)_{ba}\,,
\end{equation}
where $\alpha$ is the tree level decay constant in the chiral\index{chiral expansion!in heavy meson mixing} expansion, and $\varkappa _{1,2}$ are the counterterm coefficients. Together with the strong coupling $g$, these parameters are not predicted within  HM$\chi$PT\index{HM$\chi$PT!parameters}. Instead, they are expected to be fixed by matching\index{matching!HM$\chi$PT to HQET} the HM$\chi$PT\index{HM$\chi$PT!matching to HQET} expressions with the results of lattice QCD\index{lattice QCD!results} for a given quantity (see reviews in ref.~\cite{Sharpe:2006pu,Aubin:2006gp,Bernard:2006zp}). The notation used above is the same as in ref.~\cite{Becirevic:2002sc}.
\begin{figure}
\vspace*{-0.3cm}
\begin{center}
\begin{tabular}{@{\hspace{-0.25cm}}c}
\psfrag{0-}[cc]{$\color{red}  H_a(v)$}                                                 \psfrag{1-}[cc]{$\color{red}  H_b(v)$}                                                 \psfrag{0+}[cc]{$\color{red}  H_b(v)$}
\epsfxsize4cm\epsffile{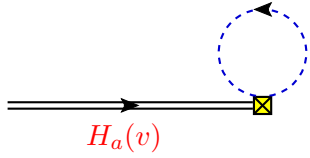}    \\
\end{tabular}
\caption{\small\it\label{fig0}{The diagram which gives non-vanishing chiral logarithmic corrections to the pseudoscalar heavy-light meson decay constant.}}
\end{center}
\end{figure}
The chiral\index{chiral corrections!to heavy meson decay constant} logarithmic corrections to the decay\index{decay constant!of heavy meson} constant come from the diagrams shown in fig.~\ref{fig0}
\begin{eqnarray}\label{fq}
&&\hat f_d = \alpha \left[ 1 - {1\over (4\pi f)^2 }
 \left(
 {3\over 4}m_\pi^2\log{m_\pi^2\over \mu^2} + {1\over 2}m_K^2\log{m_K^2\over \mu^2} +{1\over 12}  m_\eta^2\log{m_\eta^2\over \mu^2}\right) \right.\nn\\
&&\hspace*{1.7cm} \left.+ \varkappa _1(\mu) m_d + \varkappa _2(\mu)(m_u+m_d+m_s) + {1\over 2}\delta Z_d\right]\,,\nn\\
&&\hat f_s = \alpha \left[ 1 - {1\over (4\pi f)^2 }\left( m_K^2\log{m_K^2\over \mu^2} +{1\over 3}  m_\eta^2\log{m_\eta^2\over \mu^2}\right) \right.\nn\\
 &&\hspace*{1.7cm} \left.+ \varkappa _1(\mu) m_s + \varkappa _2(\mu)(m_u+m_d+m_s) + {1\over 2}\delta Z_s\right]\,,
\end{eqnarray}
where it should be stressed that we work in the exact isospin limit ($m_u=m_d$) so that the index $d$ means either $u$- or $d$-quark.
Only explicit in the above expressions is the term arising from the tadpole diagram (in fig.~\ref{fig0}), whereas $Z_{d,s}$, the heavy meson field renormalization factors, come from the self energy diagram  (left in fig.~\ref{fig0}) and are given in eq.~(\ref{eq_Z_2_P}). We write out the leading order contributions of negative parity heavy states only
\begin{eqnarray}\label{Zq}
&&Z_d  = 1 - {3g^2\over (4\pi f)^2 } \left(
 {3\over 2}m_\pi^2\log{m_\pi^2\over \mu^2} + m_K^2\log{m_K^2\over \mu^2} +{1\over 6}  m_\eta^2\log{m_\eta^2\over \mu^2} \right)\nn \\
 &&\hspace{3.1cm}+ k_1(\mu) m_d + k_2(\mu)(m_u+m_d+m_s) ,\nn\\
&&Z_s = 1 - {3 g^2\over (4\pi f)^2 }\left(2 m_K^2\log{m_K^2\over \mu^2} +{2 \over 3}  m_\eta^2\log{m_\eta^2\over \mu^2}\right) + k_1(\mu) m_s + k_2(\mu)(m_u+m_d+m_s)\ .\nn\\
\end{eqnarray}
In both eqs.~(\ref{fq}) and~(\ref{Zq}) the $\mu$ dependence in the logarithm cancels against the one in the local counterterms.
\begin{figure}
\vspace*{-0.3cm}
\begin{center}
\begin{tabular}{@{\hspace{-0.25cm}}c}
\psfrag{0-}[cc]{$\color{red}  H_a(v)$}                                                 \psfrag{1-}[cc]{$\color{red}  H_b(v)$}                                                 \psfrag{0+}[cc]{$\color{red}  H_b(v)$}
\epsfxsize12.cm\epsffile{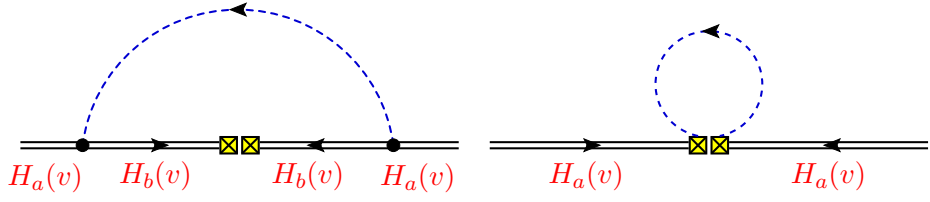}    \\
\end{tabular}
\caption{\small\it\label{fig1}{The diagrams relevant to the chiral corrections to the SM bag parameter $\widetilde B_{1a}$. In the text we refer to the left one as ``sunset", and to the right one as ``tadpole". Only the tadpole diagram gives a non-vanishing contribution to the bag parameters $\widetilde B_{2,4a}$.} }
\end{center}
\end{figure}
With these ingredients in hands it is now easy to deduce that the only diagrams which contribute to the SM\index{bag parameter of heavy meson} bag parameter,
$\widetilde B_{1a}$, are the two shown in fig.~\ref{fig1}.  They arise from the two terms  in
$\widetilde O_1 = 8\widehat \beta_1 [ (\xi^{\dagger} P^{\ast-}_\mu)_a (\xi^{\dagger} P^{\ast + \mu})_a  + (\xi^{\dagger} P^-)_a (\xi^{\dagger}  P^+)_a]$ and  yield
\begin{subequations}
\begin{eqnarray}
{\rm ``sunset"} :&& 4\widehat \beta_1 {3g^2\over (4\pi f)^2 }\sum_i (\lambda^i_{aa})^2 m_{i}^2\log{m_{i}^2\over \mu^2}\,,\\
{\rm ``tadpole"} : &-&4\widehat \beta_1  {1\over (4\pi f)^2} \sum_i (\lambda^i_{aa})^2 m_{i}^2\log{m_{i}^2\over \mu^2}\,,
\end{eqnarray}
\end{subequations}
respectively,  where $\lambda^i$ are the SU(3) generators and $m_i$ masses of the pseudo-Goldstone\index{pseudo-Goldstone boson} bosons. The SM bag parameters now read\index{bag parameter of heavy meson}\index{chiral corrections!to heavy meson bag parameters}\index{counterterm contributions!to heavy meson bag parameters}
\begin{eqnarray} \label{B1}
&&\widetilde B_{1d}=\widetilde B_1^{\rm Tree}\left[ 1 - {1-3g^2\over (4\pi f)^2}\left(
{1\over 2}m_\pi^2\log{m_\pi^2\over \mu^2} + {1\over 6}m_\eta^2\log{m_\eta^2\over \mu^2} \right)\right.\nn\\
 &&\hspace{4.2cm} \biggl.+ b_1(\mu) m_d +b_1^\prime(\mu) (m_u+m_d+m_s) \biggr]\,,\nn\\
&&\widetilde B_{1s}=\widetilde B_1^{\rm Tree}\left[ 1 - {1-3g^2\over (4\pi f)^2}
 {2\over 3}m_\eta^2\log{m_\eta^2\over \mu^2} + b_1(\mu) m_s +b_1^\prime(\mu) (m_u+m_d+m_s)
\right]\,,
\end{eqnarray}
where we also wrote the counterterm contributions and, for short, we wrote $\widetilde B_1^{\rm Tree}=3\widehat \beta_1/\alpha^2$.  The above results agree  with the ones
presented in refs.~\cite{Grinstein:1992qt,Hiorth:2003ci}, in which the pion loop contribution \index{loop corrections!to heavy meson bag parameter} was left out, and with the ones  recently
presented in ref.~\cite{Arndt:2004bg}.

\par

As for the bag parameters $\widetilde B_{2q}$ and $\widetilde B_{4q}$ we obtain\index{bag parameter of heavy meson}\index{chiral corrections!to heavy meson bag parameters}\index{counterterm contributions!to heavy meson bag parameters}
\begin{eqnarray} \label{B24}
\widetilde B_{2,4 d}&=&\widetilde B_{2,4}^{\rm Tree}\left[ 1 + {3 g^2 Y\mp 1\over (4\pi f)^2}\left(
{1\over 2}m_\pi^2\log{m_\pi^2\over \mu^2} + {1\over 6}m_\eta^2\log{m_\eta^2\over \mu^2} \right) \right. \nn\\
&& \Biggl. \hspace*{30mm} + b_{2,4}(\mu) m_d +b_{2,4}^\prime(\mu) (m_u+m_d+m_s) \Biggr] \,,\nn\\
\widetilde B_{2,4s}&=&\widetilde B_{2,4}^{\rm Tree}\left[ 1 +{2\over 3} {3 g^2 Y \mp 1\over (4\pi f)^2}
 m_\eta^2\log{m_\eta^2\over \mu^2} + b_{2,4}(\mu) m_s +b_{2,4}^\prime(\mu) (m_u+m_d+m_s)\right]\,,\nn\\
\end{eqnarray}
where
$Y=(\widehat \beta^*_{2,4}/\widehat \beta_{2,4})$, with
$\widehat \beta^\ast_2 = \beta_{2\gamma_{\nu}} +
\beta_{2\gamma_{\nu}\gamma_5} - 4 \beta_{2\sigma_{\nu\rho}}$, and $ \widehat
\beta^\ast_4 = - \beta_{4\gamma_{\nu}} + \beta_{4\gamma_{\nu}\gamma_5}$. The sign difference in the second terms of eq.~(\ref{B24}) is due to the different chiral structure of the $\widetilde O_{2,4}$ operators so that the right diagram in fig.~\ref{fig1} receives a minus sign in the case of $\widetilde O_{4}$.

\section{Impact of the $1/2^+$-mesons}
\index{resonance contribution!in heavy meson mixing}
\index{resonance contribution!in chiral loops}

In this section we examine the impact of the heavy-light mesons belonging to the
$1/2^+$ doublet when propagating in the loops onto the chiral logarithmic corrections derived in the previous section. We use the full Lagrangian\index{HM$\chi$PT!Lagrangian} of eq.~(\ref{eq_2_13}) and also the weak current operators~(\ref{eq_2_17}), which we also extend for the $1/2^+$ heavy meson chiral \index{counterterm contributions!to heavy meson decay constants} counterterm contributions:
\begin{equation}\label{VAscalar}
J^{(1)\mu}_{(V-A)\mathrm{HM}\chi\mathrm{PT}} += \widetilde\varkappa _2 \mathrm{Tr} [ \gamma^{\mu}(1-\gamma_5) (\xi^{\dagger} S)_a]  (m_q)_{cc} +
 \widetilde\varkappa _1 {\rm Tr}\left[ \gamma^\mu(1-\gamma_5) (\xi^{\dagger} S)_b\right]  (m_q)_{ba} \,,
\end{equation}
where  $\widetilde \varkappa _{1,2}$ are the coefficients of two new counterterms.

\subsection{Decay constants}
\index{chiral corrections!to heavy meson decay constant}
\index{decay constant!of heavy meson}

\begin{figure}
\vspace*{-0.3cm}
\begin{center}
\begin{tabular}{@{\hspace{-0.25cm}}c}
\psfrag{0-}[cc]{$\color{red}  H_a(v)$}                                                 \psfrag{1-}[cc]{$\color{red}  H_b(v)$}                                                 \psfrag{0+}[cc]{$\color{red}  H_b(v)$}
\epsfxsize4cm\epsffile{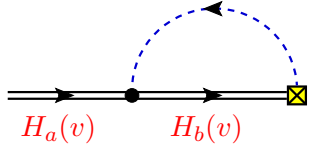}    \\
\end{tabular}
\caption{\small\it\label{fig3}{ In addition  to the diagram shown
in fig.~\ref{fig0}, this diagram  contributes the loop corrections to the
pseudoscalar meson decay constant after the $1/2^+$ mesons are included in HM$\chi$PT.}}
\end{center}
\end{figure}

Since we focus on the pseudoscalar meson decay constant, it should be clear that only the scalar meson from the $1/2^+$ doublet can propagate in the loop \index{loop corrections!to heavy meson decay constant}.
The diagrams that give non-vanishing contributions are shown in fig.~\ref{fig3} and the corresponding expressions now read
\begin{eqnarray}
\hat f_{a} &=& \alpha \left\{
1 + { \lambda^i_{ab}\lambda^{i\dagger}_{ba} \over 2 (4\pi f)^2}  \biggl[
- 3 g^2 C'_1\left(\frac{\Delta_{ba}}{m_i},m_i\right) - C_0(m_i) \right.\nonumber\\*
&&\left.+h^2
C'\left( \frac{\Delta_{\widetilde ba}+\Delta_{SH}}{m_i}, m_i\right)
+ 2 h{\alpha'\over {\alpha \Delta_{SH}}}
C\left( \frac{\Delta_{\widetilde ba} + \Delta_{SH}}{m_i}, m_i\right)
\biggl]\right\}\,,
\end{eqnarray}
where the summation over ``$i$" is implicit, and we omit the counterterm contributions to make the expressions simpler. The integral functions $C_i$ can be found in Appendix~\ref{ch_loop}. The last term in the decay constant (the one proportional to $h$) comes from the right graph in fig.~\ref{fig3}  which was absent before the inclusion of the scalar meson. Notice that due to the chiral \index{chiral behavior!of heavy meson decay constant} behavior of the $C_i$ functions in eq.~(\ref{eq_4.13}), the presence of the nearby $1/2^+$ state does not affect the pionic logarithmic behavior of the decay constant. It does, however, affect the kaon\index{meson!$K$!in chiral loops} and $\eta$-meson\index{meson!$\eta$!in chiral loops} loops because those states are heavier than $\Delta_{SH}$  and the coefficients of their logarithms cease to be the predictions of this approach since those logarithms are competitive in size with the terms proportional to $\Delta_{SH}^2 \log(4\Delta_{SH}^2/ \mu^2)$.
Stated equivalently, the relevant chiral logarithmic corrections are those coming from the $SU(2)_L\otimes SU(2)_R \to SU(2)_V$ theory, and the pseudoscalar decay constant reads
\begin{eqnarray}\label{fB-correct}
\hat f_{q} = \alpha \left[ 1-  {1+3g^2\over 2 (4\pi f)^2} {3\over 2} m_\pi^2 \log{m_\pi^2\over \mu^2} + c_f(\mu) m_\pi^2 \right]\,,\end{eqnarray}
where $c_f(\mu)$ stands for the combination of the counterterm coefficients considered in the
previous section.~\footnote{More specifically, $2B_0 c_f(\mu) + \displaystyle{{3h^2 \over 4(4\pi f)^2}} \left[
3+\log(4\Delta_S^2/\mu^2)\right]+\displaystyle{{3 h \alpha' \over 2\alpha (4\pi f)^2}} [1+\log(4\Delta_S^2/\mu^2)] = \frac{1}{2}k_1(\mu)+ \frac{1}{2}k_1^\prime (\mu)+k_2(\mu)+k_2^\prime (\mu)+\varkappa _1(\mu)+2\varkappa _2(\mu)$, where we use the Gell-Mann--Oakes--Renner formula, $m_\pi^2 = 2 B_0 m_d$ from eq.~(\ref{eq_2.6}). We stress again that the exact isospin symmetry\index{isospin symmetry!in chiral extrapolation} ( $m_u=m_d$) is assumed throughout this chapter.}
At this point we also note that we checked that the chiral logarithms in the scalar heavy-light meson decay constant, which has recently been computed on the lattice\index{lattice QCD!in heavy meson mixing} in ref.~\cite{Herdoiza:2006qv,McNeile:2004rf}, are the same as for the pseudoscalar meson,
with the coupling $g$ being replaced by $\widetilde g$, i.e.,
\begin{eqnarray}\label{su2f}
\widetilde f_{q} = \alpha' \left[ 1-  {1+3\widetilde g^2\over 2 (4\pi f)^2} {3\over 2} m_\pi^2
\log{ m_\pi^2\over \mu^2} + \widetilde  c_f(\mu) m_\pi^2 \right]\,.\end{eqnarray}
Since, as we already checked in chapter~\ref{ch_strong} (see table~\ref{table_summary}) $\widetilde g^2/g^2 \ll 1$ the deviation from the linear behavior in $m_\pi^2$
is less pronounced for  $\widetilde f_{q}$ than it is for $\hat f_{q}$.

\subsection{Bag parameters}
\index{bag parameter of heavy meson}
\index{chiral corrections!to heavy meson bag parameters}

In this subsection we show that the situation with the bag parameters is similar to the one with decay constant, namely the pion loop chiral logarithms remain unchanged when the nearby scalar meson is included in HM$\chi$PT\index{HM$\chi$PT!in heavy meson mixing}.
To that end, besides eq.~(\ref{eq_2_13}), we should include the contributions of $1/2^+$-mesons to the  operators~(\ref{bosbase1}-\ref{bosbase3}). Generically the operators $\widetilde{\cal O}_{1,2,4}$ now become\index{bosonization!of mixing amplitudes}\index{chiral expansion!in heavy meson mixing}\index{HM$\chi$PT!matching to HQET}
\begin{eqnarray}\label{eq-9}
\widetilde O_{1}&= & \sum_X \beta_{1 X} {\rm Tr}\left[ \left(\xi^{\dagger} H\right)_a \gamma_{\mu }(1-\gamma_5) X\right]
 {\rm Tr}\left[ \left(\xi^{\dagger} H\right)_a \gamma^{\mu }(1-\gamma_5) X \right] \nonumber \\
&&\hspace*{5mm}  +
 \beta_{1 X}^\prime
 {\rm Tr}\left[ \left(\xi^{\dagger} H\right)_a \gamma_{\mu }(1-\gamma_5) X  \right]
 {\rm Tr}\left[ \left(\xi^{\dagger} S\right)_a  \gamma^{\mu }(1-\gamma_5) X  \right]    \nonumber \\
&&\hspace*{5mm}+  \beta_{1 X}^{\prime\prime}  {\rm Tr}\left[ \left(\xi^{\dagger} S\right)_a \gamma_{\mu }(1-\gamma_5) X \right]
 {\rm Tr}\left[ \left(\xi^{\dagger} S\right)_a  \gamma^{\mu }(1-\gamma_5) X\right] \,,
\end{eqnarray}
where $\beta_{1X}^{\prime}$ are the couplings of the operator  $\widetilde O_{1}$
to both $1/2^-$ and $1/2^+$ mesons, while $\beta_{1X}^{\prime\prime}$ come from the coupling to the
$1/2^+$ mesons only.  Similarly, the operators $\widetilde O_{2,4}$ now read:\index{bosonization!of mixing amplitudes}\index{chiral expansion!in heavy meson mixing}
\begin{eqnarray}\label{eq-92}
\widetilde O_{2}&= & \sum_X \beta_{2 X} {\rm Tr}\left[ \left(\xi^{\dagger}  H\right)_a  (1-\gamma_5) X\right]
 {\rm Tr}\left[ \left(\xi^{\dagger} H\right)_a  (1-\gamma_5) X \right] \nonumber \\
&&\hspace*{5mm}  +
 \beta_{2 X}^\prime
 {\rm Tr}\left[ \left(\xi^{\dagger} H\right)_a  (1-\gamma_5) X  \right]
 {\rm Tr}\left[ \left(\xi^{\dagger} S\right)_a   (1-\gamma_5) X  \right]    \nonumber \\
&&\hspace*{5mm}+  \beta_{2 X}^{\prime\prime}  {\rm Tr}\left[ \left(\xi^{\dagger}  S\right)_a  (1-\gamma_5) X \right]
 {\rm Tr}\left[ \left(\xi^{\dagger} S\right)_a  (1-\gamma_5) X\right] \,,
\end{eqnarray}
\begin{eqnarray}\label{eq-93}
\widetilde O_{4}
&= & \sum_X \beta_{4 X} {\rm Tr}\left[ \left(\xi^{\dagger} H\right)_a  (1-\gamma_5) X\right]
 {\rm Tr}\left[ \left(\xi H\right)_a  (1+\gamma_5) X \right] \nonumber \\
&&\hspace*{5mm}  +
 \beta_{4 X}^\prime
 {\rm Tr}\left[ \left(\xi^{\dagger} H\right)_a  (1-\gamma_5) X  \right]
 {\rm Tr}\left[ \left(\xi S\right)_a   (1+\gamma_5) X  \right]   \nonumber \\
&&\hspace*{5mm}  +
\overline \beta_{4 X}^\prime
 {\rm Tr}\left[ \left(\xi^{\dagger}  S\right)_a  (1-\gamma_5) X  \right]
 {\rm Tr}\left[ \left(\xi H\right)_a   (1+\gamma_5) X  \right]   \nonumber \\
&&\hspace*{5mm}+  \beta_{4 X}^{\prime \prime} {\rm Tr}\left[ \left(\xi^{\dagger}  S\right)_a  (1-\gamma_5) X\right]
 {\rm Tr}\left[ \left(\xi S\right)_a  (1+\gamma_5) X \right] \,.
\end{eqnarray}
After evaluating the traces in eqs.~(\ref{eq-9}), and keeping in mind that the external states are the pseudoscalar mesons, we have
\begin{eqnarray}
\widetilde O_1&= & 8\widehat \beta_1\left[   \left(\xi^{\dagger} P^{\ast - \mu} \right)_a  \left(\xi^{\dagger} P^{\ast+}_\mu \right)_a
 +   \left(\xi^{\dagger} P^- \right)_a  \left(\xi^{\dagger} P^+  \right)_a \right]  \nn\\
&& + 4\widehat \beta_1^\prime \left[
    \left(\xi^{\dagger} P^- \right)_a  \left(\xi^{\dagger} P^+_0  \right)_a  +\left(\xi^{\dagger} P^-_0 \right)_a  \left(\xi^{\dagger} P^+  \right)_a \right] + 8\widehat \beta_1^{\prime\prime}  \left(\xi^{\dagger} P^-_0 \right)_a  \left(\xi^{\dagger} P^+_0  \right)_a  \,,
\end{eqnarray}
where $\widehat \beta_1^{(\prime,\prime\prime)}$ have forms analogous to the ones written
in eq.~(\ref{betas}).
\begin{figure}
\vspace*{-0.3cm}
\begin{center}
\psfrag{0-}[cc]{$\color{red}  H_a(v)$}                                                 \psfrag{1-}[cc]{$\color{red}  H_b(v)$}                                                 \psfrag{0+}[cc]{$\color{red}  H_b(v)$}
\psfrag{1-0+}[cc]{{$ \color{red}{\quad H_b(v)}$}}
\epsfxsize12cm\epsffile{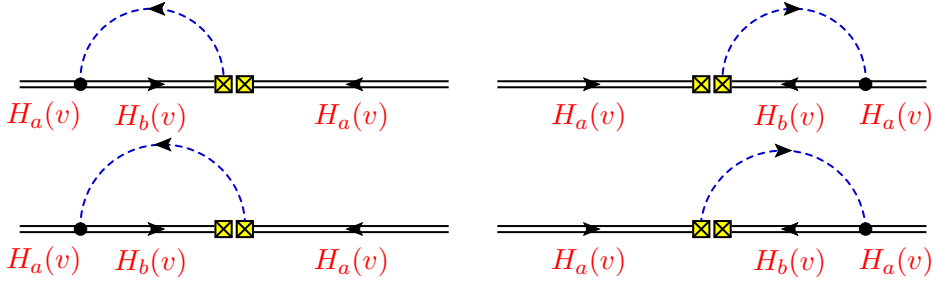}
\caption{\small\it\label{fig4}{Additional diagrams which enter in the calculation of the chiral corrections to the operators
$\langle \widetilde {\cal O}_{1,2,4}\rangle$ once positive parity heavy states are taken into account.}}
\end{center}
\end{figure}
The corresponding additional 1-loop chiral diagrams are shown in fig.~\ref{fig4}.  Since the couplings of the four-quark operators\index{four quark operator!in heavy meson mixing} to the scalar meson(s) are proportional to $\beta_i^{\prime,\prime\prime}$ and of the pseudoscalar decay\index{decay constant!of heavy meson} constant to $\alpha'$, the cancellation between
the chiral loop corrections in the operators  $\widetilde O_{i}$ and in the decay constant is not automatic. For that reason, instead of writing the chiral logarithmic corrections to the bag-parameter, we will write them for the  full operator, namely \index{bag parameter of heavy meson}\index{chiral corrections!to heavy meson bag parameters}
\begin{eqnarray}\label{eqO1}
\widetilde B_{1a} \hat f_a^2 &=& {3}\widehat \beta_1
 \left\{
1 - { \lambda^i_{ab}\lambda^{i\dagger}_{ba} \over 2 (4\pi f)^2}  \biggl[
6 g^2  C'_1\left(\frac{\Delta_{ba}}{m_i},m_i\right) + 2 C_0(m_i)\biggr.\right.
\nn\\
&&-2h^2
 C'\left(\frac{\Delta_{\widetilde b a}+\Delta_{SH}}{m_i},m_i\right)
 \biggl. -4 h{\widehat \beta_1^\prime\over {\Delta_{SH}\widehat  \beta_1}}C\left(\frac{\Delta_{\widetilde b a}+\Delta_{SH}}{m_i},m_i\right)
 \biggr]
  \nn\\
&&  - { \lambda^i_{aa}\lambda^{i\dagger}_{aa} \over 2 (4\pi f)^2} \biggl[
- 6 g^2 C'_1\left(\frac{\Delta_{b a}}{m_i},m_i\right)  +   2 C_0(m_i)  \biggr.
 \nn\\
&& \biggl.  \biggl.
- 4 h {\widehat \beta_1^{\prime}\over {\Delta_{SH}\widehat \beta_1}}
C\left(\frac{\Delta_{\widetilde b a}+\Delta_{SH}}{m_i},m_i\right)  +
h^2 {\widehat \beta_1^{\prime\prime}\over {\Delta_{SH}\widehat \beta_1}} \sum_{s=\pm 1}C\left(s \frac{\Delta_{\widetilde b a}+ \Delta_{SH}}{m_i},m_i\right)
\biggr]\biggr\}\,,\nn\\
\end{eqnarray}
To keep the above expression simpler we do not write
the counterterms  since their structure remains the same as before.
The similar formulae for $\widetilde B_{2,4q}\hat f_a^2$ are  \index{bag parameter of heavy meson}\index{chiral corrections!to heavy meson bag parameters}
\begin{eqnarray}\label{eqO2}
\widetilde B_{2a} \hat f_a^2 &=& \frac{24}{5}\widehat \beta_2
 \left\{
1 - { \lambda^i_{ab}\lambda^{i\dagger}_{ba} \over 2 (4\pi f)^2}  \biggl[
6 g^2  C'_1\left(\frac{\Delta_{ba}}{m_i},m_i\right) + 2 C_0(m_i)\biggr.\right.
\nn\\
&&-2h^2
 C'\left(\frac{\Delta_{\widetilde b a}+\Delta_{SH}}{m_i},m_i\right)
 \biggl. -4 h{\widehat \beta_2^\prime\over {\Delta_{SH}\widehat  \beta_2}}C\left(\frac{\Delta_{\widetilde b a}+\Delta_{SH}}{m_i},m_i\right)
 \biggr]
  - { \lambda^i_{aa}\lambda^{i\dagger}_{aa} \over 2 (4\pi f)^2} \biggl[
 2 C_0(m_i)  \biggr.
 \nn\\
&& \biggl.  \biggl.
- 4 h {\widehat \beta_2^{\prime}\over {\Delta_{SH}\widehat \beta_2}}
C\left(\frac{\Delta_{\widetilde b a}+\Delta_{SH}}{m_i},m_i\right)  +
h^2 {\widehat \beta_2^{\prime\prime}\over {\Delta_{SH}\widehat \beta_2}} \sum_{s=\pm 1}C\left(s \frac{\Delta_{\widetilde b a}+ \Delta_{SH}}{m_i},m_i\right)
\biggr]\biggr\}\,,\nn\\
\end{eqnarray}
\begin{eqnarray}\label{eqO4}
\widetilde B_{4a} \hat f_a^2 &=& \frac{24}{7}\widehat \beta_4
 \left\{
1 - { \lambda^i_{ab}\lambda^{i\dagger}_{ba} \over 2 (4\pi f)^2}  \biggl[
6 g^2  C'_1\left(\frac{\Delta_{ba}}{m_i},m_i\right) + 2 C_0(m_i)\biggr.\right.
\nn\\
&&-2h^2
 C'\left(\frac{\Delta_{\widetilde b a}+\Delta_{SH}}{m_i},m_i\right)
 \biggl. -4 h{\widehat \beta_4^\prime\over {\Delta_{SH}\widehat  \beta_4}}C\left(\frac{\Delta_{\widetilde b a}+\Delta_{SH}}{m_i},m_i\right)
 \biggr]
  - { \lambda^i_{aa}\lambda^{i\dagger}_{aa} \over 2 (4\pi f)^2} \biggl[
- 2 C_0(m_i)  \biggr.
 \nn\\
&& \biggl.  \biggl.
- 4 h {\widehat \beta_4^{\prime}\over {\Delta_{SH}\widehat \beta_4}}
C\left(\frac{\Delta_{\widetilde b a}+\Delta_{SH}}{m_i},m_i\right)  -
h^2 {\widehat \beta_4^{\prime\prime}\over {\Delta_{SH}\widehat \beta_4}} \sum_{s=\pm 1}s C\left(s \frac{\Delta_{\widetilde b a}+ \Delta_{SH}}{m_i},m_i\right)
\biggr]\biggr\}\,.\nn\\
\end{eqnarray}


\par

We now turn to the case $m_\pi \ll \Delta_{SH}$  and study the behavior of eq.~(\ref{eqO1}) around $m_\pi^2\to 0$.  We shall proceed similarly to what has been done in section~\ref{sec_dsh}, namely  we expand the integrand in $E_\pi/\Delta_{SH}$\index{$1/\Delta_{SH}$ expansion}. We see that after expanding  eq.~(\ref{eqO1}) around $m_\pi^2=0$, the leading chiral logarithms arising from the pion loops remain  unchanged even when the coupling to the scalar meson is included
in the loops. On the other hand,  as discussed in the previous subsection,
the logarithms arising from the kaon\index{meson!$K$!in chiral loops} and the $\eta$-meson\index{meson!$\eta$!in chiral loops} are competitive in size with those arising from the coupling to the heavy-light scalar meson, which is the consequence of the  smallness of $\Delta_{SH}$ .
Therefore, like for the decay constants,  the relevant chiral expansion
is the one derived in the $SU(2)_L \otimes SU(2)_R \to SU(2)_V$ theory, i.e.,\index{bag parameter of heavy meson}\index{chiral corrections!to heavy meson bag parameters}
\begin{subequations}
\begin{eqnarray}
\widetilde B_{1a} \hat f_a^2 &=& \widetilde B_{1}^{\rm Tree} \alpha^2
 \left[
1 - {3 g^2 +2 \over  (4\pi f)^2} m_\pi^2\log{m_\pi^2\over \mu^2} + c_{{\cal O}_1}(\mu)m_\pi^2\right]\,,\\
\widetilde B_{2,4a} \hat f_a^2 &=&  \widetilde B_{2,4}^{\rm Tree} \alpha^2
 \left[
1 - {3g^2(3 -Y) +3\pm 1 \over 2 (4\pi f)^2} m_\pi^2\log{m_\pi^2\over \mu^2}
+ c_{{\cal O}_{2,4}}(\mu)m_\pi^2\right] \,,
\end{eqnarray}
\end{subequations}
or by using eq.~(\ref{fB-correct}), for the bag parameters we obtain\index{chiral corrections!to heavy meson bag parameters}
\begin{subequations}
\begin{eqnarray}\label{B1-correct}
\widetilde B_{1q}  &=&\widetilde B_1^{\rm Tree}
 \left[
1 - {1-3 g^2  \over 2 (4\pi f)^2} m_\pi^2\log{m_\pi^2\over \mu^2} + c_{{B}_1}(\mu)m_\pi^2\right]\,,\\
\label{B24-correct}
 \widetilde B_{2,4q}  & = & \widetilde B_{2,4}^{\rm Tree}  \left[
 1 + {3 g^2 Y \mp 1 \over 2 (4\pi f)^2} m_\pi^2\log{m_\pi^2\over \mu^2}
 + c_{{B}_{2,4}}(\mu)m_\pi^2\right]\,,
\end{eqnarray}
\end{subequations}
which coincide with the pion loop \index{loop corrections!to heavy meson bag parameter} contributions shown in eqs.~(\ref{B1}) and (\ref{B24}), as they should.

\section{Relevance to the analysis of the lattice QCD data}
\index{chiral extrapolation!of heavy meson bag parameter}

It should be stressed that the consequence of the discussion in the previous section
is mainly important to the phenomenological approaches in which the sizable kaon\index{meson!$K$!in chiral loops} and $\eta$-meson\index{meson!$\eta$!in chiral loops} logarithmic corrections  are taken as predictions, whereas the counterterm coefficients are fixed by
matching\index{matching!to large $N_c$} to large $N_c$ expansion or some other model.
We showed that the contributions of the nearby heavy-light scalar states are competitive in size and thus they cannot
be ignored nor separated from the discussion of the kaon and/or $\eta$-meson loops.
However, the fact that the nearby scalar heavy-light mesons do not spoil the dominant pion logarithmic correction to the decay constant and the bag-parameters is most welcome from the lattice practitioners' point of view, because the formulae derived in HM$\chi$PT can still (and should) be used to guide the chiral\index{chiral extrapolation!of heavy meson bag parameter} extrapolations of the lattice\index{lattice QCD!extrapolation} results, albeit for the pion masses lighter than $\Delta_{SH}$ .

\section{Conclusion}
\index{conclusions}
\index{summary}

In this chapter we revisited the computation of the $\bbarq$ mixing\index{mixing!of heavy neutral mesons} amplitudes in the framework
of HM$\chi$PT\index{HM$\chi$PT!in heavy meson mixing}.  Besides the SM bag parameter, we also provided the expressions for the chiral
logarithmic correction to the SUSY bag parameters.
More importantly, we study the impact of the near scalar mesons to the predictions derived in HM$\chi$PT in which these
contributions were previously ignored. We showed that while the corrections due to the nearness of the scalar mesons
are competitive in size with the kaon and $\eta$ meson loop corrections, they do not alter the pion chiral logarithms.
In other words the valid (pertinent) $\chi$PT\index{$\chi$PT} expressions for the quantities discussed in this chapter are those involving pions only.
This is of major importance for the chiral\index{chiral extrapolation!of heavy meson bag parameter} extrapolations of the results obtained from the QCD simulations on
the lattice\index{lattice QCD!in heavy meson mixing}, because precisely the pion chiral logarithms provide the most important guideline in those extrapolations.
As a  side-result we verified that the chiral logarithmic corrections to the scalar meson decay\index{decay constant!of heavy meson} constant
are the same as to the pseudoscalar one, modulo replacement $g\to \widetilde g$ (c.f. eq.~(\ref{su2f})).

\chapter[Rare hadronic decays of heavy mesons]{$\bf \Delta S=2$ and $\bf \Delta S = -1$ rare hadronic decays of $\bf B_c$ mesons}

\index{R-parity violation!see{RPV}}

\index{$\Delta S=2$ transitions}
\index{$\Delta S=-1$ transitions}
\index{meson!$B_c$!decay}

Rare decays of $B$\index{meson!$B$!decay} mesons are considered to be one of the
promising areas for the discovery of new physics beyond the SM~\cite{Grossman:2003qi,Isidori:2004rd, Buras:2004sc}.
This is based on the expectation
that virtual new particles will affect these decays, in
particular in processes induced by FCNCs\index{FCNC}; such processes are suppressed in the SM since they
proceed via loop diagrams. This venue, typified by transitions
like $b\to s(d) \gamma$, $b \to s(d) l \bar l $
has been investigated intensively in recent years~\cite{Grossman:2003qi,
Isidori:2004rd, Buras:2004sc}. In particular the experimental results on
decay rates and the parameters describing CP-violation\index{CP violation} in the $B$\index{meson!$B$!decay}
meson nonleptonic two-body weak decays\index{meson!$B$!decay} such as $B \to \pi K$\index{decay!$B \to K \pi$}\index{decay!$B
\to \phi K_S$} and $B
\to \phi K_S$ have attracted a lot of attention during the last few
years~(see e.g.~\cite{Silvestrini:2005zb} and references
 therein).  In the theoretical explanation of these decay  rates and
CP\index{CP violation} violating parameters it is usually assumed that an interplay of the
SM contributions and new physics\index{new physics!contributions} occurs. Grossman et al.~\cite{Grossman:1999av} have investigated the decay mechanisms of $B \to K \pi$ decays\index{decay!$B \to K \pi$} and found that new
physics might give important contributions to the relevant
observables.
In their search for the explanation of the $B \to K \pi$ puzzle, the authors
of~\cite{Barger:2004hn} have investigated the $B \to K \pi$ decay\index{decay!$B \to K \pi$} mode
within a model with an extra flavor changing $Z'$\index{gauge boson!$Z'$} boson, making
predictions for the CP\index{CP violation} violating asymmetries in these decays . $Z'$\index{gauge boson!$Z'$}
mediated penguin\index{penguin operators} operators\index{penguin operators} have also been considered in many other
scenarios.  Contributions of SUSY models with and without
R-parity violation (RPV)\index{RPV} in the same decay  channel were also discussed
in~\cite{Arnowitt:2005qz}. The difficulty with this decay mode is that
the SM contribution is the dominant one. The use of QCD\index{QCD} in the treatment of the weak hadronic $B$\index{meson!$B$!decay} meson decays\index{decay of $B$ meson}
is not a straightforward procedure.  Numerous theoretical studies have
been attempted to obtain the most appropriate framework to describe
nonleptonic $B$\index{meson!$B$!decay} meson decays to two light meson states.  But even the
most sophisticated approaches such as QCD factorization\index{QCD factorization} (BBNS\index{BBNS} and
SCET\index{SCET})~\cite{Bauer:2000yr,Beneke:1999br,Beneke:2000ry,Beneke:2001ev,Beneke:2003zv,Keum:2003js,Bauer:2000ew,Bauer:2001ct,Bauer:2001yt,Bauer:2004tj,Bauer:2005kd,Williamson:2006hb}
still have parameters which are difficult to obtain from ``first
principles''. Consequently, searches for new physics\index{new physics!searches} in decay modes
dominated by SM contributions suffer from large uncertainties.

\par

A related, though alternative approach is the search for rare
$b$ decays\index{decay of $b$ quark} which have extremely small rates in the SM, and
their mere detection would be a sign for new physics\index{new physics!searches}. Several
years ago, Huitu, Lu, Singer and Zhang suggested
\cite{Huitu:1998vn, Huitu:1998pa} the study of $b \to s
s \bar d$ and $b \to d d \bar s$ as prototypes of the
alternative method. This proposal is based~\cite{Huitu:1998vn}
on the fact that these transitions are indeed exceedingly small
in the SM, where they occur via box\index{box diagrams} diagrams with up-type quarks
and $W$'s in the box. The matrix elements of these transitions
are approximately proportional in SM to $\lambda^7$ and
$\lambda^8$ ($\lambda$ is the sine of the Cabibbo angle),
resulting in branching ratios  of approximately $10^{-11}$
and $10^{-13}$ respectively, probably too small for detection
even at LHC\index{LHC}. Further discussions on these ($\Delta S=2$) and
($\Delta S= -1$) transitions are given in Refs.
\index{$\Delta S=2$ transitions}
\index{$\Delta S=-1$ transitions}
\cite{Grossman:1999av, Fajfer:2001ht, Chun:2003rg, Wu:2003kp}.

\par

Huitu et al. have then investigated~\cite{Buras:2004sc,Huitu:1998vn}
the $b \to s s \bar d$ transition in several models of
physics beyond SM\index{new physics!in $B_c$ decays}, namely the MSSM\index{MSSM}, the MSSM with
RPV couplings \index{RPV} and the THDM\index{THDM}.
Within a certain range of allowed parameters, the MSSM\index{MSSM} predicts
\cite{Huitu:1998vn} a branching ratio  as high as $10^{-7}$ for
$b \to s s \bar d$. On the other hand, Higgs models may
lead~\cite{Huitu:1998pa} to a branching ratio  at the $10^{-8}$
level. Most interestingly, in RPV\index{RPV} the process can proceed as a
tree diagram~\cite{Huitu:1998vn} and limits on $\lambda'_{ijk}$
RPV\index{RPV} superpotential couplings that existed at the time did not
constrain at all the $b \to s s \bar d$ transition.

\par

In Ref.~\cite{Huitu:1998vn} the hadronic decay\index{decay!$B^- \to K^- K^- \pi^+$}
$B^- \to K^- K^- \pi^+$, proceeding either directly or
through a $\overline K^{*0}$, has been selected as a convenient
signal for the $b \to s s \bar d$ transition.
A rough estimate~\cite{Huitu:1998vn} has shown that the
semi-inclusive decays\index{semi-inclusive decay}
$B^- \to K^- K^- + \text{(nonstrange)}$  \index{decay!$B^- \to K^- K^- + \text{(nonstrange)}$} \index{meson!$B$!decay}\index{decay of $B$ meson}may account for
about $1/4$ of the $b \to s s \bar d$ transitions.
The OPAL\index{OPAL} collaboration has undertaken  the search for this
decay  establishing~\cite{Abbiendi:1999st} the first upper limit
for it, subsequently constrained by both B-factories\index{B-factories}~\cite{Garmash:2003er,Abe:2002av,Aubert:2003xz} to
$\mathcal{BR}(B^- \to K^- K^- \pi^+) <$ $2.4 \times 10^{-6}$. In an experiment planned~\cite{Damet:2000ct} by ATLAS\index{ATLAS} at LHC\index{LHC}, a two orders of magnitude improvement is expected.

\par

In order to use these results for restricting the size of the $b \to s s \bar d$ and $b \to d s \bar s$ transitions one needs also an estimate
for the long distance contributions\index{long distance effects} to such ($\Delta S=2$ and $\Delta S=-1$)
\index{$\Delta S=2$ transitions}
\index{$\Delta S=-1$ transitions}
processes. A calculation~\cite{Fajfer:2000ax} of such
contributions provided by $D D$ and by $D \pi$ intermediate states
 as well as those induced by virtual $D$\index{meson!$D$} and $\pi$ mesons lead
 to a branching ratio  $\mathcal{BR}^{LD}
 (B^- \to K^- K^- \pi^+ ) = 6 \times 10^{-12}$,
 only slightly larger than the short-distance result of SM for this transition.
 Thus, this decay\index{decay!$B^- \to K^- K^- \pi^+$} and similar ones are indeed suitable for the
 search of new physics\index{new physics!searches}, the LD contribution not overshadowing the
new physics, if it leads to rates of the order of $10^{-10}$  or larger.
 A survey~\cite{Fajfer:2000ny} of various
 possible two-body ($\Delta S=2$) decays\index{decay of $B$ meson}\index{$\Delta S=2$ transitions}\index{meson!$B$!decay} of $B$ mesons to
 $V V$, $V P$, $P P$ states has singled out
 $B^- \to K^{*-} \overline K^0$ and
 $B^- \to K^- \overline K^0$ as the most likely
 ones for the detection of the presence of RPV\index{RPV} transitions
 at the $10^{-7}$ level.

\par

The $b \to d d \bar s$ transition has not been
subject of such intensive theoretical studies although experimental
information on the upper bound for the $B^- \to \pi^- \pi^- K^+$ decay \index{decay!$B^- \to \pi^- \pi^- K^+$}
rate already exists. Namely, the BaBar\index{BaBar} collaboration has reported that
${\rm BR} (B^- \to \pi^- \pi^- K^+) < 1.8\E{-6}$~\cite{Aubert:2003xz}
while the Belle\index{Belle} collaboration found ${\rm BR} (B^- \to \pi^- \pi^- K^+
)< 4.5\E{-6}$~\cite{Garmash:2003er}.  The LHC-b\index{LHC}\index{LHC-b} is expected to give
even better constraints. Therefore we shall consider possible
candidate nonleptonic decay  channels proceeding with the $b \to d d \bar s$ transition for the experimental searches.

\par

At the forthcoming LHC\index{LHC} accelerator one expects about
$5 \times 10^{10}$ $B_c$\index{meson!$B_c$} events/year, at a luminosity of
$10^{34}\mathrm{~cm}^{-2} s^{-1}$~\cite{Gouz:2002kk}.
 Even if the actual number will be a couple of orders
of magnitude lower, the search for rare decays\index{decay of $B_c$ meson} of $B_c$\index{meson!$B_c$!decay} exhibiting possible
features of physics beyond SM\index{new physics!in $B_c$ decays} will become possible.
The effect of such physics on radiative decays\index{radiative decay} of $B_c$, as
caused for example by $c$-quark decay\index{decay of $c$ quark} via the
$c \to u + \gamma$ transition has already been
investigated~\cite{Fajfer:1999dq, Aliev:1999tg}.
Here, we will address the effects of new physics\index{new physics!contributions} on rare $b$-decays\index{decay of $b$ quark} caused by
the ($\Delta S=2$ and $\Delta S=-1$)\index{$\Delta S=2$ transitions}
\index{$\Delta S=-1$ transitions}
 transitions, which we already mentioned to be very rare in SM~\cite{Huitu:1998vn, Huitu:1998pa, Grossman:1999av}. Specifically, we will calculate two-body decay modes\index{meson!$B_c$!decay} $B_c^- \to D^{*-}_s \overline K^{*0}$\index{decay!$B_c^- \to D^{*-}_s \overline K^{*0}$}, $B_c^- \to D^{*-}_s \overline K^0, D_s^- \overline K^{*0}$\index{decay!$B_c^- \to D^{*-}_s \overline K^0, D_s^- \overline K^{*0}$} and $B_c^- \to D_s^- \overline K^0$\index{decay!$B_c^- \to D_s^- \overline K^0$} as well as three body modes  $B_c^- \to D_s^- K^- \pi^+$\index{decay!$B_c^- \to D_s^- K^- \pi^+$}, $B_c^- \to D^{*-}_s K^- \pi^+$\index{decay!$B_c^- \to D^{*-}_s K^- \pi^+$}, $B_c^- \to D_s^-  D^{*-}_s D^+$\index{decay!$B_c^- \to D_s^-  D^{*-}_s D^+$}, $B_c^- \to D_s^-  D_s^- D^{*+}$\index{decay!$B_c^- \to D_s^-  D_s^- D^{*+}$}, $B_c^- \to D_s^-  D_s^- D^+$\index{decay!$B_c^- \to D_s^-  D_s^- D^+$}, $B_c^- \to D^0 \overline K^0 K^-$\index{decay!$B_c^- \to D^0 \overline K^0 K^-$} and $B_c^- \to D^{*0} \overline K^0 K^-$\index{decay!$B_c^- \to D^{*0} \overline K^0 K^-$}. We expect these modes to be most likely candidates for the experimental observation.


\section[Inclusive $\Delta S = 2$ and $\Delta S = -1$ transitions]{Inclusive $b\to s s \bar{d}$ and $b \to d d \bar{s}$ transitions in SM and beyond}
\index{new physics!in $B_c$ decays}
 \index{new physics!MSSM|see{MSSM}}
 \index{new physics!RPV|see{RPV}}
 \index{new physics!THDM|see{THDM}}
 \index{new physics!$Z'$|see{gauge boson!$Z'$}}

\index{transition!$b\to d d \bar s$}\index{transition!$b\to s s \bar d$}

\subsection{Operator basis and mixing}
\index{mixing!of operators}

The effective weak Hamiltonian \index{effective weak Hamiltonian} encompassing the $b \to dd \bar s$
\index{transition!$b\to d d \bar s$}\index{transition!$b\to s s \bar d$}process has been introduced by the authors of~\cite{Grossman:1999av}
in the case of $B \to K \pi$ decays\index{decay!$B \to K \pi$}.  Following their notation we
also include the $b \to ss \bar d$ transitions and write it as
\begin{equation}
\mathcal H_{\mathrm{eff.}} = \sum_{n=1}^5 \left[ C^s_n \mathcal O^s_n + \widetilde C^s_n \widetilde {\mathcal O}^s_n + C^d_n \mathcal O^d_n + \widetilde C^d_n \widetilde {\mathcal O}^d_n\right],
\end{equation}
where $C^q_i$ and $\widetilde C^q_i$ denote effective Wilson coefficients\index{Wilson coefficient}
multiplying the complete operator basis
 of all the four quark
operators\index{four quark operator!in rare $B$ decays}
 which can contribute to the processes $b\to d d \bar s$ (for $q=s$) and $b\to s s \bar d$ (for $q=d$). We
choose
\begin{equation}
\begin{array}{c}
  \mathcal O^s_1 = \bar d^i_L \gamma^{\mu} b^i_L \bar d^j_R \gamma_{\mu} s^j_R,\quad
  \mathcal O^s_2 = \bar d^i_L \gamma^{\mu} b^j_L \bar d^j_R \gamma_{\mu} s^i_R,\quad
  \mathcal O^s_3 = \bar d^i_L \gamma^{\mu} b^i_L \bar d^j_L \gamma_{\mu} s^j_L, \\
  \\
  \mathcal O^s_4 = \bar d^i_R b^i_L \bar d^j_L s^j_R,\quad
  \mathcal O^s_5 = \bar d^i_R b^j_L \bar d^j_L s^i_R,

\end{array}
\end{equation}
plus additional operators ${\mathcal {\widetilde O}^s_{1,2,3,4,5}}$, with
the chirality  exchanges $L\leftrightarrow R$, plus the same operators with $s$ and $d$ quark flavors interchanged. In these expressions,
the superscripts $i,j$ are $SU(3)$ color indices. All other operators

with the correct Lorentz and color structure can be related to these
by operator\index{operator identities}
 identities and Fierz rearrangements. We perform our
calculations of inclusive and exclusive decays at the scale of the $b$
quark mass ($\mu=m_b$), therefore we have to take into account the
RGE\index{RGE!running of operators} running of these operators from the interaction
scale $\Lambda$. At leading log order in the strong coupling, the
operators $\mathcal O^q_{1,2}$ mix\index{mixing!of operators} with the anomalous dimension\index{anomalous dimension} matrix\index{$\alpha_s$!corrections}
\begin{equation}
\gamma(\mathcal O^q_1\mathcal O^q_2) = \frac{\alpha_s}{2\pi}
\left(\begin{array}{cc}
-8 & 0 \\
-3 & 1
\end{array}\right).
\end{equation}
The same holds for operators $\mathcal O^q_{4,5}$ ($\gamma(\mathcal
O^q_1\mathcal O^q_2)=\gamma(\mathcal O^q_4\mathcal O^q_5)$) due to Fierz
identities, while the operator $\mathcal O^q_3$ has anomalous dimension\index{anomalous dimension}
$\gamma(\mathcal O^q_3) = \alpha_s/\pi$\index{$\alpha_s$!corrections}. Anomalous dimension\index{anomalous dimension} matrices for chirally
flipped operators $\mathcal{\widetilde O}^q_{1,2,3,4,5}$ are identical to
these.

\subsection{SM}

\begin{figure}[!t]
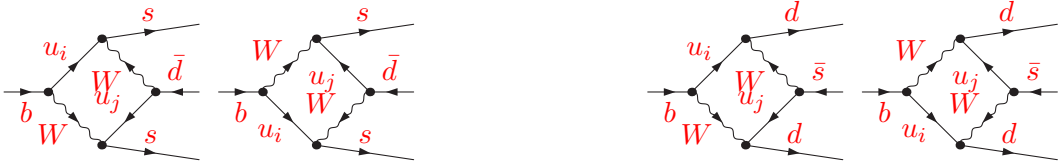

\begin{center}
\unitlength=1bp%

\begin{feynartspicture}(400,80)(5,1)

\FADiagram{}
\FAProp(0.,10.)(4.5,10.)(0.,){/Straight}{1}
\FALabel(2.25,8.93)[t]{$\Red b$}
\FAProp(20.,17.)(10.,15.5)(0.,){/Straight}{-1}
\FALabel(14.7701,17.3029)[b]{$\Red s$}
\FAProp(20.,3.)(10.,4.5)(0.,){/Straight}{-1}
\FALabel(15.0371,4.76723)[b]{$\Red s$}
\FAProp(20.,10.)(15.5,10.)(0.,){/Straight}{1}
\FALabel(17.5,11.27)[b]{$\Red {\bar d}$}
\FAProp(4.5,10.)(10.,15.5)(0.,){/Straight}{1}
\FALabel(6.63398,13.366)[br]{$\Red {u_i}$}
\FAProp(4.5,10.)(10.,4.5)(0.,){/Sine}{1}
\FALabel(6.63398,6.63398)[tr]{$\Red W$}
\FAProp(10.,15.5)(15.5,10.)(0.,){/Sine}{-1}
\FALabel(12.134,12.134)[tr]{$\Red W$}
\FAProp(10.,4.5)(15.5,10.)(0.,){/Straight}{-1}
\FALabel(12.134,7.86602)[br]{$\Red {u_j}$}
\FAVert(4.5,10.){0}
\FAVert(10.,15.5){0}
\FAVert(10.,4.5){0}
\FAVert(15.5,10.){0}

\FADiagram{}
\FAProp(0.,10.)(4.5,10.)(0.,){/Straight}{1}
\FALabel(2.25,8.93)[t]{$\Red b$}
\FAProp(20.,17.)(10.,15.5)(0.,){/Straight}{-1}
\FALabel(14.7701,17.3029)[b]{$\Red s$}
\FAProp(20.,3.)(10.,4.5)(0.,){/Straight}{-1}
\FALabel(15.0371,4.76723)[b]{$\Red s$}
\FAProp(20.,10.)(15.5,10.)(0.,){/Straight}{1}
\FALabel(17.5,11.27)[b]{$\Red {\bar d}$}
\FAProp(4.5,10.)(10.,15.5)(0.,){/Sine}{1}
\FALabel(6.63398,13.366)[br]{$\Red W$}
\FAProp(4.5,10.)(10.,4.5)(0.,){/Straight}{1}
\FALabel(6.63398,6.63398)[tr]{$\Red {u_i}$}
\FAProp(10.,15.5)(15.5,10.)(0.,){/Straight}{-1}
\FALabel(12.134,12.134)[tr]{$\Red {u_j}$}
\FAProp(10.,4.5)(15.5,10.)(0.,){/Sine}{-1}
\FALabel(12.134,7.86602)[br]{$\Red W$}
\FAVert(4.5,10.){0}
\FAVert(10.,15.5){0}
\FAVert(10.,4.5){0}
\FAVert(15.5,10.){0}

\FADiagram{}

\FADiagram{}
\FAProp(0.,10.)(4.5,10.)(0.,){/Straight}{1}
\FALabel(2.25,8.93)[t]{$\Red b$}
\FAProp(20.,17.)(10.,15.5)(0.,){/Straight}{-1}
\FALabel(14.7701,17.3029)[b]{$\Red d$}
\FAProp(20.,3.)(10.,4.5)(0.,){/Straight}{-1}
\FALabel(15.0371,4.76723)[b]{$\Red d$}
\FAProp(20.,10.)(15.5,10.)(0.,){/Straight}{1}
\FALabel(17.5,11.27)[b]{$\Red {\bar s}$}
\FAProp(4.5,10.)(10.,15.5)(0.,){/Straight}{1}
\FALabel(6.63398,13.366)[br]{$\Red{u_i}$}
\FAProp(4.5,10.)(10.,4.5)(0.,){/Sine}{1}
\FALabel(6.63398,6.63398)[tr]{$\Red W$}
\FAProp(10.,15.5)(15.5,10.)(0.,){/Sine}{-1}
\FALabel(12.134,12.134)[tr]{$\Red W$}
\FAProp(10.,4.5)(15.5,10.)(0.,){/Straight}{-1}
\FALabel(12.134,7.86602)[br]{$\Red{ u_j} $}
\FAVert(4.5,10.){0}
\FAVert(10.,15.5){0}
\FAVert(10.,4.5){0}
\FAVert(15.5,10.){0}

\FADiagram{}
\FAProp(0.,10.)(4.5,10.)(0.,){/Straight}{1}
\FALabel(2.25,8.93)[t]{$\Red b$}
\FAProp(20.,17.)(10.,15.5)(0.,){/Straight}{-1}
\FALabel(14.7701,17.3029)[b]{$\Red d$}
\FAProp(20.,3.)(10.,4.5)(0.,){/Straight}{-1}
\FALabel(15.0371,4.76723)[b]{$\Red d$}
\FAProp(20.,10.)(15.5,10.)(0.,){/Straight}{1}
\FALabel(17.5,11.27)[b]{$\Red {\bar s}$}
\FAProp(4.5,10.)(10.,15.5)(0.,){/Sine}{1}
\FALabel(6.63398,13.366)[br]{$\Red W$}
\FAProp(4.5,10.)(10.,4.5)(0.,){/Straight}{1}
\FALabel(6.63398,6.63398)[tr]{$\Red {u_i}$}
\FAProp(10.,15.5)(15.5,10.)(0.,){/Straight}{-1}
\FALabel(12.134,12.134)[tr]{$\Red {u_j}$}
\FAProp(10.,4.5)(15.5,10.)(0.,){/Sine}{-1}
\FALabel(12.134,7.86602)[br]{$\Red W$}
\FAVert(4.5,10.){0}
\FAVert(10.,15.5){0}
\FAVert(10.,4.5){0}
\FAVert(15.5,10.){0}

\end{feynartspicture}
\caption{\small\it\label{fig_bssd_SM}{Dominant contributions to the $b \to s s \bar d$ (left) and $b \to d d \bar s$ (right) transitions in the SM . Straight lines denote quarks while wavy lines represent $W$ bosons. Filled dots stand for weak vertex insertion.}}
\end{center}
\end{figure}

Within the SM only the operators
 $\mathcal O^{s(d)}_3$ contribute
to the $b \to dd \bar s$ ($b \to ss \bar d$)\index{transition!$b\to d d \bar s$}\index{transition!$b\to s s \bar d$} transitions at one loop. The dominating contributions to the Wilson
coefficients\index{Wilson coefficient} come from the up quark and $W$ boson box\index{box diagrams} diagrams in fig.~\ref{fig_bssd_SM}.
The top and charm quark loop\index{loop contributions} contributions dominate and lead to \index{CKM!matrix elements!$V_{cd}$}\index{CKM!matrix elements!$V_{cs}$}\index{G$_F$}
\begin{subequations}
\begin{eqnarray}
    C^{d,SM}_3 &=& \frac{G_F^2}{4\pi^2} m_W^2 V_{tb} V^*_{ts} \Bigg[ V_{td} V_{ts}^* f\left( \frac{m_W^2}{m_t^2} \right)  + V_{cd} V_{cs}^* \frac{m_c^2}{m_W^2} g\left( \frac{m_W^2}{m_t^2},\frac{m_c^2}{m_W^2} \right) \Bigg], \\*
   C_3^{s,SM} &=& \frac{G_F^2}{4 \pi^2} m_W^2 V_{tb} V_{td}^* \Bigg[ V_{ts} V_{td}^*
  f\left(\frac{m_W^2}{m_t^2}\right) + V_{cs} V_{cd}^* \frac{m_c^2}{m_W^2} g\left(\frac{m_W^2}{m_t^2},\frac{m_c^2}{m_W^2}\right)\Bigg],
\end{eqnarray}
\end{subequations}
where
\begin{subequations}
\begin{eqnarray}
f(x)&=&\frac{1-11x+4x^2}{4x(1-x)^2}
-\frac{3}{2(1-x)^3}{\rm ln}x,\\
~g(x,y)&=& \frac{4x-1}{4(1-x)} +\frac{8x-4x^2-1}{4(1-x)^2}\ln x -\ln y .
\end{eqnarray}
\end{subequations}
Using numerical values of the relevant
CKM\index{CKM!matrix elements} matrix elements from
PDG~\cite{Eidelman:2004wy} and possibly including the CKM\index{CKM!phase} phase in $V_{td}$ one always finds $\left|C_3^{s,SM}\right| \leq 3\E{-13}\e{GeV}^{-2}$ and $\left|C^{d,SM}_{3}\right|\leq 4\times10^{-12} \mathrm{~GeV}^{-2}$.
Renormalization group running from the weak interaction scale to the bottom quark mass
scale, due to anomalous dimension\index{anomalous dimension} of the operator
 $\mathcal O_3$,
induces only a small correction factor of $0.8$ which can be safely neglected.
The inclusive \index{transition!$b\to d d \bar s$}\index{transition!$b\to s s \bar d$}$b\to d d \bar s$ amd $b\to ss \bar d$ decay widths within SM  can then be calculated. By accounting for the colors of final state quarks in combination with the crossing symmetry\index{crossing symmetry} one obtains
\begin{equation}
  \Gamma^{q,SM}_{\mathrm{inc.}} = \frac{\left|C_3^{q,SM}\right|^2 m_b^5}{48 (2\pi)^3},
\label{eq_Gamma_SM}
\end{equation}
which leads to the branching ratios  of the order $10^{-12}$ (for $b\to ss \bar d$) to $10^{-14}$ (for $b\to dd \bar s$).

\subsection{Beyond SM}
\index{new physics!in $B_c$ decays}

Next we discuss contributions of several models containing physics
beyond the SM : the MSSM\index{MSSM} with and without RPV\index{RPV} and models with an extra
$Z'$\index{gauge boson!$Z'$} boson. For the THDM\index{THDM} on the other hand, the charged Higgs\index{charged Higgs boson}\index{Higgs boson!charged} box\index{box diagrams} diagram contributions were found to be negligible in the \index{transition!$b\to d d \bar s$}\index{transition!$b\to s s \bar d$}$b\to s s \bar d$ case~\cite{Huitu:1998pa}. Due to higher CKM\index{CKM!suppression} suppression, the argument holds also for the $b\to dd \bar s$ case. In addition, the tree level neutral Higgs exchange amplitude\index{amplitude!in $\Delta S=2$ and $\Delta S=-1$ decays} for $b\to dd \bar s$ is proportional to
$|\xi_{db}\xi_{ds}|/m_H^2$, where $\xi_{db}$ and $\xi_{ds}$ are flavor
changing Yukawa couplings and $m_H$ is a common Higgs mass scale. This
ratio is constrained from the neutral meson mixing~\cite{Huitu:1998pa}\index{mixing!of neutral mesons}. Using presently known values of $\Delta m_K$ and $\Delta m_B$~\cite{Eidelman:2004wy} one can obtain an upper bound of $|\xi_{db}\xi_{ds}|/m_H^2<10^{-13}~\mathrm{GeV}^{-2}$
rendering this contribution negligible. Similarly for the  $b\to s s \bar d$ process the relevant effective THDM\index{THDM} coupling $|\xi_{sb}\xi_{sd}|/m_H^2$ is bounded from above by the upper limit on $\Delta m_{B_s}$\index{meson!$B_s$!oscillations}. With the recent two-sided bound on the $B_s$ oscillation\index{oscillations!$B^0_s-\overline B^0_s$}\index{meson!$B_s$!oscillations} frequency from the D0 and CDF collaborations~\cite{Abulencia:2006mq,Abazov:2006dm} it is now possible to constrain this contribution to $|\xi_{sb}\xi_{sd}|/m_H^2<10^{-12}$. This value is two orders of magnitude smaller than the one used in existing studies. Correspondingly, all the decay  rate predictions for the THDM\index{THDM} are diminished by four orders of magnitude and thus rendered negligible.

\subsubsection{MSSM}

\begin{figure}[!t]
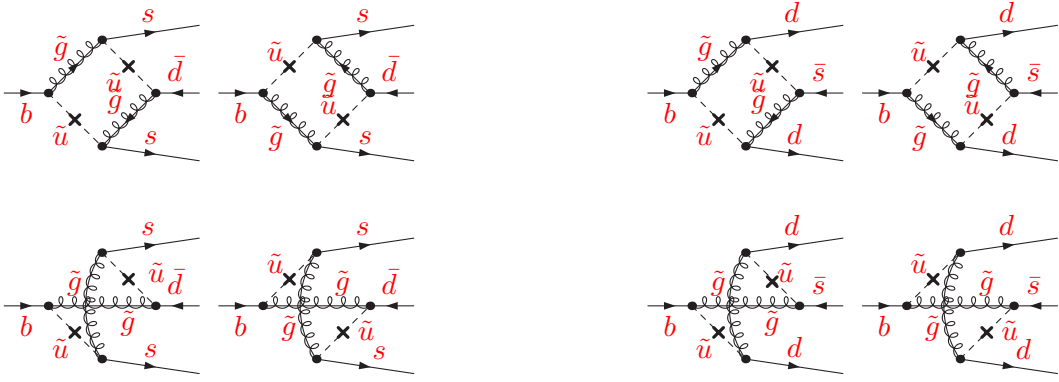

\begin{center}
\unitlength=1bp%

\begin{feynartspicture}(400,160)(5,2)

\FADiagram{}
\FAProp(0.,10.)(4.5,10.)(0.,){/Straight}{1}
\FALabel(2.25,8.93)[t]{$\Red b$}
\FAProp(20.,17.)(10.,15.5)(0.,){/Straight}{-1}
\FALabel(14.7701,17.3029)[b]{$\Red s$}
\FAProp(20.,3.)(10.,4.5)(0.,){/Straight}{-1}
\FALabel(15.0371,4.76723)[b]{$\Red s$}
\FAProp(20.,10.)(15.5,10.)(0.,){/Straight}{1}
\FALabel(17.5,11.27)[b]{$\Red {\bar d}$}
\FAProp(4.5,10.)(10.,15.5)(0.,){/Straight}{1}
\FAProp(4.5,10.)(10.,15.5)(0.,){/Cycles}{0}
\FALabel(6.63398,13.366)[br]{$\Red {\tilde g}$}
\FAProp(4.5,10.)(10.,4.5)(0.,){/ScalarDash}{0}
\FAVert(7.25,7.25){1}
\FALabel(6.63398,6.63398)[tr]{$\Red {\tilde u}$}
\FAProp(10.,15.5)(15.5,10.)(0.,){/ScalarDash}{0}
\FAVert(12.75,12.75){1}
\FALabel(12.134,12.134)[tr]{$\Red {\tilde u}$}
\FAProp(10.,4.5)(15.5,10.)(0.,){/Straight}{-1}
\FAProp(10.,4.5)(15.5,10.)(0.,){/Cycles}{0}
\FALabel(12.134,7.86602)[br]{$\Red {\tilde g}$}
\FAVert(4.5,10.){0}
\FAVert(10.,15.5){0}
\FAVert(10.,4.5){0}
\FAVert(15.5,10.){0}

\FADiagram{}
\FAProp(0.,10.)(4.5,10.)(0.,){/Straight}{1}
\FALabel(2.25,8.93)[t]{$\Red b$}
\FAProp(20.,17.)(10.,15.5)(0.,){/Straight}{-1}
\FALabel(14.7701,17.3029)[b]{$\Red s$}
\FAProp(20.,3.)(10.,4.5)(0.,){/Straight}{-1}
\FALabel(15.0371,4.76723)[b]{$\Red s$}
\FAProp(20.,10.)(15.5,10.)(0.,){/Straight}{1}
\FALabel(17.5,11.27)[b]{$\Red {\bar d}$}
\FAProp(4.5,10.)(10.,15.5)(0.,){/ScalarDash}{0}
\FAVert(7.25,12.75){1}
\FALabel(6.63398,13.366)[br]{$\Red {\tilde u}$}
\FAProp(4.5,10.)(10.,4.5)(0.,){/Straight}{1}
\FAProp(4.5,10.)(10.,4.5)(0.,){/Cycles}{0}
\FALabel(6.63398,6.63398)[tr]{$\Red {\tilde g}$}
\FAProp(10.,15.5)(15.5,10.)(0.,){/Straight}{-1}
\FAProp(10.,15.5)(15.5,10.)(0.,){/Cycles}{0}
\FALabel(12.134,12.134)[tr]{$\Red {\tilde g}$}
\FAProp(10.,4.5)(15.5,10.)(0.,){/ScalarDash}{0}
\FAVert(12.75,7.25){1}
\FALabel(12.134,7.86602)[br]{$\Red {\tilde u}$}
\FAVert(4.5,10.){0}
\FAVert(10.,15.5){0}
\FAVert(10.,4.5){0}
\FAVert(15.5,10.){0}

\FADiagram{}

\FADiagram{}
\FAProp(0.,10.)(4.5,10.)(0.,){/Straight}{1}
\FALabel(2.25,8.93)[t]{$\Red b$}
\FAProp(20.,17.)(10.,15.5)(0.,){/Straight}{-1}
\FALabel(14.7701,17.3029)[b]{$\Red d$}
\FAProp(20.,3.)(10.,4.5)(0.,){/Straight}{-1}
\FALabel(15.0371,4.76723)[b]{$\Red d$}
\FAProp(20.,10.)(15.5,10.)(0.,){/Straight}{1}
\FALabel(17.5,11.27)[b]{$\Red {\bar s}$}
\FAProp(4.5,10.)(10.,15.5)(0.,){/Straight}{1}
\FAProp(4.5,10.)(10.,15.5)(0.,){/Cycles}{0}
\FALabel(6.63398,13.366)[br]{$\Red{\tilde g}$}
\FAProp(4.5,10.)(10.,4.5)(0.,){/ScalarDash}{0}
\FAVert(7.25,7.25){1}
\FALabel(6.63398,6.63398)[tr]{$\Red {\tilde u}$}
\FAProp(10.,15.5)(15.5,10.)(0.,){/ScalarDash}{0}
\FAVert(12.75,12.75){1}
\FALabel(12.134,12.134)[tr]{$\Red {\tilde u}$}
\FAProp(10.,4.5)(15.5,10.)(0.,){/Straight}{-1}
\FAProp(10.,4.5)(15.5,10.)(0.,){/Cycles}{0}
\FALabel(12.134,7.86602)[br]{$\Red{ \tilde g} $}
\FAVert(4.5,10.){0}
\FAVert(10.,15.5){0}
\FAVert(10.,4.5){0}
\FAVert(15.5,10.){0}

\FADiagram{}
\FAProp(0.,10.)(4.5,10.)(0.,){/Straight}{1}
\FALabel(2.25,8.93)[t]{$\Red b$}
\FAProp(20.,17.)(10.,15.5)(0.,){/Straight}{-1}
\FALabel(14.7701,17.3029)[b]{$\Red d$}
\FAProp(20.,3.)(10.,4.5)(0.,){/Straight}{-1}
\FALabel(15.0371,4.76723)[b]{$\Red d$}
\FAProp(20.,10.)(15.5,10.)(0.,){/Straight}{1}
\FALabel(17.5,11.27)[b]{$\Red {\bar s}$}
\FAProp(4.5,10.)(10.,15.5)(0.,){/ScalarDash}{0}
\FAVert(7.25,12.75){1}
\FALabel(6.63398,13.366)[br]{$\Red {\tilde u}$}
\FAProp(4.5,10.)(10.,4.5)(0.,){/Straight}{1}
\FAProp(4.5,10.)(10.,4.5)(0.,){/Cycles}{0}
\FALabel(6.63398,6.63398)[tr]{$\Red {\tilde g}$}
\FAProp(10.,15.5)(15.5,10.)(0.,){/Straight}{-1}
\FAProp(10.,15.5)(15.5,10.)(0.,){/Cycles}{0}
\FALabel(12.134,12.134)[tr]{$\Red {\tilde g}$}
\FAProp(10.,4.5)(15.5,10.)(0.,){/ScalarDash}{0}
\FAVert(12.75,7.25){1}
\FALabel(12.134,7.86602)[br]{$\Red {\tilde u}$}
\FAVert(4.5,10.){0}
\FAVert(10.,15.5){0}
\FAVert(10.,4.5){0}
\FAVert(15.5,10.){0}

\FADiagram{}
\FAProp(0.,10.)(4.5,10.)(0.,){/Straight}{1}
\FALabel(2.25,8.93)[t]{$\Red b$}
\FAProp(20.,17.)(10.,15.5)(0.,){/Straight}{-1}
\FALabel(14.7701,17.3029)[b]{$\Red s$}
\FAProp(20.,3.)(10.,4.5)(0.,){/Straight}{-1}
\FALabel(15.0371,4.76723)[b]{$\Red s$}
\FAProp(20.,10.)(15.5,10.)(0.,){/Straight}{1}
\FALabel(17.5,11.27)[b]{$\Red {\bar d}$}
\FAProp(4.5,10.)(15.5,10.)(0.,){/Straight}{0}
\FAProp(4.5,10.)(15.5,10.)(0.,){/Cycles}{0}
\FALabel(8,11)[br]{$\Red {\tilde g}$}
\FAProp(4.5,10.)(10.,4.5)(0.,){/ScalarDash}{0}
\FAVert(7.25,7.25){1}
\FALabel(6.63398,6.63398)[tr]{$\Red {\tilde u}$}
\FAProp(10.,15.5)(15.5,10.)(0.,){/ScalarDash}{0}
\FAVert(12.75,12.75){1}
\FALabel(16.5,14.5)[tr]{$\Red {\tilde u}$}
\FAProp(10.,4.5)(10.,15.5)(-0.3,){/Straight}{0}
\FAProp(10.,4.5)(10.,15.5)(-0.3,){/Cycles}{0}
\FALabel(13.5,9.5)[tr]{$\Red {\tilde g}$}
\FAVert(4.5,10.){0}
\FAVert(10.,15.5){0}
\FAVert(10.,4.5){0}
\FAVert(15.5,10.){0}

\FADiagram{}
\FAProp(0.,10.)(4.5,10.)(0.,){/Straight}{1}
\FALabel(2.25,8.93)[t]{$\Red b$}
\FAProp(20.,17.)(10.,15.5)(0.,){/Straight}{-1}
\FALabel(14.7701,17.3029)[b]{$\Red s$}
\FAProp(20.,3.)(10.,4.5)(0.,){/Straight}{-1}
\FALabel(16.5,4.5)[b]{$\Red s$}
\FAProp(20.,10.)(15.5,10.)(0.,){/Straight}{1}
\FALabel(17.5,11.27)[b]{$\Red {\bar d}$}
\FAProp(4.5,10.)(10.,15.5)(0.,){/ScalarDash}{0}
\FAVert(7.25,12.75){1}
\FALabel(6.63398,13.366)[br]{$\Red {\tilde u}$}
\FAProp(4.5,10.)(15.5,10.)(0.,){/Straight}{0}
\FAProp(4.5,10.)(15.5,10.)(0.,){/Cycles}{0}
\FALabel(8,9)[tr]{$\Red {\tilde g}$}
\FAProp(10.,15.5)(10.,4.5)(0.3,){/Straight}{0}
\FAProp(10.,15.5)(10.,4.5)(0.3,){/Cycles}{0}
\FALabel(13.5,11)[br]{$\Red {\tilde g}$}
\FAProp(10.,4.5)(15.5,10.)(0.,){/ScalarDash}{0}
\FAVert(12.75,7.25){1}
\FALabel(16,6.5)[br]{$\Red {\tilde u}$}
\FAVert(4.5,10.){0}
\FAVert(10.,15.5){0}
\FAVert(10.,4.5){0}
\FAVert(15.5,10.){0}

\FADiagram{}

\FADiagram{}
\FAProp(0.,10.)(4.5,10.)(0.,){/Straight}{1}
\FALabel(2.25,8.93)[t]{$\Red b$}
\FAProp(20.,17.)(10.,15.5)(0.,){/Straight}{-1}
\FALabel(14.7701,17.3029)[b]{$\Red d$}
\FAProp(20.,3.)(10.,4.5)(0.,){/Straight}{-1}
\FALabel(15.0371,4.76723)[b]{$\Red d$}
\FAProp(20.,10.)(15.5,10.)(0.,){/Straight}{1}
\FALabel(17.5,11.27)[b]{$\Red {\bar s}$}
\FAProp(4.5,10.)(15.5,10.)(0.,){/Straight}{0}
\FAProp(4.5,10.)(15.5,10.)(0.,){/Cycles}{0}
\FALabel(8,11)[br]{$\Red {\tilde g}$}
\FAProp(4.5,10.)(10.,4.5)(0.,){/ScalarDash}{0}
\FAVert(7.25,7.25){1}
\FALabel(6.63398,6.63398)[tr]{$\Red {\tilde u}$}
\FAProp(10.,15.5)(15.5,10.)(0.,){/ScalarDash}{0}
\FAVert(12.75,12.5){1}
\FALabel(15,14.5)[tr]{$\Red {\tilde u}$}
\FAProp(10.,4.5)(10.,15.5)(-0.3,){/Straight}{0}
\FAProp(10.,4.5)(10.,15.5)(-0.3,){/Cycles}{0}
\FALabel(13.5,9.5)[tr]{$\Red {\tilde g}$}
\FAVert(4.5,10.){0}
\FAVert(10.,15.5){0}
\FAVert(10.,4.5){0}
\FAVert(15.5,10.){0}

\FADiagram{}
\FAProp(0.,10.)(4.5,10.)(0.,){/Straight}{1}
\FALabel(2.25,8.93)[t]{$\Red b$}
\FAProp(20.,17.)(10.,15.5)(0.,){/Straight}{-1}
\FALabel(14.7701,17.3029)[b]{$\Red d$}
\FAProp(20.,3.)(10.,4.5)(0.,){/Straight}{-1}
\FALabel(16.5,4.5)[b]{$\Red d$}
\FAProp(20.,10.)(15.5,10.)(0.,){/Straight}{1}
\FALabel(17.5,11.27)[b]{$\Red {\bar s}$}
\FAProp(4.5,10.)(10.,15.5)(0.,){/ScalarDash}{0}
\FAVert(7.25,12.75){1}
\FALabel(6.63398,13.366)[br]{$\Red {\tilde u}$}
\FAProp(4.5,10.)(15.5,10.)(0.,){/Straight}{0}
\FAProp(4.5,10.)(15.5,10.)(0.,){/Cycles}{0}
\FALabel(8,9)[tr]{$\Red {\tilde g}$}
\FAProp(10.,15.5)(10.,4.5)(0.3,){/Straight}{0}
\FAProp(10.,15.5)(10.,4.5)(0.3,){/Cycles}{0}
\FALabel(13.5,11)[br]{$\Red {\tilde g}$}
\FAProp(10.,4.5)(15.5,10.)(0.,){/ScalarDash}{0}
\FAVert(12.75,7.25){1}
\FALabel(16,6.5)[br]{$\Red {\tilde u}$}
\FAVert(4.5,10.){0}
\FAVert(10.,15.5){0}
\FAVert(10.,4.5){0}
\FAVert(15.5,10.){0}

\end{feynartspicture}

\caption{\small\it\label{fig_b_ssd_MSSM}{Dominant contributions to the $b \to s s \bar d$ (left) and $b \to d d \bar s$ (right) transitions in the MSSM. Dashed lines denote squarks\index{squark} while curly-straight lines represent gluinos. Filled dots stand for strong vertex insertion, while crosses denote off-diagonal squark\index{squark} mass insertions.}}
\end{center}
\end{figure}

In the MSSM\index{MSSM}, like in the SM , the main contributions come from the
$\mathcal O^q_3$ operators
, while the corresponding Wilson coefficients\index{Wilson coefficient} are
here\index{$\alpha_s$!corrections}
\begin{subequations}
\begin{eqnarray}
C_3^{s,MSSM} &=& -\frac{\alpha_s^2 \delta_{21}^* \delta_{13}}{216
    m_{\widetilde{d}}^2} \left[24 x f_6(x) + 66 \widetilde{f}_6 (x)\right],\\
C_3^{d,MSSM} &=& -\frac{\alpha_s^2 \delta^{*}_{12}
  \delta_{23}}{216m_{\widetilde
  d}^2}\left[ 24x f_6 (x) + 66 \widetilde f_6 (x) \right],
\label{eq:Cmssm}
\end{eqnarray}
\end{subequations}
as found in analyses~\cite{Gabbiani:1996hi} taking into account only
contributions from the left-handed squarks\index{squark} in the loop\index{loop contributions}\index{box diagrams} (see fig.~\ref{fig_b_ssd_MSSM}).
Recently it has been also verified~\cite{Wu:2003kp} that the chargino contribution in MSSM\index{MSSM} to this process is indeed smaller by an order of magnitude than contributions calculated in~\cite{Huitu:1998vn}. The functions $f_6(x)$ and $\widetilde f_6 (x)$ read~\cite{Gabbiani:1996hi}
\begin{subequations}
\begin{eqnarray}
f_6(x)=\frac{6(1+3x)\ln x +x^3-9x^2-9x+17}{6(x-1)^5}\; ,  \\
\widetilde{f}_6(x)=\frac{6x(1+x)\ln x -x^3-9x^2+9x+1}{3(x-1)^5}\;,
\end{eqnarray}
\end{subequations}
with $x=m_{\widetilde g}^2/m_{\widetilde d}^2$. We take $\alpha_s(m_W)\simeq 0.12$~\cite{Eidelman:2004wy}\index{$\alpha_s$!value}, while couplings $\delta^d_{ij}$ parametrize the mixing\index{mixing!of quark flavors} between the down-type left-handed squarks\index{squark}. The value of $\delta_{12}$ is determined from the $K^0-\overline K^0$ mixing~\cite{Gabbiani:1996hi}\index{mixing!of neutral mesons} and is currently bounded by $m_{K_L}-m_{K_S}=3.49\times10^{-15}\mathrm{~GeV}$~\cite{Eidelman:2004wy}. We follow
ref.~\cite{Ciuchini:2005kp} and take $x=m_{\widetilde g^2}/m_{\widetilde
  d^2}=1$; using results from~\cite{Gabbiani:1996hi} we estimate the absolute value of $\delta_{12}$ to be below $3\times 10^{-2}$ at average squark\index{squark} mass $m_{\widetilde d}=350\mathrm{~GeV}$. The strongest bounds on $\delta_{23}$ come from the radiative $b\to s\gamma$ decay\index{decay!$b\to s\gamma$} ~\cite{Gabbiani:1996hi, Ciuchini:2003sq,Ciuchini:2003rg}. These studies give at $x=1$ and for $m_{\widetilde d}=350\mathrm{~GeV}$ the stronger bound on $|\delta_{23}(x=1)|\lesssim 0.4$ which results in $|C_3^{d,MSSM}| \lesssim 5 \times 10^{-12} \mathrm{~GeV}^{-2}$. This updated value for $C_3^{d,MSSM}$ is somewhat smaller than those used in~\cite{Huitu:1998vn,Fajfer:2000ny}. Similarly, the recent limits on $ \delta_{21}^* \delta_{13}$~\cite{Ciuchini:2005kp,Khalil:2005qg,Ciuchini:2006dx}
disallow significant contributions from the mixed\index{mixing!of squarks}\index{squark} and the right-handed
squark\index{squark} mass insertion terms. Therefore, we only include the dominant
contributions given in the above expression.  We follow
ref.~\cite{Ciuchini:2005kp} and take $x=m_{\widetilde g^2}/m_{\widetilde
  d^2}=1$ and the corresponding values of
$\left|\delta_{13}(x=1)\right| \leq 0.14$ and
$\left|\delta_{21}(x=1)\right| \leq 0.042$~\cite{Gabbiani:1996hi}. We
take for the average mass of squarks\index{squark} $m_{\widetilde d} = 500\e{GeV}$, and find
$\left|C_3^{s,MSSM}\right| \leq 2 \E{-12}\e{GeV}^{-2}$.  Using
expression~(\ref{eq_Gamma_SM}) and substituting for the correct Wilson
coefficient\index{Wilson coefficient} one finds the MSSM\index{MSSM} prediction for the inclusive $b\to d d
\bar s$ and $b\to ss
\bar d$ decay  branching ratios of the order of $10^{-12}$.
\index{transition!$b\to d d \bar s$}\index{transition!$b\to s s \bar d$}

\subsubsection{RPV\label{sec:suppressedO5}}
\index{RPV}

If RPV\index{RPV} interactions are included in the MSSM\index{MSSM}, the part of the
superpotential which becomes relevant here is $W =
\lambda^{\prime}_{ijk} L_i Q_j d_k$, where $i,j,$ and $k$ are family
indices, and $L$, $Q$ and $d$ are superfields for the lepton doublet,
the quark doublet, and the down-type quark singlet, respectively.
\begin{figure}[!t]
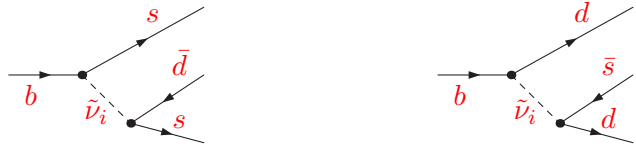

\begin{center}
\unitlength=1bp%

\begin{feynartspicture}(240,80)(3,1)

\FADiagram{}
\FAProp(0.,10.)(7.5,10.)(0.,){/Straight}{1}
\FALabel(2.25,8.93)[t]{$\Red b$}
\FAVert(7.5,10.){0}
\FAProp(20.,17.)(7.5,10.)(0.,){/Straight}{-1}
\FALabel(14.7701,15.5)[b]{$\Red s$}
\FAProp(20.,3.)(12.5,5.)(0.,){/Straight}{-1}
\FALabel(17.5,4.5)[b]{$\Red s$}
\FAProp(20.,10.)(12.5,5.)(0.,){/Straight}{1}
\FALabel(17.5,10.)[b]{$\Red {\bar d}$}
\FAVert(12.5,5.){0}
\FAProp(7.5,10.)(12.5,5.)(0.,){/ScalarDash}{0}
\FALabel(9.,5.)[b]{$\Red {\tilde \nu_i}$}

\FADiagram{}

\FADiagram{}
\FAProp(0.,10.)(7.5,10.)(0.,){/Straight}{1}
\FALabel(2.25,8.93)[t]{$\Red b$}
\FAVert(7.5,10.){0}
\FAProp(20.,17.)(7.5,10.)(0.,){/Straight}{-1}
\FALabel(14.7701,15.5)[b]{$\Red d$}
\FAProp(20.,3.)(12.5,5.)(0.,){/Straight}{-1}
\FALabel(17.5,4.5)[b]{$\Red d$}
\FAProp(20.,10.)(12.5,5.)(0.,){/Straight}{1}
\FALabel(17.5,10.)[b]{$\Red {\bar s}$}
\FAVert(12.5,5.){0}
\FAProp(7.5,10.)(12.5,5.)(0.,){/ScalarDash}{0}
\FALabel(9.,5.)[b]{$\Red {\tilde \nu_i}$}

\end{feynartspicture}

\caption{\small\it\label{fig_bssdRPV}{Dominant contributions to the $b \to s s \bar d$ (left) and $b \to d d \bar s$ (right) transitions in the RPV\index{RPV} model. Dashed lines denote sneutrinos\index{sneutrino}, while filled dots stand for RPV vertex insertions.}}
\end{center}
\end{figure}
The tree level effective Hamiltonian \index{effective weak Hamiltonian} due to sneutrino\index{sneutrino} exchanges in fig.~\ref{fig_bssdRPV} receives contributions from the
operators
 $\mathcal O^q_{4}$ and $\mathcal{\widetilde O}^q_{4}$ with the
Wilson coefficients\index{Wilson coefficient} defined at the interaction scale $\Lambda \sim
m_{\widetilde \nu}$
\begin{eqnarray}
  C_4^{s,RPV} = -\sum_{n=1}^3 \frac{\lambda_{n31}'\lambda_{n12}'^*}{m_{\widetilde{\nu}_n}^2}, &&
  \widetilde C_4^{s,RPV} = -\sum_{n=1}^3 \frac{\lambda_{n21}'\lambda_{n13}'^*}{m_{\widetilde{\nu}_n}^2},\nonumber\\
  C_4^{d,RPV} = -\sum_{n=1}^3 \frac{\lambda_{n32}'\lambda_{n21}'^*}{m_{\widetilde{\nu}_n}^2}, &&
  \widetilde C_4^{d,RPV} = -\sum_{n=1}^3 \frac{\lambda_{n12}'\lambda_{n23}'^*}{m_{\widetilde{\nu}_n}^2}.\nonumber\\
  \label{eq:RPVcouplings}
\end{eqnarray}
The QCD\index{QCD corrections} corrections were found to be important for this
transition~\cite{Bagger:1997gg}. For our purpose it suffices
to follow~\cite{Huitu:1998vn} retaining the leading order QCD\index{QCD}
result. Namely, the RGE\index{RGE!running of operators} running of the operators
 induces a common
correction factor for $C_4^{q,RPV}(\mu) = f_{QCD}(\mu) C_4^{q,RPV}$ and
$\widetilde C_4^{q,RPV}(\mu) = f_{QCD}(\mu) \widetilde C_4^{q,RPV}$:\index{$\alpha_s$!corrections}
\begin{equation}
  f_{QCD}(\mu) = \left\{
\begin{array}{ll}
  \left[\frac{\alpha_s(\mu)}{\alpha_s(\Lambda)}\right]^{24/23}, & \Lambda < m_t \\
  \left[\frac{\alpha_s(\mu)}{\alpha_s(m_t)}\right]^{24/23}\left[\frac{\alpha_s(m_t)}{\alpha_s(\Lambda)}\right]^{24/21}, & \Lambda > m_t \\
\end{array}\right\},
\end{equation}
which evaluates to $f_{QCD}(m_b)\simeq 2$ for a range of sneutrino\index{sneutrino}
masses between $100~\mathrm{GeV}\lesssim m_{\widetilde \nu} \lesssim
1~\mathrm{TeV}$. 
In addition the mixing\index{mixing!of operators} with the operators $\mathcal O^q_5$ and
$\widetilde{\mathcal O^q}_5$ induces a small contribution to the Wilson
coefficients\index{Wilson coefficient} $C_5^{q,RPV}(\mu) = \widetilde f_{QCD}(\mu) C_4^{q,RPV}$ and
$\widetilde C_5^{q,RPV}(\mu) = \widetilde f_{QCD}(\mu) \widetilde C_4^{q,RPV}$:\index{$\alpha_s$!corrections}
\begin{equation}
\widetilde f_{QCD}(\mu) =  \frac{1}{3}\left\{
  \begin{array}{ll}
    \left[\frac{\alpha_s(\mu)}{\alpha_s(\Lambda)}\right]^{24/23}-\left[\frac{\alpha_s(\mu)}{\alpha_s(\Lambda)}\right]^{-3/23}, & \Lambda < m_t \\
    \left[\frac{\alpha_s(\mu)}{\alpha_s(m_t)}\right]^{24/23}\left[\frac{\alpha_s(m_t)}{\alpha_s(\Lambda)}\right]^{24/21}-\left[\frac{\alpha_s(\mu)}{\alpha_s(m_t)}\right]^{-3/23}\left[\frac{\alpha_s(m_t)}{\alpha_s(\Lambda)}\right]^{-3/21}, & \Lambda > m_t \\
  \end{array}\right\}
\end{equation}
which is of the order $\widetilde f_{QCD}(m_b) \simeq 0.4$ for the chosen
sneutrino mass\index{sneutrino} range. The relevant part of the effective Hamiltonian \index{effective weak Hamiltonian}
we use in this scenario is then 
\begin{eqnarray}
  \mathcal H_{\mathrm{eff.}}^{RPV} &=& \sum_{q=s,d} \left\{ f_{QCD}(\mu) \left[C_4^{q,RPV}
    \mathcal O^q_4(\mu) + \widetilde C_4^{q,RPV} \widetilde {\mathcal O}^q_4(\mu)
  \right]\right.\nonumber\\
  && \left.\hskip1cm+\widetilde{f}_{QCD}(\mu) \left[C_4^{q,RPV} \mathcal O^q_5(\mu) + \widetilde
    C_4^{q,RPV} \widetilde {\mathcal O}^q_5(\mu) \right]\right\}.
\end{eqnarray}
We neglect the $\widetilde f_{QCD}$ suppressed contributions of $\mc O^q_5$,
$\widetilde{\mc O}^q_5$ to the amplitudes\index{amplitude!in $\Delta S=2$ and $\Delta S=-1$ decays} in the cases where the operators
$\mc O^q_4$, $\widetilde{\mc O}^q_4$ give non-zero
contribution. The inclusive $b\to d d \bar s$\index{transition!$b\to d d \bar s$}\index{transition!$b\to s s \bar d$}
and $b\to ss \bar d$ decay  rates induced by the RPV\index{RPV} model become
\begin{equation}
  \Gamma^{q,RPV}_{\mathrm{inc.}} = \frac{m_b^5 f^2_{QCD}(m_b)}{256 (2\pi)^3}\left(| C_4^{q,RPV}|^2 + | \widetilde C_4^{q,RPV}|^2\right).
\label{eq_Gamma_RPV}
\end{equation}

\par

The most recent upper bound on the specific combination of
couplings entering Wilson coefficients\index{Wilson coefficient} $C_4^{q,RPV}$ and
$\widetilde C_4^{q,RPV}$ can be obtained from Belle's\index{Belle}
search for the  and $B^+\to K^+ K^+ \pi^-$ and $B^+\to \pi^+ \pi^+ K^-$ decays\index{decay!$B^+ \to K^+ K^+ \pi^-$}\index{decay!$B^+ \to \pi^+ \pi^+ K^-$}
\cite{Garmash:2003er,Abe:2002av} which we shall explore in the next section.

\subsubsection{Z'}
\index{gauge boson!$Z'$}

\begin{figure}[!t]
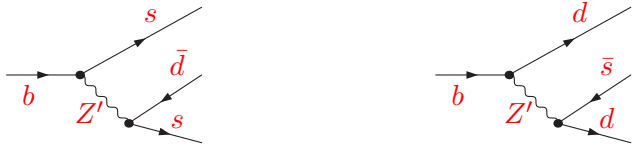

\begin{center}
\unitlength=1bp%

\begin{feynartspicture}(240,80)(3,1)

\FADiagram{}
\FAProp(0.,10.)(7.5,10.)(0.,){/Straight}{1}
\FALabel(2.25,8.93)[t]{$\Red b$}
\FAVert(7.5,10.){0}
\FAProp(20.,17.)(7.5,10.)(0.,){/Straight}{-1}
\FALabel(14.7701,15.5)[b]{$\Red s$}
\FAProp(20.,3.)(12.5,5.)(0.,){/Straight}{-1}
\FALabel(17.5,4.5)[b]{$\Red s$}
\FAProp(20.,10.)(12.5,5.)(0.,){/Straight}{1}
\FALabel(17.5,10.)[b]{$\Red {\bar d}$}
\FAVert(12.5,5.){0}
\FAProp(7.5,10.)(12.5,5.)(0.,){/Sine}{0}
\FALabel(8.5,5.)[b]{$\Red {Z'}$}

\FADiagram{}

\FADiagram{}
\FAProp(0.,10.)(7.5,10.)(0.,){/Straight}{1}
\FALabel(2.25,8.93)[t]{$\Red b$}
\FAVert(7.5,10.){0}
\FAProp(20.,17.)(7.5,10.)(0.,){/Straight}{-1}
\FALabel(14.7701,15.5)[b]{$\Red d$}
\FAProp(20.,3.)(12.5,5.)(0.,){/Straight}{-1}
\FALabel(17.5,4.5)[b]{$\Red d$}
\FAProp(20.,10.)(12.5,5.)(0.,){/Straight}{1}
\FALabel(17.5,10.)[b]{$\Red {\bar s}$}
\FAVert(12.5,5.){0}
\FAProp(7.5,10.)(12.5,5.)(0.,){/Sine}{0}
\FALabel(8.5,5.)[b]{$\Red {Z'}$}

\end{feynartspicture}

\caption{\small\it\label{fig_bssd_Z}{Dominant contributions to the $b \to s s \bar d$ (left) and $b \to d d \bar s$ (right) transitions in the $Z'$ model\index{gauge boson!$Z'$}. Wavy lines denote $Z'$\index{gauge boson!$Z'$} propagation, while filled dots stand for effective flavor violating $Z'$-fermion vertex insertions.}}
\end{center}
\end{figure}

In many extensions\index{Standard Model!extensions|see{new physics}} of the SM ~\cite{Langacker:2000ju} an additional
neutral gauge\index{gauge boson} boson appears. Heavy neutral bosons are also present in
many extensions of the SM such as grand unified, superstring theories
and theories with large extra dimensions~\cite{Erler:1999nx}.  This
induces contributions in fig.~\ref{fig_bssd_Z} to the effective tree level Hamiltonian \index{effective weak Hamiltonian} from the operators
 $\mathcal O^q_{1,3}$ as well as $\mathcal {\widetilde O}^q_{1,3}$.
Following \cite{Langacker:2000ju,Erler:1999nx}, the Wilson
coefficients\index{Wilson coefficient} for the corresponding operators read at the interaction
scale $\Lambda \sim m_{Z'}$\index{G$_F$}
\begin{equation} \label{eq:zPrimeOperators}
\begin{array}{cc}
C_1^{s,Z'} = -\frac{4G_F y}{\sqrt{2}} B_{12}^{d_L} B_{13}^{d_R}, &
\widetilde C_1^{s,Z'} = -\frac{4G_F y}{\sqrt{2}} B_{12}^{d_R} B_{13}^{d_L}, \\
C_3^{s,Z'} = -\frac{4G_F y}{\sqrt{2}} B_{12}^{d_L} B_{13}^{d_L}, &
\widetilde C_3^{s,Z'} = -\frac{4G_F y}{\sqrt{2}} B_{12}^{d_R} B_{13}^{d_R}, \\
C_1^{d,Z'} = -\frac{4G_F y}{\sqrt{2}} B_{21}^{d_L} B_{23}^{d_R}, &
\widetilde C_1^{d,Z'} = -\frac{4G_F y}{\sqrt{2}} B_{21}^{d_R} B_{23}^{d_L}, \\
C_3^{d,Z'} = -\frac{4G_F y}{\sqrt{2}} B_{21}^{d_L} B_{23}^{d_L}, &
\widetilde C_3^{d,Z'} = -\frac{4G_F y}{\sqrt{2}} B_{21}^{d_R} B_{23}^{d_R},

\end{array}
\end{equation}
where $y = (g_2/g_1)^2 (\rho_1 \sin^2 \theta+ \rho_2 \cos^2 \theta)$
and $\rho_i = m_W^2/m_i^2 \cos^2\theta_W$. In this expression $g_1$,
$g_2$, $m_1$ and $m_2$ stand for the gauge\index{gauge coupling} couplings and masses of the
$Z$\index{gauge boson!$Z$} and $Z'$\index{gauge boson!$Z'$}\index{gauge boson!$Z'$} bosons, respectively, while $\theta$ is their mixing
angle\index{mixing!of neutral gauge bosons}.  Again renormalization group running induces corrections and
mixing\index{mixing!of operators} between the operators.  As already mentioned, the mixing of
operators $\mathcal O^q_{1,2}$ and their chirally  flipped counterparts
is identical to that of operators $\mathcal O^q_{4,5}$ since these
operators are connected via Fierz rearrangement. Thus the same scaling
and mixing\index{mixing!of operators} factors $f_{QCD}$ and $\widetilde f_{QCD}$ apply. For the
operator $\mathcal O_3$ on the other hand the renormalization can be
written as $C^{q,Z'}_3(\mu) = f'_{QCD}(\mu) C^{q,Z'}_3$ with\index{$\alpha_s$!corrections}
\begin{equation}
  f'_{QCD}(\mu) = \left\{
    \begin{array}{ll}
      \left[\frac{\alpha_s(\mu)}{\alpha_s(\Lambda)}\right]^{-6/23}, & \Lambda < m_t \\
      \left[\frac{\alpha_s(\mu)}{\alpha_s(m_t)}\right]^{-6/23}\left[\frac{\alpha_s(m_t)}{\alpha_s(\Lambda)}\right]^{-6/21}, & \Lambda > m_t \\
    \end{array}\right\}.
\end{equation}
In particular for a common $Z'$ boson scale of $m_{Z'}\simeq
500~\mathrm{GeV}$~\cite{Langacker:2000ju} one gets numerically
$f_{QCD} (m_b) \simeq 2$, $\widetilde f_{QCD} (m_b) \simeq 0.4 $ and
$f'_{QCD} (m_b) \simeq 0.8$. The full contributing part of the
effective Hamiltonian \index{effective weak Hamiltonian} in this case is
\begin{eqnarray}
  \mathcal H_{\mathrm{eff.}}^{Z'} &=& \sum_{q=s,d} \left\{ f_{QCD}(\mu) \left[C_1^{q,Z'} \mathcal O^q_1(\mu) + \widetilde C_1^{q,Z'} \widetilde {\mathcal O}_1(\mu) \right] \right.\nonumber\\
  &&\hskip1cm + \widetilde f_{QCD}(\mu) \left[C_1^{q,Z'} \mathcal O^q_2(\mu) + \widetilde C_1^{q,Z'} \widetilde {\mathcal O}^q_2(\mu) \right] \nonumber\\
  &&\hskip1cm\left.+ f'_{QCD}(\mu) \left[C_3^{q,Z'} \mathcal O^q_3(\mu) + \widetilde C_3^{q,Z'} \widetilde {\mathcal O}^q_3(\mu) \right]\right\}.\label{eq:zHamiltonian}
\end{eqnarray}
For the inclusive $b\to d d \bar s$ and $b\to s s \bar d$ \index{transition!$b\to d d \bar s$}\index{transition!$b\to s s \bar d$}decay  rates the $\mathcal O^q_2$ and
$\widetilde {\mathcal O^q_2} $ are numerically suppressed due to the $\widetilde
f_{QCD}$ factor and we write
\begin{equation}
  \Gamma^{q,Z'}_{\mathrm{inc.}} = \frac{m_b^5}{192 (2\pi)^3}\Big[ 3 f^2_{QCD}(m_b) \left(|C^{q,Z'}_1|^2+|\widetilde C^{q,Z'}_1|^2\right) + 4 f'^2_{QCD}(m_b)\left(|C^{q,Z'}_3|^2+|\widetilde C^{q,Z'}_3|^2\right)\Big].
  \label{eq_Gamma_Z}
\end{equation}
In Section~\ref{sec:discussion} we discuss bounds on Wilson coefficients\index{Wilson coefficient}
$C_{1,3}^{q,Z'}$ and $\widetilde{C}_{1,3}^{q,Z'}$ which might be estimated
from the $B^- \to \pi^- \pi^- K^+$ and $B^- \to K^- K^- \pi^+$ decay  rates.

\section[Nonleptonic decays of $B_c$ mesons]{Two- and three-body non-leptonic decays of $B_c$ mesons}
\index{meson!$B_c$!decay}
\index{$\Delta S=2$ transitions}
\index{$\Delta S=-1$ transitions}
In calculating decay
 rates of various $B_c$ meson decay
 modes based on the $b \to dd\bar{s}$ and $b \to ss\bar{d}$ quark transition\index{transition!$b\to d d \bar s$}\index{transition!$b\to s s \bar d$}, one has to calculate matrix elements of the effective Hamiltonian\index{effective weak Hamiltonian} operators
 between meson states.
As a first approximation, the calculation will be performed in the factorization approximation\index{VSA}. In the $B$ meson decays\index{meson!$B$!decay}
 this works in a good number of cases, while in other cases a more sophisticated approach is needed (for a recent review see~\cite{Wei:2003tv}). In this first calculation we consider that factorization approximation\index{VSA} is sufficient for obtaining the correct features of the decays
 of the various channels considered. An exception is the case in which matrix elements vanish as a result of factorization\index{VSA}, which in a better approximation can be improved.

\subsection{Preliminaries}

In out na\"ive factorization\index{VSA}\index{na\"ive factorization|see{VSA}} of two- and three-body
amplitudes\index{amplitude!in $\Delta S=2$ and $\Delta S=-1$ decays}, we express the resulting one- and two-point transition amplitudes
between mesons in terms of the standard weak transition decay\index{decay constant}
 constants and form factors  (\ref{eq_3_19} - \ref{def-ff_0}), as dictated by the Lorentz
covariance. Some general formulae can be devised to assist our calculation.

\subsubsection{Factorization and Kinematics}

\begin{figure}[!t]
\begin{center}
\psfrag{B}[cc]{$\color{red}  B(p_B)$}
\psfrag{P1}[cl]{$\color{red}  M_1(p_1)$}                                                 \psfrag{P2}[br]{$\color{red}  M_2(p_2)$}
\epsfxsize10cm\epsffile{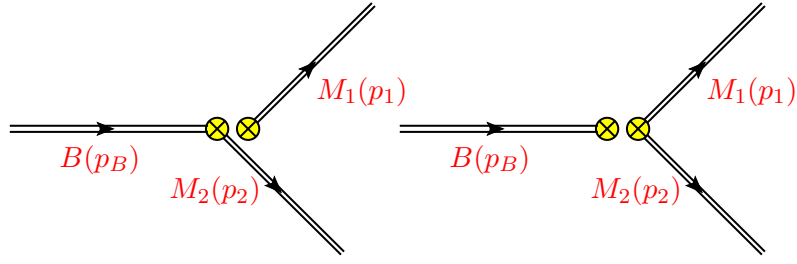}    \\
\caption{\small\it\label{fig_BtoPP}{Diagrams contributing to the factorized matrtix elements of two body nonleptonic decays of $B$ mesons\index{decay of $B$ meson}\index{meson!$B$!decay}. Double lines represent meson propagation, while crossed circles represent factorized weak current insertions.}}
\end{center}
\end{figure}

For the decays\index{decay of $B$ meson}\index{meson!$B$!decay}
 of a pseudoscalar meson $B$ containing a $b$ quark to two mesons $M_1$ and $M_2$ two diagram topologies are possible in the factorization approximation\index{VSA} (in fig.~\ref{fig_BtoPP}). The right ``anhilation'' diagram does not contribute in spectator processes, where the light quark flavor of the $B$\index{meson!$B$} meson ($\bar u$ in $B^-$\index{meson!$B$} and $\bar c$ in $B_c^-$\index{meson!$B_c$}) does not feature in the effective Hamiltonian\index{effective weak Hamiltonian}. All processes that we shall consider are of this type.
We first derive a general expression for the
factorized matrix element of the $\mathcal O^q_3$ operators
, relevant in
the framework of SM~(MSSM)\index{MSSM}. In case the final state mesons are pseudoscalars (we deonte them by $P1$ and $P2$) we get for the $\mathcal O^s_3$\index{amplitude!in $\Delta S=2$ and $\Delta S=-1$ decays}
\begin{equation}
  \Braket{P_1 (p_1) | \bar{d} \gamma_\mu\gamma_5 s |0} \Braket{P_2 (p_2) | \bar{d} \gamma^\mu b| B(p_B) } = i (m_B^2-m_{P_2}^2) f_{P_1} F_0^{P_2 B} (m_{P_1}^2)
  \label{eq:2bodySM}
\end{equation}
Taking care of sign difference due to different chirality\index{chiral structure} structure, the same expression also applies for operators
 $\widetilde {\mc O}^q_3$, $\mc O_1^q$ and $\widetilde
{\mc O}_1^q$.

\par

Similarly, when one of the final state mesons is a vector (we denote it as $V$), we have two possibilities: either (I) the vector state is paired with the vacuum or (II) with the initial $B$\index{meson!$B$} meson. The two possibilities give\index{amplitude!in $\Delta S=2$ and $\Delta S=-1$ decays}
\begin{subequations}
\begin{eqnarray}
\label{eq_BPV}
&\mathrm{(I)}& \bra{V(\epsilon,p_V)}\bar{d} \gamma_\mu s\ket{0} \bra{P(p_P)} \bar{d} \gamma^\mu b \ket{B(p_B)} = 2 m_V f_V F_+^{P B}(m_V^2) \epsilon \cdot p_B\\
&\mathrm{(II)}& \bra{P(p_P)}\bar{d} \gamma_\mu\gamma_5 s\ket{0} \bra{V(\epsilon,p_V)} \bar{d} \gamma^\mu\gamma_5 b \ket{B(p_B)} = - 2 m_V f_P A_0^{V B}(m_P^2) \epsilon \cdot p_B
\label{eq_BPVb}
\end{eqnarray}
\end{subequations}
\par
Because only vector currents contribute in the expression (I), it
also applies for operators
 $\widetilde {\mc O}^q_3$, $\mc O_1^q$ and $\widetilde
{\mc O}_1^q$, while in the case (II) the expression is valid for these operators up to a sign difference.

\par

In two-body decays with a vector meson $V$ and a pseudoscalar meson
$P$ in the final state we also sum over the polarizations\index{polarization of vector meson} of $V$. The sum
in our case reduces to
\begin{equation}
  \label{eq:polSum} \sum_{\epsilon_V} \left|\epsilon_V^*(p_V)\cdot p_B \right|^2
  = \frac{\lambda(m_B^2,m_V^2,m_P^2)}{4m_V^2},
\end{equation}
where $\epsilon_V$ is the polarization\index{polarization of vector meson} vector of $V$ and $\lambda$ is
defined as $\lambda(x,y,z) = (x+y+z)^2-4(xy+yz+zx)$.

\par

For decay to two vector mesons in the final state we use the helicity
amplitudes\index{helicity amplitudes} formalism as described in ref.~\cite{Kramer:1991xw}.
Non-polarized decay  rate is expressed as an incoherent sum of helicity
amplitudes
\begin{equation}
  \Gamma=\frac{|\vek{p}_1|}{8 \pi m_B^2} \left(\left|H_0\right|^2+\left|H_{+1}\right|^2+\left|H_{-1}\right|^2 \right),
\end{equation}
where $\vek{p}_1$ is momentum of the vector meson in $B$\index{meson!$B$} meson rest
frame and helicity amplitudes\index{helicity amplitudes} are expressed as
\begin{equation}
  H_{\pm 1} = a \pm \frac{\sqrt{\lambda(m_B^2,m_1^2,m_2^2)}}{2 m_1 m_2} c,\qquad
  H_0 = -\frac{m^2-m_1^2-m_2^2}{2m_1 m_2} a
  -\frac{\lambda(m_B^2,m_1^2,m_2^2)}{4 m_1^2 m_2^2}b.
\end{equation}
Vector meson masses are denoted by $m_{1,2}$, while definition of the
constants $a$, $b$ and $c$ is given by general Lorentz decomposition
of the polarized amplitude
\begin{equation}
\label{eq:helicityDecomposition}
  H_\lambda=\epsilon_{1\mu}^*(\lambda) \epsilon_{2\nu}^*(\lambda) \left(a g^{\mu\nu} + \frac{b}{m_1 m_2}p_B^\mu p_B^\nu\right.
  + \left.\frac{ic}{m_1 m_2} \epsilon^{\mu\nu\alpha\beta}p_{1\alpha}
    p_{2\beta} \right),
\end{equation}
where $\epsilon_{1,2}$ and $p_{1,2}$ are the vector mesons polarizations and momenta.

\par

\begin{figure}[!t]
\begin{center}
\psfrag{B}[cc]{$\color{red}  B(p_B)$}
\psfrag{P1}[cl]{$\color{red}  M_1(p_1)$}
\psfrag{P2}[br]{$\color{red}  M_2(p_2)$}
\psfrag{P3}[br]{$\color{red}  M(p)$}
\epsfxsize16cm\epsffile{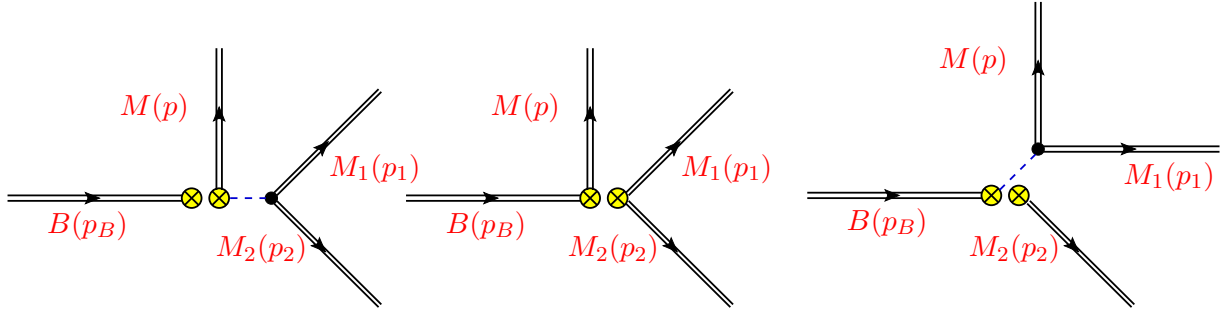}    \\
\caption{\small\it\label{fig_BtoPPP}{Diagrams contributing to the factorized matrix elements of three body nonleptonic decays\index{decay of $B$ meson}\index{meson!$B$!decay} of $B$ mesons. Dashed lines represent intermediate (resonant) state propagation while filled circles represent effective strong vertex insertions.}}
\end{center}
\end{figure}

Situation is more complicated in the case of three body nonleptonic decays.
Beside the simple factorized topology in the center of figure~\ref{fig_BtoPPP}, we also need to consider possible dominant contributions comming from (resonant) intermediate states, such as those pictured in the left and right diagrams in fig.~\ref{fig_BtoPPP}.
Here again the left diagram does not contribute in our chosen ``spectator'' channels.
For the decays of $B$\index{decay of $B$ meson}\index{meson!$B$!decay} to three pseudoscalar mesons $P$, $P_1$ and $P_2$ the factorized matrix element of $\mathcal O^s_3$ due to the topology in center fig.~\ref{fig_BtoPPP} reads\index{amplitude!in $\Delta S=2$ and $\Delta S=-1$ decays}
\begin{align}
  \Braket{P_2 (p_2) P_1 (p_1) | \bar{d} \gamma_\mu s |0} &\Braket{P(p)
    | \bar{d} \gamma^\mu b| B(p_B) }
  = (t-u) F_+^{P_2 P_1} (s) F_+^{PB} (s)\nonumber\\
  &+ \frac{(m_{P_1}^2-m_{P_2}^2)(m_B^2-m_P^2)}{s} \left[ F_+^{P_2 P_1}
    (s) F_+^{PB} (s) -F_0^{P_2 P_1} (s) F_0^{PB} (s) \right].
  \label{eq:3bodySM}
\end{align}
Because only vector currents contribute in the above expression, it
also applies for operators
 $\widetilde {\mc O}^q_3$, $\mc O_1^q$ and $\widetilde
{\mc O}_1^q$.  The Mandelstam kinematical variables are as before $s=(p_B-p)^2$,
$t=(p_B-p_1)^2$ and $u=(p_B-p_2)^2$. We assume these contributions to be the dominant ones, where present. Possible contributions from the right diagram of fig.~\ref{fig_BtoPPP} are to be considered where eq.~(\ref{eq:3bodySM}) does not contribute. We employ resonance\index{resonance dominance approximation} dominance approximation and saturate the intermediate state with the lowest lying resonances\index{resonance} coupling to the weak current and the pair of final state mesons.
The lowest resonance coupling to a pair of pseudoscalar mesons in a parity conserving way is of vector type (we denote it as $V^*$) and only contributes via the (pseudo)scalar part of the current. Therefore we may write such contribution as\index{amplitude!in $\Delta S=2$ and $\Delta S=-1$ decays}
\begin{eqnarray}
&&\bra {P(p) P_1(p_1) P_2 (p_3)} \bar s \gamma_{\mu} \gamma_5 d \bar s \gamma^{\mu} \gamma_5 b \ket{B(p_B)}  = \nonumber\\
&&\hskip-1cm \bra{P(p)} \bar s \gamma_{\mu} \gamma_5 d \ket{ 0} \braket{P_1(p_1) P_2 (p_3) | V^*(p_V,\epsilon)} \otimes i G_{V^*}(p_V) \otimes \bra {V^*(p_V,\epsilon)} \bar s \gamma^{\mu} b \ket{B(p_B)},
\end{eqnarray}
where $G_{V^*}(q)=(-g_{\mu\nu} + q_{\mu}q_{\nu}/m_{V^*}^2)/(q^2 - m_{V^*}^2 + i \Gamma(V^{*}) m_{V^*})$ is the $V^{*}$ meson propagator of Breit-Wigner form, while  $\braket{P_1(p_1) P_2 (p_2) | V^*} = G_{P_1,P_2,V^*} \epsilon \cdot p_1$ is the strong transition matrix element defined in eq.~(\ref{eq_strong_v}) and needs to be calculated in a suitable QCD effective theory description or model. Contraction of propagator Lorentz indices with vector state polarizations is understood. The amplitude\index{amplitude!in $\Delta S=2$ and $\Delta S=-1$ decays}  thus becomes
\begin{eqnarray}
&&\bra {P(p) P_1(p_1) P_2 (p_3)} \bar s \gamma_{\mu} \gamma_5 d \bar s \gamma^{\mu} \gamma_5 b \ket{B(p_B)}  = \nonumber\\
&&\hskip-1cm= f_{P} m_P^2 A^{BV}_0(m_{P}^2) G_{P_1,P_2,V^*} \frac{m_{V}^2(u-m_{P_1}^2-m_{P}^2) + \frac{1}{2} (s-m_{P_2}^2+m_{P_1}^2)(s-m_{P}^2+m_{B}^2)}{s~m_{V}^2
(s + i \Gamma(V) m_{V} - m_{V}^2)} .\nn\\
\label{eq:resonance}
\end{eqnarray}
Other possible three-body decay  channels with vector mesons in final states will not be considered as they are very difficult to reconstruct experimentaly.

\par

In the context of RPV, the contributions of the operators
 $\mc{O}^q_4$
and $\widetilde {\mc O}^q_4$ to hadronic amplitudes are dominant.  One can
use the Dirac equation to express scalar~(pseudoscalar) density
operators in terms of derivatives of vector~(axial-vector) currents
\begin{equation}
\label{eq:diracTrick}
    \bar{q}_i q_j = \frac{i \partial_\mu(\bar{q}_i \gamma^{\mu} q_j)}{m_{q_j} - m_{q_i}},\qquad
    \bar{q}_i \gamma^5 q_j = -\frac{i \partial_\mu (\bar{q}_i
      \gamma^{\mu} \gamma^5 q_j)}{m_{q_j} + m_{q_i}}.
\end{equation}
Using these relations we can derive expressions for the factorized matrix
element of the $\mc O_4^s$ and $\widetilde{\mc O}^s_4$ operators
 in all two- and threebody channels considered above. For the two-body $B\to P_1 P_2$ case we obtain
\begin{equation}
  \Braket{P_1 (p_1) | \bar{d} \gamma_5 s |0} \Braket{P_2 (p_2) | \bar{d} b| B(p_B) } = \frac{i (m_B^2-m_{P_2}^2) m_{P_1}^2}{(m_b-m_d)(m_s+m_d)} f_{P_1} F_0^{P_2 B} (m_{P_1}^2).
\label{eq:RPV2Body}
\end{equation}
In $B\to P V$ channels, only case (II) (eq.~(\ref{eq_BPVb})) contributes in the factorization approximation\index{VSA} due to the vector polarization transverzality condition $\epsilon_V \cdot p_V=0$ in case (I). This gives
\begin{equation}
  \Braket{P (p_P) | \bar{d} \gamma_5 s |0} \Braket{V (\epsilon,p_V) | \bar{d}\gamma_5 b| B(p_B) } = - \frac{2 m_V m_P^2}{(m_b+m_d)(m_s+m_d)} f_{P} A_0^{V B} (m_{P}^2) \epsilon^* \cdot p_B.
\end{equation}
Vector transversality condition also kills any $\mc O_4^s$ and $\widetilde{\mc O}^s_4$ contributions in $B\to V_1 V_2$ channels. For the three-body $B\to P_1 P_2 P_3$ channel topology in fig.~\ref{fig_BtoPPP} on the other hand one obtains
\begin{equation}
  \Braket{P_2 (p_2) P_1 (p_1) | \bar{d} s |0} \Braket{P(p) | \bar{d} b | B(p_B) } =
  \frac{(m_{P_1}^2-m_{P_2}^2)(m_B^2-m_P^2)}{(m_b - m_d)(m_s - m_d)}
  F_0^{P_2 P_1} (s) F_0^{PB} (s),
  \label{eq:3bodyO4}
\end{equation}
and for the resonance\index{resonance contribution} contribution (fig.~\ref{fig_BtoPPP})
\begin{eqnarray}
&&\bra {P(p) P_1(p_1) P_2 (p_3)} \bar s \gamma_5 d \bar s \gamma_5 b \ket{B(p_B)}  = \nonumber\\
&&\hskip-1cm= \frac{f_{P} m_P^4 A^{BV}_0(m_{P}^2) G_{P_1,P_2,V^*}}{(m_b+m_s)(m_d+m_s)} \frac{m_{V}^2(u-m_{P_1}^2-m_{P}^2) + \frac{1}{2} (s-m_{P_2}^2+m_{P_1}^2)(s-m_{P}^2+m_{B}^2)}{s~m_{V}^2
(s + i \Gamma(V) m_{V} - m_{V}^2)} .\nn\\
\label{DKK_R}
\end{eqnarray}

\par

Finally, the color non-singlet operators
 $\mc
O_2^q$ and $\widetilde {\mc O}^q_2$ can be Fierz rearranged to $\mc O^q_4$ and
$\widetilde {\mc O}^q_4$ and then same expressions apply as well. Remaining operators are all of the $V\pm A$ form and their forms are therefore given above.

\subsubsection{Modeling Form Factors}

In our calculations we need the $B_c\to D_s^{(*)}$\index{meson!$B_c$!decay} transition form factors\index{form factor!$B_c\to D_s^{(*)}$} $F_{\pm}$, $V$ and $A_{0,1,2}$. Since HQET\index{HQET} and the whole discussion of chapter~\ref{chapter_semileptonic} is not directly applicable to the decays\index{decay of $B_c$ meson}\index{meson!$B_c$!decay} of the $B_c$ meson, we assume pole dominance for these form factors ~\cite{Wirbel:1985ji,Kiselev:2002vz}:
\begin{eqnarray}
F(s) &=& \frac{F(0)}{(1 - s/m_{\mathrm{pole}}^2)}
\end{eqnarray}
and take numerical values for $F(0)$ and $m_{\mathrm{pole}}$ from from QCD\index{QCD sum rules} sum rules calculations~\cite{Kiselev:2002vz} (see tables \ref{F0_table} and \ref{M_table}).
\begin{table}[!t]
\begin{center}
\begin{tabular}{|r|cccccc|}
    \hline
    Decays & $F_+(0)$ & $F_-(0)$ & $V(0)$ & $A_0(0)$ & $A_1(0)$ & $A_2(0)$\\\hline
    $B_c\to D^{(*)}$ & $0.32$ & $-0.34$ & $0.20$ & $3.7$ & $-0.062$ & $0.10$\\
    $B_c\to D_s^{(*)}$ & $0.45$ & $-0.43$ & $0.24$ & $4.7$ & $-0.077$ & $0.13$\\\hline
\end{tabular}
\end{center}
\caption{\small\it\label{F0_table} Numerical values of $B_c\to D^{(*)}_{(s)}$ transition form factors at $s=0$ by Kiselev.}
\end{table}
\begin{table}[!t]
 \begin{center}
 \begin{tabular}{|r|cccccc|}
 \hline
    Decays & $F_+ [\mathrm{GeV}]$ & $F_- [\mathrm{GeV}]$ & $V [\mathrm{GeV}]$ & $A_0 [\mathrm{GeV}]$ & $A_1 [\mathrm{GeV}]$ & $A_2 [\mathrm{GeV}]$\\\hline
    $B_c\to D^{(*)}$ & $5.0$ & $5.0$ & $6.2$ & $\infty$ & $6.2$ & $6.2$\\
    $B_c\to D_s^{(*)}$ & $5.0$ & $5.0$ & $6.2$ & $\infty$ & $6.2$ & $6.2$\\\hline
 \end{tabular}
 \end{center}
 \caption{\small\it\label{M_table} Pole masses used in $B_c\to D^{(*)}_{(s)}$ transition form factors by Kiselev.}
 \end{table}

\par

For the $B^-\to \pi^-$ and $B^-\to K^-$ transitions used to constrain new physics\index{new physics!constraints} model parameters we use the form factors  calculated in the relativistic constituent quark model with
numerical input from lattice QCD\index{lattice QCD} at high $s$~\cite{Melikhov:2000yu}
\begin{subequations}
\begin{align}
  F_1^{\pi B}(s) &= \frac{F^{\pi
      B}_1(0)}{(1-s/m_{B^*}^2)[1-\sigma_1 s/m_{B^*}^2]},
  \quad F^{\pi B}_1(0) = 0.29,\quad \sigma_1 = 0.48,\\
  F_0^{\pi B}(s) &= \frac{F^{\pi B}_0(0)}{1-\sigma_1
    s/m_{B^*}^2+\sigma_2 s^2/m_{B^*}^4}, \quad F^{\pi
    B}_0(0) = 0.29,\quad \sigma_1 = 0.76, \quad\sigma_2=0.28,\\
  F_1^{K B}(s) &= \frac{F^{K
      B}_1(0)}{(1-s/m_{B_s^*}^2)[1-\sigma_1 s/m_{B_s^*}^2]},
  \quad F^{K B}_1(0) = 0.29,\quad \sigma_1 = 0.48,\\
  F_0^{K B}(s) &= \frac{F^{K B}_0(0)}{1-\sigma_1
    s/m_{B_s^*}^2+\sigma_2 s^2/m_{B_s^*}^4}, \quad F^{K
    B}_0(0) = 0.29,\quad \sigma_1 = 0.76, \quad\sigma_2=0.28,
\end{align}
\end{subequations}

\par

In the three body decay modes involving pairs of $D$\index{meson!$D$!decay} and
$D_s$\index{meson!$D_s$!decay} mesons, we also need the form factors  for the
$D_s^{} \to D^{}$ transitions. These are not available in the literature and we calculate them by utilizing HM$\chi$PT\index{HM$\chi$PT!inspired model!of light scalar interactions}, including the light scalar meson interactions with heavy mesons as it has been done recently~\cite{Bardeen:2003kt}, and presuming the main contributions from exchange of light scalar meson resonance $K^{*0}(1430)$.
Interactions of heavy mesons are described by the HM$\chi$PT\index{HM$\chi$PT!Lagrangian} Lagrangian~\ref{eq_2_13}, to which we add interactions of
scalar mesons, which we put into an $SU(3)$ nonet ($1 \oplus 8$) representation
\begin{eqnarray}
    \mathcal L^{(0)}_{HM\chi PT} += - \frac{g_{\pi}}{4} \mathrm{Tr} \left[ H \widetilde \sigma \overline H \right] + \ldots
\end{eqnarray}
Here the light scalar mesons are introduced \index{resonance!of light meson}
through the $\widetilde \sigma=\sqrt{2/3} \hat\sigma$ field, where
$\hat\sigma$ is the light scalar meson matrix
\begin{equation}
\hat \sigma =
   \begin{pmatrix}
    \frac{1}{\sqrt 2} (\sigma(600) + f^0(980)) & f^+ & K^{'+} \\
    f^- & \frac{1}{\sqrt 2} (\sigma(600) - f^0(980)) & K^{*0}(1430) \\
    K^{'-} & \overline K^{*0}(1430) & a^0(980)
   \end{pmatrix}.
\end{equation}
The ellipses indicate further terms involving only light meson
fields, chiral and $1/m_H$ corrections. At the leading order in heavy quark mass and chiral\index{chiral expansion!in $D_s \to D$ transitions} expansion $F_+^{D_sD}$, 
is found to vanish, so the only contributions come from the $F_0^{D_sD}$
form factor . We then use resonance\index{resonance dominance approximation} dominance appoximation to obtain
\begin{equation}
  F^{D_sD}_0(s) = \frac{s}{m_{D_s}^2-m_{D}^2} \frac{(g_\pi/4)
    f_{K(1430)} \sqrt{m_{D_s}m_D}}{s-m_{K(1430)}^2+i\sqrt{s}
    \Gamma_{K(1430)}}.
\end{equation}
From our analyses in Chapter~\ref{chapter_semileptonic} this result comes as no suprise since our choice of form factor\index{form factor!parameterization} parameterization dictates which resonances\index{resonance contribution} may contribute. Taking account of only (pseudo)scalar resonance exchange thus singles out the $F_0$ form factor . In the numerical calculation we use $g_\pi =\simeq 3.73$~\cite{Bardeen:2003kt}.
The rest of parameters are taken from PDG~\cite{Eidelman:2004wy}.

\par

A similar method has been used to obtain the light to light $K^- \to
\pi^-$ meson transition\index{transition!$K^- \to
\pi^-$} form factors\index{form factor!$K^- \to
\pi^-$} in ref.~\cite{Fajfer:1999hh}
\begin{subequations}
\begin{align}
F_1^{\pi K}(s) &=\frac{2 g_{VK^*} g_{K^*}}{s-m_{K^*}^2+i \sqrt{s} \Gamma_{K^*}},\\
F_0^{\pi K}(s) &= F_1^{\pi K}(s)\left(1-\frac{s}{m_{K^*}^2}\right) + \frac{s}{m_{K}^2-m_{\pi}^2}\frac{f_{K(1430)} g_{SK(1430)}}{s-m_{K(1430)}^2+i \sqrt{s}
\Gamma_{K(1430)}}.
\end{align}
\end{subequations}
In our numerical calculations we use the following values: $f_{K (1430)} \simeq 0.05\mathrm{~GeV}$, $g_{K^*}= f_{K^*} m_{K^*} =  0.196\mathrm{~GeV}^2$, $g_{V K^*}=4.59  $ and $g_{S K(1430)} =
3.67 \pm 0.3\mathrm{~GeV}$ taken from~\cite{Fajfer:1999hh}.

\subsection{Amplitudes}
\index{amplitude!in $\Delta S=2$ and $\Delta S=-1$ decays}

\subsubsection{$B^- \to \pi^- \pi^- K^+$ and $B^- \to K^- K^- \pi^+$}
\index{decay!$B^- \to \pi^- \pi^- K^+$}
\index{decay!$B^- \to K^- K^- \pi^+$}

Experimentally, these are the only constrained processes proceeding through the $b\to d d \bar s$ and $b\to s s \bar d$ transitions\index{transition!$b\to d d \bar s$}\index{transition!$b\to s s \bar d$}. Therefore we may constrain new physics\index{new physics!constraints} model parameters and use these constraints to predict other viable decay channels .

\par

Hadronic matrix element entering in the amplitudes for $B^- \to \pi^-
\pi^- K^+$ ($B^- \to \pi^-
\pi^- K^+$) in SM~(MSSM)\index{MSSM} is readily given by eq.~(\ref{eq:3bodySM})
after identifying $P = \pi^-(K^-)$, $P_1 = K^+(K^-)$, $P_2 = \pi^-(\pi^+)$ and using
appropriate form factors . Eq.~(\ref{eq:3bodyO4}) is used instead for RPV, while the $Z'$\index{gauge boson!$Z'$}
amplitude incorporates both eqs.~(\ref{eq:3bodySM}) and
(\ref{eq:3bodyO4}).  There are two contributions in each model to this
mode, with an additional term with the $u \leftrightarrow s$ ($t \leftrightarrow s$)
replacement in eqs.  (\ref{eq:3bodySM}) and (\ref{eq:3bodyO4}),
representing an interchange of the two pions (kaons\index{meson!$K$}) in the final state.
After phase space integration\index{phase space integration}, the decay  rates can be written very
compactly with only Wilson coefficients\index{Wilson coefficient} left in symbolic form in table~\ref{table:bpipik}.
\begin{table}[!t]
 \begin{center}
 \begin{tabular}{|r|ll|}
 \hline
    Model & $\Gamma_{\pi\pi K}~[10^{-3}~\mathrm{GeV}^5]$
            & $\Gamma_{K K \pi}~[10^{-3}~\mathrm{GeV}^5]$\\\hline\hline
    (MS)SM & $2.1 \times\left|C_3^{s,(MS)SM}\right|^2$
             & $3.1 \times\left|C_3^{d,(MS)SM}\right|^2$\\
    RPV & $7.8 \times\left|C_4^{s,RPV}+\widetilde C_4^{s,RPV}\right|^2 $
         &  $11 \times\left|C_4^{d,RPV}+\widetilde C_4^{d,RPV }\right|^2 $\\
    Z' & $ 9.0 \times\left|C_1^{s,Z'} + \widetilde C_1^{s,Z'}\right|^2  $
        &  $16 \times\left|C_1^{d,Z'} + \widetilde C_1^{d,Z'}\right|^2$ \\
    & $ + 2.1\times \left|C_3^{s,Z'} + \widetilde C_3^{s,Z'}\right|^2$
    & $ +3.1\times \left|C_3^{d,Z'} + \widetilde C_3^{d,Z'}\right|^2$  \\
    & $ + 8.3 \times\Re\left[\left(C_1^{s,Z'} + \widetilde C_1^{s,Z'}\right)\right.$
    &  $ +13\times\Re\left[\left(C_1^{d,Z'} + \widetilde C_1^{d,Z'}\right)\right.$\\
    &   \hskip1.8cm $\left.\left(C_3^{s,Z'} + \widetilde C_3^{s,Z'}\right)^*\right]$
     & \hskip1.7cm$\left.\left(C_3^{d,Z'} + \widetilde C_3^{d,Z'}\right)^*\right]$  \\\hline
 \end{tabular}
 \end{center}
 \caption{\small\it\label{table:bpipik} $B^- \to \pi^-
\pi^- K^+$  and $B^- \to \pi^-
\pi^- K^+$ decay rates in various models and in terms of the relevant Wilson coefficients.}
 \end{table}
Assuming as in~\cite{Huitu:1998pa,Fajfer:2000ny} that interference
between the two chiral contributions in RPV\index{RPV} and $Z'$\index{gauge boson!$Z'$} models is small,
the decay  rates in these models become approximately
\begin{subequations}
\begin{align}
  &\Gamma^{RPV}_{\pi\pi K} = \left(|C_4^{s,RPV}|^2+ |\widetilde C_4^{s,RPV
    }|^2\right) \times
  7.8\E{-3}\e{GeV}^5,\label{eq:3bodyRPVrate}\\
  &\Gamma^{Z'}_{\pi\pi K} = \left(|C_1^{s,Z'}|^2 + |\widetilde C_1^{s,Z'}|^2\right) \times 9.0\E{-3}\e{GeV}^5+\left(|C_3^{s,Z'}|^2 + |\widetilde C_3^{s,Z'}|^2\right) \times 2.1\E{-3}\e{GeV}^5
\end{align}
\end{subequations}
and
\begin{subequations}
\begin{align}
  &\Gamma^{RPV}_{KK\pi} = \left(|C_4^{d,RPV}|^2+ |\widetilde C_4^{d,RPV
    }|^2\right) \times
  10.6\E{-3}\e{GeV}^5,\label{eq:3bodyRPVrate1}\\
  &\Gamma^{Z'}_{KK\pi} = \left(|C_1^{d,Z'}|^2 + |\widetilde C_1^{d,Z'}|^2\right) \times 15.6\E{-3}\e{GeV}^5+\left(|C_3^{d,Z'}|^2 + |\widetilde C_3^{d,Z'}|^2\right) \times 3.1\E{-3}\e{GeV}^5.
\end{align}.
\end{subequations}

\subsubsection{$B_c^- \to D^- D^- D_s^+$ and $B_c^- \to D_s^- D_s^- D^{+}$}
\index{decay!$B_c^- \to D^- D^- D_s^+$}
\index{decay!$B_c^- \to D_s^- D_s^- D^{+}$}

In calculation of the $B_c^- \to D^- D^- D_s^+$ ($B_c^- \to D_s^- D_s^- D^{+}$) decay  rates again we
use eqs.~(\ref{eq:3bodySM}) and (\ref{eq:3bodyO4}) now with
substitutions $P_3=D^-(D_s^-)$, $P_1 = D_s^+(D^+)$ and $P_2 = D^-(D_s^-)$. Numerical results are presented in table~\ref{table:bddds}.
\begin{table}[!t]
 \begin{center}
 \begin{tabular}{|r|ll|}
 \hline
    Model & $\Gamma_{D D D_s}~[10^{-5}~\mathrm{GeV}^5]$
            & $\Gamma_{D_s D_s D}~[10^{-5}~\mathrm{GeV}^5]$\\\hline\hline
    (MS)SM & $1.9\E{-3}\left|C_3^{s,(MS)SM}\right|^2$
             & $3.1\E{-3}\left|C_3^{d,(MS)SM}\right|^2$\\
    RPV & $11 \times \left|C_4^{s,RPV}+\widetilde C_4^{s,RPV}\right|^2 $
         &  $18 \times\left|C_4^{d,RPV}+\widetilde C_4^{d,RPV }\right|^2 $\\
    Z' & $ 1.5  \times \left|C_1^{s,Z'} + \widetilde C_1^{s,Z'}\right|^2  $
        &  $3.3 \times\left|C_1^{d,Z'} + \widetilde C_1^{d,Z'}\right|^2$ \\
    & $ + 1.9\E{-3} \left|C_3^{s,Z'} + \widetilde C_3^{s,Z'}\right|^2$
    & $ +3.1\E{-3} \left|C_3^{d,Z'} + \widetilde C_3^{d,Z'}\right|^2$  \\
    & $ + 0.1 \times \Re\left[\left(C_1^{s,Z'} + \widetilde C_1^{s,Z'}\right)\right.$
    & $ +0.2\times\Re\left[\left(C_1^{d,Z'} + \widetilde C_1^{d,Z'}\right)\right.$\\
     &\hskip1.8cm$\left.\left(C_3^{s,Z'} + \widetilde C_3^{s,Z'}\right)^*\right]$
     &\hskip1.8cm$\left.\left(C_3^{d,Z'} + \widetilde C_3^{d,Z'}\right)^*\right]$  \\\hline
 \end{tabular}
 \end{center}
 \caption{\small\it\label{table:bddds} $B_c^- \to D^- D^- D_s^+$ and $B_c^- \to D_s^- D_s^- D^{+}$ decay  rates in various models and in terms of the relevant Wilson coefficients.}
 \end{table}
These decay  rates are suppressed\index{phase space supression} due to the small phase space in
comparison to the rates of the $B^- \to \pi^- \pi^- K^+$ and $B^- \to K^- K^- \pi^+$ decays\index{decay!$B^- \to \pi^- \pi^- K^+$}\index{decay!$B^- \to K^- K^- \pi^+$}. In numerical analysis we will again and in all following cases neglect all the interference terms appearing in RPV\index{RPV} and $Z'$\index{gauge boson!$Z'$} models.

\subsubsection{$B_c^- \to D^- \pi^- K^+$ and $B_c^- \to D_s^- K^- \pi^+$}
\index{decay!$B_c^- \to D^- \pi^- K^+$}
\index{decay!$B_c^- \to D_s^- K^- \pi^+$}

Here we identify $P_3=D^-(D_s^-)$, $P_1 = \pi^-(K^-)$ and $P_2 = K^+(\pi^+)$ and obtain results in table~\ref{table:bcdkpi}.
\begin{table}[!t]
 \begin{center}
 \begin{tabular}{|r|ll|}
 \hline
    Model & $\Gamma_{D\pi K}~[10^{-3}~\mathrm{GeV}^5]$
            & $\Gamma_{D_s K \pi}~[10^{-3}~\mathrm{GeV}^5]$\\\hline\hline
    (MS)SM & $3.3 \times\left|C_3^{s,(MS)SM}\right|^2$
             & $6.4 \times\left|C_3^{d,(MS)SM}\right|^2$\\
    RPV & $6.5 \times\left|C_4^{s,RPV}+\widetilde C_4^{s,RPV}\right|^2 $
         &  $13 \times\left|C_4^{d,RPV}+\widetilde C_4^{d,RPV }\right|^2 $\\
    Z' & $ 14 \times\left|C_1^{s,Z'} + \widetilde C_1^{s,Z'}\right|^2  $
        &  $29 \times\left|C_1^{d,Z'} + \widetilde C_1^{d,Z'}\right|^2$ \\
    & $ + 3.3\times \left|C_3^{s,Z'} + \widetilde C_3^{s,Z'}\right|^2$
    & $ +6.4\times \left|C_3^{d,Z'} + \widetilde C_3^{d,Z'}\right|^2$  \\
    & $ + 13 \times\Re\left[\left(C_1^{s,Z'} + \widetilde C_1^{s,Z'}\right)\right.$
    & $ +27\times\Re\left[\left(C_1^{d,Z'} + \widetilde C_1^{d,Z'}\right)\right.$\\
     &\hskip1.7cm$\left.\left(C_3^{s,Z'} + \widetilde C_3^{s,Z'}\right)^*\right]$
     &\hskip1.7cm$\left. \left(C_3^{d,Z'} + \widetilde C_3^{d,Z'}\right)^*\right]$  \\\hline
 \end{tabular}
 \end{center}
 \caption{\small\it\label{table:bcdkpi} $B_c^- \to D^- \pi^- K^+$ and $B_c^- \to D_s^- K^- \pi^+$ decay rates in various models and in terms of the relevant Wilson coefficients.}
 \end{table}

\subsubsection{$B_c^- \to K^0 D^0 \pi^{-}$ and  $B_c^- \to \overline K^0 D^0 K^{-}$}
\index{decay!$B_c^- \to K^0 D^0 \pi^{-}$}
\index{decay!$B_c^- \to \overline K^0 D^0 K^{-}$}

This transition only proceeds through the resonance\index{resonance contribution} contribution and eq.~(\ref{eq:resonance}) applies with the identification $P_1=\pi^-(K^-)$, $P_2=D^0$, $P=K^0 (\overline K^0)$ and $V^*=D^{*-}(D_s^{*-})$. We use HM$\chi$PT\index{HM$\chi$PT} eq.~(\ref{eq_4.11}) for the evaluation of the $D^{*-} D^0 \pi^-$ ($D_s^{*-} D^0 K^-$) vertices $G_{D^{*-} D^0 \pi^-}= 2 g \sqrt {m_{D^*}m_D}/f$ ($G_{D_s^{*-} D^0 K^-}= 2 g \sqrt {m_{D_s^*}m_D}/f$). The decay rates are then given in table~\ref{table:bckdk}.
\begin{table}[!t]
 \begin{center}
 \begin{tabular}{|r|ll|}
 \hline
    Model & $\Gamma_{K D\pi}~[10^{-5}~\mathrm{GeV}^5]$
            & $\Gamma_{K D K}~[10^{-5}~\mathrm{GeV}^5]$\\\hline\hline
    (MS)SM & $0.06 \times\left|C_3^{s,(MS)SM}\right|^2$
             & $0.04 \times\left|C_3^{d,(MS)SM}\right|^2$\\
    RPV & $23 \times\left|C_4^{s,RPV}+\widetilde C_4^{s,RPV}\right|^2 $
         &  $14 \times\left|C_4^{d,RPV}+\widetilde C_4^{d,RPV }\right|^2 $\\
    Z' & $ 2.1 \times\left|C_1^{s,Z'} + \widetilde C_1^{s,Z'}\right|^2  $
        &  $1.3 \times\left|C_1^{d,Z'} + \widetilde C_1^{d,Z'}\right|^2$ \\
    & $ + 0.06\times \left|C_3^{s,Z'} + \widetilde C_3^{s,Z'}\right|^2$
    & $ +0.04\times \left|C_3^{d,Z'} + \widetilde C_3^{d,Z'}\right|^2$  \\
    & $ + 0.7 \times\Re\left[\left(C_1^{s,Z'} + \widetilde C_1^{s,Z'}\right)\right.$
    & $ +0.5\times\Re\left[\left(C_1^{d,Z'} + \widetilde C_1^{d,Z'}\right)\right.$\\
     &\hskip1.8cm$\left. \left(C_3^{s,Z'} + \widetilde C_3^{s,Z'}\right)^*\right]$
     &\hskip1.8cm$\left. \left(C_3^{d,Z'} + \widetilde C_3^{d,Z'}\right)^*\right]$  \\\hline
 \end{tabular}
 \end{center}
 \caption{\small\it\label{table:bckdk} $B_c^- \to K^0 D^0 \pi^{-}$ and  $B_c^- \to \overline K^0 D^0 K^{-}$ decay  rates in various models and in terms of the relevant Wilson coefficients.}
 \end{table}
Note that the analogous decay  channels $B_c^- \to K^0 D^- \pi^{0}$ and  $B_c^- \to \overline K^0 D^- \overline K^{0}$ will not be analyzed, since they contain two neutral light mesons in the final state which are notoriously difficult to detect.


\subsubsection{$B_c^- \to D^- K^0$ and $B_c^- \to D_s^- \overline K^0$}
\index{decay!$B_c^- \to D^- K^0$}
\index{decay!$B_c^- \to D_s^- \overline K^0$}

The operators
 $\mc O^s_{1,3}$ and $\widetilde{\mc O}^s_{1,3}$ that are present in SM~(MSSM)\index{MSSM} and $Z'$\index{gauge boson!$Z'$} model obtain contributions in the form given by eq.~(\ref{eq:2bodySM}) with identification $P_1=K^0(\overline K^0)$ and $P_2=D^-(D_s^-)$. Operators $\mc O_4$ and $\mc{\widetilde O}_4$, relevant for the RPV\index{RPV} and
$Z'$\index{gauge boson!$Z'$} models result in expressions of the form~(\ref{eq:RPV2Body}). However, in the latter two models, the two chirally  flipped contributions to the amplitude have opposite signs,
resulting in a slightly different combination of Wilson coefficients\index{Wilson coefficient} (in table~\ref{table:BcDK})
in comparison with the $B^- \to \pi^- \pi^- K^+$ ($B^- \to K^- K^- \pi^+$) decay rates .
\begin{table}[!t]
 \begin{center}
 \begin{tabular}{|r|ll|}
 \hline
    Model & $\Gamma_{DK}~[10^{-3}~\mathrm{GeV}^5]$
            & $\Gamma_{D_s K}~[10^{-3}~\mathrm{GeV}^5]$\\\hline\hline
    (MS)SM & $0.7 \times\left|C_3^{s,(MS)SM}\right|^2$
             & $1.3 \times\left|C_3^{d,(MS)SM}\right|^2$\\
    RPV & $0.6 \times\left|C_4^{s,RPV}-\widetilde C_4^{s,RPV}\right|^2 $
         &  $1.2 \times\left|C_4^{d,RPV}-\widetilde C_4^{d,RPV }\right|^2 $\\
    Z' & $ 1.8 \times\left|C_1^{s,Z'} - \widetilde C_1^{s,Z'}\right|^2  $
        &  $3.4 \times\left|C_1^{d,Z'} - \widetilde C_1^{d,Z'}\right|^2$ \\
    & $ + 0.7\times \left|C_3^{s,Z'} - \widetilde C_3^{s,Z'}\right|^2$
    & $ +1.3\times \left|C_3^{d,Z'} - \widetilde C_3^{d,Z'}\right|^2$  \\
    & $ + 2.2 \times\Re\left[\left(C_1^{s,Z'} - \widetilde C_1^{s,Z'}\right)\right.$
    & $ +4.2\times\Re\left[\left(C_1^{d,Z'} - \widetilde C_1^{d,Z'}\right)\right.$\\
     &\hskip1.8cm$\left. \left(C_3^{s,Z'} - \widetilde C_3^{s,Z'}\right)^*\right]$
     &\hskip1.8cm$\left. \left(C_3^{d,Z'} - \widetilde C_3^{d,Z'}\right)^*\right]$  \\\hline
 \end{tabular}
 \end{center}
 \caption{\small\it\label{table:BcDK} $B_c^- \to D^- K^0$ and  $B_c^- \to D_s^- \overline K^0 $ decay rates in various models and in terms of the relevant Wilson coefficients.}
 \end{table}

\subsubsection{$B_c^- \to D^{*-} K^0$ and $B_c^- \to D_s^{*-} \overline K^0$}
\index{decay!$B_c^- \to D^{*-} K^0$}
\index{decay!$B_c^- \to D_s^{*-} \overline K^0$}

Scenario (II) (eq.~(\ref{eq_BPVb})) applies here with the identification $P=K^0(\overline K^0)$ and $V = D^{*-}(D_s^{*-})$. We sum over polarizations\index{polarization sum} of the $D^*$ meson using
eq.~(\ref{eq:polSum}), and the unpolarized decay  rates are given in table~\ref{table:BcDstarK}.
\begin{table}[!t]
 \begin{center}
 \begin{tabular}{|r|ll|}
 \hline
    Model & $\Gamma_{D^*K}~[10^{-3}~\mathrm{GeV}^5]$
            & $\Gamma_{D^*_s K}~[10^{-3}~\mathrm{GeV}^5]$\\\hline\hline
    (MS)SM & $0.7 \times\left|C_3^{s,(MS)SM}\right|^2$
             & $1.2 \times\left|C_3^{d,(MS)SM}\right|^2$\\
    RPV & $0.6 \times\left|C_4^{s,RPV}+\widetilde C_4^{s,RPV}\right|^2 $
         &  $1.0 \times\left|C_4^{d,RPV}+\widetilde C_4^{d,RPV }\right|^2 $\\
    Z' & $ 1.8 \times\left|C_1^{s,Z'} + \widetilde C_1^{s,Z'}\right|^2  $
        &  $3.2 \times\left|C_1^{d,Z'} + \widetilde C_1^{d,Z'}\right|^2$ \\
    & $ + 0.7\times \left|C_3^{s,Z'} + \widetilde C_3^{s,Z'}\right|^2$
    & $ +1.2\times \left|C_3^{d,Z'} + \widetilde C_3^{d,Z'}\right|^2$  \\
    & $ + 2.2 \times\Re\left[\left(C_1^{s,Z'} + \widetilde C_1^{s,Z'}\right)\right.$
    & $ +3.9\times\Re\left[\left(C_1^{d,Z'} + \widetilde C_1^{d,Z'}\right)\right.$\\
     &\hskip1.8cm$\left. \left(C_3^{s,Z'} + \widetilde C_3^{s,Z'}\right)^*\right]$
     &\hskip1.8cm$\left. \left(C_3^{d,Z'} + \widetilde C_3^{d,Z'}\right)^*\right]$  \\\hline
 \end{tabular}
 \end{center}
 \caption{\small\it\label{table:BcDstarK} $B_c^- \to D^{*-} K^0$ and  $B_c^- \to D_s^{*-} \overline K^0 $ decay rates in various models and in terms of the relevant Wilson coefficients.}
 \end{table}


\subsubsection{$B_c^- \to D^- K^{*0}$ and $B_c^- \to D_s^- \overline K^{0*}$}
\index{decay!$B_c^- \to D^- K^{*0}$}
\index{decay!$B_c^- \to D_s^- \overline K^{0*}$}

Factorized matrix element is here of type~(\ref{eq_BPV}) (I) with the identification $V=K^{*0}(\overline K^{*0})$ and $P=D^-(D_s^-)$. The density operators
 $\mc O^s_4$ and $\widetilde{\mc O}^s_4$ do not
contribute and consequently in the RPV\index{RPV} model this mode is dominated by the operators $\mc O^s_5$ and $\widetilde{\mc O}^s_5$ which are, as mentioned in Section~\ref{sec:suppressedO5}, suppressed\index{RGE!suppression} by the renormalization group running. Using Fierz rearrangements, we
write them down as $\mc O_1$, $\widetilde{\mc O}_1$ and yield an additional
$1/2$ suppression factor. Results are presented in table~\ref{table:BcDKstar}.
\begin{table}[!t]
 \begin{center}
 \begin{tabular}{|r|ll|}
 \hline
    Model & $\Gamma_{DK^*}~[10^{-3}~\mathrm{GeV}^5]$
            & $\Gamma_{D_s K^*}~[10^{-3}~\mathrm{GeV}^5]$\\\hline\hline
    (MS)SM & $1.2 \times\left|C_3^{s,(MS)SM}\right|^2$
             & $2.3 \times\left|C_3^{d,(MS)SM}\right|^2$\\
    RPV & $4.7\E{-2} \times\left|C_4^{s,RPV}+\widetilde C_4^{s,RPV}\right|^2 $
         &  $9.1\E{-2} \times\left|C_4^{d,RPV}+\widetilde C_4^{d,RPV }\right|^2 $\\
    Z' & $ 4.8 \times\left|C_1^{s,Z'} + \widetilde C_1^{s,Z'}\right|^2  $
        &  $9.1 \times\left|C_1^{d,Z'} + \widetilde C_1^{d,Z'}\right|^2$ \\
    & $ + 1.2\times \left|C_3^{s,Z'} + \widetilde C_3^{s,Z'}\right|^2$
    & $ +2.3\times \left|C_3^{d,Z'} + \widetilde C_3^{d,Z'}\right|^2$  \\
    & $ + 4.8 \times\Re\left[\left(C_1^{s,Z'} + \widetilde C_1^{s,Z'}\right)\right.$
    & $ +9.1\times\Re\left[\left(C_1^{d,Z'} + \widetilde C_1^{d,Z'}\right)\right.$\\
     &\hskip1.8cm$\left. \left(C_3^{s,Z'} + \widetilde C_3^{s,Z'}\right)^*\right]$
     &\hskip1.8cm$\left. \left(C_3^{d,Z'} + \widetilde C_3^{d,Z'}\right)^*\right]$  \\\hline
 \end{tabular}
 \end{center}
 \caption{\small\it\label{table:BcDKstar} $B_c^- \to D^{-} K^{*0}$ and  $B_c^- \to D_s^{-} \overline K^{*0} $ decay rates in various models and in terms of the relevant Wilson coefficients.}
 \end{table}

\subsubsection{$B_c^- \to D^{*-} K^{*0}$ and $B_c^- \to D_s^{*-} \overline K^{*0}$}
\index{decay!$B_c^- \to D^{*-} K^{*0}$}
\index{decay!$B_c^- \to D_s^{*-} \overline K^{*0}$}

Like in the previous case, this mode only receives contributions from
the RGE\index{RGE!suppression} suppressed RPV\index{RPV} terms. We calculate
unpolarized hadronic amplitudes of the operators
 $\mc O^s_{1,3}$ and
$\widetilde{\mc O}^s_{1,3}$ by utilizing the helicity amplitudes\index{helicity amplitudes} formalism.
Using form factor\index{form factor!decomposition} decomposition~(\ref{eq_3_19b},
\ref{eq:PtoV}), we write down the expression for the polarized
amplitude~(\ref{eq:helicityDecomposition}) and identify
constants $a$, $b$ and $c$:
\begin{subequations}
\begin{align}
a&=-\frac{i}{4}(m_{B_c}+m_{D_{(s)}^*})g_{K^*} A_1^{{D_{(s)}^*} {B_c}} (m_{K^*}^2)(C-\widetilde C),\\
b&=\frac{i}{2} \frac{m_{K^*}m_{D_{(s)}^*}}{m_{B_c}+m_{D_{(s)}^*}}g_{K^*} A_2^{{D_{(s)}^*} {B_c}}(m_{K^*}^2)(C -\widetilde C),\\
c&=-\frac{i}{2} \frac{m_{K^*}m_{D_{(s)}^*}}{m_{B_c}+m_{D_{(s)}^*}}g_{K^*} V^{{D_{(s)}^*} {B_c}}(m_{K^*}^2)(C +\widetilde C).
\end{align}
\end{subequations}
$C$ and $\widetilde C$ are combinations of the Wilson coefficients\index{Wilson coefficient} present
in a considered model. We have $C=C_3^{s,(MS)SM}$, $\widetilde C=0$ in the
SM~(MSSM)\index{MSSM}, $C=-\widetilde{f}_{QCD}(m_b) C_4^{s,RPV}/2$, $\widetilde
C=-\widetilde{f}_{QCD}(m_b) \widetilde C_4^{s,RPV}/2$ in the case of the RPV\index{RPV}
model and $C=f_{QCD}(m_b) C_1^{s,Z'} + f_{QCD}'(m_b) C_3^{s,Z'}$, $\widetilde
C = f_{QCD}(m_b) \widetilde C_1^{s,Z'} + f_{QCD}'(m_b) \widetilde C_3^{s,Z'}$ in
the $Z'$\index{gauge boson!$Z'$} model. Decay  rates are then given in table~\ref{table:BcDsKs}.
\begin{table}[!t]
 \begin{center}
 \begin{tabular}{|r|ll|}
 \hline
    Model & $\Gamma_{D^*K^*}~[10^{-4}~\mathrm{GeV}^5]$
            & $\Gamma_{D^*_s K^*}~[10^{-4}~\mathrm{GeV}^5]$\\\hline\hline
    (MS)SM & $0.9 \times\left|C_3^{s,(MS)SM}\right|^2$
             & $1.6 \times\left|C_3^{d,(MS)SM}\right|^2$\\
    RPV & $3.3\E{-2} \times\left|C_4^{s,RPV}-\widetilde C_4^{s,RPV}\right|^2 $
         &  $6.0\E{-2} \times\left|C_4^{d,RPV}-\widetilde C_4^{d,RPV }\right|^2 $\\
         &  $4.4\E{-3} \times\left|C_4^{s,RPV}+\widetilde C_4^{s,RPV }\right|^2 $
         &  $6.0\E{-3} \times\left|C_4^{d,RPV}+\widetilde C_4^{d,RPV }\right|^2 $\\
    Z' & $ 3.3 \times\left|C_1^{s,Z'} - \widetilde C_1^{s,Z'}\right|^2  $
        &  $6.0 \times\left|C_1^{d,Z'} - \widetilde C_1^{d,Z'}\right|^2$ \\
        & $ 0.4 \times\left|C_1^{s,Z'} + \widetilde C_1^{s,Z'}\right|^2  $
        &  $0.6 \times\left|C_1^{d,Z'} + \widetilde C_1^{d,Z'}\right|^2$ \\
    & $ + 0.8\times \left|C_3^{s,Z'} - \widetilde C_3^{s,Z'}\right|^2$
    & $ +1.5\times \left|C_3^{d,Z'} - \widetilde C_3^{d,Z'}\right|^2$  \\
    & $ + 0.1\times \left|C_3^{s,Z'} + \widetilde C_3^{s,Z'}\right|^2$
    & $ +0.2\times \left|C_3^{d,Z'} + \widetilde C_3^{d,Z'}\right|^2$  \\
    & $ + 3.3 \times\Re\left[\left(C_1^{s,Z'} - \widetilde C_1^{s,Z'}\right)\right.$
    & $ +6.0\times\Re\left[\left(C_1^{d,Z'} - \widetilde C_1^{d,Z'}\right)\right.$\\
     &\hskip1.8cm$\left. \left(C_3^{s,Z'} - \widetilde C_3^{s,Z'}\right)^*\right]$
     &\hskip1.8cm$\left. \left(C_3^{d,Z'} - \widetilde C_3^{d,Z'}\right)^*\right]$  \\
    & $ + 0.4 \times\Re\left[\left(C_1^{s,Z'} + \widetilde C_1^{s,Z'}\right)\right.$
    & $ +0.6\times\Re\left[\left(C_1^{d,Z'} + \widetilde C_1^{d,Z'}\right)\right.$\\
     &\hskip1.8cm$\left. \left(C_3^{s,Z'} + \widetilde C_3^{s,Z'}\right)^*\right]$
     &\hskip1.8cm$\left. \left(C_3^{d,Z'} + \widetilde C_3^{d,Z'}\right)^*\right]$  \\\hline
 \end{tabular}
 \end{center}
 \caption{\small\it\label{table:BcDsKs} $B_c^- \to D^{*-} K^{*0}$ and  $B_c^- \to D_s^{*-} \overline K^{*0} $ decay rates in various models and in terms of the relevant Wilson coefficients.}
 \end{table}
In numerical analysis we shall neglect mixing\index{mixing!of operators} terms between the chirally  flipped Wilson
coefficients\index{Wilson coefficient} in the RPV\index{RPV} and the $Z'$\index{gauge boson!$Z'$} models and also omit the last two
terms in the $Z'$\index{gauge boson!$Z'$} model decay  rate.

\section{Constraining new physics\label{sec:discussion}}
\index{new physics!constraints}
\index{new physics!in $B_c$ decays}

The usefulness of $\Delta S = 2$ decays\index{$\Delta S=2$ transitions}\index{decay of $B$ meson}\index{meson!$B$!decay} of $B$ mesons in the
search for new physics\index{new physics!searches} has been discussed in several
publications~\cite{Huitu:1998vn,Huitu:1998pa,Grossman:1999av,Fajfer:2001ht,Chun:2003rg,Wu:2003kp,Fajfer:2000ax,Fajfer:2000ny}. From the models considered so far it appears that these decays  are particularly relevant in the search for SUSY, with and without $\mathcal R$-parity violation.
\par
The results obtained in the MSSM\index{MSSM} framework depend on the values of the $\delta_{ij}^{d}$ parameters of the mass-insertion approximation which we use. The constraints on these parameters have been improved in recent years~\cite{Ciuchini:2003rg,Ciuchini:2002uv} vs. the values which were used in the first calculation~\cite{Huitu:1998vn} of the $b \to s s \bar d$\index{transition!$b\to d d \bar s$}\index{transition!$b\to s s \bar d$}.
For the RPV\index{RPV} MSSM\index{MSSM} and the $Z'$\index{gauge boson!$Z'$} model may obtain the stringest limits on the effective couplings using the experimental upper limits on the
$B^+ \to K^+ K^+ \pi^-$\index{decay!$B^+ \to K^+ K^+ \pi^-$} and $B^{+} \to \pi^+ \pi^+ K^-$\index{decay!$B^+ \to \pi^+ \pi^+ K^-$} decay  rates from Belle\index{Belle}~\cite{Garmash:2003er}. For this purpose we use the explicit calculation of
these decay  rates in Table~\ref{table:bpipik}. Normalizing the masses of
sneutrinos\index{sneutrino} to a common mass scale of $100\e{GeV}$ we derive bounds on
the RPV\index{RPV} terms given in eq.~(\ref{eq:RPVcouplings})
\begin{subequations}
\begin{eqnarray}
\label{eq:par-R1}
  \left| \sum_{n=1}^3 \left(\frac{100~\mathrm{GeV}}{m_{\widetilde\nu_n}}\right)^2 \left(\lambda_{n31}' \lambda_{n12}'^* + \lambda_{n21}' \lambda_{n13}'^*\right)\right| &<& 9.5\E{-5},\\
  \left| \sum_{n=1}^3 \left(\frac{100~\mathrm{GeV}}{m_{\widetilde\nu_n}}\right)^2 \left(\lambda_{n32}' \lambda_{n21}'^* + \lambda_{n21}' \lambda_{n13}'^*\right)\right| &<& 9.5\E{-5}.
\label{eq:par-R}
\end{eqnarray}
\end{subequations}
Assuming that new physics\index{new physics!in $B_c$ decays} arises due to an extra $Z'$\index{gauge boson!$Z'$} gauge boson\index{gauge boson} we
derive bounds on the parameters given in
Eq.~(\ref{eq:zPrimeOperators}).  We neglect interference between
Wilson coefficients\index{Wilson coefficient}, namely the last lines in Table~\ref{table:bpipik}. Experimental bound of this simplified
expression now confines $\left(|C_1^{q,Z'}+\widetilde C_1^{q,Z'}|,
  |C_3^{q,Z'}+\widetilde C_3^{q,Z'}|\right)$ to lie within an ellipse with
semiminor and semimajor axes as upper limits
\begin{subequations}
  \begin{align}
    y^2 \left| B_{12}^{s_L}\, B_{13}^{s_R}+
      B_{12}^{s_R}\,  B_{13}^{s_L}\right| &<  2.7\E{-4},\\
    y^2 \left| B_{12}^{s_L}\, B_{13}^{s_L}+ B_{12}^{s_R}\,
      B_{13}^{s_R}\right| &< 5.6\E{-4},
\end{align}
\end{subequations}
and
\begin{subequations}
  \begin{align}
    y^2 \left| B_{21}^{d_L}\, B_{23}^{d_R}+
      B_{21}^{d_R}\,  B_{32}^{d_L}\right| &<  2.4\E{-4},\\
    y^2 \left| B_{21}^{d_L}\, B_{32}^{d_L}+ B_{21}^{d_R}\,
      B_{32}^{d_R}\right| &< 5.3\E{-4}.
  \label{eq:par-Z}
\end{align}
\end{subequations}
The bounds (\ref{eq:par-R1}-\ref{eq:par-Z}) are interesting since they
offer an independent way of constraining the particular combination of
the parameters, which are not constrained by the $B^0_d - \overline B^0_d$,
$B^0_s-\overline B^0_s$, $K^0-\overline K^0$ oscillations\index{oscillations!$B^0_d-\overline B^0_d$}\index{oscillations!$B^0_s-\overline B^0_s$}\index{meson!$B$!oscillations}\index{oscillations!$K^0-\overline K^0$}\index{meson!$K$!oscillations} or $b\to s\gamma$ decay \index{transition!$b\to s\gamma$} rates (see e.g.~\cite{Silvestrini:2005zb}).

\par

Using these inputs we predict the branching ratios  for the various
possible two- and three-body decay\index{decay of $B_c$ meson} modes of the $B_c$\index{meson!$B_c$!decay}.
The results are summarized in Table~\ref{tab1}.
\begin{table}[!t]
\begin{center}
\begin{tabular}{|l|cccc|}
  \hline
  Decay  & SM & MSSM & RPV & $Z'$ \\\hline
  \hline
  $B_c^- \to D^- D^- D_s^+$ & $1\E{-21}$ & $5\E{-20}$ & $7\E{-9}$ & $9\E{-10}$\\
  $B_c^- \to D_s^- D_s^- D^+$ & $4\E{-19}$ & $5\E{-19}$ & $1\E{-8}$ & $1\E{-9}$\\
  $B_c^- \to D^- K^+ \pi^-$ & $2 \times 10^{-16}$ & $5\times 10^{-15}$ & $4 \times 10^{-7}$ & $2\E{-6}$ \\
  $B_c^- \to D_s K^- \pi^{+}$ & $7 \times 10^{-14}$ & $1\times 10^{-13}$ & $8 \times 10^{-7}$ & $3\E{-6}$ \\
  $B_c^- \to \overline D^0 \pi^- K^0$ & $4 \times 10^{-20}$ & $2\times 10^{-18}$ & $2 \times 10^{-8}$ & $1\times 10^{-9}$ \\
  $B_c^- \to \overline D^0 K^- \overline K^0$ & $4 \times 10^{-18}$ & $7\times 10^{-18}$ & $9 \times 10^{-9}$ & $6\times 10^{-10}$ \\
  $B_c^- \to D^- K^0$ & $4\E{-17}$ & $2\E{-15}$ & $4\E{-8}$ & $3\E{-7}$\\
  $B_c^- \to D_s^- \overline K_0$ & $1 \times 10^{-14}$ & $2 \times 10^{-14}$ & $7 \times 10^{-8}$ & $4 \times 10^{-7}$ \\
  $B_c^- \to D^{*-} K^0$ & $4\E{-17}$ & $2\E{-15}$ & $4\E{-8}$ & $3\E{-7}$\\
  $B_c^- \to D_s^{*-} \overline K_0$ & $1 \times 10^{-14}$ & $2\times 10^{-14}$ & $6 \times 10^{-8}$ & $4 \times 10^{-7}$ \\
  $B_c^- \to D^- K^{*0}$ & $8\E{-17}$ & $3\E{-15}$ & $3\E{-9}$ & $5\E{-7}$\\
  $B_c^- \to D_s^- \overline K_0^*$ & $3 \times 10^{-14}$ & $4 \times 10^{-14}$ & $6\E{-9}$ & $9\E{-7}$ \\

  $B_c^- \to D^{*-} K^{*0}$ & $6\E{-18}$ & $3\E{-16}$ & $2\E{-10}$ & $4\E{-8}$\\
  $B_c^- \to D_s^{*-} \overline K_0^*$ & $2 \times 10^{-15}$ & $3\times 10^{-15}$ & $4\E{-10}$ & $5\E{-8}$ \\
  \hline
\end{tabular}
\end{center}
\caption{\small\it The branching ratios  for the $\Delta S= -1$ and $\Delta S= 2$ decays of the
  $B_c^-$ meson calculated within SM, MSSM, RPV\index{RPV} and $Z'$ \index{gauge boson!$Z'$}models.  The
  experimental upper bounds for the $BR(B^- \to \pi^- \pi^- K^+)$ $<
  1.8 \times 10^{-6}$ and $BR(B^- \to K^- K^- \pi^+)$ $<
  2.4 \times 10^{-6}$ have been used as an input parameters to fix the
  unknown combinations of the RPV terms (IV column) and the model with
  an additional $Z'$ boson (V column).
}
\label{tab1}
\end{table}
The SM and MSSM\index{MSSM} give negligible contributions. Using constraints for the
particular combination of the RPV\index{RPV} parameters present in the $B^- \to
\pi^- \pi^- K^+$ and $B^- \to K^- K^- \pi^+$ decays\index{decay!$B^- \to
\pi^- \pi^- K^+$}\index{decay!$B^- \to K^- K^- \pi^+$} we obtain the largest possible branching ratios \index{meson!$B_c$!decay}
for the three-body decays
$B_c^- \to D^- K^+ \pi^-$\index{decay!$B_c^- \to D^- K^+ \pi^-$} and
$B_c^- \to D_s^- K^- \pi^+$\index{decay!$B_c^- \to D_s^- K^- \pi^+$},
and two-body decays of
$B_c^- \to D^- K^0$\index{decay!$B_c^- \to D^- K^0$},
$B_c^- \to D_s^- \overline K^0$\index{decay!$B_c^- \to D_s^- \overline K^0$},
$B_c^- \to D^{*-} K^0$\index{decay!$B_c^- \to D^{*-} K^0$} and
$B_c^- \to D_s^{*-} \overline K^0$\index{decay!$B_c^- \to D_s^{*-} \overline K^0$},
while for the $B_c^- \to D^- K^{*0}$\index{decay!$B_c^- \to D^- K^{*0}$} and $B_c^- \to D^{*-} K^{*0}$\index{decay!$B_c^- \to D^{*-} K^{*0}$}
the RPV\index{RPV} contribution is suppressed\index{RGE!suppression} by renormalization group running\index{RGE!running of operators}.
Their order of magnitude is $10^{-9}$ and thus still experimentally
unreachable. However, these two decay  channels are besides the ones already mentioned, most likely to be
observed in the model with an additional $Z'$\index{gauge boson!$Z'$} boson, if we assume that
interference terms are negligible.

\par

Since in the experimental measurements only $K_S$ or $K_L$ are
detected and not $K^0$ or $\overline K^0$, it might be difficult to observe
new physics\index{new physics!in $B_c$ decays} in decay modes containing neutral final state kaons\index{meson!$K$} due to pollution of SM penguin dominated decays\index{penguin dominated decays}~\cite{Grossman:1999av}. Therefore, the
decay modes with charged kaons\index{meson!$K$} as well as $K^{*0}$ or $\overline K^{*0}$ in the final state seem to be
better candidates for the experimental searches of new physics\index{new physics!searches} in the
$b \to d d \bar s$ and $b \to s s \bar d$\index{transition!$b\to d d \bar s$}\index{transition!$b\to s s \bar d$} transitions.

\par

In our calculation we have relied on the na\"ive
factorization approximation\index{VSA}, which is as a first approximation
sufficient to obtain correct gross features of new physics\index{new physics!contributions} effects.
One might think that the nonfactorisable contributions might induce
large additional uncertainties, but we do not expect them to change
the order of magnitude of our predictions.  However, since in SM the
basic decays\index{decay of $b$ quark}\index{transition!$b\to s s \bar d$}\index{transition!$b\to d d \bar s$} $b \to ss \bar d$ and $b \to dd \bar s$\index{transition!$b\to d d \bar s$}\index{transition!$b\to s s \bar d$} have branching ratios  of the  order $10^{-12}-10^{-14}$ and one expects that the rates for exclusive decays should be even smaller, the gap between this and the predictions of beyond SM\index{new physics!in $B_c$ decays} is so large, that it makes the
search for these modes a useful tool. Additional uncertainties
might originate in the poor knowledge of the input parameters such as
form factors\index{form factor}. However, we do not expect these to invalidate our order of magnitude estimates. All these decays  should be looked for, when sizable samples of $B_c$'s\index{meson!$B_c$} will be available.

\chapter{Concluding Remarks}
\index{conclusions}
\index{summary}

The nonperturbative nature of QCD is a persisting problem of calculations in hadronic physics. One of its manifestations is the appearance of resonances\index{resonance} in the hadronic spectrum. In processes where the exchanged momenta are small compared to the chiral symmetry breaking scale\index{chiral symmetry breaking scale} $\sim 1~\mathrm{GeV}$, one may employ the effective theory approach based on the approximate chiral symmetry\index{chiral symmetry} of light quarks and the approximate spin-flavor symmetry\index{heavy quark symmetry} of heavy quarks (both compared to the chiral scale\index{chiral symmetry breaking scale}). In such framework, the impacts of the nearest resonances\index{resonance contribution!in processes of heavy mesons} in the processes of heavy mesons can be systematically studied.

\index{spin symmetry|see{heavy quark symmetry}}
\index{heavy flavor symmetry|see{heavy quark symmetry}}
\index{heavy quark symmetry}

\par

The HM$\chi$PT\index{HM$\chi$PT} has been applied to strong, semileptonic and rare processes of heavy mesons. The lowest lying positive and negative parity heavy meson multiplets were included systematically into the framework at leading order in heavy quark and at the next to leading order in the chiral expansion\index{chiral expansion}.

\par

At LO it was found that the nearby heavy meson excited states may help explain certain features of the heavy-to-light semileptonic form factors\index{form factor!heavy-to-light} . Namely, using a constrained form factor parameterization based on approximate effective theory limits, it was possible to saturate the whole tower of intermediate states beyond $t$-channel production threshold with just the nearest resonances\index{resonance} of suitable quantum numbers. The parametrization was matched\index{matching} onto HM$\chi$PT\index{HM$\chi$PT} calculation at small momentum exchanges, were predictions were most reliable. Such model reproduced most $H\to P$ and $H\to V$ form factor\index{form factor!$H\to P$}\index{form factor!$H\to V$} shapes successfully within current experimental errors and compatible with existing lattice QCD\index{lattice QCD} calculations.

\par

In other processes considered, the excited heavy meson resonances\index{resonance!of heavy meson} contribute only at the NLO in HM$\chi$PT\index{HM$\chi$PT} through chiral\index{loop corrections} loop corrections. Considering strong decays\index{strong decay} of heavy measons, the effective strong couplings between pairs of heavy positive or negative parity mesons and light pseudoscalar mesons were calculated at NLO in chiral\index{chiral expansion} expansion. From the measured $D^*\to D\pi$ and $D'_0\to D \pi$ decay \index{decay!$D^* \to D \pi$} rates the LO effective couplings were extracted. The effects of the large number of unknown counterterms entering NLO calculation were estimated by varying the renormalization scale and by scanning the parameter space with the experimental fit. Then the chiral\index{chiral extrapolation} extrapolation of the couplings was studied in limit where the light pseudoscalar masses tend to zero. It was found that in the naive calculation of chiral\index{loop corrections} loop corrections involving excited heavy states, the chiral\index{chiral limit} limit is ill-defined. One can instead perform an expansion in the inverse mass splitting between the ground and excited heavy meson states to recover a well behaved chiral limit\index{chiral limit}. Such expansion is reliable for light pseudoscalar masses, smaller then the heavy meson parity splitting scale. Then the effects of excited heavy mesons are formally expressed as higher order chiral\index{loop corrections} corrections to a theory without dynamical excited states. The result is especially important for lattice QCD studies\index{lattice QCD} where chiral\index{chiral extrapolation} extrapolation is used in order to reach the physical limit of light quark masses used in simulations. It means that the relevant chiral\index{chiral limit}\index{chiral symmetry} symmetry limit for such expansions is the $SU(2)$ isospin limit and that chiral\index{chiral expansion} expansions may only be reliable for pion masses smaller then the heavy meson parity mass splitting. At the same time the reliability of the leading log order extrapolations in this limit can be estimated using the leading higher order contributions due to excited states.

\par

The decoupling of excited resonances\index{resonance decoupling} and their leading order effects was probed also in the case of heavy-to-heavy semileptonic form factors\index{form factor!heavy-to-light} were the chiral corrections\index{chiral corrections} to Isgur-Wise functions in weak transitions among heavy mesons of both parities were calculated. The very accurate determination of the decay  rates from experiments and the form factors from lattice QCD\index{lattice QCD} requires detailed knowledge of the chiral limit\index{chiral limit} in order to extract CKM\index{CKM!matrix elements} matrix element $V_{bc}$. It was found that the effects of the excited heavy meson resonances\index{resonance!of heavy meson} may be comparable in size to current theoretical error estimates and therefore should be taken into account in future studies.

\par

The prime interest in the rare heavy meson processes is the search for new physics signatures beyond the SM\index{new physics}. But in order to be successful, hadronic effects have to be well understood and under control. For this purpose the chiral behavior was studied for the full SUSY\index{SUSY!basis of $\Delta B = 2$ operators} basis of effective $\Delta B=2$ operators, which mitigate oscillations\index{oscillations!of heavy neutral mesons} of heavy neutral mesons. Chiral\index{loop corrections}\index{chiral loop corrections|see{loop corrections}} loop corrections were calculated in the NLO in the chiral\index{chiral expansion} and LO in heavy quark expansion including effects of positive parity heavy mesons. The decoupling of the excited states was confirmed and the leading log order chiral extrapolation\index{chiral extrapolation} formulae for the whole basis were given, to be used by future lattice QCD studies\index{lattice QCD} of these transitions. As an auxiliary result also the leading chiral log corrections to the positive parity heavy meson decay\index{decay constant} constants were calculated.

\par

Finally the very rare $b\to ss\bar d$ and $b \to dd\bar s$ transitions\index{transition!$b\to d d \bar s$}\index{transition!$b\to s s \bar d$} of the $B_c$\index{meson!$B_c$} meson were evaluated in the effective theory approach. The hadronic decay  amplitudes were estimated using factorization\index{VSA} and resonance\index{resonance saturation approximation} saturation approximations. The transitions were analyzed in several new physics\index{new physics} models. Based on existing experimental limits on $B\to K K \pi$ and $B \to \pi \pi K$ decay  rates the relevant new physics parameter combinations could be constrained. Finally, based on these limits the most promising two- and three-body nonleptonic decays of the $B_c$\index{meson!$B_c$!decay} meson were identified, were signals of the rare transitions could be searched for in future colliders.

\par

To obtain our results, several technical details had to be resolved as well. The complete set of NLO counterterms contributing to strong transitions among heavy positive and negative parity heavy mesons, and light pseudoscalar mesons had to be identified. The inclusion of excited heavy meson states also spoiled the chiral\index{chiral limit} limit of the leading log order calculations. The issue was resolved using a truncated loop\index{loop integral!expansion} integral expansion in the inverse powers of the heavy meson parity mass splitting, which however reduced the scale of validity of HM$\chi$PT  calculations. In $H\to P,V$ transitions the HQET\index{HQET!static limit} and SCET\index{SCET} limits had to be correctly reproduced in order to obtain a valid form factor\index{form factor!parameterization} parameterization. Also, the bases of QCD\index{QCD} and HQET\index{HQET!matching to QCD} form factors had to be matched\index{matching} correctly and identified with the results of the HM$\chi$PT  calculation. It was found that only such correct matching\index{matching} faithfully reproduces resonances\index{resonance contribution} contributions of correct quantum numbers the the form factors\index{form factor!resonance contributions}. Also, in order to reproduce the pole structure of the form factor\index{form factor!parameterization} parameterizations, heavy meson radial excitations had to be introduced into HM$\chi$PT\index{HM$\chi$PT}. In the calculation of chiral\index{chiral corrections} corrections to the heavy meson mixing operators\index{mixing!of heavy neutral mesons} a correct operator bosonization\index{bosonization!of mixing amplitudes} prescription had to be identified. It turns out that the large general basis of HM$\chi$PT operators contributing to the matching\index{matching} can be greatly reduced using heavy quark spin symmetry\index{heavy quark symmetry} and $4 \times 4$ matrix identities. Similarly in $b\to ss\bar d$ and $b \to dd\bar s$ transitions\index{transition!$b\to d d \bar s$}\index{transition!$b\to s s \bar d$}, a complete basis of quark operators and their LO RGE running and mixing\index{mixing!of operators} had to identified in order to have control over leading order QCD\index{QCD corrections} corrections in the UV. Finally, several hadronic amplitudes entering two- and three body nonleptonic decays of the $B_c$\index{meson!$B_c$!decay} meson required HM$\chi$PT input calculations including light vector and scalar meson contributions and correct ressonace saturation prescriptions in order to yield sensible phenomenological results.

\appendix

\chapter{HM$\chi$PT Feynman rules}
\index{HM$\chi$PT!Feynman rules}

In deriving the Feynman rules from the leading order HM$\chi$PT\index{HM$\chi$PT!Lagrangian} Lagrangian~(\ref{eq_2_13}) we set the overall heavy quark mass scale to a common scale for all processes and states under study inducing a mass gap $\Delta_S$ ($\Delta_H$) terms in the propagators of the positive (negative) parity doublet states due to the relevant residual mass counterterms\index{counterterms} in the Lagrangian~(\ref{eq_L_1}). The same approach could be taken with regards to the chiral\index{chiral symmetry breaking} symmetry breaking contributions, which also induce mass gaps $\Delta_a$ in the heavy meson propagators due to relevant $\mathcal O(m_q)$ counterterm\index{counterterms} contributions in Lagrangian~(\ref{eq_L_1}). However, their non-analyitic contributions to the chiral\index{chiral corrections} corrections are of higher order in the chiral\index{chiral power counting} power counting and we can safely neglect them in our calculations. Likewise, we neglect hyper-fine splittings within individual spin-parity heavy meson doublets. These are degenerate at zeroth order in the $1/m_H$ expansion at which we are working due to heavy quark spin symmetry\index{heavy quark symmetry}.

\par

Following is a list of derived Feynman rules used in the calculations in the text. The standard $+i0$ -- prescriptions are implicitly understood in the propagators.
\psfrag{P}[cc]{$\Red{P_a(v)}$}
\psfrag{Pa}[bc]{$\Red{P_a(v)}$}
\psfrag{Pstarb}[bc]{$\Red{P^*_b(v)}$}
\psfrag{Pstara}[bc]{$\Red{P^{*\mu}_a(v)}$}
\psfrag{Pstar}[cc]{$\Red{P^*_a(v)}$}
\psfrag{Sb}[bc]{$\Red{P_{0b}(v)}$}
\psfrag{S}[cc]{$\Red{P_{0a}(v)}$}
\psfrag{Sstar}[cc]{$\Red{P^*_{1a}(v)}$}
\psfrag{Sstarb}[bc]{$\Red{P^{*\nu}_{1 b}(v)}$}
\psfrag{Sa}[bc]{$\Red{P_{0 a}(v)}$}
\psfrag{Sstara}[bc]{$\Red{P^{*\mu}_{1 a}(v)}$}
\begin{longtable}{rlcl}
$P_a$ propagator: &
\psfrag{Ha}[cc]{$\Red{P_a(v)}$}
\epsfxsize3cm\epsffile{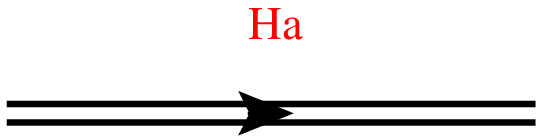}
& $=$ & $\frac{i}{2(k\cdot v - \Delta_H -\Delta_a)}$ \\
&&\\
$P^*_a$ propagator: &
\psfrag{Ha}[cc]{$\Red{P^*_a(v)}$}
\epsfxsize3cm\epsffile{propagator_P.eps}
 & $=$ & $\frac{-i(g^{\mu\nu}-v^{\mu}v^{\nu})}{2(k\cdot v - \Delta_H -\Delta_a)}$ \\
&&\\
$P_{0a}$ propagator: &
\psfrag{Ha}[cc]{$\Red{P_{0a}(v)}$}
\epsfxsize3cm\epsffile{propagator_P.eps}
& $=$ & $\frac{i}{2(k\cdot v-\Delta_S-\widetilde \Delta_a)}$ \\
&&\\
$P^*_{1a}$ propagator: &
\psfrag{Ha}[cc]{$\Red{P^*_{1a}(v)}$}
\epsfxsize3cm\epsffile{propagator_P.eps}
 & $=$ & $\frac{-i(g^{\mu\nu}-v^{\mu}v^{\nu})}{2(k\cdot v -\Delta_S - \widetilde \Delta_a)}$ \\
&&\\
$\pi^i$ propagator: & \includegraphics{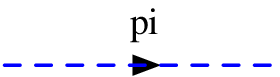} & $=$ & $\frac{i}{k^2-m_i^2}$ \\
&&\\
$P_a P_b^*\pi^i$ coupling: &
\psfrag{Ha}[bc]{$\Red{P_a(v)}$}
\psfrag{Hstarb}[bc]{$\Red{P^*_b(v)}$}
\epsfxsize3cm\epsffile{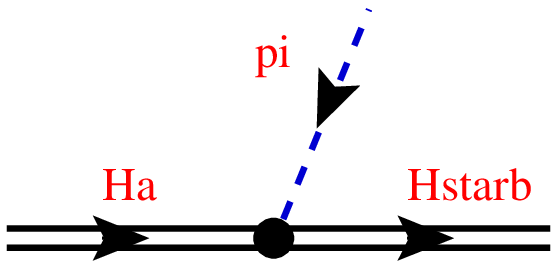}
& $=$ & $\frac{2g}{f} k^{\nu} \lambda^i_{ab}$ \\
&&\\
$P^*_a P^*_b \pi^i$ coupling: &
\psfrag{Ha}[bc]{$\Red{P^*_a(v)}$}
\psfrag{Hstarb}[bc]{$\Red{P^*_b(v)}$}
\epsfxsize3cm\epsffile{vertex_PpiP.eps}
 & $=$ & $\frac{2 i g}{f} \epsilon^{\mu\nu\alpha\beta} k^{\alpha} v_{\beta} \lambda^i_{ab}$ \\
&&\\
$P_{0a} P^*_{1b}\pi^i$ coupling: &
\psfrag{Ha}[bc]{$\Red{P_{0a}(v)}$}
\psfrag{Hstarb}[bc]{$\Red{P^*_{1b}(v)}$}
\epsfxsize3cm\epsffile{vertex_PpiP.eps}
 & $=$ & $\frac{2\widetilde g}{f} k^{\nu} \lambda^i_{ab}$ \\
&&\\
$P^*_{1a} P^*_{1b} \pi^i$ coupling: &
\psfrag{Ha}[bc]{$\Red{P^*_{1a}(v)}$}
\psfrag{Hstarb}[bc]{$\Red{P^*_{1b}(v)}$}
\epsfxsize3cm\epsffile{vertex_PpiP.eps}
 & $=$ & $\frac{2 i \widetilde g}{f} \epsilon^{\mu\nu\alpha\beta} k^{\alpha} v_{\beta} \lambda^i_{ab}$ \\
&&\\
$P_a P_{0b}\pi^i$ coupling: &
\psfrag{Ha}[bc]{$\Red{P_a(v)}$}
\psfrag{Hstarb}[bc]{$\Red{P_{0b}(v)}$}
\epsfxsize3cm\epsffile{vertex_PpiP.eps}
 & $=$ & $\frac{-2h}{f} (k\cdot v) \lambda^i_{ab}$ \\
&&\\
$P^*_a P^*_{1b} \pi^i$ coupling: &
\psfrag{Ha}[bc]{$\Red{P^*_a(v)}$}
\psfrag{Hstarb}[bc]{$\Red{P^*_{1b}(v)}$}
\epsfxsize3cm\epsffile{vertex_PpiP.eps}
 & $=$ & $\frac{2h}{f} (k\cdot v) g^{\mu\nu} \lambda^i_{ab}$ \\
  &  &  \\
\end{longtable}

%

\chapter{\label{ch_loop}One loop scalar and tensor functions, special cases}
\index{loop functions}
\index{loop integral}

Following is a list of loop\index{loop integral!expressions|see{loop functions}} integral expressions used in the text. Our notation follows roughly that of Ref.~\cite{Colangelo:1995ph}. We employ dimensional regularization\index{dimensional regularization} in the renormalization scheme where the subtracted divergences $2/\varepsilon -\gamma +\log 4\pi + 1$ are absorbed into the appropriate counterterms\index{counterterms}. All expressions below already have these infinite parts of the integrals subtracted.
\begin{equation}
I_0(m) = \mu^{(4-D)} \int \frac{\mathrm{d}^D q}{(2\pi)^D} \frac{1}{(q^2-m^2)} = - \frac{i}{16\pi^2} m^2 \log\left(\frac{m^2}{\mu^2}\right),
\end{equation}
\begin{equation}
I_2^{\mu\nu}(m) = \mu^{(4-D)} \int \frac{\mathrm{d}^D q}{(2\pi)^D} \frac{q^{\mu}q^{\nu}}{(q^2-m^2)} =\frac{i}{16\pi^2} C_0(m) g^{\mu\nu}
\end{equation}
\begin{eqnarray}
I_2^{\mu\nu}(m,\Delta) &=& \mu^{(4-D)} \int \frac{\mathrm{d}^D q}{(2\pi)^D} \frac{q^{\mu} q^{\nu}}{(q^2-m^2)(v\cdot q - \Delta )} \nn\\
&=& \frac{i}{16\pi^2} \left[ C_1 \left(\frac{\Delta}{m},m\right) g^{\mu\nu} + C_2 \left(\frac{\Delta}{m},m\right) v^{\mu} v^{\nu}\right],
\end{eqnarray}
\begin{equation}
I_1^{\mu}(m,\Delta) = \mu^{(4-D)} \int \frac{\mathrm{d}^D q}{(2\pi)^D} \frac{q^{\mu} }{(q^2-m^2)(v\cdot q - \Delta)} = \frac{i}{16\pi^2}  C \left(\frac{\Delta}{m},m\right) \frac{v^{\mu}}{\Delta},
\end{equation}
\begin{eqnarray}
I_2^{\mu\nu}(m,\Delta_1,\Delta_2) &=& \mu^{(4-D)} \int \frac{\mathrm{d}^D q}{(2\pi)^D} \frac{q^{\mu} q^{\nu}}{(q^2-m^2)(v\cdot q - \Delta_1)(v\cdot q - \Delta_2)}\nonumber\\
&=& \frac{1}{\Delta_1-\Delta_2} \left[ I_2^{\mu\nu}(m,\Delta_1)  - I_2^{\mu\nu}(m,\Delta_2) \right],
\end{eqnarray}
where
\begin{equation}
I_2^{\mu\nu}(m,\Delta,\Delta) = \frac{\mathrm{d}}{\mathrm{d}\Delta} I_2^{\mu\nu}(m,\Delta),
\end{equation}
\begin{eqnarray}
I_2^{\mu\nu}(v,v',m,\Delta_1,\Delta_2) &=& \mu^{(4-D)} \int \frac{\mathrm{d}^D q}{(2\pi)^D} \frac{q^{\mu} q^{\nu}}{(q^2-m^2)(v\cdot q - \Delta_1)(v'\cdot q - \Delta_2)}\nonumber\\
&=& \frac{i}{16\pi^2} \Big[ C_1(w,m,\Delta_1,\Delta_2) g^{\mu\nu} + C_2(w,m,\Delta_1,\Delta_2) (v^{\mu} {v'}^{\nu} + v^{\nu} {v'}^{\mu}) \nonumber\\
&& + C_3(w,m,\Delta_1,\Delta_2){v'}^{\mu}{v'}^{\nu} + C_4(w,m,\Delta_1,\Delta_2){v}^{\mu}{v}^{\nu} \Big].
\end{eqnarray}

In the text we then make use of the following expressions
\begin{equation}
C_0(m)= -\frac{1}{4} m^4 \log \left(\frac{m^2}{\mu^2}\right),
\end{equation}
\begin{equation}
C(x,m) =\frac{m^3}{9} \left[ -18x^3 + (18x^3 -9x) \log\left(\frac{m^2}{\mu^2}\right) + 36 x^3 F\left(\frac{1}{x}\right) \right],
\end{equation}
\begin{equation}
C_1(x,m) =\frac{m^3}{9} \left[ -12x + 10 x^3 + (9x-6x^3) \log\left(\frac{m^2}{\mu^2}\right) - 12 x (x-1) F\left(\frac{1}{x}\right)  \right],
\end{equation}
\begin{equation}
C_2(x,m) = C(x,m) - C_1(x,m),
\end{equation}
with
\begin{equation}
C'_{1,2} \left(x,y,m\right) = \frac{1}{m}\frac{1}{x-y} [C_{1,2}(y,m)-C_{1,2}(x,m)],
\end{equation}
\begin{equation}
C'_{1,2}(x,m)  = C'_{1,2} \left(x,x,m\right) = \frac{1}{m}\frac{\mathrm{d}}{\mathrm{d}x} C_{1,2}(x,m).
\end{equation}
The function $F(x)$ was calculated in Ref.~\cite{Stewart:1998ke}
\begin{equation}
F\left(\frac{1}{x}\right) = \left\{
\begin{array}{lc}
\frac{\sqrt{x^2-1}}{x} \log\left( x+\sqrt{x^2-1} \right), & |x| \geq 1, \\
-\frac{\sqrt{1-x^2}}{x} \left[\frac{\pi}{2} - \tan^{-1}\left( \frac{x}{\sqrt{1-x^2}} \right) \right], & |x| \leq 1.
\end{array}
\right.
\end{equation}

We also make use of the $C_i(v,v',m,\Delta_1,\Delta_2)$ loop integral  functions which have been defined in~\cite{Boyd:1995pq}.
The $1/\Delta$ expansion for $C_i(x,m)$ has been given in sec.~\ref{sec_dsh}, while for $C_i(v,v',m,\Delta_1,\Delta_2)$ it follows as
\begin{eqnarray}
C_1(v,v',m,\Delta,0) = C_1(v',v,m,0,\Delta) &\to& -\frac{1}{\Delta} C_1(m,0) - \frac{1}{\Delta^2} C_0(m) w + \mathcal O(1/\Delta^3), \nonumber\\
C_2(v,v',m,\Delta,0) = C_2(v',v,m,0,\Delta) &\to& - \frac{1}{\Delta^2} C_0(m) + \mathcal O(1/\Delta^3), \nonumber\\
C_3(v,v',m,\Delta,0) = C_4(v,v',m,0,\Delta) &\to& - \frac{1}{\Delta} C_1(m,0) + \frac{2}{\Delta^2} C_0(m) w + \mathcal O(1/\Delta^3), \nonumber\\
C_4(v,v',m,\Delta,0),~C_3(v,v',m,0,\Delta) &\to& \mathcal O(1/\Delta^3), \nonumber\\
C_1(v,v',m,\Delta,\Delta) = - C_1(v,v',m,\Delta,-\Delta) &\to& \frac{1}{\Delta^2} C_0(m) + \mathcal O(1/\Delta^3), \nonumber\\
C_2(v,v',m,\Delta,\Delta),~C_2(v,v',m,\Delta,-\Delta) &\to& \mathcal O(1/\Delta^3), \nonumber\\
C_3(v,v',m,\Delta,\Delta),~C_3(v,v',m,\Delta,-\Delta) &\to& \mathcal O(1/\Delta^3), \nonumber\\
C_4(v,v',m,\Delta,\Delta),~C_4(v,v',m,\Delta,-\Delta) &\to& \mathcal O(1/\Delta^3).
\end{eqnarray}

\backmatter

\chapter{List of abbreviations}

\begin{tabular}{l|l}
{\bf$\bi\chi$PT} & Chiral Perturbation Theory \\
{\bf BBNS} & Beneke-Buchalla-Neubert-Sachrajda \\
{\bf BSM} & Beyond the Standard Model \\
{\bf CKM} & Cabibbo-Kobayashi-Maskawa \\
{\bf CPT} & Charge-Parity-Time conjugation \\
{\bf FCNC} & Flavor Changing Neutral Currents \\
{\bf GUT} & Great Unified Theory \\
{\bf HM$\bi\chi$PT} & Heavy Meson Chiral Perturbation Theory \\
{\bf HQET} & Heavy Quark Effective Theory \\
{\bf ILC} & International Linear Collider \\
{\bf IW} & Isgur-Wise \\
{\bf LEET} & Large Energy Effective Theory \\
{\bf LD} & Long distance \\
{\bf LHC} & Large Hadron Collider \\
{\bf LO} & Leading order \\
{\bf MFV} & Minimal Flavor Violation \\
{\bf MSSM} & Minimal Supersymmetric Standard Model \\
{\bf NLO} & Next-to-leading order \\
{\bf NRQCD} & Non-relativistic Quantum Chromodynamics \\
{\bf OPE} & Operator Product Expansion \\
{\bf PDG} & Particle Data Group \\
{\bf RG} & Renormalization Group \\
{\bf RGE} & Renormalization Group Equations \\
{\bf RPV} & R-parity violation \\
{\bf QCD} & Quantum Chromodynamics \\
{\bf QFT} & Quantum Field Theory \\
{\bf SCET} & Soft Collinear Effective Theory \\
{\bf SM} & Standard Model \\
{\bf THDM} & Two Higgs Doublets Model \\
{\bf VSA} & Vacuum Saturation Approximation \\
\end{tabular}

\chapter{List of publications}

\begin{list}{}{\setlength{\itemsep}{0.2cm}\setlength{\listparindent}{0pt}\setlength{\leftmargin}{0pt}}
\item\large{\bf Published articles}
\item
  J.~O.~Eeg, S.~Fajfer and J.~Kamenik,
  ``Chiral loop corrections to weak decays of B mesons to positive and
  negative parity charmed mesons,''
  JHEP {\bf 0707}, 078 (2007)
  arXiv:0705.4567 [hep-ph].
\item
  D.~Becirevic, S.~Fajfer and J.~Kamenik,
  ``Chiral behavior of the $B_{s,d}^0 - \overline B_{s,d}^0$ mixing amplitude in the
  standard model and beyond,''
  JHEP {\bf 0706}, 003 (2007)
  [arXiv:hep-ph/0612224].
\item
  S.~Fajfer and J.~Kamenik,
  ``Chiral loop corrections to strong decays of positive and negative  parity
  charmed mesons,''
  Phys.\ Rev.\  D {\bf 74}, 074023 (2006)
  [arXiv:hep-ph/0606278].
\item
  S.~Fajfer, J.~Kamenik and N.~Kosnik,
  ``$b \to d d \bar s$ transition and constraints on new physics in B- decays,''
  Phys.\ Rev.\  D {\bf 74}, 034027 (2006)
  [arXiv:hep-ph/0605260].
\item
  S.~Fajfer and J.~Kamenik,
  ``Note on helicity amplitudes in $D \to V$ semileptonic decays,''
  Phys.\ Rev.\  D {\bf 73}, 057503 (2006)
  [arXiv:hep-ph/0601028].
\item
  S.~Fajfer and J.~Kamenik,
  ``Charm meson resonances and $D \to V$ semileptonic form factors,''
  Phys.\ Rev.\  D {\bf 72}, 034029 (2005)
  [arXiv:hep-ph/0506051].
\item
  S.~Fajfer and J.~Kamenik,
  ``Charm meson resonances in $D \to P \ell \nu$ decays,''
  Phys.\ Rev.\  D {\bf 71}, 014020 (2005)
  [arXiv:hep-ph/0412140].
\item
  S.~Fajfer, J.~Kamenik and P.~Singer,
  ``New-physics scenarios in Delta(S) = 2 decays of the B/c meson,''
  Phys.\ Rev.\  D {\bf 70}, 074022 (2004)
  [arXiv:hep-ph/0407223].
\end{list}

\vskip1cm

\begin{list}{}{\setlength{\itemsep}{0.2cm}\setlength{\listparindent}{0pt}\setlength{\leftmargin}{0pt}}
\item\large{\bf Proceedings}

\item
  S.~Fajfer, J.~Kamenik and S.~Prelovsek,
  ``D physics,''
{\it In the Proceedings of International Conference on Heavy Quarks and Leptons (HQL 06), Munich, Germany, 16-20 Oct 2006, pp 018}
  [arXiv:hep-ph/0702172].
\item
  S.~Fajfer and J.~Kamenik,
  ``Charm meson resonances in D semileptonic decays,''
  AIP Conf.\ Proc.\  {\bf 806}, 203 (2006)
  [arXiv:hep-ph/0509166].
\end{list}

\bibliographystyle{JHEP.bst}

\addcontentsline{toc}{chapter}{Bibliography}
\bibliography{article}

\addcontentsline{toc}{chapter}{Index}
\printindex

\pagestyle{empty}

\end{document}